\documentstyle[prl,aps,preprint,epsf]{revtex}
\tighten
\begin{document}
%
\tightenlines
\title{Quantum Phase Transitions and Vortex Dynamics in Superconducting 
	Networks}
\author{
	Rosario Fazio$^{(1,2)}$ and Herre van der Zant$^{(3)}$}
\address{
$^{(1)}$Dipartimento di Metodologie Fisiche e Chimiche (DMFCI), 
        Universit\`a di Catania,\\
        viale A. Doria 6, 95125 Catania, Italy\\
$^{(2)}$Istituto Nazionale per la Fisica della Materia (INFM),
        Unit\`a di Catania, Italy\\
$^{(3)}$Department of Applied  Sciences and DIMES, 
        Delft University of Technology,\\ 
	Lorentzweg 1, CJ 2628, Delft, The Netherlands}
\maketitle

\begin{abstract}
Josephson-junction arrays are ideal model systems to study a variety
of phenomena such as phase transitions, frustration effects, vortex dynamics and
chaos.
In this review, we focus on the quantum dynamical properties of low-capacitance 
Josephson-junction arrays.
The two characteristic energy scales in these systems are the Josephson
energy, associated with the tunneling of Cooper pairs between neighboring
islands, and the charging energy, which is the energy needed to add an
extra electron charge to a neutral island. 
The phenomena described in  
this review stem from the competition between single-electron 
effects with the Josephson effect. They give rise to (quantum) 
Superconductor-Insulator phase transitions that occur when the ratio 
between the coupling constants is varied or when the external fields are varied. 
We describe the dependence of the various control parameters on the phase 
diagram and the transport properties close to the quantum critical
points. On the superconducting side of the transition, vortices are the 
topological excitations. In low-capacitance junction arrays these vortices 
behave as massive particles that exhibit quantum behavior. We review the 
various quantum-vortex experiments and theoretical treatments of their quantum
dynamics. 
\end{abstract}



\newpage

\tableofcontents

\newpage

\section{Introduction}
\subsection{Josephson-junction arrays}
The first artificially fabricated Josephson-Junctions Arrays (JJAs)
were realized twenty years ago at IBM~\cite{IBM} as part of their
effort to develop an electronics based on superconducting devices. In the 
first ten years of their existence, Josephson arrays were intensively 
studied to explore a wealth of 
classical phenomena~\cite{nato86,delft88,trieste95,frascati95,simanek94}.
JJAs proved to be an ideal model system in which classical phase
transitions, frustration effects, classical vortex dynamics, non-linear 
dynamics and chaos could be studied in a controlled way.
The observation of the Berezinskii-Kosterlitz-Thouless (BKT)
transition~\cite{Berezinskii70,Kosterlitz73} in Josephson
arrays is probably one of the most spectacular experiments~\cite{btkexp}
in this respect.

All classical phenomena can be successfully explained by studying 
the classical (thermo)dynamics of the phases of the superconducting order 
parameter on each island. This approach is justified because experiments are 
usually carried out at temperatures well below the BCS transition temperature.
 Each island
is then superconducting with a well defined gap, but phase fluctuations are 
still allowed. Under these conditions, classical JJAs are a 
physical realization of the two-dimensional XY-model and above the BKT 
transition temperature, phase fluctuations destroy global phase coherence 
preventing the system to reach the superconducting state. 
Global phase coherence is only restored below temperatures corresponding to 
the Josephson coupling energy $E_J$, which is the energy scale
associated with Cooper pair tunneling between neighboring islands.

As we now understand, the relatively large junctions at that time had 
resistances too low~\footnote{
A simple estimate for value of the junction resistance above which clear 
quantum effects become visible can be 
obtained by using the Heisenberg relation $\Delta E \Delta \tau \ge \hbar$. 
By taking the charging energy ($e^2/2C$) for $\Delta E$ and $\tau$ of the 
order 
of the junction $RC$-time one finds that the junction resistance $R_N$ 
should satisfy the inequality $R_N > R_Q = h/4e^2$ for quantum effects 
to be observable.}
to observe clear quantum effects.
By the end of the eighties semiconductor technology had pushed device
dimensions well below the micron size. It became possible to fabricate
arrays with Josephson tunnel junctions of sizes $100 \times 100$~nm$^2$.
Circuits with such small junctions showed single-electron effects 
when cooled down to temperatures corresponding to the charging energy 
$E_C$, the energy needed to add an extra electron charge to a neutral island. 
It was soon realized that the competition between single-electron 
effects~\cite{schoen90,averin91} and the Josephson effect would lead to new,
exciting physics. 

An appealing feature of JJAs already emerges at this stage as they can be
visualized as model systems to investigate quantum (zero-temperature) phase 
transitions~\cite{sondhi97,sachdev99}. In recent years, this field of 
research has attracted the attention of many physicists. Experiments on 
thin, superconducting films, high-temperature superconductors, spin
systems and two-dimensional electron gases have all shown the existence
of quantum critical points.
In arrays made of submicron junctions, the quantum fluctuations drive the
system through a variety of quantum phase transitions. A quantum JJA may 
be insulating at zero temperature even though each island is still 
superconducting. 
In the classical limit $E_J \gg E_C$, the system turns superconducting 
at low temperatures 
since the fluctuations of the phases are weak and the system is globally 
phase coherent. In the opposite limit, $E_{J}
\ll E_{C}$, the array becomes a Mott insulator since the charges on each 
islands are localized and an activation energy of the order of $E_C$ is 
required to transport charges through the system (Coulomb blockade of
Cooper pairs). Strong quantum fluctuations of the phases prevent the 
system from reaching long-range phase coherence in this regime. 

Granular superconducting thin films are closely related to arrays.
In granular films, superconducting islands of various sizes and with various 
coupling energies are connected together. Therefore, disorder plays a 
crucial role in these granular materials while it is
virtually absent in JJAs (or it can be introduced in a controlled way). 
Models based on the behavior of 
Josephson arrays also form the starting point to describe the physics of 
ultra-thin, amorphous superconducting films in which superconductivity is 
quenched by disorder or by an applied magnetic field. 
In these two-dimensional homogeneous films it is believed that, 
although the order parameter is suppressed, phase fluctuations are still 
responsible for driving the system through the Superconductor-Insulator 
(S-I) transition~\cite{goldman00}.

Another important field of investigation addressed with JJAs,
is the study of the quantum dynamics of macroscopic objects. In the classical
limit vortices are the topological excitations that determine the
(thermo)dynamic properties of JJAs. In the opposite situation ($E_{J}
\ll E_{C}$) the charges on each island are the relevant degrees of freedom.
Vortices and charges play a dual role and many features of JJAs can be
observed in the two limits if the role of charges and vortices are 
interchanged. By fabricating arrays with different geometries, vortices 
can be manipulated to a great extent. Quantum dynamics of 
macroscopic objects requires knowledge of the coupling to the surrounding 
environment~\cite{weiss99}. To a certain degree, the dissipative environment 
can be modeled and therefore JJAs are prototype systems to study 
macroscopic quantum mechanics as well.
Born as a problem related to the foundations of quantum mechanics, 
macroscopic coherence in superconducting nanocircuits is acquiring 
increasing attention since the advent of quantum computation.

\subsection{Phase-number relation}

Throughout this review, the interplay between the phase $\phi$ of an island 
and number of charge carriers $Q$ on it plays a crucial role. Together they 
determine the properties of quantum Josephson networks. The competition 
between these two canonically conjugated variables is captured by the 
following Heisenberg relation~\cite{anderson64}: 
$$
 [\phi_{i},Q_{j}]=2e \,i \; \delta _{ij}
$$
where the subscripts $i$ and $j$ label the island positions. 

\begin{figure}
\centerline{{\epsfxsize=14cm\epsfysize=14cm\epsfbox{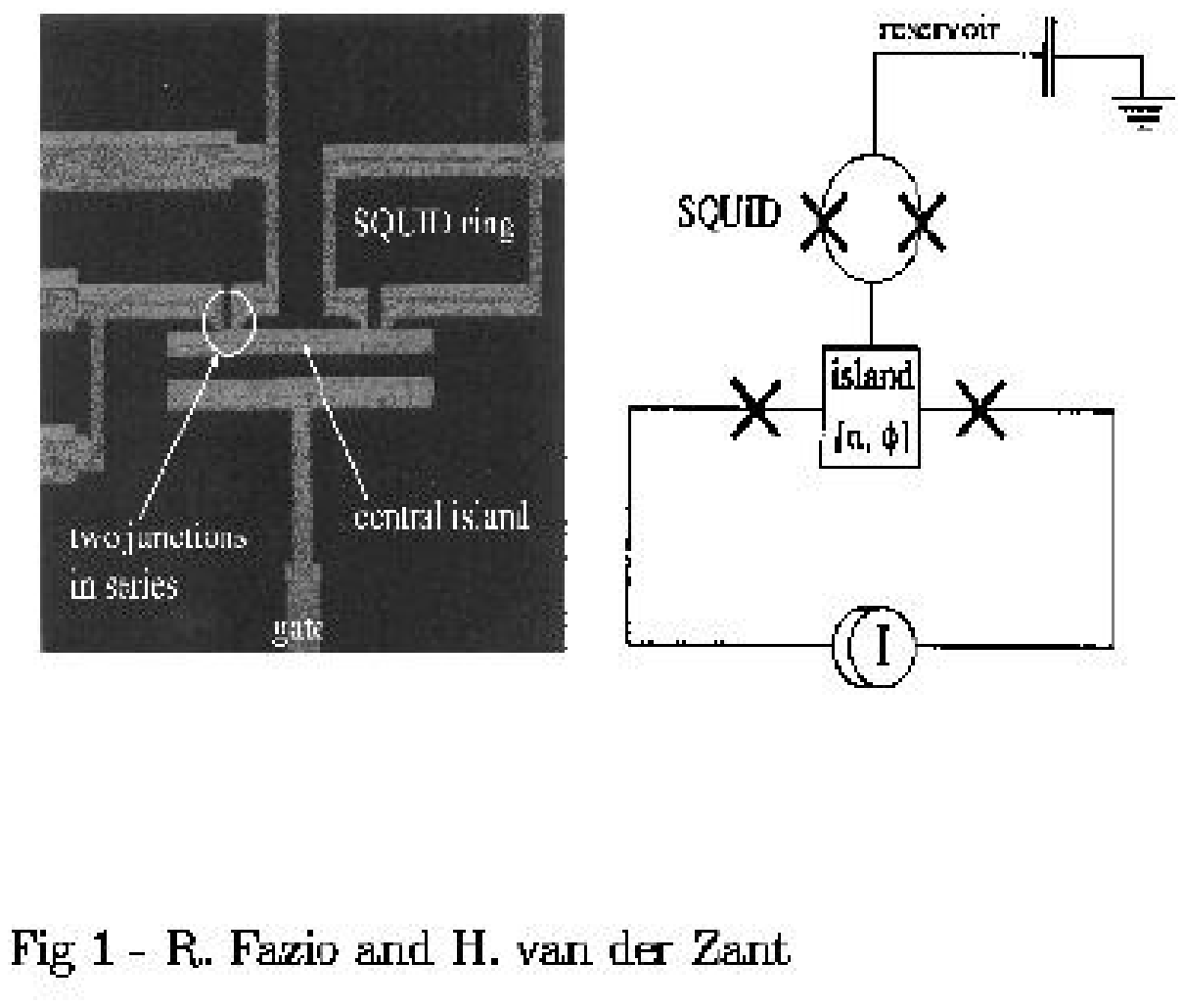}}}
\caption{Left hand side: a Scanning-Electron-Microscope (SEM) photograph of the 
	Heisenberg transistor. The leads in upper, left corner are used to perform a 
	four-terminal  measurement on the two junctions in series. Offset-charges on 
	the central island are nulled out by the gate capacitor situated below the 
	central island. (Picture taken by W.J. Elion.)
	Right hand side: a schematic drawing of the Heisenberg 
	transistor \protect\cite{elion94}. The phase and the charge on 
	the island are quantum 
	mechanical conjugated variables. By varying the flux through
	the SQUID ring the effective Josephson energy is tuned and, as a 
	result, phase fluctuations on the central island can be varied. 
	(Reprinted by permission from  Nature {\bf 371}, 594 (1994) copyright 1994
	Macmillan Magazines Ltd.)}
\label{jjaintrofig}
\end{figure}

An elegant illustration of this competition is presented by what became 
known as the Heisenberg transistor~\cite{elion94,matters95}. The aim of the 
experiment was to control and measure the quantum phase fluctuations through 
a modulation of the critical current of the system. 
In Fig.\ref{jjaintrofig} the layout of the device
is shown. Two junctions in series (indicated by crosses) are connected to 
a current source. The junction parameters are such that $E_J \sim E_C$, i.e., 
quantum mechanical 
fluctuations of the number of Cooper pairs and of the phase of the central 
island are comparable.
A large superconducting reservoir is coupled to the island through a
Superconducting QUantum Interference Device (SQUID). 

In the experiment the critical current was measured as a function of the 
applied flux through the SQUID ring. It shows a periodic modulation with 
a period equal to the superconducting flux quantum ($\Phi_0 = h/2e$). The 
role of the SQUID is to provide a tunable coupling to the reservoir of 
Cooper pairs. When the flux equals an integer times half a flux quantum 
($n Phi_0/2$) the coupling is turned off and fluctuations in the number 
of Cooper pairs are suppressed. At the same time, phase fluctuations reach 
their maximum as indicated by the Heisenberg relation. At fields equal to 
zero or an integer number of flux quanta, the coupling is maximum so that 
the amount of charge fluctuations reaches a maximum as well. 
In the experiment, the critical current probes the scale of charge 
fluctuations: The situation with large charge fluctuations corresponds to 
favorable Cooper-pair tunneling and a high critical current.
Thus, for zero applied field a high critical current is measured because 
charge fluctuations are at their maximum. At half a flux quantum applied, 
the critical current reaches its minimum value because phase fluctuations 
are at their maximum. 

\subsection{Structure of the review}
This review is organized as follows. In Chapter~\ref{qpt} the basic
physical properties and models are introduced. Some theoretical tools to 
study the phase diagram including the boundary of the S-I transition are 
briefly discussed: the mean-field approximation, the coarse-graining
approach to derive a Ginzburg-Landau effective free energy, and the
Villain transformation that leads to a description in terms of charges and
vortices. These approaches capture most of the essential physics. 
Sections~\ref{magfrus} and \ref{supersolid} are devoted to a description 
of the phase diagram when charge and/or magnetic
frustration are included. 
Since the number of control parameters that can be varied is
large, the phase diagram is discussed in some limiting cases only.
Section~\ref{dissip} describes the various sources of dissipation
in JJAs and their effect on the phase diagram. 
The final three sections of Chapter~\ref{qpt} report on the transport 
properties
close to the S-I transition, the S-I transition in one-dimensional 
arrays and the physics of the field-tuned S-I transitions. 
In all sections, comparison of the theoretical phase diagrams with the 
experimental results is discussed as well.

Chapter~\ref{qvd} deals with quantum vortex dynamics. After introducing
the important vortex properties (vortex mass, pinning potential,...) and 
its classical equation of motion, a theoretical
description of quantum corrections to the classical equation of motion is
presented. The remainder of the chapter concerns the description of 
quantum vortex-experiments. 
We start with the single vortex properties (tunneling, interference,
and Bloch oscillations) and then proceed with collective vortex motion in
quasi-one dimensional samples (Mott insulation of vortices and Anderson
vortex localization).

In the last chapter, some future directions are explored. The theoretical 
description of two new experiments is outlined: persistent vortex currents 
in Corbino geometries and the Quantum Hall effect for vortices/charges. 
The experimental technicalities for the observation
of these phenomena are described as well. We end this review with a brief 
discussion on Josephson qubits in which fundamental aspects of quantum 
mechanics and quantum information theory can be studied.

We tried to keep this review self-contained and, at the same time, to
give a comprehensive overview of the quantum properties of Josephson
networks. For each of the sections, we present the main
ideas without going into a detailed enumeration of all the results obtained
in the field. There are some
topics which are not discussed here. Probably the most important,
which would require a review by itself, is the effect of disorder which
seems to be more important for granular materials and
ultra-thin films. 

Basics in superconductivity and Josephson physics can be found in 
the books by Tinkham~\cite{tinkham96} and by Barone and Patern\`o
~\cite{barone82}. Since many ideas (e.g. persistent current, localization) 
were born in the field of more traditional mesoscopic physics we refer 
for these topics to 
the books by Beenakker~\cite{beenakker91} and Imry~\cite{imry97} and 
to the conference proceedings~\cite{karlsruhe94,curacao97}.
Various other aspects of JJAs have already been discussed in previous 
reviews devoted to this topic~\cite{doniach83,mooij92,katsumoto95}.

Throughout the review we put $\hbar=k_B=c=1$. Distances are expressed in 
units of the lattice constant $a$. We restore S.I. units in the formulas 
expressing measurable quantities.

\section{Quantum Phase Transitions}
\label{qpt}
A quantum Josephson array consists of a regular network of superconducting 
islands weakly coupled by tunnel junctions. Thanks to submicron lithography, 
array's parameters (associated to the shape of the islands, the 
thickness of the oxide barrier,...) can be made uniform (virtually identical) 
across the whole array. With present-day technology variations in junction 
parameters are below 20~$\%$ across the  array. The dimensions of the unit
cell are of the other of few $\mu$m$^{2}$ while the superconducting
islands have an area of about 1$\thinspace \mu$m$^{2}$.
The largest samples consist of about 10000 junctions (e.g. 100 by 100 or 1000 
by 7). 

Quantum arrays are fabricated of all-aluminum high-quality Josephson 
tunnel junctions with a shadow-evaporation technique. The evaporation mask 
is made of electron sensitive resists in which the bottom resist layer 
has an undercut to ensure a proper lift-off after evaporation.
Junctions are formed by evaporating a thin aluminum layer (25~nm) as the 
bottom electrode followed by in situ oxygen oxidation and evaporation 
of the counter electrode of about 50 nm from an opposite angle. 
A more detailed discussion of the fabrication techniques is presented in the 
Appendix~\ref{fabrication}.

\subsection{The model of a Josephson Junction Array}

\subsubsection{Quantum Phase Model}
\begin{figure}
\centerline{{\epsfxsize=14cm\epsfysize=16cm\epsfbox{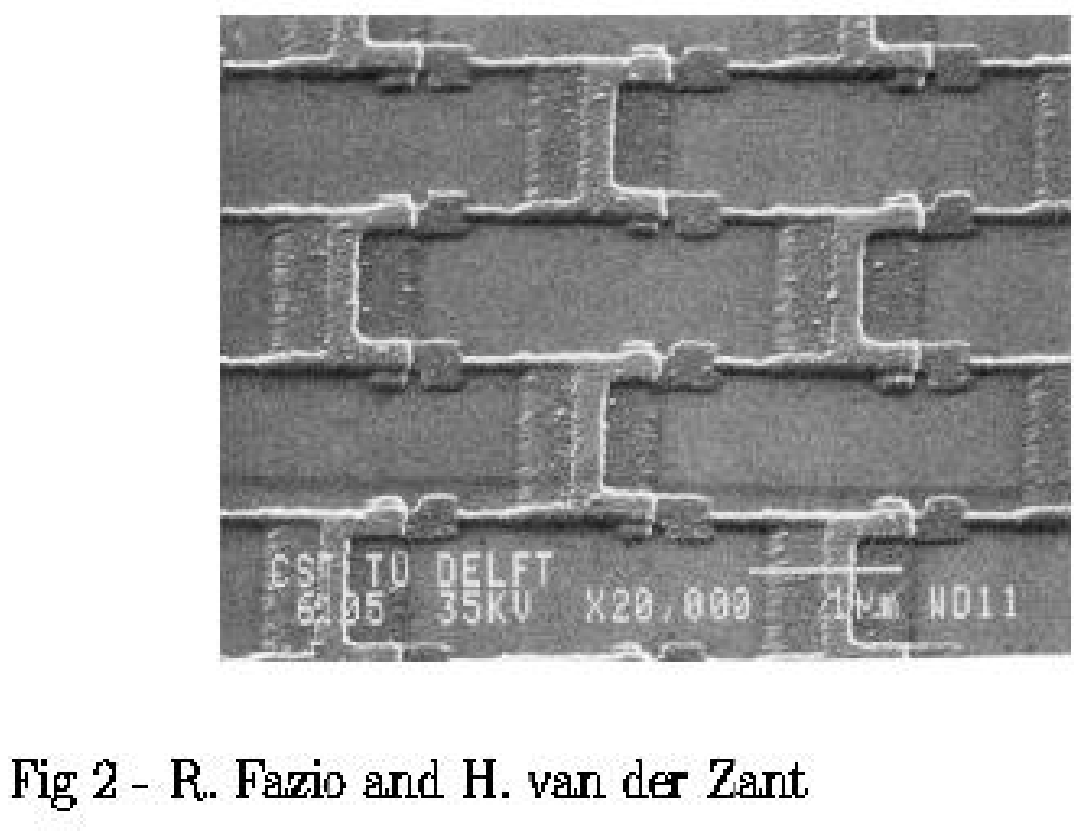}}}
\caption{A Scanning-Electron-Microscope (SEM) photograph of a two-dimensional, square 
	array of small tunnel junctions produced by shadow evaporation. The white 
	bar is 1~$\mu$m long. (Picture taken from the Ph.D. thesis of L.J. Geerligs, 
	Delft 1990 (unpublished).)}
\label{jjafig}
\end{figure}

\begin{figure}
\centerline{{\epsfxsize=14cm\epsfysize=16cm\epsfbox{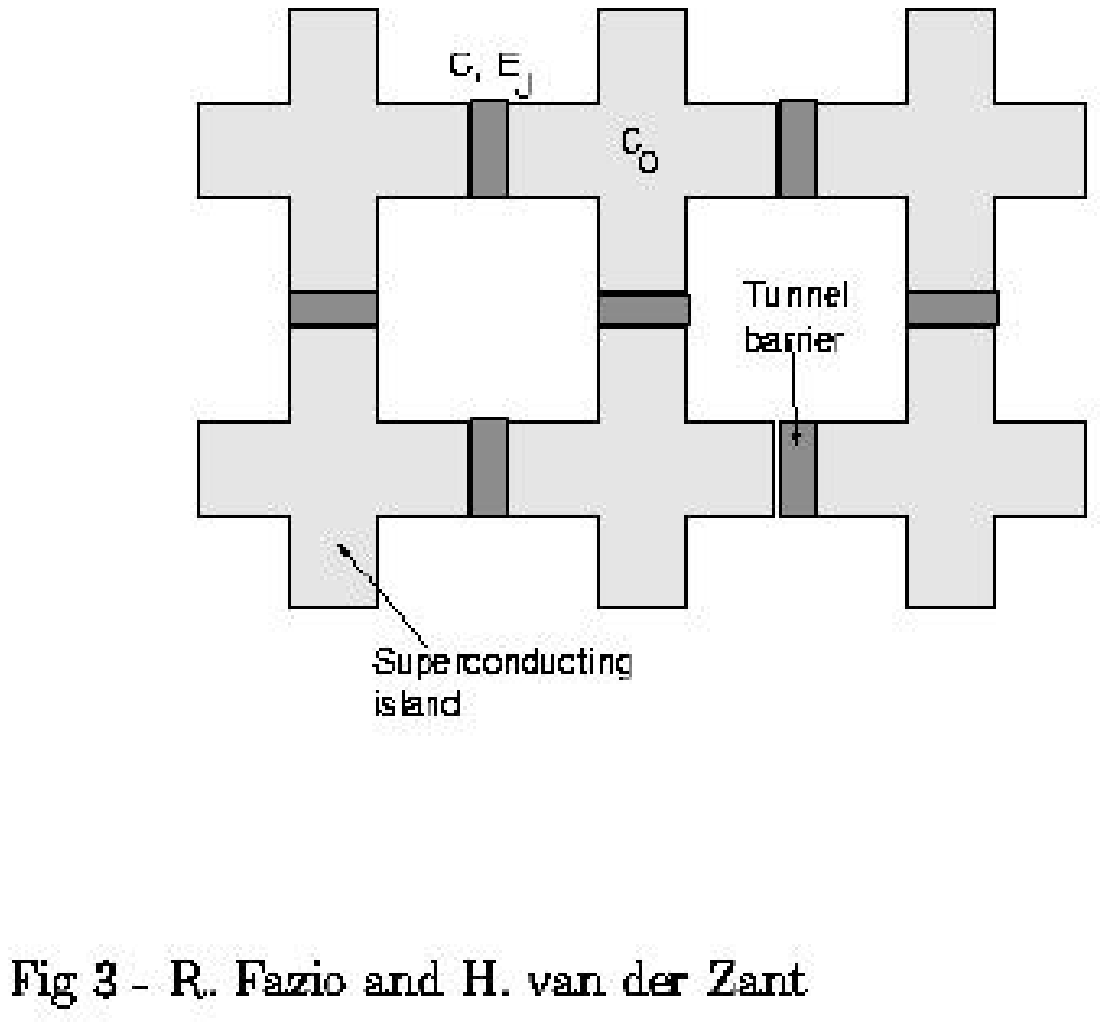}}}
\caption{A Josephson array consists of a regular network of superconducting islands weakly 
	coupled by tunnel junctions. 
	Each junction is characterized by the Josephson coupling $E_J$ and the junction capacitance 
	$C$; each island by the capacitance to the ground $C_0$.}
\label{jjamodel}
\end{figure}

In Fig.~\ref{jjafig}, we show a scanning-electron-microscope (SEM) picture of a 
Josephson junction array. Its schematic representation is given in 
Fig.~\ref{jjamodel}. 
In this square geometry the coordination number, ${\rm z}=4$. The coupling strength
between adjacent islands is determined by the Josephson energy
$E_J=\Phi_{0}I_{c}/(2\pi)$.
This coupling energy is inversely proportional to the normal-state junction
resistance $R_{N}$. Experimentally $R_N$ is determined from the normal-state
array resistance $r_N$ measured at 4.2 K, $R_N=(N+1)r_N M_y/M_x$, where $M_y$
is the number of cells across the array and $M_x$ is the number of cells along
its length. The maximum junction critical current $I_{c}$, in the absence of 
charging effects and thermal fluctuations, is assumed to be given 
by the Ambegaokar-Baratoff value~\cite{ambegaokar63} 
$$
	I_{c}R_N = \pi \frac{\Delta}{2e}
$$
with the measured critical temperature $T_c$. For a BCS critical temperature 
$T_c=1.35$ $\thinspace$K one gets $I_{c}R_{N}$=0.32$ \thinspace$mV at 
low temperatures. 

Quantum effects in Josephson arrays come into play when the charging energy 
(associated with non-neutral charge configurations of the islands) is comparable with the Josephson coupling (the physics associated with 
charging effects in single normal and superconducting junctions has been 
reviewed in Refs.~\cite{schoen90,averin91}).  In addition, as explained 
in the introduction, the junction resistance should be of the order of 
(or larger than) $R_Q$\cite{juncres}.

Arrays are made in a planar geometry, in which each island is coupled to 
each of the other islands and to a far-away ground by its self-capacitance 
$C_0$. The junctions are made of two overlapping superconducting 
layers separated by a thin oxide layer and the main contribution to the 
capacitance therefore comes from the junction capacitance $C$. 
An estimate of the 
total island capacitance $C_{\Sigma}$ is obtained from measuring the 
voltage offset ($V_{offset}$) in the I - V characteristics at high bias 
currents at $T$=10$\thinspace$mK 
in a magnetic field of 2$\thinspace$T. Using the so-called local 
rule~\cite{schoen90,local}, $C_{\Sigma} =Me^2/2V_{offset}$. For 
junctions of 0.01~$\mu$m$^2$, $C_{\Sigma}$ is found to be 1.1$\thinspace$fF. 
If one identifies $C$ with $C_{\Sigma}$, the specific capacitance is 
110~fF/$\mu$m$^2$. 
Measurements on large-area junctions have yielded a 
specific capacitance that is about a factor of two lower. This discrepancy 
shows that stray capacitance (capacitance between next-nearest and further
neighbors) may play a role in Josephson circuits as pointed out by Lu 
{\em et al.}~\cite{lu98}. However, for simplicity one often identifies the measured 
$C_{\Sigma}$ as the junction capacitance $C$. 

Reliable estimates of this self-capacitance ($C_0$)are obtained from separate 
measurements on small series arrays with high $E_C/E_J$ ratio. A magnetic field 
of 2~Tesla is applied so that the series arrays are in the normal state; 
$C_0$ is then measured by varying the potential of the circuit with respect 
to the ground potential. Recording the current through the circuit yields 
a periodic signal with period $e/C_0$. For islands of $1~\mu$m by 
$1~\mu$m, $C_0 \approx 1.2$x$10^{-17}$~F which is much smaller than 
$C$~\cite{conduct}. 

As already mentioned, the electrostatic energy can be determined once 
the capacitance matrix $C_{ij}$ and the  gate voltages (if present) are 
known~\cite{local,jackson75}. Generally one only considers 
the junction capacitance $C$ and the capacitance to the ground $C_0$. 
In this case the capacitance matrix has the form $C_{ii} = C_0 + {\rm z}C$, 
$C_{ij} = -C$ (if $i,j$ nearest neighbors) and zero in all other cases. 
Consequently the charging energy (for two charges placed in islands $i$ 
and $j$ of coordinates ${\bf r}_i$ and ${\bf r}_j$ respectively) 
is given by
\begin{equation}
E_{ij}^{(ch)}  =  \frac{e^2}{2}C_{ij}^{-1} =
	      \frac{e^2}{2} \int \frac{d{\bf k}}{4\pi^2}
	      \frac{e^{i{\bf k}\cdot ({\bf r}_i- {\bf r}_j)}}
	      {C_0 + 2 C (1-\cos k_x) + 2 C (1-\cos k_y)} \;\;\; ,
\end{equation}
which is well approximated by the expression:
\begin{equation}
	E_{ij}^{(ch)}  
	\sim  
	\frac{e^2}{4\pi C} K_0 \left( \frac{|{\bf r}_i-{\bf r}_j |}{\lambda}\right)
\label{charging_interaction}   
\end{equation}
Here, $K_0$ is the modified Bessel function. The charging interaction 
increases logarithmically up to distances of the order of the screening 
length $\lambda$ and then dies out exponentially. The characteristic energy scale is 
$$
		E_C = \frac{e^2}{2C} \;\;\; .
$$
Equation~(\ref{charging_interaction}) assumes three-dimensional screening and 
the range of the electrostatic interaction between Cooper pairs is given by (
in units of the lattice spacing):
$$
	\lambda =\sqrt{C/C_{0}} \;\;\; .
$$. 
If the two-dimensional limit is considered (if e.g. the array is sandwiched 
between two media with large dielectric coefficients) the screening length 
scales linearly with $C/C_{0}$ yielding a longer ranged interaction. 

From all these considerations, we arrive at the following Hamiltonian
which describes Cooper pair tunneling in superconducting quantum networks 
(quasi-particle tunneling is ignored at this stage). This model is frequently 
called the Quantum Phase Model (QPM) and is its most general form it is given by:
\begin{eqnarray}
	H &=& H_{ch} + H_{J} \nonumber \\  
	  &=& \frac{1}{2} \sum_{{i} , {j} }  
	(Q_{i}-Q_{{\rm x},j}) \; C_{ij}^{-1} \; (Q_{j}-Q_{{\rm x},j}) 
	-E_{J} \sum_{< {i} , {j} >} 
	\cos \left (  \phi_{i}-\phi_{j} - A_{ij} \right ) \; . 
\label{QPM}
\end{eqnarray}
The first term in the Hamiltonian is the charging energy in which the $C_{ij}^{-1}$ is 
the capacitance matrix; the second is due to 
the Josephson tunneling. An external gate voltage $V_{{\rm x},i}$ gives the 
contribution to the energy via the induced charge $Q_{\rm x,i} = 2e q_{\rm x} 
= \sum _j C_{ij} V_{\rm x,j}$. This external voltage can be either applied to 
the ground plane or it may be (unintentionally) induced by trapped charges in 
the substrate. In this latter case $Q_{\rm x,i}$ will be a random variable. 

A perpendicular magnetic field with vector potential ${\bf A}$ enters the 
Hamiltonian of Eq.~(\ref{QPM}) through  $A_{ij}= 2 e \int_{i}^{j} {\bf A} 
{\bf \cdot} d {\bf l}$. 
The relevant parameter that describes the magnetic frustration is  
$$
f= (1/ 2 \pi) \; \sum_{P} A_{ij} 
$$ 
where the summation runs over an elementary plaquette. 
In quantum arrays, the 2D flux penetration depth 
$\lambda_{\perp}(T)=\Phi_{0}/ 2\pi \mu_{0}I_{c}(T)$ 
is much larger than the array size so that the magnetic 
field is essentially uniform over the whole array, i.e., $f$ is position independent. 
A similar conclusion can be 
drawn by considering the ratio of the geometric inductance (we estimate 
it to be of the order of 1~pH) to the Josephson inductance (larger than $1~$nH).

Throughout this review, two limits of the QPM are frequently discussed: 
$C \gg C_{0}$ and $C \ll C_{0}$.  The former limit has already been discussed 
in detail as it is the appropriate regime of Josephson arrays. The latter limit 
is more appropriate for granular films that have a short-range Coulomb interaction. 
To describe these systems we use the following notation.
When the on-site contribution is dominant, the characteristic energy is  
$$
		E_0 = \frac{e^2}{2}C_{00}^{-1} \;\;\; .
$$
However, some properties (see e.g. discussion in Section~\ref{supersolid}) are 
crucially dependent 
on the details of the electrostatic energy at small distances, i.e., on weather 
the nearest-neigbor interaction is also included or not. $E_1$ represents the 
variable denoting this nearest-neigbor interaction; 
$E_2$ the interaction between next nearest-neighors and so on. 

The two contributions in the Hamiltonian of Eq.(\ref{QPM}) favor different 
types of ground states. The Josephson energy tends to establish phase 
coherence which can be achieved if supercurrents flow through the array. 
On the other hand the charging energy favors charge localization on each 
island and therefore tends to suppress superconducting coherence. 
This interplay becomes evident if one recalls the Josephson relation 
(which here can be obtained at the operator level by calculating the 
Heisenberg equation of motion for the phase)
\begin{equation}
	\frac{d\phi_i}{dt} 
	= 
	\frac{2e}{\hbar}V_i = \frac{2e}{\hbar}C_{ij}^{-1}Q_j
\label{Jorel}
\end{equation}
A constant (in time) charge on the islands implies strong fluctuations 
in the phases. On the other hand phase coherence leads to strong fluctuations 
in the charge.

The low-lying excitations of the model defined in Eq.(\ref{QPM}), are 
long wave-length phase waves whose dispersion relation can be obtained 
by considering the QPM in the harmonic approximation
\begin{equation}
	H \sim
	  \frac{1}{2}\sum_{{i} , {j} }  
	  Q_{i} \; C_{ij}^{-1} \; Q_{j} 
	- \frac{E_J}{2}\sum_{< {i} , {j} >} 
	\left (  \phi_{i}-\phi_{j} \right ) ^2 \; . 
\end{equation}
The dispersion relation of these modes (usually named spin-waves from 
the magnetic analogy of the Josephson coupling with the XY model) 
depends on the form of the capacitance matrix 
(see Section~\ref{ball}). The QPM possesses topological excitations as 
well, charges and vortices, that will be discussed in Section~\ref{duality}.

A qualitative understanding of the phase diagram can be obtained by 
considering the two limiting cases in which one of the two coupling energies 
is largest. For simplicity we look at the ground state of the system ignoring external 
voltages and magnetic field. If the Josephson term is dominant, the array minimizes 
its energy by aligning all the phases, i.e. it is in the superconducting state. If 
instead the charging energy is dominant, each island has a zero charge 
in the ground state. In order to put an extra charge on the island one 
has to overcome a Coulomb gap of the order of the charging energy (max{ $E_C,E_0$}). 
The array behaves as an insulator although each island is still in the 
superconducting state.  

\subsubsection{Dissipative Models}

Since the seminal paper by Caldeira and Leggett~\cite{caldeira81,caldeira83}
it became clear that dissipation changes the quantum dynamics of macroscopic systems. 
One can therefore ask the question to what extent dissipation plays a role in 
Josepson-junction arrays and what its role is on the SI transition in quantum arrays.  
At low temperatures one expects quasi-particle tunneling not to be present
since the charging energy is smaller than the superconducting gap 
$\Delta$. Experiments on small arrays indicate 
that even at mK-temperatures a small but finite amount of quasi-particles 
is always present, although it  has not been possible to discriminate 
the exact details of dissipation. Therefore, we treat the various models that 
have been proposed to describe dissipation in superconducting networks. 

The QPM defined in Eq.(\ref{QPM}) only accounts for Cooper pair tunneling 
between neighboring islands and needs to be generalized. The 
appropriate description is formulated in terms of an effective action 
by the authors of Ref.~\cite{schoen90,ambegaokar82}:
\begin{equation} 
	Z = \int \prod_{i} D\phi_i (\tau) \exp [-S\{\phi\}]
\label{dissiZ}
\end{equation}
The Euclidean effective action $S\{\phi\}$, corresponding to 
the Hamiltonian of Eq.(\ref{QPM}) has the form 
(for simplicity we ignore charge and magnetic frustration for the time being) 
\begin{equation}
	S[\phi] =
	\int_{0}^{\beta } d\tau \left\{ \vphantom{\frac{}{{}_{}}} 
	\frac{C_{0}}{8e^2}\sum_i
	(\dot{\phi}_{i})^2 
	+  \frac{C}{8e^2}\sum_{\langle ij \rangle} 
	(\dot{\phi}_{i} - \dot{\phi}_{j})^2 
	- E_J \sum_{\langle ij \rangle} 
	\cos(\phi_{i} - \phi_{j})  \vphantom{\frac{}{{}_{}}} \right\}
 \label{action} 
\end{equation}
with $\beta$ being the inverse temperature.
The first two terms are easily recognized as charging energies
expressed in terms of voltages (see Eq.(\ref{Jorel})). In the presence 
of dissipative tunneling, the effective action has a Caldeira-Leggett 
form and acquires an additional term
\begin{equation}
	S_{D}[\phi]
	= 
	\frac{1}{2}\int_{0}^{\beta } d\tau  d\tau'\sum_{<ij>} 
	\alpha (\tau-\tau')F(\phi_{ij}(\tau)-\phi_{ij}(\tau'))
\label{dissiS}
\end{equation}
where $\phi_{ij} = \phi_{i} - \phi_{j}$.
Both the dissipative kernel $\alpha(\tau)$ related to the $I-V$ characteristic of tunnel junctions~\cite{schoen90}, and the function 
$F(\left\{\phi \right\})$ depend on the nature of the dissipation. 
Various mechanisms: tunneling of quasi-particles and/or the flow of Ohmic currents through the substrate or between the junctions themselves~\cite{barone82,mccumber}.

For an Ohmic junction, as is the case when the bath is formed by 
quasi-particle excitations in normal metals (or gapless superconductors), 
the kernel is 
$$
	\alpha (\tau)
	= 
	\frac{\pi}{2e^2R_N}\frac{1}{ \beta^2}
	\frac{1}{\sin^2(\pi \tau /  \beta)}
$$
Here, the normal-state resistance $R_N$ controls the coupling to the 
environment~\cite{beloborodov00}.  
For ideal superconducting islands on the other hand, the BCS gap inhibits 
leakage currents at small voltages and as a consequence the dissipative kernel 
is short range in imaginary time. Therefore, quasi-particle tunneling results 
in a renormalization of the junction capacitance~\cite{ambegaokar82}
\begin{equation}
	C \longrightarrow C + \frac{3\pi }{32\Delta R_N} \;\; .
\label{capren}
\end{equation}

The dissipation mechanism also affects the form of the function $F(\phi)$ in 
Eq.~(\ref{dissiS}). If normal-electron tunneling occurs via discrete charge 
transfer, as it is for a quasi-particle current, $F(\phi)$ is a periodic function 
of the phase $\phi$
\begin{equation}
	F_{QP}(\phi_{ij}(\tau)-\phi_{ij}(\tau')) 
	= 
	1 - 
	\cos \left(\frac{\phi_{ij}(\tau)-\phi_{ij}(\tau')}{2}\right) \;\; .
\label{qpdissi}
\end{equation}
If, on the contrary, dissipation is due to normal shunts or more 
generally to the interaction with the electromagnetic environment, 
$F$ is quadratic in $\phi$
\begin{equation}
	F_{N}(\phi_{ij}(\tau)-\phi_{ij}(\tau')) 
	=  
	\frac{1}{2} \left(\frac{\phi_{ij}(\tau)-\phi_{ij}(\tau')}{2}\right)^2
\label{ohmdissi}
\end{equation} 
indicating that the charge at a junction can assume continuous values. 

One can also consider dissipation due to currents flowing to the substrate. 
This local-damping model plays an important role in classical, proximity coupled 
Josephson arrays. Voltage fluctuations compared to the ground instead of voltage 
differences between junctions are the crucial variables:  
the dissipative part of the action now depends on the phase 
$\phi_i$ of each island and not on the phase difference $\phi_{ij}$
\begin{equation}
	F_{LD}(\phi_{i}(\tau)-\phi_{i}(\tau')) 
	= 
	\frac{1}{2}
	 \left(\frac{\phi_{i}(\tau)-\phi_{i}(\tau')}{2}\right)^2 \;\;\;\; .
\label{loc}
\end{equation}

The path integration in Eq.(\ref{dissiZ}) depends on the nature of 
dissipation mechanism and it is related to the charge on the 
islands being a continuous or discrete variable~\cite{schoen90}.
In the first case, the phase is considered an extended variable and in the
path integral $ \phi_{i}(0) =  \phi_{i}(\beta ) $. If the charge is a
discrete variable, a summation over winding number is implied in
Eq.(\ref{dissiZ}) 
\begin{equation} 
	\int D\phi
	\longrightarrow
	\prod_{i} \int_{0}^{2\pi} d\phi_{i0} 
	\sum_{\{m_i = 0,\pm1,\dots\}} 
 	\int_{\phi_{i0}}^{\phi_{i0} + 2\pi m_{i}} 
	D\phi_i (\tau) \;.
\label{discrete}
\end{equation}
These non-trivial boundary conditions express the fact that the charges 
of the grains are integer multiples of $2e$.

The dissipative coupling strength is usually expressed in the form 
$\alpha = R_e/R_Q$. The exact value of the effective
resistance $R_e$ is not a priori clear. Consider an array of unshunted tunnel junctions.
 If thermally excited quasi-particles were the only source of damping, a measure of 
$R_e$ would be the subgap resistance which is many orders of magnitude larger than 
the normal-state resistance $R_N$. However, measurements hint at a much smaller $R_e$ 
that is closer to $R_n$. The exact mechanism is not clear but one should always keep 
in mind that Josephson junctions are highly nonlinear elements. Some coupling to higher 
energy scales may therefore occur and $R_e$ is smaller than the subgap resistance but
not lower than $R_N$. This coupling may for instance occur 
when a vortex crosses a single junction thereby producing voltage
spikes. Throughout this review, we keep the notation simple and use $\alpha$ to 
denote the dissipation strength regardless of the underlying dissipation mechanism.
Its origin will be specified from case to case.

\subsubsection{Related Models}
 
The S-I transition has been investigated by studying model Hamiltonians, 
the so called Bose-Hubbard and XXZ models, closely related to the QPM of 
Eq.(\ref{QPM}). 
We follow the notation used currently in the literature and point out the connection 
with the couplings used to defined 
the QPM. The Bose-Hubbard (BH) model~\cite{fisher89b} is defined as
\begin{equation}
	H=\frac{1}{2}\sum_{i}n_{i}U_{ij}n_{i} -\mu\sum_{i}n_{i}
	-\frac{t}{2}\sum_{\langle ij\rangle}(e^{-iA_{ij}} \;
	b^{\dagger}_{i}b_{j}+ h.c.)
\label{bh}
\end{equation}
Here, $b^{\dagger}, b$ are the creation and annihilation
operators for bosons and $n_i =b^{\dagger}_i b_i $ is the number operator. $U_{ij}$ 
describes the Coulomb interaction between bosons
($U_{ij} \longrightarrow E_{ij}^{ch}$),  
$\mu$ is the chemical potential, and $t$ the hopping matrix element.
The connection between the Bose-Hubbard model and the QPM is easily seen
by writing the field $b_i$ in terms of its amplitude and phase and by subsequently 
approximating the amplitude by its average. This procedure leads to the 
identification $b_i \sim e^{i\phi _i}$. 
The hopping term is then associated with the Josephson tunneling
($<n> t \longrightarrow E_J$) while the chemical potential plays 
the same role as the external charge in the QPM ($\mu \longrightarrow Q_x$). 
The mapping becomes more accurate as the average number of bosons per 
sites increases.

In the case of strong  on-site Coulomb interaction 
$U_{ii} \rightarrow \infty$ and very low temperatures 
only few charge states are important. If the gate
voltage is tuned close to a degeneracy point, the relevant physics is 
captured by considering only two charge states for each island, and
the QPM is equivalent to an anisotropic XXZ
spin-$1/2$ Heisenberg model\cite{liu73,bruder93} 
\begin{equation}
	H_{S} = - h\sum_{i} \, S^{z}_{i} + \sum_{ {i} \ne {j} } 
 	\, S^{z}_{i} \; U_{ij} \; S^{z}_{j} 
	- E_J \sum_{\langle {i} , {j} \rangle}  
	\left ( e^{iA_{ij}} \; S^{+}_{i} \; S^{-}_{j} +
	e^{-iA_{ij}} \; S^{+}_{j} \; S^{-}_{i} \right ) \;.
\label{Sham}
\end{equation}
The operators $ S^{z}_{i}, \; S^{+}_{i},\;S^{-}_{j}  $ are the 
spin-$1/2$ operators, $ S^{z}_{i}$ being related to the 
charge on each island ($q_{i}=S^{z}_{i}+ \frac{1}{2}$),
and the raising and lowering $S^{\pm}_{i}$ operators 
corresponding to the "creation" and "annihilation" operators 
$e^{\pm i \phi_{j}}$ of the QPM. The  "external" field $h$ is related 
to the external charge by $h=\left (q_{\rm x}-1/2\right ) \sum_{j} U_{ij}$.  
Different magnetic ordered phases of the XXZ Hamiltonian correspond to the
different phases in the QPM. Long-range order in $\langle S^+ \rangle$ 
indicates superconductivity in the QPM while  long-range order in 
$ \langle S^z  \rangle $ describes order in the  charge configuration 
(Mott insulator).

The three models are equivalent in the sense that they belong to the same 
universality class (they lead to the same Ginzburg-Landau effective free 
energy). The non-universal features like  the location of the phase 
transitions, the shape of the phase diagram (and sometimes the very existence 
of some intermediate phase) depend quantitatively on the specific choice 
of the model. A more rigorous discussion of the different dynamical algebras 
realized by the three models can be found in 
Refs.\cite{amico00,das99,herbut99,kopec00,amico00a}.

A generalization of these models to the case in which amplitude fluctuations 
are coupled to phase fluctuations was discussed in Ref.~\cite{fazio88}. 
Very recetly Yurkevich and Lerner~\cite{lerner00} developed a 
nonlinear sigma model description of granular superconductors that reduces to 
the BH model in the limit in which amplitude fluctuations are ignored. 

\subsection{The zero-field phase diagram}
\label{phase0}
\begin{figure}
\centerline{{\epsfxsize=16cm\epsfysize=16cm\epsfbox{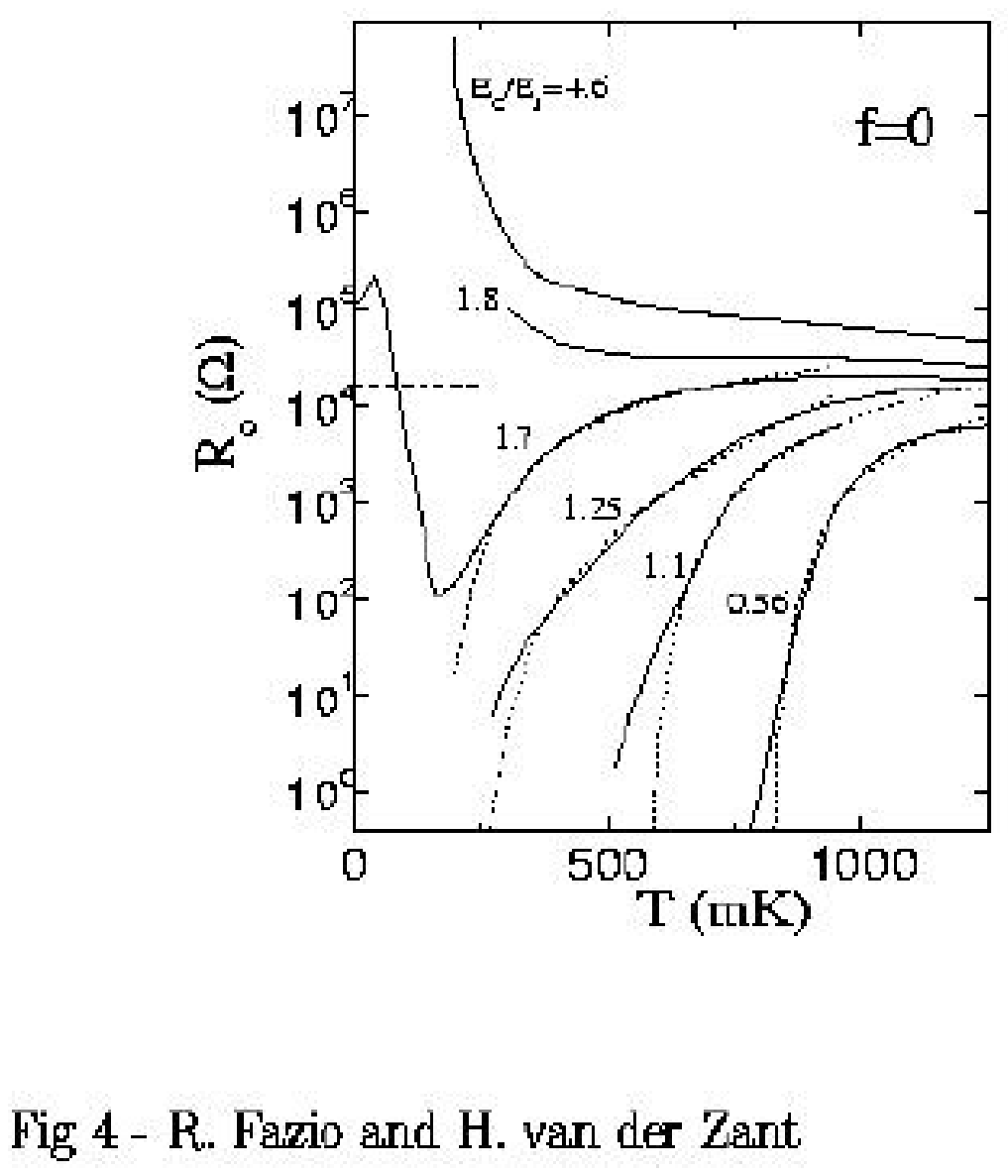}}}
\caption{The zero-field linear resistance per square measured as a function 
 of the temperature for six different square arrays. The curved dashed lines 
 are fits to the vortex-BKT square root cusp formula. The dashed 
	horizontal line indicates the zero-temperature universal resistance at 
	the S-I transition calculated in Ref.\protect\cite{Fisher90b}. 
	(From Ref.\protect\cite{zant96}.)}
\label{ktb_zero}
\end{figure}

Superconductor-Insulator transition has been investigated in great details 
in JJAs~\cite{geerligs89,zant96,chen95,chen96} as well as in granular
systems~\cite{strongin70,deutscher80,kobayashi80,orr86,jaeger89,kobayashi92,yagi96}
and uniform ultra-thin films~\cite{haviland89,liu91,yazdani95}.
The first controlled measurements on the S-I transition in junction 
arrays have been carried out by Geerligs {\em et al.}~\cite{geerligs89}.
Part of the data of Ref\cite{geerligs89} together with the new data of 
Ref.~\cite{zant96} are presented in Fig.~\ref{ktb_zero}. It shows the 
resistive behavior of six different square arrays in zero magnetic field. 
The zero-bias resistance per junction $R_0(T)$
has been measured with a very small transport bias (current per junction 
smaller than $<10^{-3}I_c$) in the linear part of the current-voltage 
characteristics. 
Three arrays become superconducting; two arrays insulating and one
array that lies very close to the S-I transition  
shows a doubly reentrant dependence.
The horizontal dashed line in Fig.\ref{ktb_zero} is the critical
resistance value of $8R_Q/\pi$ (see Eq.(\ref{univres})).

For the three arrays that become superconducting, 
the data are fitted to the predicted BKT square-root cusp dependence 
on temperature, 
$$
	R_0(T)/R_N=c \exp(-b[E_J/(T-T_J)]^{1/2})
$$
with $b$ and $c$ constants of order one. 
In order to compensate for the temperature dependence of $E_J$, it 
is convenient to define a normalized temperature is defined as 
$T/E_J(T)$. From the fits the normalized BKT transition
temperature $T_J$ is determined.
Near the S-I transition $T_J$ is substantially smaller than the
classical value of $0.90 E_J$. Note that at low resistance levels 
($R_0(T) <10^{-3}R_N$), deviations from the square-root cusp 
dependence are found and that the resistance decreases exponentially. 
This is indicative of thermal activation of single vortices across 
the whole array width~\cite{zant90}.

The two arrays with a ratio of $E_J/E_C \le 0.55$
show a continuous increase of the resistance as the temperature is lowered,
i.e., the arrays become insulating at zero temperature despite the fact that
each island is still superconducting with a well developed BCS gap! 
It has been proposed that, due to the long range interaction between the
charges, the conductance will follow a square-root cusp
dependence on temperature in a similar way as the resistance for 
the superconducting samples. This square-root cusp dependence
characteristic for a charge-BKT transition is generally not observed. 
Instead the conductance decreases exponentially as temperature is lowered.
This issue will be discussed in some details in the section devoted to the
BKT transition (section~II.B.4).

The resistance of sample with $E_C/E_J=1.7$ has a very remarkable 
dependence on temperature.
Starting at high temperatures, $R_0(T)$ first decreases when the temperature 
is lowered. Below $T$=150$\thinspace$mK, however, $R(T)$ increases by more 
than three orders of magnitude and at the same time a charging gap 
develops in the $I$-$V$ curve. Finally at 40$\thinspace$mK, $R_0(T)$ 
starts to decrease again. 
The second reentrant transition at 40~mK seems to be a more general feature
of arrays near the S-I transition which is also present in a magnetic field. 

Reentrant behavior in the resistance has also been observed in granular
superconductors. Already in the early theoretical works on the QPM
various explanations have been proposed. We will summarize some of the 
ideas. Efetov~\cite{efetov80} suggested that the thermally excited 
quasi-particles could screen the Coulomb energy thereby lowering the 
threshold for the onset of phase coherence. Stimulated by Efetov's work
a number of theoretical papers showed that a reentrant
phase boundary can be obtained in the QPM however it turned out that 
the very existence of the re-entrance was sensitive to the approximation 
scheme used. Moreover it was shown that even if there is no re-entrance 
in the phase diagram, the QPM leads to a fluctuation dominated
region~\cite{fazekas84,fazio86,fishman88a} which may account for the 
observed re-entrance. Many physical ingredients not contained in the QPM 
(random offset, dissipation,...) may play a role as well. 
Recently Feigelman {\em et al.}\cite{feigelman97} proposed that the 
parity effects may be responsible. At intermediate 
temperatures the screening of quasi-particles would decrease the effective 
Coulomb interaction (and therefore the resistance). At lower temperatures 
screening disappears due to the excess free energy associated with odd 
grains, leading to an upturn of the resistance curve.

\begin{figure}
\centerline{{\epsfxsize=14cm\epsfysize=16cm\epsfbox{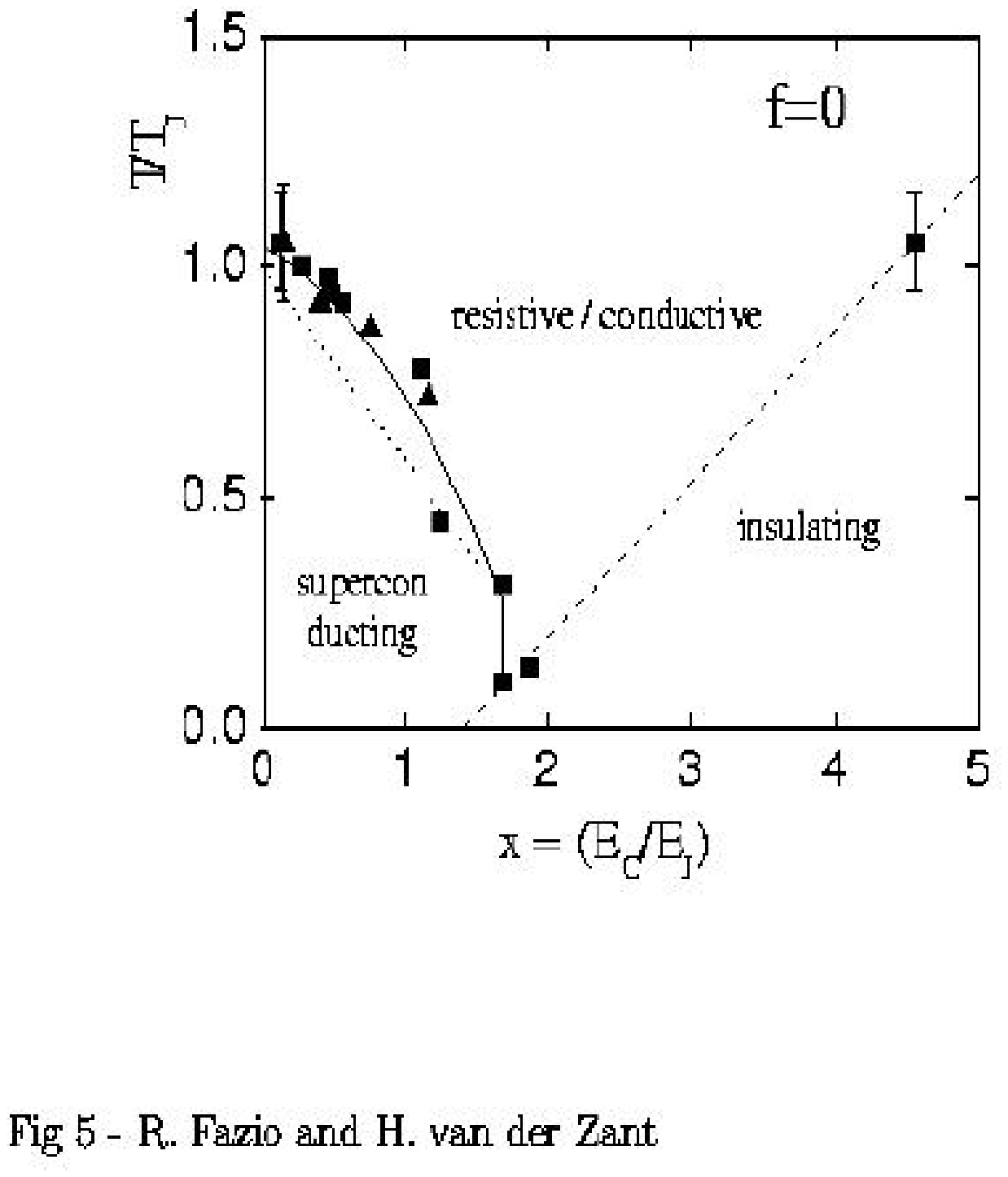}}}
\caption{Measured phase diagram in zero magnetic field for square 
	(solid squares) and triangular (solid triangles) arrays showing the 
	S-I transition at $E_C/E_J \sim 1.7$. The solid line is a 
	guide to the eye connecting the data points and the dotted line 
	on the superconducting side represents the result of the calculation 
	of Ref.\protect\cite{jose94}.
	(From Ref.\protect\cite{zant96}.)}
\label{phase_zero}
\end{figure}

The data of Fig.\ref{ktb_zero} can be used to construct a phase diagram
for phase transition of Josephson arrays in zero-magnetic field~\cite{zant96}
as shown in Fig.~\ref{phase_zero}. 
In this figure the superconducting-to-normal phase boundary is 
the vortex-BKT phase transition. Temperature on the vertical axis in
this figure is given in units of $E_J$ and scaled to the classical
(in absence of charging effects) BKT transition $T_J^{(0)}$.
The experimental value $T_J^{(0)}=0.95E_J$ is close to the theoretical 
value of 0.90 determined from Monte Carlo simulations.
On the insulating side of the figure no strict phase transition was
observed. The dashed line therefore is somewhat arbitrary. It represents
the crossover to the low-temperature region with $R_0 >10^3 R_N$.
Fig.\ref{phase_zero} indicates that at zero temperature the S-I transition
takes place at $E_C/E_J \approx 1.7$. 

The existence of a zero temperature (quantum) phase transition can be 
understood by simple arguments as already discussed in the previous section. 
Mean-field~\cite{simanek81}, variational approaches~\cite{lozovik81,wood82,cuccoli99}, 
$1/{\rm z}$ expansion~\cite{fishman88b}, Monte Carlo simulations~\cite{jacobs88,jose94} 
and cluster expansions~\cite{fazekas80} were also applied to the QPM model of 
Eq.~(\ref{QPM}). We refer the reader to the various articles for a discussion 
and comparison. Here, we present a few approaches that give
a self-contained description of the phase diagram. We 
start with the mean-field calculation. Inaccurate in determining the 
critical behavior, this approach is yet capable to capture most of the 
features measured in the experiments.  We first consider the case in  
which both charge and magnetic frustration are absent.
Note that for the time being, dissipation is not included.
It will be considered in detail in Section~\ref{dissip}. 

\subsubsection{Mean-Field approach}
The mean-field decoupling consists in approximating the Hamiltonian
of Eq.(\ref{QPM}) by~\cite{efetov80,simanek81}
$$
   H_{\rm MF} = \frac{1}{2} \sum_{i,j} Q_i C_{ij}^{-1} Q_j - {\rm z} E_J 
	\langle \cos\phi  \rangle _{\rm MF} \sum _{j} \cos \phi_{j} \; .
$$
The average 
$ \langle \cos(\phi ) \rangle _{\rm MF} $ plays the role of the order 
parameter and it should be calculated self-consistently
$$
	\langle \cos(\phi  ) \rangle _{\rm MF}
	=
	{\rm Tr} \left \{   \cos(\phi_{i} ) 
	\exp ({- \beta H_{\rm MF}}) \right \}/
	{\rm Tr} \left \{ \exp ({- \beta H_{\rm MF}}) \right \}
\:\:.
$$
Close to the transition point the thermal average on the r.h.s can be 
evaluated by expanding it in powers of $\langle \cos(\phi )\rangle _{\rm MF}$ 
(This can be done because the transition is continuous).
The transition line is determined by the equation
\begin{equation}
	1 - {\rm z}E_{J}
	\int_{0}^{\beta } d \tau  
	\langle \cos\phi_{i}(\tau ) \space \cos\phi_{i}(0) \rangle_{ch}
	\:=0
\label{selfc2}
\end{equation}
where the average   $\:   <...>_{ch} \:$ is calculated using the 
eigenstates of the charging part of the Hamiltonian only. Note that 
the (imaginary) time evolution of the phase is due to charging as well.

In the classical limit the phase correlator is unity (there are no quantum 
fluctuations) and the mean-field transition temperature is ${\rm z}E_J/2$. 
Charging effects inhibit phase fluctuations and the critical temperature 
decreases. Explicit formulas for the phase-phase correlator are given in 
Appendix~\ref{phase-correlator}.
At $T=0$ in the self-charging limit ($E_{ij}^{(ch)} = E_0 \delta_{ij}$)
the correlator reads
$$
	\langle \cos \phi_{i}(\tau ) \space \cos \phi_{i}(0) \rangle_{ch}
	= \frac{1}{2}
	\exp \{-4\tau E_0  (1 - \tau /\beta )\} \:.
$$
By substituting this expression in Eq.(\ref{selfc2}) one gets 
a value for the S-I transition which coincides with the simple estimate 
based on energy considerations. 

\begin{figure}
\centerline{{\epsfxsize=14cm\epsfysize=16cm\epsfbox{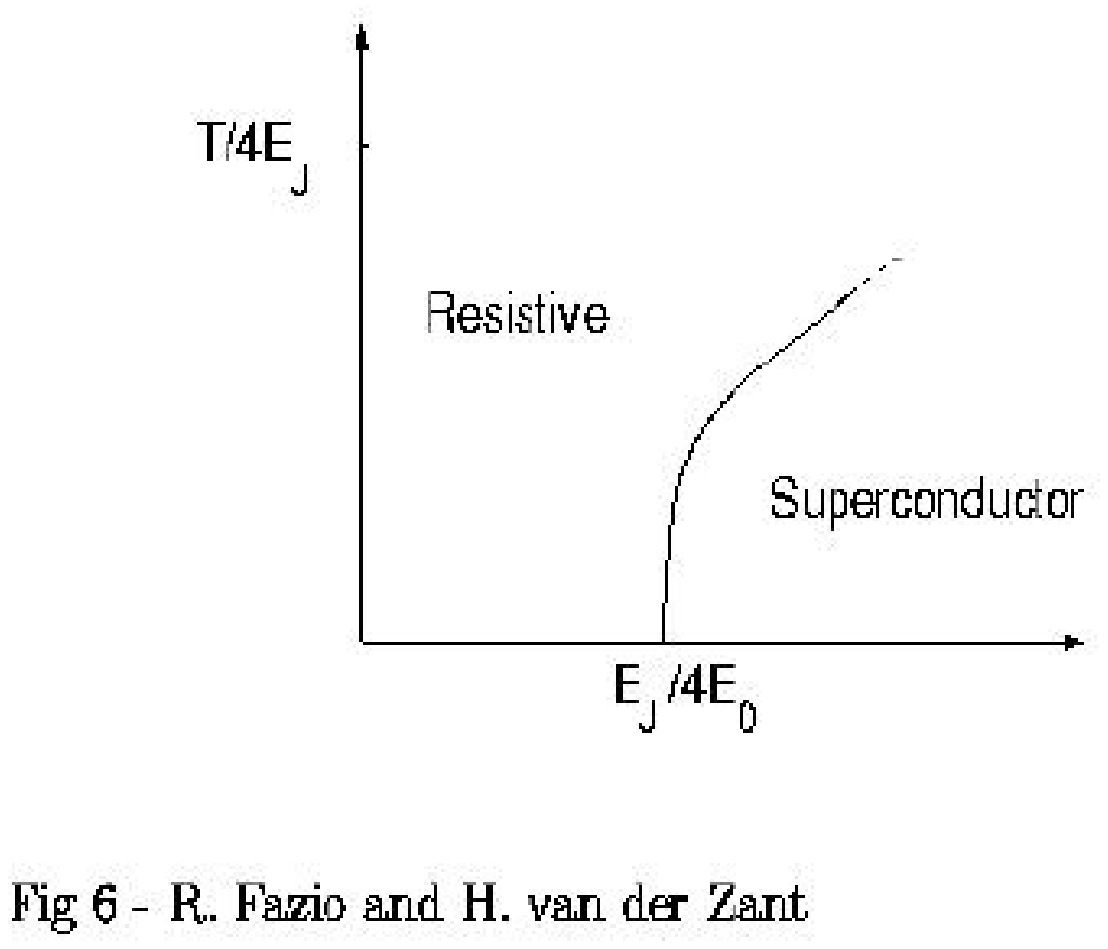}}}
\caption{Sketch of the phase diagram for short-range interaction between charges. The phase 
	incoherent state is resistive with an activated behavior of the 
	resistance. At $T=0$ the array is a insulator. The dependence of 
	the quantum critical point on the capacitance matrix is all 
	contained in $E_0$.}
\label{srdiag_fig}
\end{figure}

The detailed structure of the phase diagram depends, even in the absence of
(magnetic or electric) external frustration, on the range of the electrostatic 
energy. The phase diagram in the short-range case is sketched in 
Fig.\ref{srdiag_fig}. 
Very recently a detailed analysis of the dependence of the phase boundary 
on the form of the capacitance matrix has been performed in Ref.\cite{kim97} 
using perturbation theory and numerical simulations.

For two-dimensional arrays the transition to the superconducting state 
is of the BKT type with no 
spontaneous symmetry breaking. Quantum fluctuations renormalize the 
value of the transition temperature but do not change the universality 
class of the transition. The corrections to the BKT transition due to 
quantum fluctuations have been evaluated in a semi-classical approximation 
in Ref.~\cite{jose84}. For $E_J >> E_C$, the JJA behaves as a classical 
XY model but with a renormalized $E_J$. This approach breaks down when 
quantum fluctuation drive the transition to zero. At zero temperature 
there is a dimensional crossover and the S-I transition belongs
to the $(d+1)$-XY universality class.

\subsubsection{Coarse-Graining approach} 
\label{coarsegrain}
Although it is very useful in determining the structure of the phase 
diagram, the mean-field approach has various shortcomings. For a more 
accurate description of the quantum critical regime one has to resort to 
different approaches. Universality implies that  the critical behavior 
of the system depends only on its dimensionality and on the symmetry 
which is broken in the ordered phase. Many properties of 
JJAs can be extended to other systems that show a S-I transition.
By using the coarse-graining approximation it is possible to go from the 
QPM model to a Ginzburg-Landau model with an effective free energy which is a
function of the order parameter~\cite{doniach81,kissner93} only. Since the 
transition is governed by quantum fluctuations, the order parameter depends 
both on space and (imaginary)-time~\cite{sachdev99,herz76}.

The coarse-graining proceeds in two steps:
\begin{itemize}
\item{An auxiliary field  $\psi(x,\tau)$ (which has the role of the
	order parameter) is introduced through a 
	Hubbard-Stratonovich transformation. The partition function is then 
	expressed as a path integral over $\psi$}.
\item{The assumption that the order parameter is small close to the
	transition allows for a 
	cumulant expansion to obtain the appropriate 
	Ginzburg-Landau free energy. The coefficients depend on the details 
	of the microscopic model.} 
\end{itemize}

The partition function of the QPM is given by
\begin{equation}
	Z=Tr\{e^{-\beta H_{\rm QPM}}\}=Z_{ch}
	\langle T_{\tau}e^{-\int_0^{\beta} d\tau
	H_{J}(\tau)} \rangle_{ch}
\end{equation}
where the subscripts ${ch}$ and ${J}$ refer to the charging
and Josephson part of the Hamiltonian in Eq.(\ref{QPM}). By applying 
the Hubbard-Stratonovich transformation to the Josephson term one gets
(in absence of magnetic and charge frustration) 
$$
	{\exp \left \{ {{E_{J}}\over{2}} \int_{0}^{\beta } d
	\tau \sum_{\langle i,j\rangle}
	e^{i\phi_{i}}e^{-i\phi_{j}} + h.c. \right \}} $$
\begin{equation}
	\sim \int D \psi^*D \psi
	\exp \left\{-\int_{0}^{\beta }d \tau 
	\sum_{i,j}[E_{J}]_{ij}^{-1} 
	\psi_{i}^*(\tau) \psi_{j} (\tau)
	+ \int_{0}^{\beta} d \tau 
	\sum_{i}\left [\psi_{i}^*(\tau) e^{i\phi_i(\tau)} + h.c.
 	\right  ]\right\} 
\label{HST}
\end{equation} 
Here we introduced a matrix $[E_{ J}]_{ij}$ which is equal to 
$E_{J}$ if $i$ and $j$ are nearest neighbors and zero otherwise.
The partition function can be written as 
\begin{equation}
Z=Z_{ch}\int  D \psi^*D \psi \exp \left\{-F[\psi]\right\}
\label{partfunc}
\end{equation}
Close to the phase transition one can perform a gradient expansion:
\begin{equation}
	F =\sum_{\langle i,j\rangle}
	\int \;d\tau d\tau'
	\psi_i(\tau) 
	\left[[E_{J}]_{ij}^{-1} \delta (\tau -\tau')
	-g(\tau-\tau') \delta_{ij}\right] \psi^*_j(\tau') 
        +\kappa\sum_i\; \int d\tau|\psi_i(\tau)|^4 \; .
\label{free}
\end{equation}
The dynamics of the field $\psi$ is governed by the phase-phase correlator 
$$
	g (\tau-\tau') = 	
	\langle \exp[ \phi_{i}(\tau ) -\phi_{i}(\tau') \rangle_{ch}
$$
that was already encountered in Eq.(\ref{selfc2}). The coefficient  
$\kappa$ is related  to the  four-point phase correlator. 

The partition function in 
Eq.(\ref{partfunc}) can be calculated using a mean-field approximation for the 
phase by evaluating it in the saddle point approximation. The results 
coincide with that of the previous subsection. 
In the coarse-graining approach, however, a systematic treatment
of the fluctuations is possible and it allows to study transport as well.
In the case of zero offset charges and zero external magnetic field, 
by expanding the phase correlator around the zero-frequency and zero-momentum
value, the quadratic part of Eq.(\ref{free}) can then be rewritten as
$$
	 F^{(0)}[\psi] =T \sum_{{\bf k},\omega_n} 
	\left[
	\epsilon + \gamma k^2 + \zeta \omega_n^2
	\right]|\psi({\bf k},\omega_n)|^2
$$
where, using the expression given in Appendix~\ref{phase-correlator}, 
the coefficients can be expressed as: 
\begin{eqnarray}
	\epsilon &=& \frac{g^{-1}(\omega_n=0)}{2E_0} - \frac{E_J}{E_0} 
		\\ \nonumber
	\gamma   &=& \frac{g^{-1}(\omega_n=0)}{8E_0}
		\\ \nonumber
	\zeta    &=&  \frac{\partial^2_{\omega_n}g(\omega_n)
			\mid_{\omega_n=0}}{4E_0}   
                  \;\;\;\; .
\end{eqnarray}
At $T=0$, this system belongs to the same universality class as the $(d+1)$ 
XY model. One can readily obtain all the critical exponents from what is 
know from the XY model~\cite{parisibook}. The dynamical critical exponent 
is $z=1$ due to the symmetry between space and time. At finite temperatures 
the transition belongs to the Berezinskii-Kosterlitz-Thouless universality 
class and there is no spontaneous breaking of the symmetry. The dynamical 
critical exponent and the dimensional crossover is modified in the case of 
$1/r$-Coulomb interaction between charges~\cite{fisher88}.

\subsubsection{Duality transformations}
\label{duality}
Duality transformations have proven to be a powerful tool in field theory
and statistical mechanics~\cite{savit80}. The idea behind it is that
the weak coupling region of a particular
system can be mapped onto the strong coupling range (and vice
versa). The symmetries of the system under this transformation lead to important 
insight into the structure of the model, especially in the region 
of intermediate couplings which is usually elusive to standard treatments.
Dual transformations constitute a powerful approach since 
it is possible to recast the partition function solely in 
terms of the topological excitations of the 
system~\cite{villain75,jose77,elitzur79,nienhuis87}. 

In this section we derive some properties of quantum JJAs derived from
duality. There is a dual transformation~\cite{fisher89a,fazio91a,fazio91b} 
relating the classical vortex limit, $E_{J} \gg E_{C}$, 
to the opposite charge limit, $E_{J} \ll E_{C}$. 
The situation is most transparent in the case 
$C_0 \ll C$, which might be more relevant for arrays. The interaction 
between charges on islands is then logarithmic, analogous to vortex interactions 
in classical, superconducting arrays. The charges form a 2D Coulomb 
gas and are expected to undergo a BKT transition at $T_{ch}^{(0)} \sim
E_{C}$\cite{mooij90} (see also next subsection). 

Using the results discussed in Appendix~\ref{coupled-coulomb-gas}, the
partition function of a JJA can be expressed as a sum over charge ${q}$
and vortex ${v}$ configurations 
\begin{equation} \label{part12}
	Z = \sum_{[q,v]} e^{ - S \{q,v\}} \; .
\end{equation}
The effective action $S \{q,v\}$  reads
\begin{eqnarray} 
	S\left\{ q, v \right\} =  
	\int_0^{\beta } d\tau \sum_{ij} & &\left\{ 2e^2 q_{i}(\tau)
	C^{-1}_{ij} q_{j}(\tau)  + \pi E_J v_{i}(\tau) G_{ij}
	v_{j}(\tau)   \right. \nonumber \\
	 &+& i \left. q_{i}(\tau) \Theta_{ij} \dot{v}_{j}(\tau) 
	+  \frac{1}{4\pi E_J}
	\dot{q}_{i}(\tau) G_{ij} \dot{q}_{j}(\tau) \right\}.
\label{Sqv11}
\end{eqnarray}
This action describes two coupled Coulomb gases. We have used a continuous time 
notation  for clarity. Since $q$ and $v$ are integer valued fields, the path 
integral is well defined on a discretized time expression. The
charges interact via the inverse capacitance matrix (first term). 
The interaction among the vortices (second term) is described by the kernel
$G_{ij}$, which is the Fourier transform of $k^{-2}$. 
At large distances  $r_{ij} \gg a$ between the
sites $i$ and $j$ it depends logarithmically on the distance: 
$$
	G_{ij} \sim - \frac{1}{2}\ln r_{ij}.
$$
The third term describes the coupling between the topological excitations in the 
two limits, i.e., it describes the coupling between charges and vortices. The function 
$$
\Theta_{ij} = \arctan\left(\frac{y_i - y_j}{x_i - x_j}\right)
$$ 
represents the vortex-phase configuration at site $i$ when its center is placed 
at the site $j$. The coupling has a simple physical interpretation: a
change of vorticity at site $j$ produces a voltage at site $i$ which
is felt by the charge at this location. The last term 
$\dot{q} G \dot{q}$ stems from the spin-wave contribution to 
the charge-correlation function. In the limit in which $E_J \: \rightarrow \:0$ 
or $E_C \: \rightarrow \:0$ the action for a classical system of Cooper pair 
charges or vortices is recovered.

The effective action in Eq.(\ref{Sqv11}) shows a high degree of symmetry
between the vortex and charge degrees of freedom. In particular, in the
limit $C_0 \ll C$ the inverse capacitance matrix has the same
functional form as the kernel describing vortex interactions:
$$
	e^2 C^{-1}_{ij} = \frac{E_{ C}}{\pi} G_{ij} \; .
$$
Hence charges and vortices are dual. There is a critical point for which 
the system is self-dual with respect to interchanging them:
$$
	\frac{E_{ J}}{E_{ C}} = \frac{2}{\pi ^2}
$$
The duality is strict for vanishing self-capacitance and in the absence of the 
spin-wave  duality breaking term ($\dot{q} G \dot{q}$) in Eq.(\ref{Sqv11}). 
This latter term is irrelevant at the critical point, i.e., it merely shifts 
the transition point. However, it has important implications for the dynamical 
behavior.

Duality transformations have also been applied to a three-dimensional JJA 
consisting 
of two 2D arrays placed on top of each other~\cite{blanter96,rojas95,blanter97}.
The authors of these papers assume that there is only capacitive coupling between 
them (no Josephson coupling). The most interesting situation arises when one array 
is in the quasi-classical (vortex) regime while the other is in the quantum, charge 
regime. Then, vortices in one layer and charges in the other one are well-defined.

The Euclidean effective action, in term of phases, $S\{\phi_1,\phi_2\}$ has the form
\begin{eqnarray} \label{Sphi}	
	 &&S\{\phi_1,\phi_2\} = 
	\int_{0}^{\beta} d\tau \sum_{\mu=1,2} \nonumber \\
	&&\left\{  \vphantom{\frac{}{{}_{}}}
	  \vphantom{\frac{}{{}_{}}} 
	 \sum_i \left[\frac{C_{0\mu}}{8e^2}
	(\dot{\phi}_{i\mu})^2  +  \frac{C_{\rm x}}{8e^2}
	(\dot{\phi}_{i1} - \dot{\phi}_{i2})^2 \right] +
	\sum_{\langle ij \rangle} \left[
	 \frac{C_{\mu}}{8e^2}
	(\dot{\phi}_{i\mu} - \dot{\phi}_{j\mu})^2 
	- E_{J\mu} \cos (\phi_{i\mu} - \phi_{j\mu})
	\right] \right\}
\end{eqnarray}
where $C_{0\mu}$ are the island capacitances in array $\mu$
relative to ground, $C_{\mu}$ are the junction capacitances in the
array $\mu$, and $C_{\rm x}$ are the interlayer capacitances
between islands on top of each other, while $E_{ J\mu}$ are
the Josephson coupling constants in the layers. 
Similar as in a single array, we move from a description in terms of
phases to one in terms of charges and vortices, and use the duality of
the resulting action to investigate the transition. 

Before we proceed with the calculation it is necessary 
to stress that in the regime of interest the interlayer capacitances 
$C_{\rm x}$ not only couple the layers, but also renormalize the island 
capacitances $C_{01}$ and $C_{02}$ to ground. The physical reason for this is
that due to the strong fluctuations of charges in layer 2 and
vorticities in layer 1 these variables are effectively continuous, and
hence a coupling to the other array plays the same role as a
coupling to ground. Due to screening, the interaction between charges in each 
layer has a finite range for any non-zero $C_{\rm x}$, and the BKT transition is 
replaced by a crossover. However, in the
limit $C_{01} \ll C_{\rm x} \ll C_1$ the screening length $\xi_1 \sim
(C_1/C_{\rm x})^{1/2}$ can be large enough to make it meaningful to
speak about the charge-unbinding transition (the transition is
exponentially sharp). Below we consider this case.
For not so weak coupling, on the other hand, this description becomes meaningless, 
since the crossover is strongly smeared, and the insulating phase is absent. 

It is possible to introduce vortex degrees of freedom in the same
way as for one array. We obtain
the effective action for charges $q_{i1}(\tau)$ in layer 1 and
vorticities $v_{i2}(\tau)$ in layer 2 (to be referred below as $q_i$ and
$v_i$)
$$	 S\{q,v\} = \int_{0}^{\beta} d\tau 
	\left\{ \vphantom{\frac{}{{}_{}}} \right. \frac{2E_{ C1}}{\pi}
	\sum_{ij} q_i(\tau) G_{ij} q_j(\tau)
	 + \frac{1}{4\pi E_{ J1}} \sum_{ij} \dot{q}_i(\tau) G_{ij} 
	\dot{q}_j(\tau)
	+ \pi E_{ J2} \sum_{ij} v_i(\tau) G_{ij} v_j(\tau) 
$$
\begin{equation} \label{Sqv2}
	+ \frac{\pi }{8E_{ C2}}
	\sum_{ij} \dot{v}_i(\tau) \Big[ G_{ij} 
		- \frac{C_{\rm x}^2}{4\pi^2
	C_1C_2} \sum_{kl} \Theta_{ik} G_{kl} \Theta_{lj} \Big]
	\dot{v}_j(\tau)
	+ \frac{iC_{\rm x}}{2\pi C_1} \sum_{ijk} \dot{v}_i(\tau)
	\Theta_{ik} G_{kj} q_j(\tau)
	\left. \vphantom{\frac{}{{}_{}}}	\right\}. 
\end{equation}
This equation looks rather similar to the effective charge-vortex action in one
Josephson junction array (see Eq.(\ref{Sqv11})). 
The most important difference is that while in one layer either charges or 
vortices are well-defined degrees of freedom,
Eq. (\ref{Sqv2}) describes the system of two {\em well-defined} dynamic
variables on each site -- the charges in layer 1 and the vortices in 
layer 2. The action shows a duality 
between charges and vortices (the second term in
the square brackets is small for $C_{\rm x} \ll C_1,C_2$). Both kinetic
terms for charges and vortices violate the duality due to the
numerical coefficients. However, close enough
to the transitions these terms only produce a small renormalization of
the transition temperature, and are therefore irrelevant. 
Another interesting feature of this action is that the last term,
describing the interaction between charges and vortices, is also small, while 
in a single-layer array the interaction is always of the same order of
magnitude as the other terms.

\subsubsection{Berezinskii-Kosterlitz-Thouless transitions}

In classical arrays, it is well established that JJAs undergo a BKT phase 
transition to the superconducting state. In the opposite limit where charging 
dominates, quantum fluctuations of the phases are strong, and vortices are 
ill-defined objects. In this regime the duality transformations discussed in 
the previous section show that charges on the islands are the relevant variables. 
Similarly to vortices in classical arrays, they interact logarithmically
with each other and are expected to undergo a charge-BKT transition leading to 
insulating behavior~\cite{mooij90}. 

\begin{figure}
\centerline{{\epsfxsize=14cm\epsfysize=16cm\epsfbox{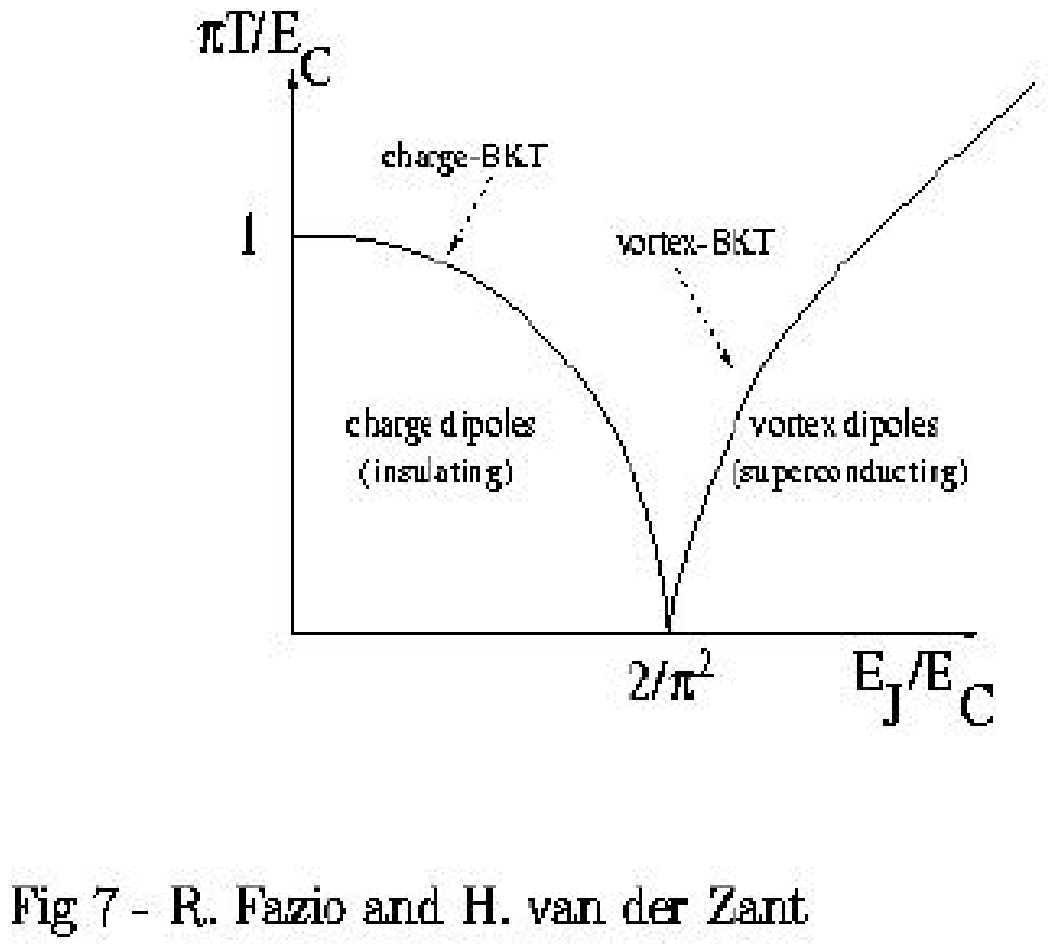}}}
\caption{The phase diagram for a quantum JJA in the limit of long-range 
	(logarithmic) interaction between charges. Similarly to vortices 
	the charges undergo a BKT transition leading to insulating behavior at low 
	temperature and small $E_J/E_C$ ratio.}
\label{lrdiag_fig}
\end{figure}

A critical point separates the superconducting and insulating regime at $T=0$.
As discussed before, the various models give different estimates for the value
of the critical point, but in all cases it lies close to $E_J/E_C \sim 1$.
The theoretical phase diagram in the limit of logarithmically interacting 
charges is shown in Fig.\ref{lrdiag_fig}. For any $T \neq 0$ the
array has three phases. Next to the superconducting and insulating phase, there
is a region with normal conduction. Here, there are always some free 
vortices/charges present as they are generated by thermal or quantum 
fluctuations. Experimental 
verification of this diagram has been reported in Refs.~\cite{zant92,yagi97}
In the remainder of this subsection, we discuss the experimental aspects of BKT 
transitions in arrays.  

The BKT transition in the classical case has been studied in great detail
(see e.g. Refs.~\cite{nato86,delft88}). On approaching the critical point 
of the S-I transition, the vortex-BKT transition temperature is lowered by 
quantum fluctuations. For $C_0 = 0$ the shift of the transition temperature 
is~\cite{jose94,v-BKT-self}
\begin{equation} \label{tsh1}
	T_{J} = 
	\frac{\pi E_{J}}{2}\left(1 - \frac{4}{3\pi}\frac{E_{0}}{E_J}\right) \;\;\; .
\end{equation}
A reduction of the transition temperature is generally observed in quantum arrays. 
The Delft data (shown in Fig.~\ref{ktb_zero}), however, show that with increasing 
$E_C/E_J$ ratio the reduction of the transition temperature goes faster 
than predicted above (Eq.~(\ref{tsh1}).  

On the charging side ($E_J \ll E_C$) of the phase diagram, the charge-BKT 
transition occurs at a temperature~\cite{fazio91a} 
$$ 
	T_{ch} = \frac{E_{C}}{\pi}  - 
	0.31\frac{E_{J}^2}{E_{C}} \; .
$$ 
Note that the charge-BKT transition exists for both arrays with superconducting 
and normal islands~\cite{normal-BKT1}. 

The existence of a charge-BKT transition implies that arrays are insulating 
below the transition temperature. The conductance should vanish with a
characteristic square-root dependence according to
\begin{equation}
	G \sim 
	R_N^{-1} \exp \left\{ -2b\left[ T/T_{ch} - 1 \right]^{-1/2}\right\}
\end{equation}
where the constant $b \sim 1$. The temperature dependence of the conductance in 
the charge regime has been investigated by 
several groups~\cite{tighe93,delsing94,normal-BKT2}. Instead of the predicted 
square-root cusp behavior, an exponential  (activated) temperature dependence 
$$
	G \sim R_N^{-1} 
	\exp \left\{ -E_a/T \right\}
$$
has been observed with an activation energy 
$$
	E_a \sim \Delta + 0.24 E_C \;\;\; .
$$ 

In arrays the screening length is about $10^2$ lattice constants and therefore 
there is no a priori reason for observing such dramatic deviations from the 
BKT theory. Recently, Feigelman {\em et al.}\cite{feigelman97} re-examined the 
problem and found that parity effects together with the  screening  of the 
Coulomb interaction due to thermally activated quasi-particles is responsible 
for masking the charge-BKT transition. At temperatures above a crossover  
temperature $T^*$ where parity effects~\cite{tuominen92} set in, the presence 
of quasi-particles rules out the possibility to observe the charge-BKT 
transition.  
If the BKT transition temperature is larger than the crossover temperature, 
the array behaves as a normal one and a charge BKT transition occurs associated 
with the unbinding of quasi-particles at temperatures close $E_C/4\pi$.
The presence of free charges screens the interaction between Cooper pairs 
resulting in the unbinding of pairs. The resistance as calculated by 
Feigelman {\em et al.}\cite{feigelman97} is expected to be 
$$
	\ln \frac{R(T)}{R(E_C/4\pi)} 
	\sim \mbox{min}
	\left[ \frac{F_P(T)}{T-E_C/4\pi}, \frac{b}{\sqrt{4\pi}}\left(
	\frac{E_C}{T-E_C/4\pi} \right)^{1/2}\right]
$$
where $F_P(T)$ is the free energy difference between islands with an even 
and an odd number of electrons~\cite{tuominen92}. From analogous 
considerations one may conclude~\cite{feigelman97} that normal arrays are  better 
suited for studying the charge unbinding transition. 

\subsection{Magnetic frustration}
\label{magfrus}
When in the classical limit ($E_C<<E_J$) a perpendicular magnetic field is applied, 
vortices enter the array above a certain threshold field~\cite{zant90}.
Just as in films, the vortex density increases with increasing
magnetic field. In junction arrays the periodic lattice 
potential, however, prevents vortices to move at low temperatures. 
Only above the depinning current, vortices move (flux-flow branch).
The resistance in this branch increases approximately
linearly with $f = \Phi / \Phi_0$ up to $f \sim 0.2$ ($\Phi$ is the 
magnetic flux piercing through an elementary plaquette).
A phenomenological model, analogous to the Bardeen-Stephen model used to 
describe flux-flow in films, is in good agreement with experiments providing 
that coupling between spin waves and vortices is taken into account (see section III.C) 
The properties of arrays at low frustration (low vortex densities) are dominated by 
single-vortex properties and are the subject of the next chapter.

\begin{figure}
\centerline{{\epsfxsize=14cm\epsfysize=16cm\epsfbox{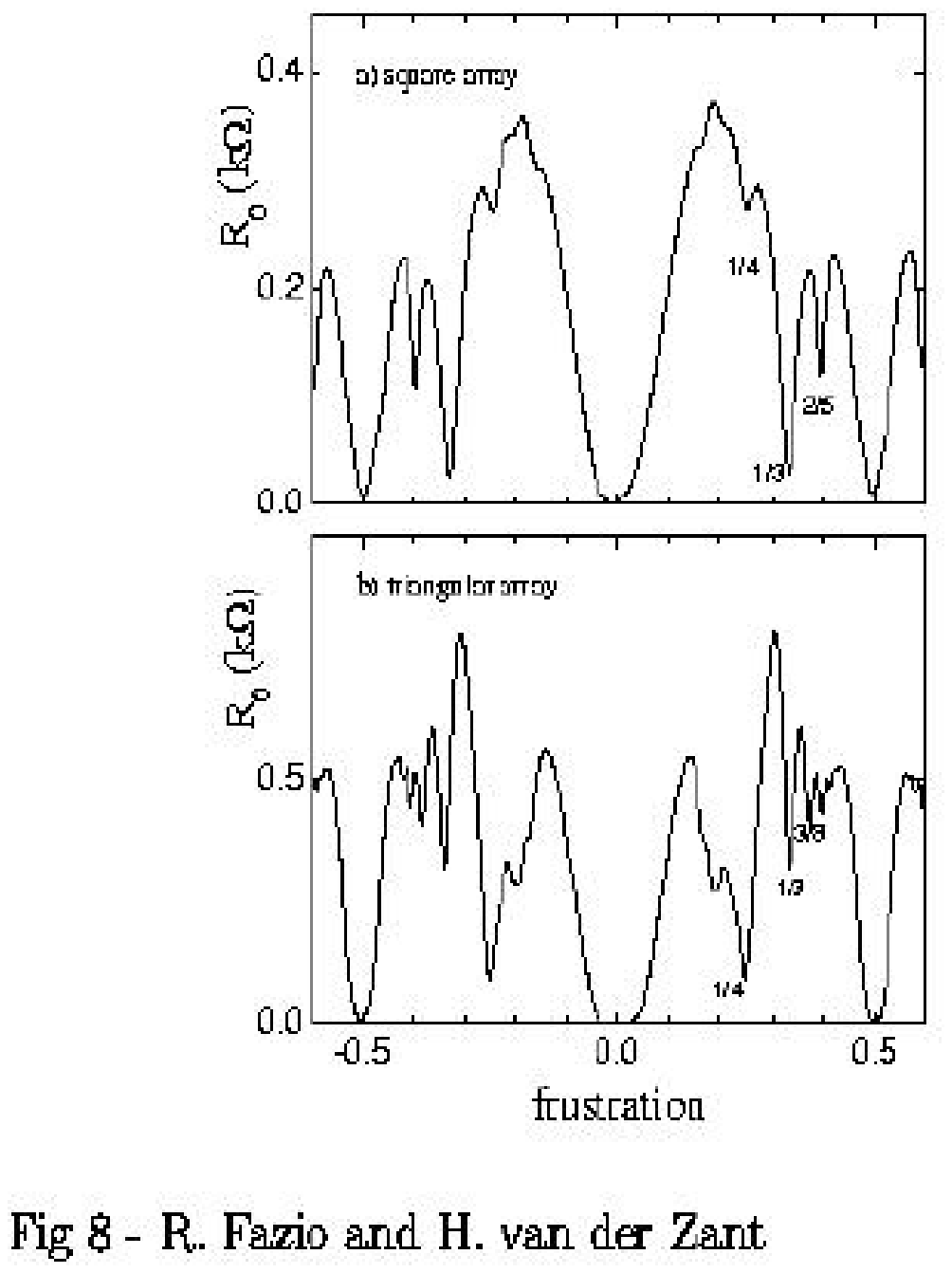}}}
\caption{Zero-bias resistance versus magnetic frustration for a square (a) and a triangular 
	(b) array. The dip at $f=1/2$ is the most pronounced one in both figures. In the 
	triangular array the dip at $f=1/4$ is  more pronounced than the one at $f=1/3$. 
	In the square array the opposite occurs. (From Ref.~\protect\cite{zant96}.)}
\label{rf_dips}
\end{figure}

In larger magnetic fields commensurability effects come into play and the 
behavior of junction arrays 
is richer than that of films. A magnetic field applied perpendicularly to the 
array leads to frustration~\cite{frustration}. The presence of the magnetic 
field  induces vortices in the system and if the frustration is a rational 
number, $f = p/q $, the ground state consists of a 
checkerboard configuration of vortices with a $q\: \times \: q$ elementary 
cell. The stability of the vortex lattice against
a bias current leads to a decrease in the small-bias resistance at finite 
fractional fillings. In order of their relative strength,  one expects 
dips at $f=1/2,1/3,1/4,2/5, ...$
in square arrays  and at $f=1/2,1/4,1/3,3/8,...$ in triangular arrays as is 
illustrated  in Fig.\ref{rf_dips}. Near these fractional values of $f$, 
defects from the ordered lattice (excess single vortices or domain walls) 
are believed to determine the array dynamics in a similar way as the field 
induced vortices determine array dynamics near $f=0$. Therefore, arrays near
commensurate values with high stability such as $f=1/2$
may qualitatively behave in a similar way as near zero magnetic field.
Because all properties are periodic in $f$ with period $f=1$ an
increase beyond $f=1/2$ does not lead to new physics.

A particularly interesting case is the fully frustrated situation
($f=1/2$) in square arrays. The two degenerate ground states consist of
a vortex lattice with a  $2 \: \times \: 2$ elementary cell. 
The current corresponding to this vortex arrangement flows
either clockwise or anti-clockwise in each plaquette (chiral ground state). 
Interaction between  domain wall excitations with 1/4 fractionally charged 
vortices ( at the corners of a domain wall) and excess single integer 
vortices are believed to trigger a combined BKT-Ising transition. 
A fully frustrated array has two critical 
temperatures related to the $Z_2$ and $U(1)$ symmetries of the problem. Their 
existence has  been investigated both by analytical methods
and Monte Carlo simulations. Even at the classical level, the complete 
scenario 
is not fully understood yet. There is numerical evidence either supporting the 
existence of two very close critical temperatures with critical behavior 
typical of
Ising and BKT transitions respectively or the existence of a single 
transition with
novel critical behavior. Further reference to classical frustrated arrays 
can be found in Ref.~\cite{martinoli00}.

\begin{figure}
\centerline{{\epsfxsize=14cm\epsfysize=16cm\epsfbox{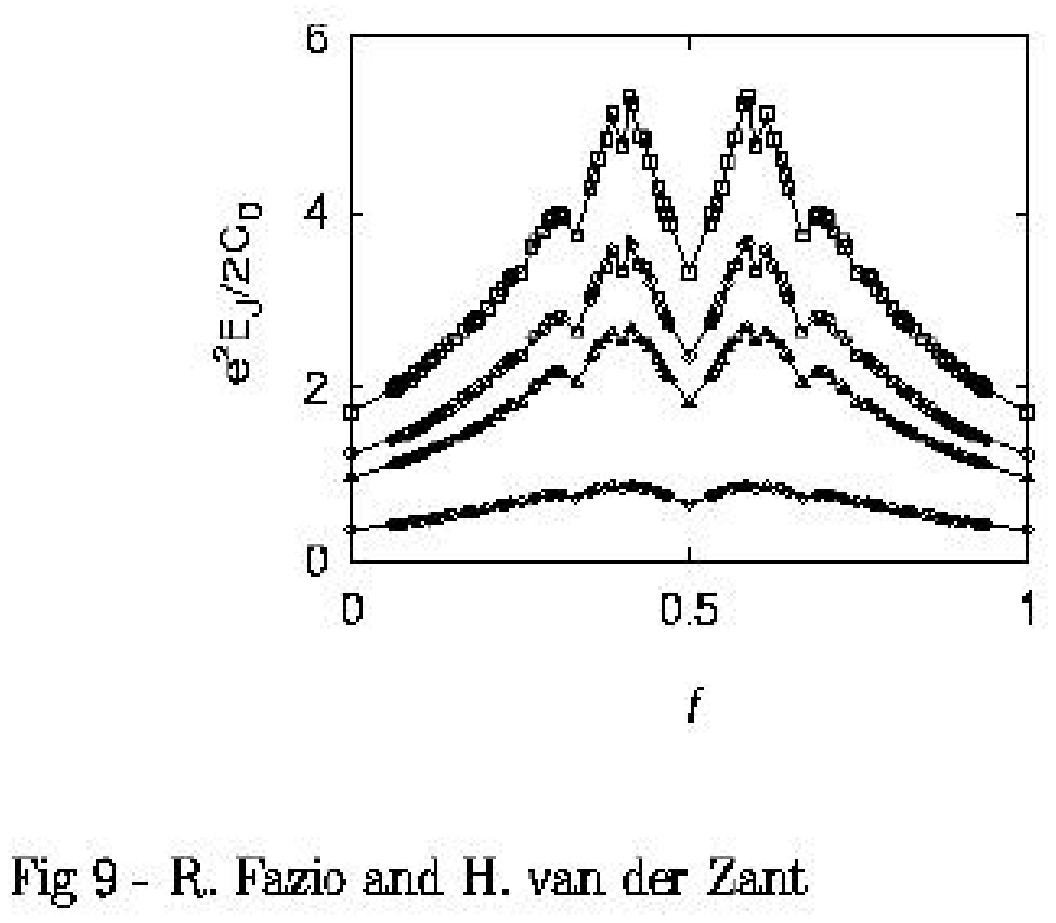}}}
\caption{Phase boundaries between the Mott insulating phase (below
	each curve) and the superconducting phase (above). 
	Boundaries for various ratios of the junction capacitance $C$ 
	to the self-capacitance $C_0$ are shown: $C/C_0$ = 0.0001($\Box$),
	0.1($\circ$), 0.2($\triangle$), and 1.0($\Diamond$). 
	(From Ref.~\protect\cite{kim98}.)}
\label{butterfly}
\end{figure}

At the mean-field level the full phase-diagram including charging effects and
magnetic fields is obtained by solving an eigenvalue problem 
equivalent to  the Hofstadter problem~\cite{hofstadter76}. 
The resulting phase boundary as a function of vorticity shows commensurability 
effects. Although the superconducting transition temperature is reduced,
the average configuration of the phases  and the supercurrent flow 
patterns are unchanged. The ground state is still chiral~\cite{fishman87}.    
More detailed calculations based on expansion in $E_J/E_0$ of the 
QPM~\cite{kim98} 
and on the BH model~\cite{niemeyer99} confirm the butterfly-like 
behavior of the S-I transition. In Fig.\ref{butterfly} the theoretical 
results obtained by Kim {\em et al.}~\cite{kim98} are shown.

\begin{figure}
\centerline{{\epsfxsize=16cm\epsfysize=16cm\epsfbox{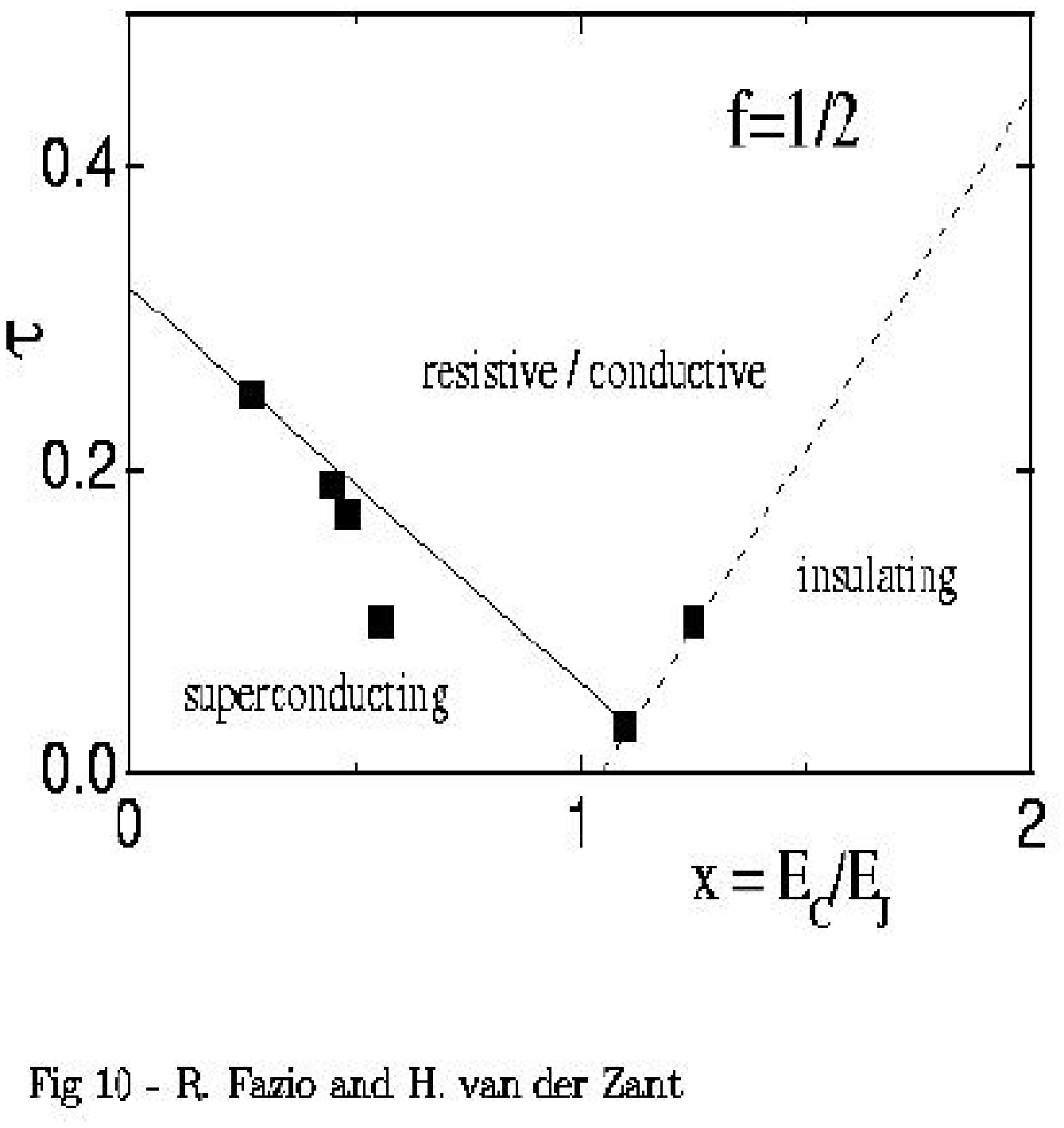}}}
\caption{Measured phase diagram for square arrays at $f=1/2$, showing the S-I transition 
	at $E_C/E_J \sim 1.2$. In the figure the temperature axis is normalized to the 
	Josephson coupling, i.e.,  $\tau = T/E_J$. The solid and dashed lines are guides 
	to the eye connecting the data points. (From Ref.\protect\cite{zant96}.)}
\label{ktb_f1/2}
\end{figure}

As in the unfrustrated case, measurements indicate a S-I
transition at $T=0$. For the same set of samples as presented 
in Fig.\ref{ktb_zero} the  S-I transition for $f=1/2$ has been studied. The
phase diagram is shown in Fig.\ref{ktb_f1/2}.
The transition takes place very close to a normal-state resistance of
11~k$\Omega$. The critical $E_C/E_J$ ratio is about 1.2, a factor 0.7 lower 
than the zero-field value.
This decrease of is consistent with the simple model that
involves a reduction of effective Josephson coupling energy at $f=1/2$: 
the interaction energy of a vortex pair is a factor $\sqrt2$ smaller
than in zero field. With this lower effective
coupling the critical value of $E_J/E_C$  of the S-I transition is  reduced 
by a factor $\sqrt{2}$, which is close to the observed reduction of 0.7. 

The experimental data agree rather well with the  
quantum Monte Carlo calculations~\cite{jose96} in the classical limit.
The experimental points of the transition 
temperatures are, however, lower than the calculated ones by entering in 
the quantum regime. At present, there is no explanation for this discrepancy. 
It would require a more detailed study and better understanding of the
phase transition at $f=1/2$.
The calculations do indicate, on the other hand, a S-I transition at 
$E_J/E_C \approx 1$, in agreement with the experiment.

\begin{figure}
\centerline{{\epsfxsize=14cm\epsfysize=16cm\epsfbox{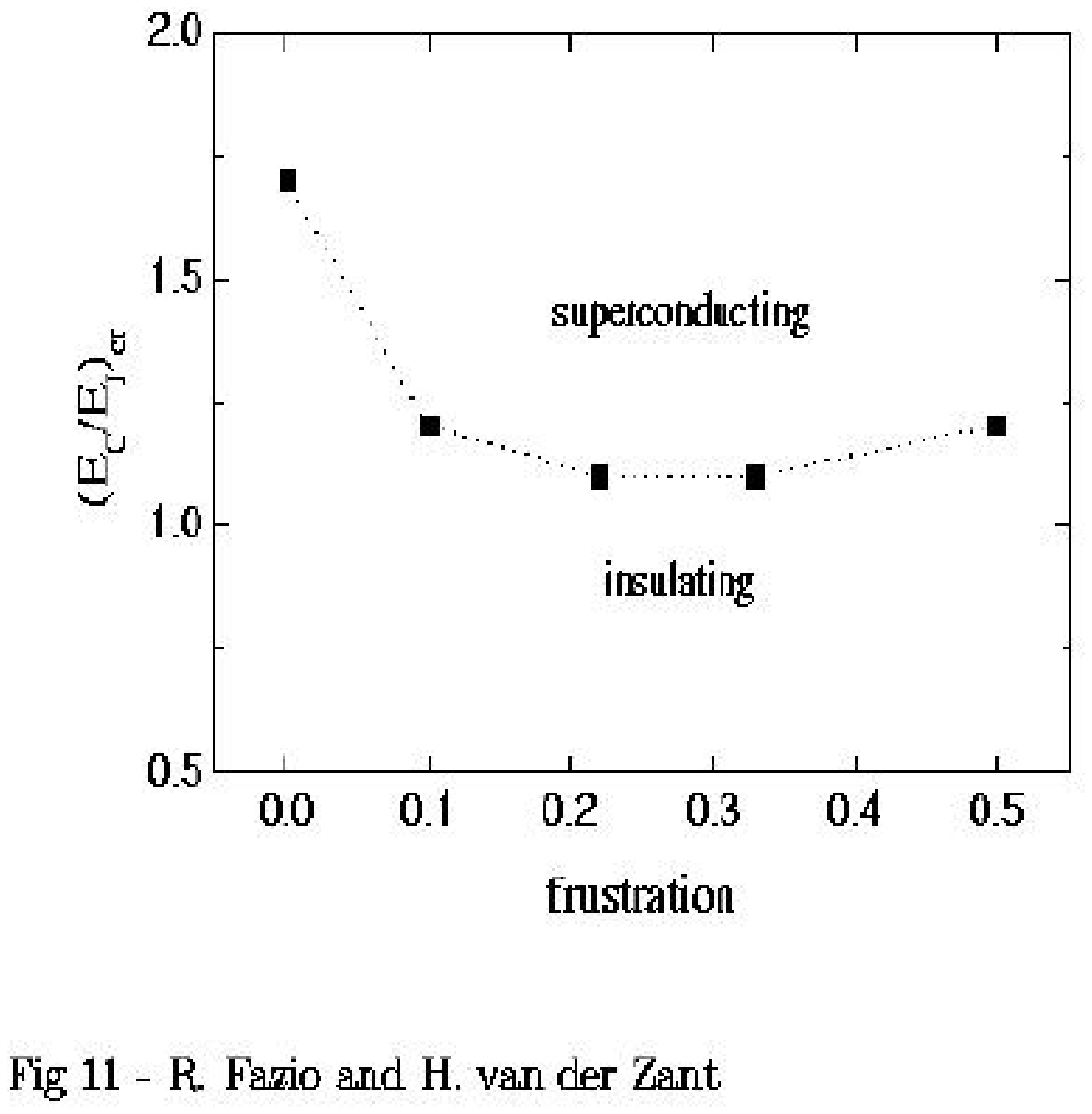}}}
\caption{Measured phase diagram for square arrays in a magnetic field.
	Below the dotted line samples become superconducting 	
	at low temperatures; above this line samples become insulating. At 
	non-commensurate magnetic fields, the S-I transition is not sharp and
	there is an additional (intermediate) metallic region not shown 
	in the figure. (From Ref.\protect\cite{zant96}.)}
\label{ecej_crit}
\end{figure}

In Fig.\ref{ecej_crit}, the critical $E_C/E_J$ 
ratio as a function of applied magnetic field for square arrays is plotted.
The three points at $f=0, \; 1/2$, and $1/3$ are combined with two data 
points of 
the field-tuned S-I transition (see Section~\ref{fieldtsection}). 
After a rapid decrease the critical ratio is almost constant for $f>0.1$. 
The critical $E_C/E_J$ ratio at $f=1/2$ is larger than at other nearby 
values of $f$, indicating once again the stability of the phase 
configuration at $f=1/2$. Fig.\ref{ecej_crit} also indicates that arrays in 
the range $1.2<E_C/E_J<1.7$ do not show special behavior at commensurate 
$f$-values (e.g. dips in the magnetoresistance); arrays are superconducting
in zero field but insulating at $f=1/2, \; 1/3, \; 1/4, ..$.

It is possible to derive a Ginzburg-Landau free energy also in the presence
of rational frustration~\cite{choi85}. The calculation proceeds along the lines outlined 
for the $f=0$ case in Section~\ref{coarsegrain}.  The important difference is 
that one should expand the free energy about the most fluctuating modes. 
In the unfrustrated case this means an expansion about ${\bf k}=(0,0)$. For the 
fully frustrated case ($f=1/2$) the expansion is carried around the two 
points: ${\bf k}=(0,0)$ and ${\bf k}=(0,\pi)$ thereby reflecting the superlattice 
structure of the ground state. The resulting free energy depends on a
multicomponent (complex) order parameter (e.g. two coupled complex fields
in the fully frustrated case).

The magnetic-field dependence of the critical exponent 
$z\nu$ ($\nu$ governs the divergence of the correlation length and 
$z$ is the dynamical critical exponent) was considered by Niemeyer 
{\em et al.}~\cite{niemeyer99}.
In zero field the mapping onto a three-dimensional XY model implies 
that $z\nu=0.67$. Their analysis hints to a dynamical exponent that increases 
with the magnetic field.
It is, however, difficult draw conclusions on the values of $z\nu$.
It could smoothly increase with magnetic field,
or immediately jump to one on once the  magnetic field has switched on.
Combining the fact that $z\nu<1$ for $f=1/2$ 
and that a  higher-order expansion predicts $z\nu=1$ there, the authors 
conclude that the answer is $z\nu=1$ for all nonzero 
magnetic fields. Monte Carlo simulations by Cha and Girvin~\cite{cha93} 
obtain for the $f=1/2$ and $f=1/3$ cases the values $z=1$ and $1/\nu =1.5$,
consistent with the analysis outlined above.

Finally we mention that, in addition to two-dimensional arrays, frustration 
effects can be studied in quantum ladders as proposed in 
Refs.~\cite{granato93-94}.

\subsection{Charge frustration and the supersolid}
\label{supersolid}
A uniform charge can be introduced in a quantum JJA by applying a 
gate voltage $V_{\rm x}$ with respect to the ground plane. This effect is 
known as {\em  charge frustration}. Although from a theoretical point 
of view charge and magnetic frustration are dual to each other, 
experimentally it 
is only possible to tune the magnetic frustration in a controlled way. 
In all arrays random offset charges, presumably caused by defects in the junctions 
or in the substrate~\cite{kruperin00}, are present. 
Electron or quasi-particle tunneling 
will partly compensate these offset charges so that their value lies
between $-e/2$ and $+e/2$. These charges can, in principle, be nulled out by
the use of a gate for each island; this procedure, however, works only 
for small networks. In large arrays they cannot be compensated 
because too many gate electrodes would be necessary, requiring too complicated 
fabrication procedures. A uniform charge frustration 
has therefore not been realized yet in 2D Josephson arrays. 
Lafarge {\em et al.}~\cite{lafarge95} have 
investigated charge frustration by placing a gate underneath a Josephson array. 
They managed 
to obtain a $40 \%$ variation of the resistance between the unfrustrated and 
the (nominally) fully frustrated array. But most importantly, in studying the 
current-voltage characteristics, it was impossible to quench the Coulomb 
blockade as it can be done in circuits with few junctions. In future arrays, 
charge frustration may be applied more uniformly if the influence of offset 
charges can be drastically lowered.
 
\subsubsection{Phase diagram}

The energy difference of two charge states in each island with $q$ and 
$q+1$ extra Cooper pairs may be reduced  by changing $V_{\rm x}$ (which means 
to change the external charge $q_{\rm x}$). Consequently the effects of a 
finite charging energy 
are weakened and the superconducting region in the phase diagram is enlarged. 
It turns out that for certain values of the gate voltages the energy 
difference vanishes implying that the Mott gap, and therefore the 
insulating behavior, is 
completely frustrated. At the degeneracy point even a small Josephson coupling 
makes the system superconducting since there is no pay in energy for moving a 
charge through the whole array. In general it is intuitive to expect that 
the extension of the insulating lobe will be maximum at integer values 
of the external charge since in this case the excitation energy is highest. 

A quantitative analysis of this phenomenon has only been obtained in models
with a short-range electrostatic potential.
Uniform charge frustration gives rise to two new effects: 
\begin{itemize}
	\item lobe structures appear in the phase diagrams 
	\item new states in the phase diagram may occur 
	      (Wigner-like  crystals and the Supersolid).
\end{itemize}
The remainder of this subsection is devoted to the lobe structures and the 
Wigner-like charge ordering; the next subsection treats the supersolid. 

\begin{figure}
\centerline{{\epsfxsize=14cm\epsfysize=16cm\epsfbox{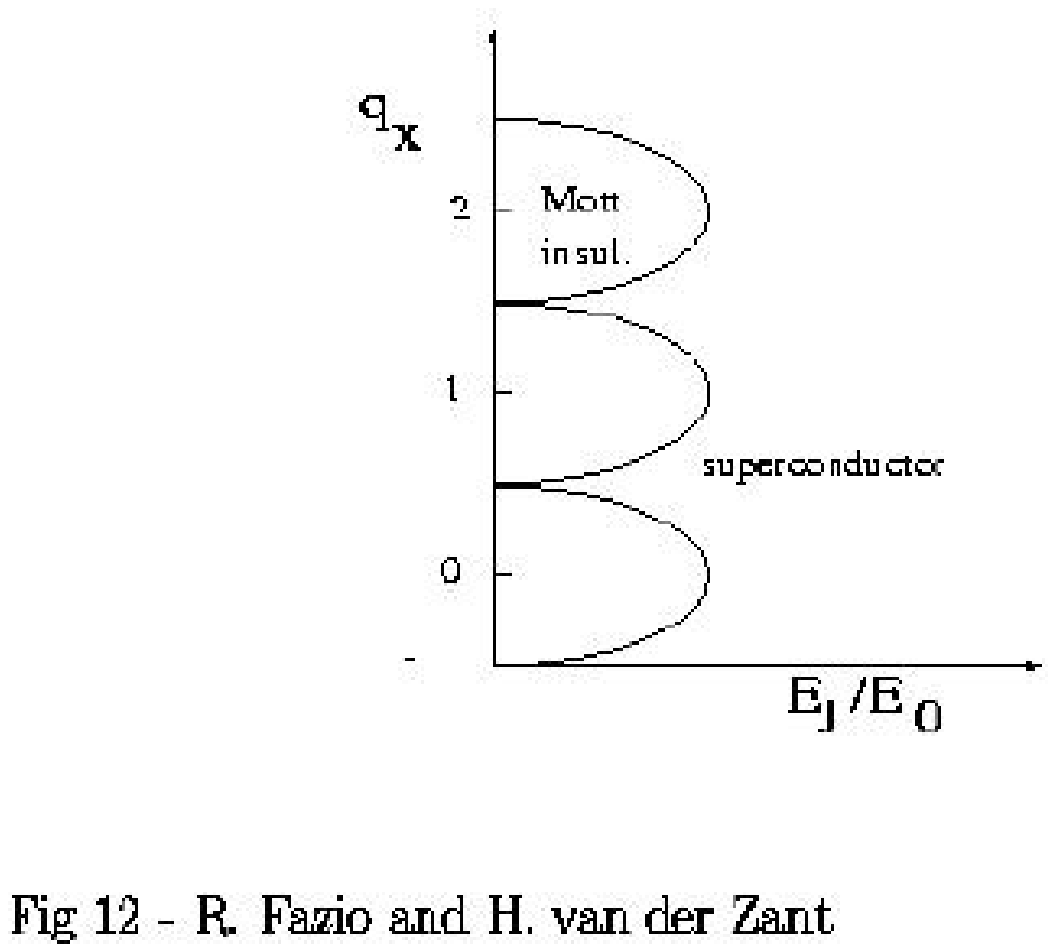}}}
\caption{The $T=0$ phase diagram in the limit of on-site interaction as a 
	function of the charge frustration. At the values of $q_{x}$ for 
	which two charge states are degenerate, the superconducting phase 
	extends to arbitrary small Josephson coupling.}
\label{lobes}
\end{figure}

The lobe-structure already follows from a mean-field analysis with on-site
interaction only. The corresponding phase diagram can be obtained by 
evaluation of the correlator given in Eq.(\ref{selfc2}) in the presence of 
an uniform charge. In Fig.\ref{lobes}, the mean-field phase boundary 
in the presence of a uniform background 
charge is shown. The detailed structure of the lobes is very sensitive to 
the model used (QPM, BH, XXZ) and on the approximation made 
~\cite{fisher89b,freericks94,batrouni90,scalettar91,bruder92,amico98,grignani00}. 
The lobes in the Monte Carlo calculation are sharper 
than predicted by mean-field, but smoother than expected from the strong 
coupling expansion. 

In the case of a finite-range interaction a number of new
classical ground states exists characterized by a crystal like ordering of
charges. The phase diagram contains extra lobes
where the charge density is pinned to a given fractional filling.
Their range of existence, in the limit of vanishing Josephson coupling, 
is found by minimizing the charging part of the Hamiltonian for a 
given charge filling. We illustrate this by considering
the simple case of short range charging energy (only on-site and next-neighbors).
In the case of $ 0 \ge q_x \ge 1/2$ (and for square lattices) only three 
different charge configuration should be considered. 
\begin{itemize}
\item All the islands are neutral 
\item A checkerboard state can be formed in which a sublattice is neutral and 
	the other is charged with one extra Cooper pair 
\item All the islands can be uniformly charged with charge $2e$. 
\end{itemize}
The corresponding energies of the different ground states are respectively
\begin{eqnarray}
E_{ch,00}/4N & = &  E_0q_x^2 +\frac{z}{2}E_1 q_x^2\\ 
E_{ch,01}/4N & = &  \frac{1}{2}E_0q_x^2 + \frac{1}{2}E_0(1- q_x)^2 - 
	\frac{z}{2}E_1q_x(1-q_x)\\ 
E_{ch,11}/4N & = & E_0 (1- q_x)^2 +\frac{z}{2}E_1(1- q_x)^2
\end{eqnarray}
where $N$ is the number of islands in the array.
The ground state energy is given by 
$E_{ch,00}$ for $0 \ge q_x \ge q_{x,1}$, $E_{ch,01}$ for 
$q_{x,1} \ge q_x \ge q_{x,2}$ 
and $E_{ch,00}$ for larger  $q_x$.
The critical values at which the ground state changes are 
\begin{eqnarray}
	q_{x,1}  &=&  \frac{1}{2} \frac{1}{1+(z/2)E_1/E_0}           \nonumber \\
	q_{x,2}  &=&  \frac{1}{2} \frac{1+zE_1/E_0}{1+(z/2)E_1/E_0}  \nonumber \; .
\end{eqnarray}

\begin{figure}
\centerline{{\epsfxsize=14cm\epsfysize=16cm\epsfbox{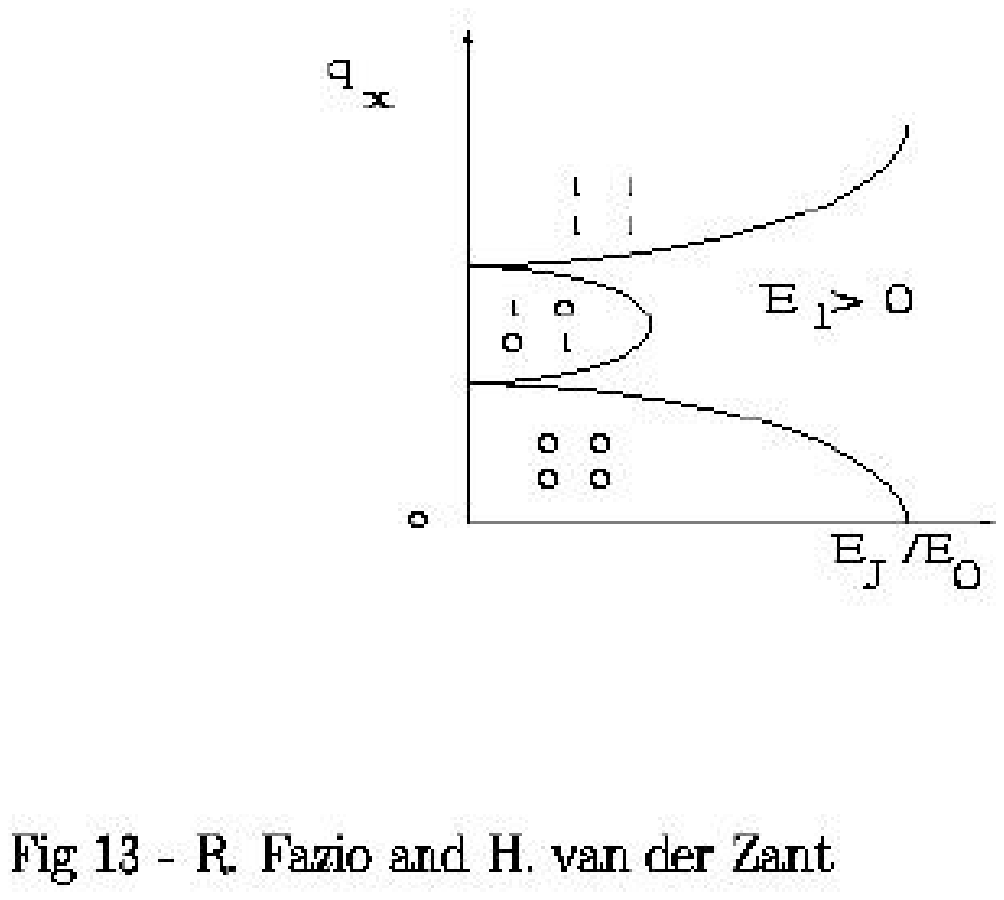}}}
\caption{ The $T=0$ phase diagram calculated with on-site interaction and a small nearest 
	neighbor charging term $E_1$. 
	Around $q_{x}=1/2$ the half-integer lobe appears. 
	Inside the lobes, each number represents the number of Cooper pairs on a particular 
	island. For example, the intermediate lobe, centered around $q_x=1/2$ has a 
	checkerboard structure.}
\label{lobes2}
\end{figure}

The S-I boundary (for finite $E_J$) can be determined, for example, using a 
mean-field approach. 
The result is presented in Fig.\ref{lobes2}. The checkerboard state for 
$q_{x,1} \ge q_x \ge q_{x,2}$ can be thought of 
as a Wigner crystal of Cooper pairs.
The role of a longer-range charging energy (next nearest neighbors,...) is to 
stabilize the crystalline phases with lower fillings ($1/4,1/8,...$). Not all the 
lobes extend down to the $E_J=0$ axis. First-order phase transitions between 
different checkerboard states are then possible.
Note, that the presence of charge ordering, characterized by a periodicity
$2\pi/k_x,2\pi/k_y$, is detected by studying the structure factor $S$ at a 
give wave-vector 
\begin{equation}
	S(k_x,k_y)=\frac{1}{L^{4}}\Big\langle\sum_{ij}e^{i{\bf k}\cdot {\bf r}_i}
	q_{i}q_{0}\Big\rangle\;.
\end{equation}
 
A uniform charge frustration changes the symmetry properties of the system. 
At $k_x=0$ the energy cost to create (or remove) a Cooper pair in a given 
island is the same. The system possesses particle-hole symmetry. For generic values 
of the external charge this symmetry is broken (i.e., $E_{ch,00} \neq E_{ch,11}$). 
In the phase diagram shown in Fig.\ref{lobes2} the tips of the lobes correspond 
to a particle-hole symmetric case while away from the tips the symmetry is broken.
This change of symmetry is reflected as a new term in the quadratic part of the 
Ginzburg-Landau free energy. This new contribution is 
\begin{equation}
	\lambda \sum_{i} \int \;d\tau \psi^*_i(\tau) 
	\partial _{\tau}  \psi_i(\tau)
\end{equation}
where
\begin{equation}
	\lambda = i \frac{\partial_{\omega_n}g(\omega_n, q_x)
	\mid_{\omega_n=0}}{2E_0} \;\;\; . 
\end{equation}
The coefficient $\lambda$ vanishes
in the particle-hole symmetric case (see Appendix~\ref{phase-correlator}).
The particle-hole symmetry has important consequences for the critical behavior of 
the system~\cite{fisher89b}.
The dynamical critical exponent $z$ changes from $z=1$ at the tip of the lobes to 
$z=2$ in the generic case.

\subsubsection{Supersolid}

A solid phase is characterized by charges being pinned on the islands whereas
a superfluid phase is characterized by phase coherence over the whole system 
(i.e.,charges are delocalized). At the end of 
60's~\cite{andreev69,leggett70,matsuda70,liu73} 
it was suggested that, in addition to the solid and superfluid phases, a new 
state should appear, characterized by the coexistence of off-diagonal 
(superfluid) and  diagonal (charge-crystalline) long-range order. 
This phase is known as the {\em supersolid}. If vacancies in a quantum 
crystal such as solid $^4$He Bose-Einstein condense, they do not necessarily 
destroy the crystal structure and they may form a superfluid solid
(or supersolid). 
Experiments have been performed on $^{4}$He, but no positive
identification of this coexistence phase has yet been made. There are,
however, hints that such a phase exists~\cite{lengua90,meisel92}. 

An exciting possibility that attracted a lot of attention was the idea to observe 
supersolids in Josephson arrays
~\cite{bruder93,roddick93,otterlo94a,otterlo95,batrouni95,frey97}. 
The supersolid phase is located in an intermediate region 
around the half-filling lobe. A simple way to understand its existence is 
to focus on a region close to the phase boundary at $q_x \sim 1/2 $.
At densities corresponding to half filling the particles form an incompressible solid. 
Away from half-filling vacancies in the charge-solid appear. As they have a 
bosonic character, they can Bose condense, and therefore they are able to move freely 
through the system. For a small enough density of vacancies one expects that the 
crystal order is not destroyed.

\begin{figure}
\centerline{{\epsfxsize=16cm\epsfysize=16cm\epsfbox{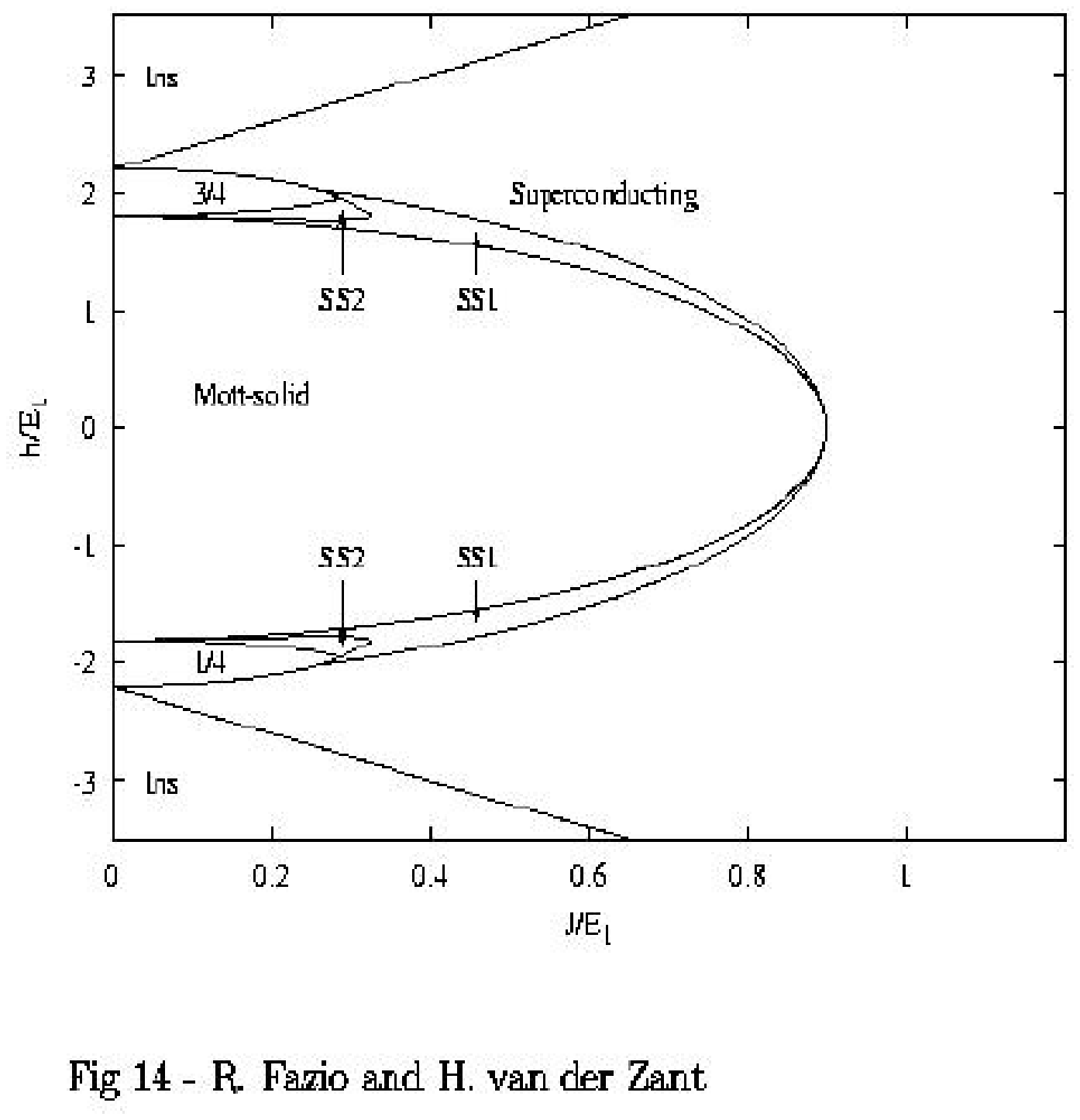}}}
\caption{Mean-field phase diagram in the presence of charge frustration 
	obtained in the hard-core limit. As discussed in the text the 
	fictitious field $h$ is related to $q_x - 1/2$ and the coupling
	$J$ corresponds to $E_J$. In the figure nn and nnn charging terms 
	are different from zero $E_2=0.1E_1$. The central lobe 
	corresponds to the half-filling case and the two small ones 
	on the side are the quarter-filling lobes. Finally SS1 and SS2 
	are different types of supersolid. 
	(From Ref.~\protect\cite{bruder93}.)}
\label{supsol1}
\end{figure}

In the limit of very large on-site charging (hard-core limit in which the 
island charge can only be zero or one)
the existence of the supersolid is related to the finite
next-nearest neighbor interaction as it does not exist for
nearest-neighbor interaction only. Furthermore, there is
no supersolid phase at exactly half-filling.
In Fig.\ref{supsol1} the mean-field phase diagram in the hard-core 
limit is shown. The supersolid region appears in a tiny region away from half filling 
between the superconducting and Mott insulating phases. 

\begin{figure}
\centerline{{\epsfxsize=16cm\epsfysize=18cm\epsfbox{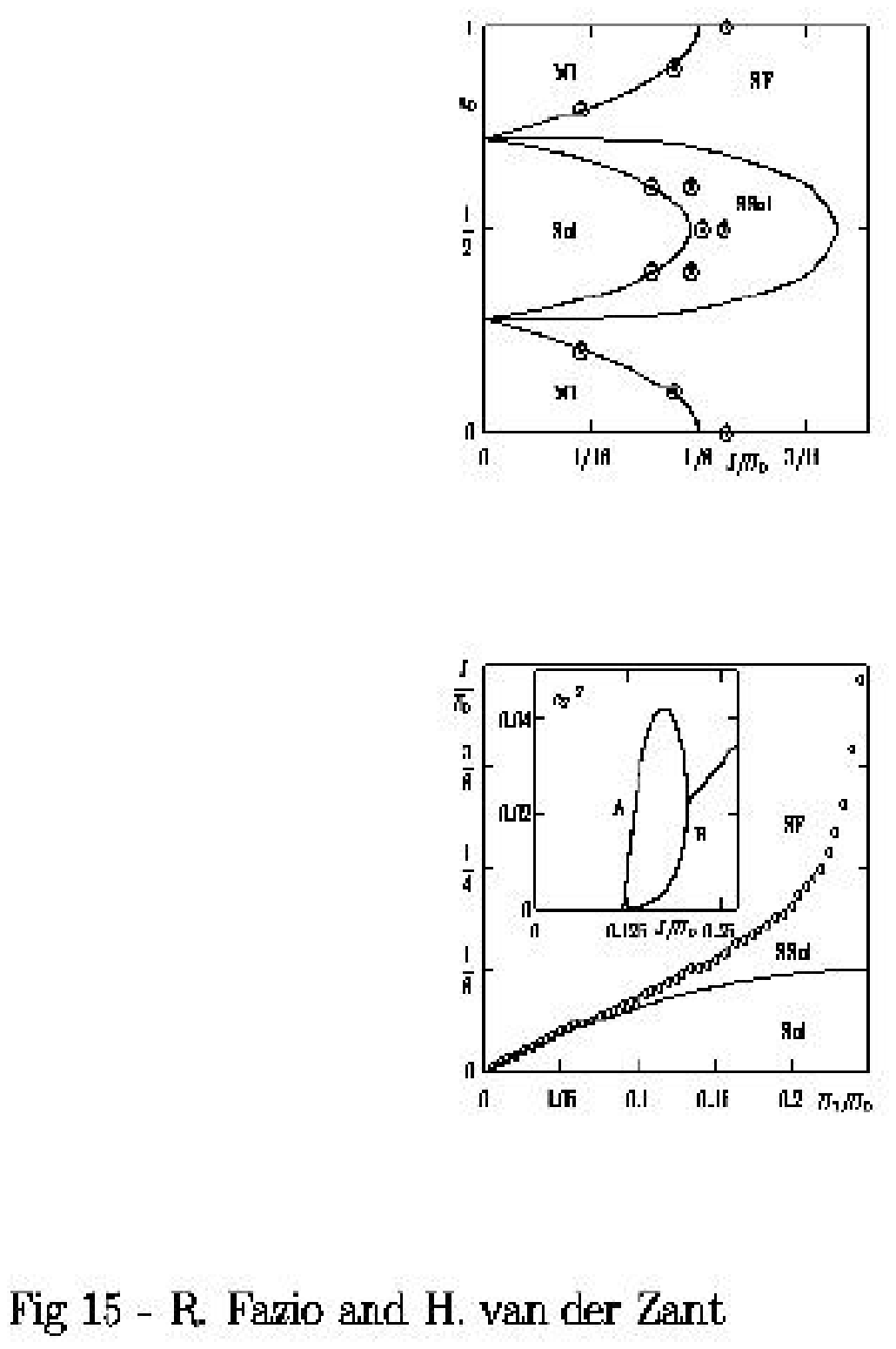}}}
\caption{Top: Mean-field phase diagram for soft-core bosons, as obtained from the 
	analysis of the QPM with on-site  and nearest-neighbor ($E_{1}/E_{0}=1/5$) interaction. 
	The symbols are the Monte Carlo data (the region between the dotted circles and the 
	crossed circles is the supersolid). 
	The checkerboard charge-density wave is denoted by ``Sol'', the supersolid phase 
	by ''SSol'', the superfluid phase is denoted by ``SF'' and the Mott-insulating 
	phase by ``MI''. Bottom: Supersolid region ``SSol'' at $q_{x}$= 0.5 as a function 
	of $U_1/U_0$ in the mean-field approximation of. 
	Inset: Occupation-number probability $|c_{2}|^{2}$ at $q_{x}$= 0.5 for the two 
	sub-lattices $A$ and $B$ at the particular value of $E_1/E_0=0.2$.
	(The notation is slightly different from that used in this review, 
	$U_{0} \rightarrow E_{0}$, $U_{1} \rightarrow E_{1}$, $J \rightarrow E_{J}$, 
	$n_{0} \rightarrow q_{x}$). (From Ref.~\protect\cite{otterlo95}.)}
\label{supsol2}
\end{figure}

If higher values of charge are allowed, the supersolid phase already 
exists for nearest-neighbor interaction and also at half-filling on the tip of 
the checkerboard lobe. This is related to excitations which are forbidden
in the hard-core limit. A large nearest-neighbor interaction
or small on-site interaction  favors the supersolid, whereas in the
hard-core limit the supersolid is suppressed. 
Thus, it seems that the system itself generates the defects
(particle-hole excitations) that Bose condense, thereby turning the solid into
the supersolid. The phase diagram for soft-core bosons is shown in 
Fig.\ref{supsol2}.

Since the supersolid phase is very sensitive to fluctuations, it was 
important to obtain independent checks of its existence. Monte Carlo 
simulations on the QPM~\cite{otterlo94a} and the BH model~\cite{batrouni95} 
have confirmed the qualitative picture discussed above. In Fig.\ref{supsol2} 
the symbols represent the phase diagram as obtained from Monte Carlo 
simulations by van Otterlo and Wagenblast~\cite{roddick93,otterlo94a}.
Note that the supersolid region is considerably reduced as compared to 
the mean-field estimates.

By changing the electrostatic interaction new phases named collinear supersolid 
were found by Scalettar {\em et al.}~\cite{batrouni95} and by Frey and 
Balents~\cite{frey97}. A detailed analysis of various supersolids including 
striped phases has recently been performed by Pich and Frey~\cite{pich98}. 
Supersolid phases in frustrated systems have been studied as well either by 
combining the effect of charge and magnetic 
frustration~\cite{amico97} or by considering arrays on Kagom\`e 
lattices~\cite{murthy97}.

Several other kinds of coexistence phases were studied.
The possibility of a spontaneous vortex anti-vortex lattice in
superfluid films was explored in Refs.~\cite{gabay93,zhang93} and a coexistence
phase of superfluidity and hexatic orientational order was proposed
in Ref.~\cite{mullen94}. Orientational order in incompressible quantum Hall fluids
is discussed in Ref.~\cite{balents95}. 
Finally, we mention the relation between 2D bosons
and 3D flux-lines in type II superconductors (high-$T_{c}$ materials)
in a magnetic field~\cite{nelson88,feigelman93}. Also in these
systems different kinds of long-range order may coexist and the equivalent of
the supersolid is discussed in Refs.~\cite{blatter94,frey94}.
Related is the question whether or not vortices may form a disentangled
liquid, which would imply a normal ground state for bosons with
long-range Coulomb interaction.

\subsection{Dissipation induced S-I transition}
\label{dissip}
The behavior of a  single Josephson junction with Ohmic dissipation has 
been discussed in a pioneering work by A. Schmid~\cite{schmid83} who 
found that there is a zero-temperature phase transition 
governed by the dissipation strength $\alpha$. Above the critical 
value  $\alpha=1$, dissipation suppresses quantum fluctuations thereby restoring
the classical behavior with a finite critical current. For weak
damping, on the other hand, quantum fluctuations destroy
global phase coherence. The supercurrent is suppressed to zero and
the junction is in the insulating state. 

Experimentally, this transition has only very recently been detected by Penttila 
{\em et al.}~\cite{penttila99}. The reason is that for a single junction the 
high-frequency coupling to the environment determines the effective damping. 
Consequently, the effective impedance is of the order of 100~$\Omega$~\cite{},
which is about two order of magnitude smaller than the 
quantum resistance. Penttila {\em et al.} increased this impedance, i.e., the 
decoupled their single junction from its environment by placing high-Ohmic, 
Chromium resistors in the leads close by. 

Dissipation plays an important role in quantum phase transitions of JJAs as 
well. Originally the interest was stimulated by the idea that dissipation could be
responsible for the observed critical resistance at the S-I transition in arrays and 
granular films. Later Fisher~\cite{Fisher90b} and Wen and Zee~\cite{Wen90}
pointed out that the observed critical resistance is a zero temperature 
property associated with the existence of a quantum phase transition and
it is not related to the presence of an ``extrinsic'' source of dissipation. 
The next section discusses the critical behavior on transport properties in more details.
Here, we discuss the influence of dissipation on the phase diagram.
 
The coupling to a dissipative bath has the effect to suppress the quantum 
fluctuations of the phase, i.e.  to quench the insulating region. 
The properties 
of the environment and the type of dissipation are important ingredients in 
Eq.(\ref{qpdissi}) and in Eq.(\ref{ohmdissi}). 
As stated before, various sources of dissipation should be considered for JJAs.
Quasi-particles may still play a role at mK as they may be generated by the 
environment or the motion of vortices themselves. From a theoretical point of view 
and in view of the recent experimental advances,
it is also possible to realize arrays in which Ohmic shunts are important. 
These shunts can be normal wires placed parallel to the junctions in a
a similar way as was done for a single junction. Ohmic shunts may also
arise because of coupling to the substrate (local damping). 
Very recently  the group of Kobayashi succeeded in fabricating a JJA in which 
each junction is shunted by a Cr resistor~\cite{takahide00}.
A different and controlled environment was investigated in the
experiments of Rimberg {\em et al.}~\cite{rimberg97} by placing a 2D electron
gas underneath the Josephson array. We briefly discuss all these 
sources of dissipation.

\subsubsection{Quasi-particle dissipation}

When the mechanism responsible for dissipation is quasi-particle 
Tunneling, the effective action is that given in Eq.(\ref{qpdissi}). 
Theoretically this  model was studied by means of mean-field 
calculation~\cite{chakravarty86,ferrell88},
variational approaches~\cite{kampf88a,kampf87,simanek88,falci91}, and Monte Carlo 
simulations~\cite{choi89}. The dominant effect coming from quasi-particle tunneling 
enters in the renormalization of the effective junction capacitance given in 
Eq.(\ref{capren}). In the mean-field calculation, this amounts to a modification of the 
capacitance matrix in the evaluation of the phase correlator. 
The zero-temperature phase boundary (for short range Coulomb interaction) obtained by 
Chakravarty {\em et al.}~\cite{chakravarty86}, is given 
by the expression
$$
	1=\frac{3\pi}{4} \alpha ^2
	\frac{1}{\ln \left( 1 +  \frac{3\pi}{4}\frac{E_0}{E_J}\alpha ^2
		\right)}	\;\;\;\; .
$$
It is important to stress that in the 
case of quasi-particle tunneling the array is either insulating or superconducting. 
The interplay between the long-range Coulomb interaction and quasi-particle 
dissipation has been discussed in Ref.~\cite{fazio91a}.

\begin{figure}
\centerline{{\epsfxsize=14cm\epsfysize=16cm\epsfbox{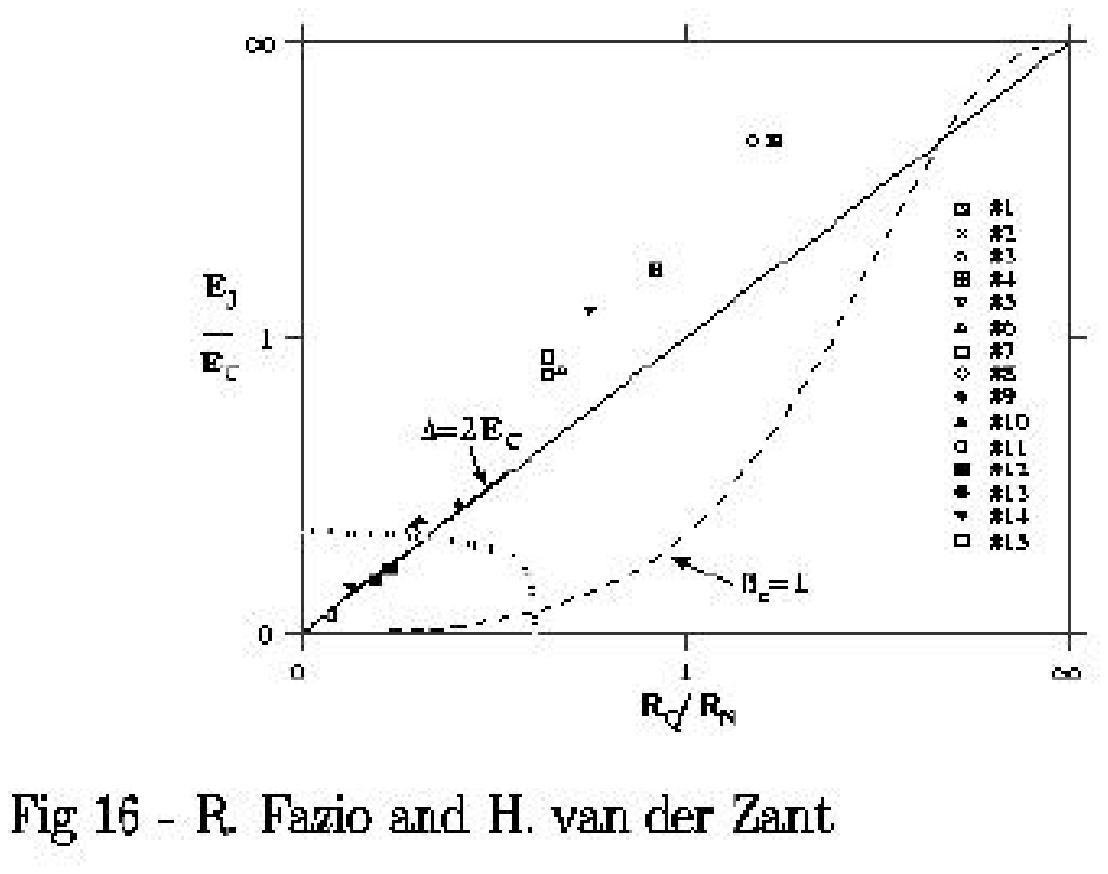}}}
\caption{Measured low-temperature phase diagram for fifteen square arrays plotted as the 
$E_J/E_C$-ratio vs. $R_Q/R_N$. 
Samples $\sharp ~1-8$ show a decreasing resistance as temperature is lowered 
(superconducting arrays). 
Samples $\sharp ~9-15$ show an increasing resistance as temperature is lowered 
(insulating arrays).
This classification roughly agrees with the theoretical prediction~\protect\cite{fazio91a} 
which states that insulating arrays are to be found in the region bounded by the dotted line. 
Only the insulating samples $\sharp ~9-11$ fall slightly outside this area.
The diagonal line represents $\Delta _0=2E_C$ and the dashed line corresponds to a 
Stewart McCumber parameter of $\beta _{c}=1$.
(From Ref.~\protect\cite{delsing97}.)}
\label{chalmers}
\end{figure}

In Section~\ref{phase0}, we have interpreted the S-I transition as being
driven solely by Coulomb interactions. However, given the uncertainty in
the damping resistance (e.g. sub-gap resistance of normal-state resistance) of 
the junctions the possibility that the transition 
is driven by quasi-particle dissipation cannot be ruled
out. The data do not exclude the possibility that
the S-I transition is influenced by the normal-state resistance.
In fact, the Chalmers group~\cite{delsing97} and the  
group of S. Kobayashi and collaborators~\cite{yagi96} have interpreted
their data in terms of a Schmid-like diagram. In Fig.\ref{chalmers} 
we show the results from the Chalmers group. The normal state resistance 
$R_N$ is used as the resistance determining the dissipation parameter.
If this resistance is used, a reasonable agreement with the theoretical 
models is obtained.

\subsubsection{Ohmic dissipation}

The influence of Ohmic shunts on the phase diagram has been intensively 
investigated as well. A new phase with local
phase coherence is possible, i.e., phase coherence only exists as a function of time. Various theoretical methods have been applied in
this case such as coarse graining~\cite{zwerger88,panyukov89,kampf88b},
variational~\cite{panyukov89,chakravarty86b,chakravarty88,falci91b,cuccoli00}
and renormalization group~\cite{panyukov89,chakravarty88,fisher87}
approaches. The general trend is, as expected, that the critical value of 
$E_J/E_0$ for the onset of phase coherence is lowered. The dependence on the 
dissipation strength is stronger as compared with the case of quasi-particle 
damping. As for a single junction, a true dissipative transition 
occurs. A rigorous analysis can be performed in various 
limits in the $E_J - \alpha$ phase diagram ~\cite{panyukov89,chakravarty88}.
For simplicity we discuss only the $T=0$ case and follow the discussion presented 
in Ref.~\cite{zimanyi88}. 
\begin{itemize}
\item In the large $\alpha$ limit 
	time-like fluctuations of the phase are strongly suppressed and they only 
	contribute to the renormalization of the effective Josephson 
	coupling. The system behaves like a classical JJA with an effective 
	Josephson coupling
$$
	E_J^{eff} = E_J \left( 1 - \frac{1}{\alpha {\rm z}} 
		    	\ln \left( 1 +  \frac{\alpha}{2\pi} \right) \right) 
			\;\;\; .		
$$
	At zero temperature the array is in the superconducting phase independent on 
	the ratio $E_J/E_0$.
\item In the case $E_J/E_0 \gg 1$ there is a phase transition at  
$$
	\alpha  = \frac{1}{{\rm z}} 
$$
which separates two phases that both exhibit long-range coherence.        
Evidence of such a phase transition could be detected by measuring 
	the voltage noise power spectrum.
\item In the limit of small damping, the critical ratio of $E_J/E_0$ is 
	renormalized to smaller values indicating that dissipation 
	enhances the superfluid phase.	 
\item If the ratio  $E_J/E_0$ is very small, a dissipative transition to 
	a phase with local order can take place at a critical value 
	of dissipation given by
$$
	\alpha  = \frac{2}{{\rm z}}  \;\;\;\; . 
$$
\end{itemize}

Very recently Takahide {\em et al.}~\cite{takahide00} fabricated a JJA in 
which each junction was shunted by a Cr resistor. By varying the resistance of 
the shunts and the ratio $E_J/E_0$ they were able to map out the phase diagram. 
The results are in good agreement with the theories of dissipation induced 
quantum phase transition discussed above.

\subsubsection{Local damping}

Dissipation may also arise from coupling with the substrate
by means of what is named as the `local damping' model.
Local damping changes the universality class of the S-I
transition~\cite{wagenblast98}, and influences the low-frequency 
dispersion of the vortex response in classical arrays
\cite{beck94,korshunov94}.
As shown in Eq.(\ref{loc})), the local damping model correlates the island phase 
in time. In proximity-coupled arrays, which consist of superconducting islands on 
top of a metallic film, the model with local-damping is appropriate to 
describe the flow of normal electrons into the substrate. 
Aluminum tunnel junction arrays are always placed on insulating substrates 
so that it is not appropriate.

\begin{figure}
\centerline{{\epsfxsize=14cm\epsfysize=16cm\epsfbox{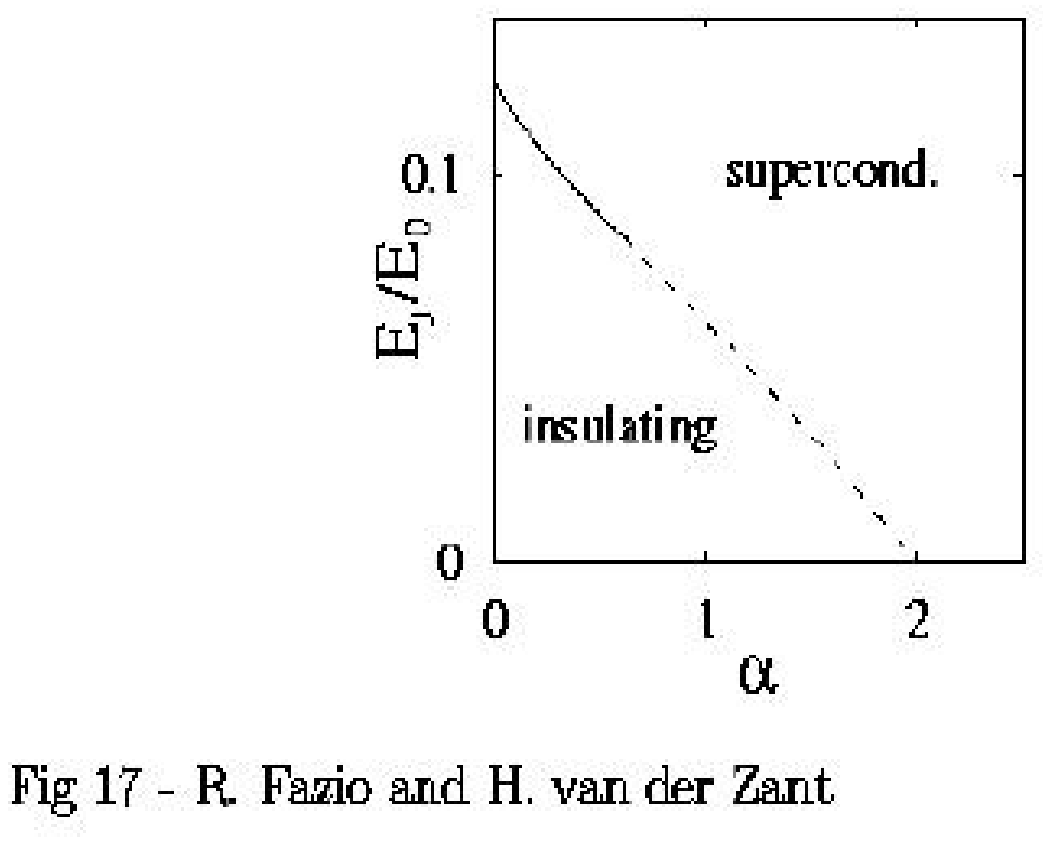}}}
\caption{The phase diagram at $T=0$ in the local damping model. 
Along the solid line the conductivity is universal, whereas it is a 	
function of the dissipation strength along the dotted line.
(From Ref.~\protect\cite{Wagenblast97}.)}
\label{localdamp}
\end{figure}

Dissipation due to local damping is associated with the phase $\phi_i$, rather with 
the phase difference $\phi_i-\phi_j$ as in the resistively shunted junctions (RSJ) 
model. 
The number of Cooper pairs in each island is allowed to decay, whereas the RSJ model 
describes only charge transfer between neighboring islands.
By going through the same steps outlined in the section on the coarse-graining 
method, it is possible to obtain also in this case an effective
Ginzburg-Landau free energy. The only difference is that now the
phase-phase correlator $g(\tau)$ has to be evaluated including the local
damping term. For small frequencies the Fourier transform reads (for more
details see Ref.~\cite{wagenblast98})
\begin{equation}
g(\omega_{\mu}) = g(0) -\eta\left|\omega_{\mu}\right|^{s}
-\zeta \omega_{\mu}^2\hspace{7mm}\mbox{with }s=\frac{2}{\alpha}-1\;.
\label{locdampg}
\end{equation}
The coefficients $\eta$ and $\zeta$ can be determined from the phase correlator
(their value is not important for our purposes).
Using this expression for $g(\omega_{\mu})$, the free energy
in Eq.(\ref{free}) contains a {\em non-Ohmic} dissipative term $(\propto
\left|\omega_{\mu}\right|^s)$ (reducing to Ohmic, or 'velocity
proportional' damping only in the special case $s=1$). This means that
{\em Ohmic} damping in the quantum phase model yields  
{\em non-Ohmic} dynamics for the coarse-grained order-parameter.
The phase boundary in the saddle point approximation
is shown in the inset of Fig.\ref{localdamp}. 
By increasing damping strength, the superconducting region is enlarged. 
At $T=0$ a quantum phase transition is ruled out beyond the critical 
value $\alpha =2$.

\subsubsection{Tunable dissipative environment}

\begin{figure}
\centerline{{\epsfxsize=14cm\epsfysize=16cm\epsfbox{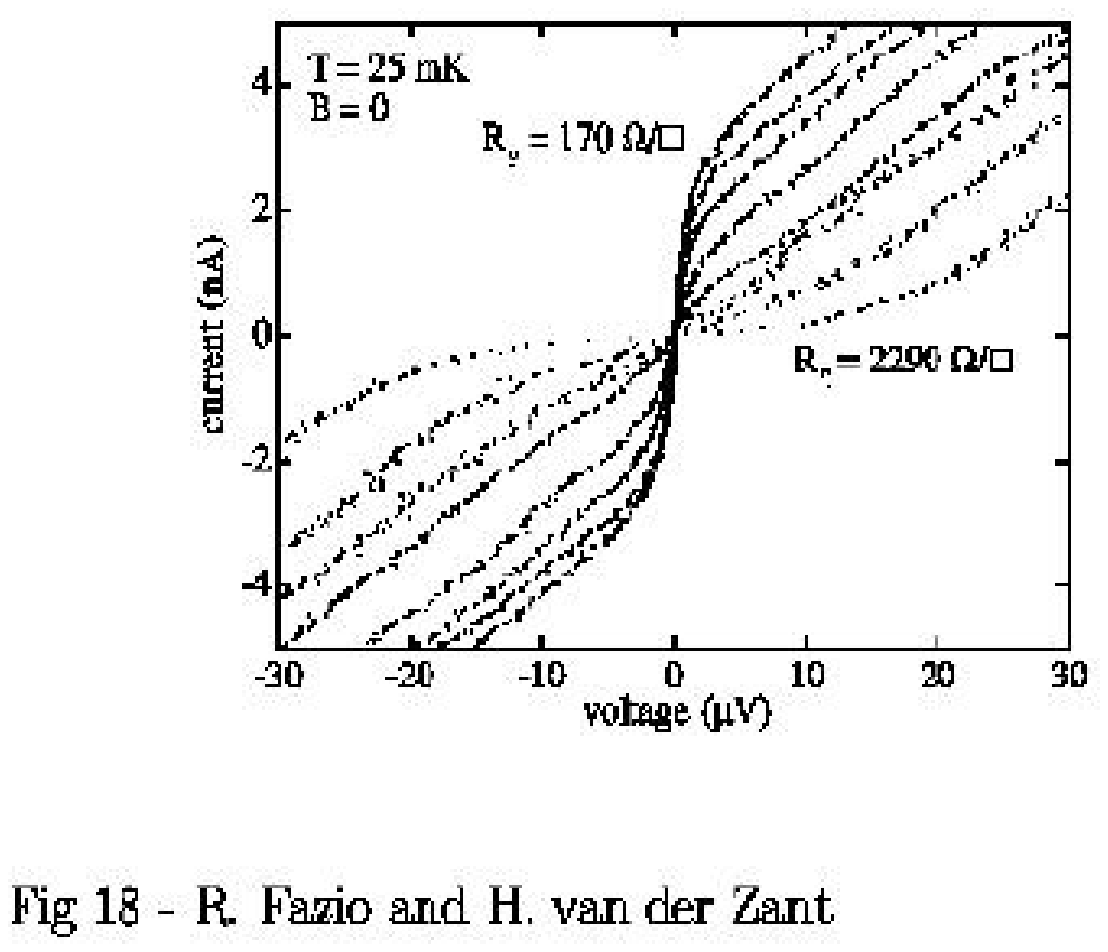}}}
\caption{Measured array $I-V$ characteristics for eight values of the back-gate voltage 
	corresponding to different ground plane resistances $R_g$. 
	The $I-V$.s change from superconducting-like to insulator-like as a function of$R_g$. 
	(From Ref.~\protect\cite{rimberg97}.)}
\label{rimberg2}
\end{figure}

\begin{figure}
\centerline{{\epsfxsize=14cm\epsfysize=16cm\epsfbox{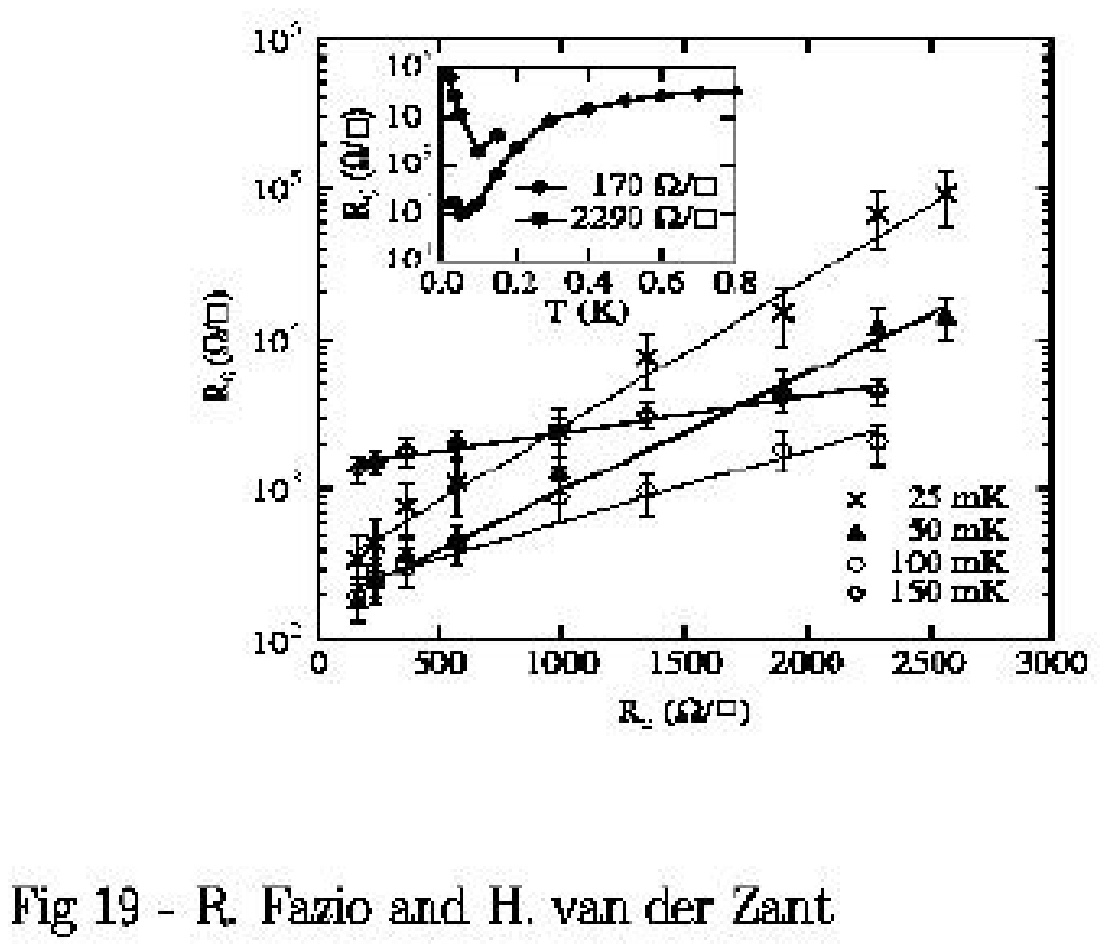}}}
\caption{Measured zero-bias array resistance $R_0$ as a function of the ground plane 
	resistance $R_g$. The inset shows the temperature dependence of the resistance. 
	(From Ref.~\protect\cite{rimberg97}.) }
\label{rimberg3}
\end{figure}

A controlled study of the dissipative S-I transition has
been performed by Rimberg {\em et al.}~\cite{rimberg97}. They placed a 
Josephson array on top of a two-dimensional electron gas (2DEG).
Junction parameters are chosen such that in the absence of
the 2DEG the array is insulating.
The array is capacitively coupled to the electron gas and its screening currents
provide a source for dissipation. 
By tuning the back-gate voltage, the electron density and the sheet resistance
$R_g$ of the 2DEG are varied without changing the array parameters.
As the resistance of the 2DEG increases the current-voltage characteristics
of the array change from superconducting to insulating with a Coulomb gap
as illustrated in Fig.\ref{rimberg2}. Moreover the  resistance of the array 
is very sensitive on $R_g$ as shown in  Fig.\ref{rimberg3}.
Note that in the experiment the island capacitance (to the 2DEG)
exceeds the estimated junction capacitance of 0.5~fF by a factor of 6.

Wagenblast {\em et al.}~\cite{Wagenblast97} analyzed 
these measurement and  modeled the experimental 
setup by capacitively coupling the array to the 2DEG. 
Assuming Ohmic dynamics of the 2DEG they obtained the following 
(Caldeira-Leggett like) effective action for the array 
\begin{equation}
	\nonumber
	S_{\text{eff}}[\varphi]=\frac{1}{2} \sum_{\omega_\mu} \int d {\bf k} D^{-1}_{0}
	(k,\omega_\mu)|\varphi_{k,\omega_\mu}|^2 + S_J
\end{equation}
($S_J$ is the action related to the Josephson coupling) with the propagator 
\begin{eqnarray}
	\label{prop}
	D^{-1}_{0}(k,\omega_\mu)=\frac{C}{4e^2} k^2\omega_\mu^2 
	+\frac{C_0}{4e^2}\frac{k^2\omega_{\mu}^2}{k^2+|\omega_\mu|/\Omega_0} ~~,
\end{eqnarray}
where $C_0$ now represents the capacitive coupling to the 2DEG and  
where $1/\Omega_0=R_{g} C_0$.
The effective action for the array is Ohmic only in an {\it intermediate} 
frequency range.  At
the lowest and highest frequencies the dynamics is capacitive.  The
two energy scales are well separated in the case $C_0\gg C$ and a 
quantum phase transition is driven by the action at
the lowest frequencies.  As the dissipative
action is cut off at the lowest frequencies,  a dissipation driven
transition cannot occur in the strict sense.  However, 
quasi-critical behavior can be observed at temperatures and voltages
exceeding the low energy scale $\Omega_0$.  In the limit
$\Omega_0\rightarrow 0$ ($C_0\rightarrow\infty$) this behavior
converges to a true dissipation-tuned transition.

\begin{figure}
\centerline{{\epsfxsize=14cm\epsfysize=16cm\epsfbox{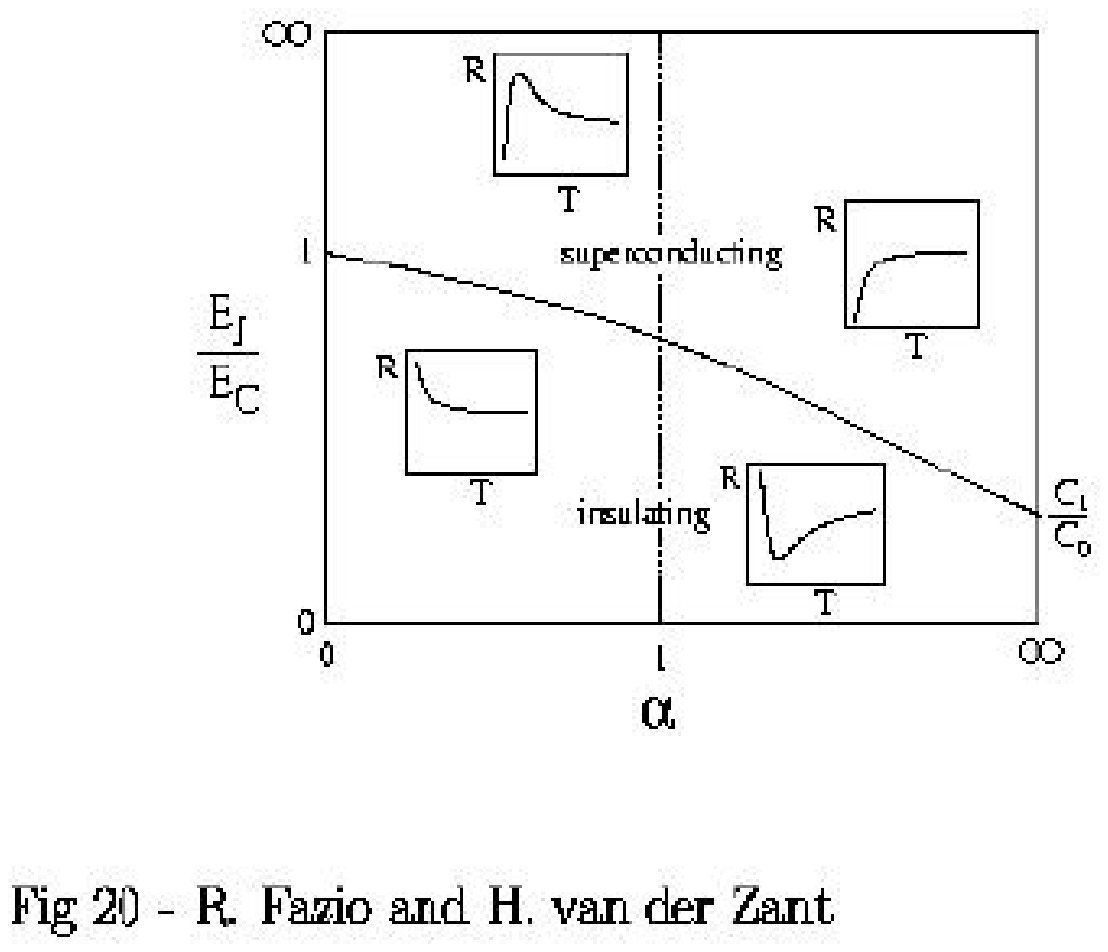}}}
\caption{Calculated phase diagram of an array coupled capacitively to a 2DEG.
	The insets show the resistance as a function of the temperature in the different 
	regions of the phase diagram. (From Ref.~\protect\cite{wagenblast98}.)}
\label{wagenblast2}
\end{figure}

An analysis of the conductivity as a function of $E_J$ and $\alpha$ suggests 
a phase diagram of the type represented in Fig.\ref{wagenblast2}.
The insets show $R_0(T)$ as a function of the temperature in
different regions. The experiments of Rimberg {\em et al.} belong to 
the right-lower sector of the Schmid diagram. 
The theoretical temperature dependence of the resistivity
$R_0(T)$ as well as the exponential
relation between $R_0$ and $R_{g}$ are in good agreement with the 
experiments~\cite{rimberg97}.

\subsection{Transport Properties}
 
The unique feature of quantum critical points is that quantum fluctuations, which 
drive the system through the transition, govern its dynamical behavior.  
The interest in understanding charge transport near a quantum phase transition 
goes beyond the study of JJAs. Important examples are the transition in 
quantized Hall systems~\cite{sondhi97}, localization in Si-MOSFETs~\cite{kravchenko94} 
and  the quantum critical point in cuprates~\cite{castellani95}.

In two dimensions right at the S-I transition 
the zero-temperature conductance~\cite{footnotetr} has been predicted to be finite 
and universal. 
This metallic behavior  is present even in the absence of extrinsic dissipation and 
it is entirely due to the presence of collective modes which become soft at 
the zero-temperature transition point. 
Universality at a quantum phase transition then implies that the properties of the 
system are governed by a set of critical exponents. 
X.G. Wen\cite{Wen92} employed a scaling theory of 
conserved currents at anisotropic critical points identifying universal amplitudes.
One of these amplitudes in two dimensions reduces to the universal 
conductance. 

A very simple argument~\cite{Girvin92} leading to a finite {\em and }
universal conductance at zero temperature can be discussed using the
duality between charges and vortices formulated in Section~\ref{phase0}.
Strictly speaking it applies to the case $C_0=0$, i.e. for logarithmic
interacting charges. From the Josephson relation the voltage drop across
an array is given by the rate of vortices crossing the sample boundary. The
current is given by the number of Cooper pairs which flow through the system
per unit time, i.e., 
$$
	V = \frac{h}{2e} \langle \dot{v} \rangle
	\;\;\;\;\;\;\;\;\;\;\;\;\;\;\;\;\;\;\;\;\;\;\;\;
	I = 2e \langle \dot{q} \rangle 	
$$
At the self-dual point $\langle \dot{v} \rangle = \langle \dot{q} \rangle$
and therefore the conductance at the transition (denoted with $\sigma^{*}$) 
is finite, universal and corresponds to the quantum of resistance for Cooper pairs:
$$
	\sigma^{*} = \frac{4e^2}{h}
$$
The value of $\sigma^*$ changes in the case of short-range charging
and/or  in the presence of disorder. Nevertheless it remains universal (independent 
on the sample parameters).  

A large amount of theoretical work has been devoted to the determination of 
the critical value $\sigma^*$ and the scaling behavior of the conductance. 
The universal conductance in a model with no disorder was considered in
Ref.~\cite{Cha91} by means of a $1/N$ expansion and Monte Carlo
simulations and in Ref.~\cite{Fazio96a} by means of an $\epsilon$-expansion.
The dirty boson system and the transition to the Bose glass phase (including
the case of  long-range Coulomb interaction) was extensively studied by Monte Carlo
simulations~\cite{Sorensen92,Batrouni93,Makivic93,Wallin94} and Lanczos
diagonalization~\cite{Runge92} as well as by using analytic calculations
~\cite{Herbut97}. The finite-frequency properties close to the transition
point were analyzed in Refs.~\cite{Otterlo93,Kampf93,Herbut98}. More recently, 
in a series of papers, Sachdev and coworkers~\cite{Damle97,Damle98,Sachdev98}
studied non-zero temperature transport properties by means of a Boltzmann equation.

A general analysis of the conductivity close to the transition can be performed 
based on scaling arguments.
The frequency dependence of the conductivity $\sigma (\omega)$ has been obtained 
from its relation with the frequency
dependent stiffness $\rho _s(\omega)$ (related to the increase of the free
energy due to a time dependent twist)
$
	\sigma (\omega ) = 4e^2 \rho  _s(-i\omega)/i \omega
$.
Close to the transition it can be shown that the conductance obeys the
scaling form~\cite{Fisher90b,Cha91,Kim91}
\begin{equation}
	\sigma(\delta,T,\omega)  =  
	\frac{h}{4e^2} f(\frac{\omega}{T} \; ,\; \frac{\delta}{T ^{z\nu}})
\label{scaling1}
\end{equation}
where $\delta$ measures the distance from the critical point, and where $f(x,y)$ is a 
dimensionless scaling function. 
In the limit of very low temperature (compared to the frequency), the temperature drops 
out of the previous expression and the two scaling variables enter in the form
\begin{equation}
	\sigma(\delta,T=0,\omega)  =  \frac{h}{4e^2}\tilde{f}(\frac{\delta}
	{\omega ^{z\nu}}).
\label{scaling2}
\end{equation}

In view of the scaling behavior of the conductance, one should consider 
two limits $\omega \ll T$ and $\omega \gg T$. This point, 
emphasized in Refs.~\cite{Damle97,Damle98,Sachdev98}, is important both from a 
conceptual point of view and for a detailed comparison 
with experiments. The two situations correspond to two 
different experimental setups for transport measurements. In the $\omega \ll T$ case, 
charge transport is governed by inelastic scattering between thermally 
excited carriers. In the opposite $\omega \gg T$ situation, collision between 
carriers can be neglected.  In Eq.(\ref{scaling1}) the two 
limiting cases correspond to $f(0,0)$ and $f(\infty,0)$ respectively.  
It turns out that both values are universal but different. 

Here, we discuss the most prominent features of  the conductivity close to the 
S-I transition by means of the Ginzburg-Landau
free energy of Eq.(\ref{free}). Connections to other models will be given. 
In the linear response regime the conductivity follows from the functional 
derivatives of the partition function. In the presence of an electromagnetic 
potential ${\bf A}$, the gradient term in the Ginzburg-Landau free energy enter
in a gauge-invariant form: 
$$
	{\bf \nabla}  \;\; \longrightarrow \;\; {\bf \nabla} - 
	\frac{2\pi }{\Phi_0}{\bf A}
$$
By noticing that the current is the derivative of the free energy with respect 
to the vector potential and that the electric field is the time derivative of 
the vector potential (with a negative sign), the conductivity, in imaginary time,
is expressed as ($a(b)=x,y$)
\begin{equation}
\sigma_{ab}(\omega_{\mu}) = \frac{1}{\omega_{\mu}}\int d^2r 
	\,d\tau\frac{\delta^2\ln Z}{\delta A_a({\bf r},\tau) \delta A_b(0)}
	\mid_{{\bf A}=0}e^{i\omega_{\mu}\tau} \; .
\end{equation}
Using Eq.(\ref{free}), the longitudinal conductivity $\sigma_{aa}= \sigma $ 
can be expressed in terms of two- and four-point Green's functions. 
In the absence of charge and magnetic frustration and by evaluating the 
correlators in the Gaussian approximation, one obtains~\cite{Cha91} 
\begin{equation}
	\sigma(\omega_{\mu})= 
	\frac{2}{R_{Q}\omega_{\mu}} \frac{1}{\beta} \sum_{\nu}\ \int dk \,
	k^3 \, G(k,\omega_{\nu})
	\left[ G(k,\omega_{\nu}) - G(k,\omega_{\nu}+\omega_{\mu})\right]\, ,
\end{equation}
where
$$
G(k,\omega_{\mu}) =
	\frac{1}{\epsilon + k^2 + \zeta\omega_{\mu}^2 } \;\;\; .
$$
This turns out to be the first term in a $1/N$ expansion~\cite{Cha91}. 
The sum over the Matsubara frequencies can be performed by contour integration.
The result is
\begin{equation}
\sigma(\omega)
  =\frac{1}{4\pi R_{Q}\omega}\int_{-\infty}^{\infty}\frac{{d}z}{1-{e}^{-\beta z}}
  \int_0^{\infty}{d}k k^3
    \Im G^{R}(k,z)
  \Big[\Re G^{R}(k,z) - \Re G^{R}(k,z+\omega)\Big],
  \label{freq} 
\end{equation}
where the retarded $G^{R}$ and advanced $G^{A}$ Green's functions are given by
$$
	G^{R/A}(k,\omega) = G(k,\omega_{\nu} \longrightarrow \omega \pm i \eta) \; .
$$ 
The previous expression can be evaluated in various important limits.

\subsubsection{Zero-temperature conductivity}
Performing the $k$-integral the real and imaginary parts of the conductivity are
\begin{equation}
\Re \, \sigma (\omega) = \frac{\pi}{8R_{Q}} \left(1 - \frac{\omega _{c} ^2}{\omega
			^2} \right) \Theta (\omega ^2 - \omega _{c}^2)
\end{equation}
\begin{equation}
\Im \,\sigma (\omega) = \frac{1}{8R_{Q}} \left[- \frac{2\omega _{c} }{\omega} + 
		\left(1 - \frac{\omega _{c} ^2}{\omega ^2} \right)
		\ln \left| \frac{\omega - \omega _{c}}{\omega + \omega _{c}}
		\right| \right]
\end{equation}
where 
$$\omega _ c = 2\sqrt{\frac{\epsilon}{\zeta}} \;\;\;\; .
$$ 
At low frequencies the real part of the conductivity
exhibits an excitation gap equal to $\omega _c$.
In the insulating region the system, as can be deduced from the behavior of the
imaginary part of the conductivity, behaves as an effective capacitor with
$$
C_{eff} = \frac{1}{6 R_{Q} \omega_c} \;\; .
$$
The previous expressions can be calculated in the lowest order in $1/N$ and obey  
the scaling law  with $z\nu =1$.
The threshold frequency vanishes at the S-I transition leading to a
finite d.c. ($\omega \rightarrow 0$) conductivity,
\begin{equation}
\sigma^\star = \frac{\pi}{8} \frac{4e^2}{h} \; .
\label{univres}
\end{equation}
As explained in the first part of this section, this corresponds to the 
evaluation of the scaling function for $\omega/T \rightarrow \infty$ (the 
collision-free regime).
Corrections to the next order in the $1/N$ expansion correct this Gaussian
result by roughly $30\%$. 

Another powerful method for evaluating critical quantities
is the $\epsilon$-expansion.  In order to set up
the $\epsilon$-expansion one should move away from two dimensions
and consider systems with $d-1$ spatial dimensions. 
This approach allows one to obtain
the scaling form of the frequency dependent conductance~\cite{Fazio96a} in $d$
dimensions. 

In the fully frustrated case ($f=1/2$) the 
conductance at the S-I transition is still finite but with a value which is 
different from the $f=0$ case. It is possible to evaluate it in a 
$1/N$-expansion~\cite{granato90} and 
to the lowest order the conductance is twice the value of the critical conductance 
in zero field. 
$$
	\sigma^\star(f=1/2) = 2 \sigma^\star(f=0)
$$
This factor of two is reminiscent of the superlattice structure
at full frustration. There are,  
however, no fundamental reasons why this ratio should hold in general.

\subsubsection{Finite-temperature conductivity}
At low  temperatures ($T \ll \omega_c$),
the real part of the conductivity
is given by~\cite{Otterlo93,Damle97} 
\begin{equation}
\Re \, \sigma (\omega)  = \frac{2\pi }{R_{Q}}T e^{-\beta \omega_c} \delta (\omega) + 
		\frac{\pi}{8R_{Q}}
		\left(1 - \frac{\omega _{c} ^2}{\omega
		^2} \right) \Theta (\omega ^2 - \omega_{c} ^2)
		\left[1+2e^{-\beta |\omega| /2}\right]
\end{equation}
The imaginary part is obtained by means of Kramers-Kronig dispersion relations
\begin{equation}
\Im \, \sigma (\omega)  =  \frac{2 }{R_{Q}\omega}T e^{-\beta \omega_c}
	 + \omega C_{eff}
\end{equation}
where
\begin{equation}
	C_{eff} = \frac{1}{6\omega_c R_Q}
	\left(1+24\frac{T^2}{\omega _c^2} e^{-\beta \omega_c}\right)
\end{equation}
The Drude peak in the real part of the conductivity arises
due to a lack of dissipation or disorder in this model. Once electron or hole
like excitations are created, they will propagate without damping thus leading to
perfect conductivity. Although the system is a perfect conductor it is not a
superconductor since it shows no Meissner effect. The response of the system to
a static ${\bf k}$-dependent magnetic field, is proportional to $k^2$, i.e. it
vanishes at long wavelengths.

The scale for the crossover to the classical behavior is set by $T \sim \omega_c$
In the high temperature limit ($T \gg \omega_c$), the real and imaginary part of the 
conductivity read
\begin{eqnarray}
\Re \, \sigma (\omega)  &=& \pi \frac{\rho_D}{R_{Q}} \delta (\omega) + 
			\frac{\pi T}{2R_{Q}\mid \omega \mid}
		\left(1 - \frac{\omega _{c} ^2}{\omega
		^2} \right) \Theta (\omega _{c} ^2 - \omega ^2) \\
\Im \, \sigma (\omega)  &=&  \frac{\rho_D}{R_{Q}\omega} + 
	\frac{T}{4R_{Q}\omega _{c} ^2}\omega 
\end{eqnarray}
where $\rho_D \sim T$ and the expression for the imaginary part is valid at 
frequencies much smaller than $\omega _{c}$.

Damle and Sachdev~\cite{Damle97,Damle98} pointed out that since the conductance
is a universal function of $\omega/T$, it makes a difference which limit is taken first 
(either $\omega \rightarrow 0$ or $T \rightarrow 0$). 
In order to study the collision-dominated regime they used a Boltzmann like approach in 
which the current is expressed in terms of distribution 
functions for the particle and hole-like excitations. 
By solving the appropriate Boltzmann 
equation (the  collision term  can be obtained by  Fermi golden rule) they 
showed that also in the collision dominated regime the conductivity is a universal 
function at the critical point. We refer to the book by Sachdev for a clear and 
comprehensive presentation of these aspects of transport close to quantum 
critical points~\cite{sachdev99}.

\subsubsection{Non universal behavior}
Despite the conceptual elegance of the theories predicting a universal conductance 
at the transition, the experiments on JJAs and two-dimensional superconducting films
show critical resistivities that are different (by
a factor up to ten) as compared to the predicted universal values. 
Wagenblast {\em et al.} ~\cite{wagenblast98}
developed a theory of this non-universal behavior using the local-damping 
model. The evaluation of the 
dynamical conductivity proceeds along the same lines discussed before 
(see Eq.~\ref{locdampg}). The advanced 
and retarded Greens functions are given by 
\begin{equation}
  \Big[G^{A/R}(k,\omega)\Big]^{-1}=\epsilon+ k^2-\zeta\omega^2
    +\eta|\omega|^s\Big[\cos\frac{s\pi}{2}
  \pm i\mbox{sign}(\omega)\sin\frac{s\pi}{2}\Big],
\end{equation}
With increasing damping the Mott gap is smeared out. For $s<2$ and
low frequencies $\omega\ll\omega_c$ one finds
\begin{equation}
  \mbox{Re}\,\sigma(\omega)=\frac{1}{R_Q}
  \frac{\eta^2\sin^2(\frac{\pi}{2}s)}{6\pi\epsilon^2} 
  \frac{[\Gamma(1+s)]^2}{\Gamma(2+2s)} |\omega|^{2s}\;.
\end{equation}
The conductivity shows a power-law behavior at low frequency, where
the power depends on the dissipation strength for $s\leq 2$.
Of particular interest is the d.c. conductivity at the transition,
which becomes a function of the strength of Ohmic
damping for $\alpha > 2/3$.  

This model with local damping was further explored by 
Dalidovich and Phillips~\cite{dalidovich00} in the case $s=1$. 
In the limit of weak damping they 
find that dissipation leads to a leveling off of the d.c. conductivity at 
intermediate temperatures. Their estimates indicate resistance 
plateaus of the order of $10$  k$\Omega$ in the mK range, compatible with 
the experiments. These results seem more applicable to uniform films rather than 
to JJAs. In any case they offer an interesting explanation for 
the experimental observation that the critical resistance is not universal.

\subsection{One-dimensional arrays}
\label{onedi}
\begin{figure}
\centerline{{\epsfxsize=14cm\epsfysize=16cm\epsfbox{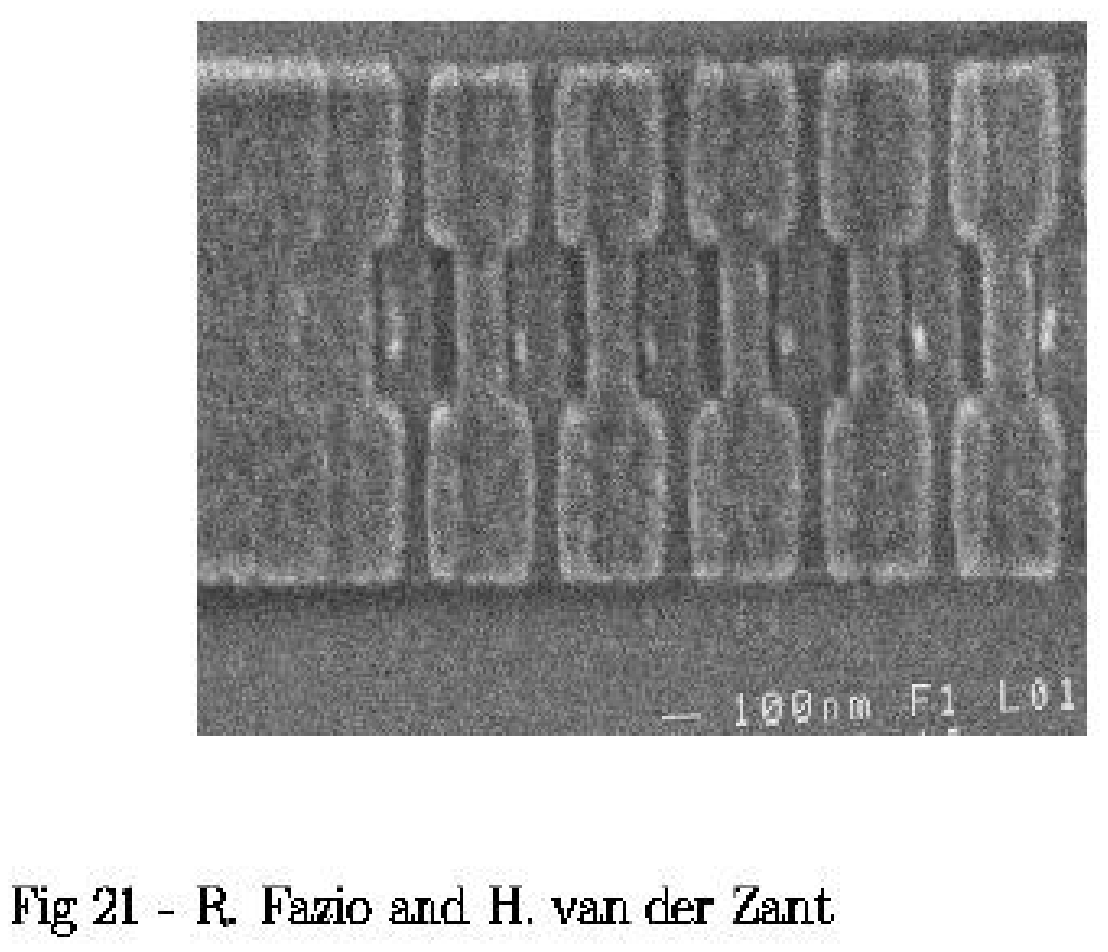}}}
\caption{A Scanning-Electron-Microscope (SEM) image of a SQUID chain. 
	(From Ref.~\protect\cite{Chow98}).}
\label{squidchain}
\end{figure}

Josephson-junction chains have been much less investigated (both theoretically and
experimentally) as compared to two-dimensional systems and only recently the S-I 
transition in one-dimensional samples has been measured~\cite{Chow98}.
In addition to the possibility to fabricate arrays with controlled couplings,
in Josephson chains the ratio of the Josephson to the charging energy
can be varied {\em in situ} by connecting mesoscopic SQUIDS in series 
(as illustrated in Fig.\ref{squidchain}). In this setup, the sample behaves as a chain 
of junctions with a tunable Josephson coupling 
$E_J(\Phi) = 2E_J \cos(\pi \Phi /\Phi_0)$ 
depending on the magnetic flux $\Phi$ piercing the SQUID. By varying $\Phi$, it is 
possible to sweep through the S-I transition while measuring on the same sample. 

At zero temperature and for short range Coulomb interaction the S-I transition of
a $d$-dimensional array is of the same universality class as a classical XY model in
$d+1$ dimensions. Therefore Josephson chains should exhibit a BKT-like transition. 
By means of duality transformation
it is possible to map the XY-model onto a gas of logarithmic interacting
vortices~\cite{Minnhagen87}. Vortices are bound
in pairs (of opposite vorticity) below the transition temperature and are in a plasma
phase in the disordered (high temperature phase). In a quantum chain the relevant
topological excitations, which correspond to vortices in space-time, are (quantum)
phase slips. The mapping of a Josephson chain onto a gas of interacting phase slips
has been performed by Bradley and Doniach~\cite{Bradley84}.

Consider, for simplicity, only the charging part of the Hamiltonian and
neglect the contribution due to the junction capacitance.
\begin{equation} 
	Z_{ch}  =  \prod_{i,} \int
	D\phi_i (\tau) Dq_i (\tau) 
	\exp \left[-\int_{0}^{\beta} d\tau 
	4E_{0}\sum_i
	q_{i}^2  + i \sum_i\int_{0}^{\beta} d\tau 
	q_{i}\dot{\phi}_{i} \right] \;\;\;\;\;.
\label{XYtime}
\end{equation}
The summation over the winding numbers fixes the charges to be integers in units of $2e$.
By discretizing the path integral (with a time slice $\tau_\epsilon$ and performing the 
summation over the  integers $q_i$ the charging contribution to the partition 
function can be recasted into the form
\begin{equation} 
	Z_{ch}  =  \prod_{i,\tau} \int  d\phi_{i,\tau} 
		\sum_{[n]} 
	\exp [- (1/8 \tau_\epsilon E_0) \sum_{i,\tau} 
	(\phi_{i,\tau} - \phi_{i,\tau+\tau_\epsilon} - 2\pi n_{i,\tau})^2]
	\;.
\label{Villaintime}
\end{equation}
Eq.(\ref{Villaintime}) is the Villain approximation of the XY potential~\cite{villain75}
if one identifies $\phi_{i,\tau} - \phi_{i,\tau+\epsilon}$
as the dynamical variable and $1/\epsilon E_0$ as
the effective coupling. The time slice $\epsilon$ can then
be chosen such that the coupling in space and time is isotropic ($\epsilon \sim
1/\sqrt{8E_J E_0}$)~\cite{1dfootnote}. The XY model in space-time has a reduced coupling
proportional to the ratio $\sqrt{E_J/8E_0}$. Therefore, all the known results for 
the classical XY model directly apply with the following replacement
\begin{equation}
	 \frac{E_J}{T}	 \;\;\; \longrightarrow \;\;\;\;  \sqrt{\frac{E_J}{8E_0}}
\end{equation}

The identification of the charging energy with the effective temperature shows the 
analogy between classical (thermal) and quantum (induced by charging effects) 
fluctuations. The partition function can now be expressed in terms
of interacting phase slips (in the same fashion as in the classical where it is
expressed in terms of vortices):
\begin{equation} 
{\cal Z} = \sum_{{p}} \exp{ \left[-\frac{2\pi^2}{N_x N_{\tau}}\sqrt{\frac{E_J}{E_0}}
	\sum_{k,\omega} p(k,\omega)G_0(k,\omega) p(-k,-\omega)\right]} \; ,
\end{equation}
where $p \pm 1$ are the "charges" associated with the occurrence of a phase slip.

The function 
$$
	G_0(k,\omega) \sim (k^2 + \omega^2)^{-1}
$$ 
implies that phase slips interact logarithmically in space-time. 
The chain undergoes a BKT phase transition at a critical value 
$$
\sqrt{E_J/E_0} \sim 2/\pi \;\;\;\;.
$$  
When a Josephson coupling is larger than the critical value the phase
correlator decays algebraically (quasi-long range order) and the chain is
superconducting. Phase slips are bound in pairs of opposite sign and therefore 
they do not lead to any dissipation over a macroscopic region (the Josephson 
relation implies that the occurrence of phase slip leads to a voltage drop).
In the opposite regime the chain is in the insulating phase. 
Phase slips are not paired and any current leads to a voltage. 
The correlation length is then given by  
$$
	\xi \sim \exp \left\{-\frac{b}{\sqrt{1 - [\pi^2E_J/16 E_0]^{1/2}}}\right\} 
$$
($b \sim 1$). Due to the isotropy in the space-time direction, 
one can now define an effective Coulomb gap $\sim E_0 \xi^{-1}$.
As long as there is particle-hole symmetry a finite range Coulomb interaction does 
not change the universality class of the transition.
A detailed analysis of the phase diagram, for realistic Coulomb interactions,
as a function of the charge frustration has been performed by
Odintsov~\cite{Odintsov95}.

\begin{figure}
\centerline{{\epsfxsize=14cm\epsfysize=16cm\epsfbox{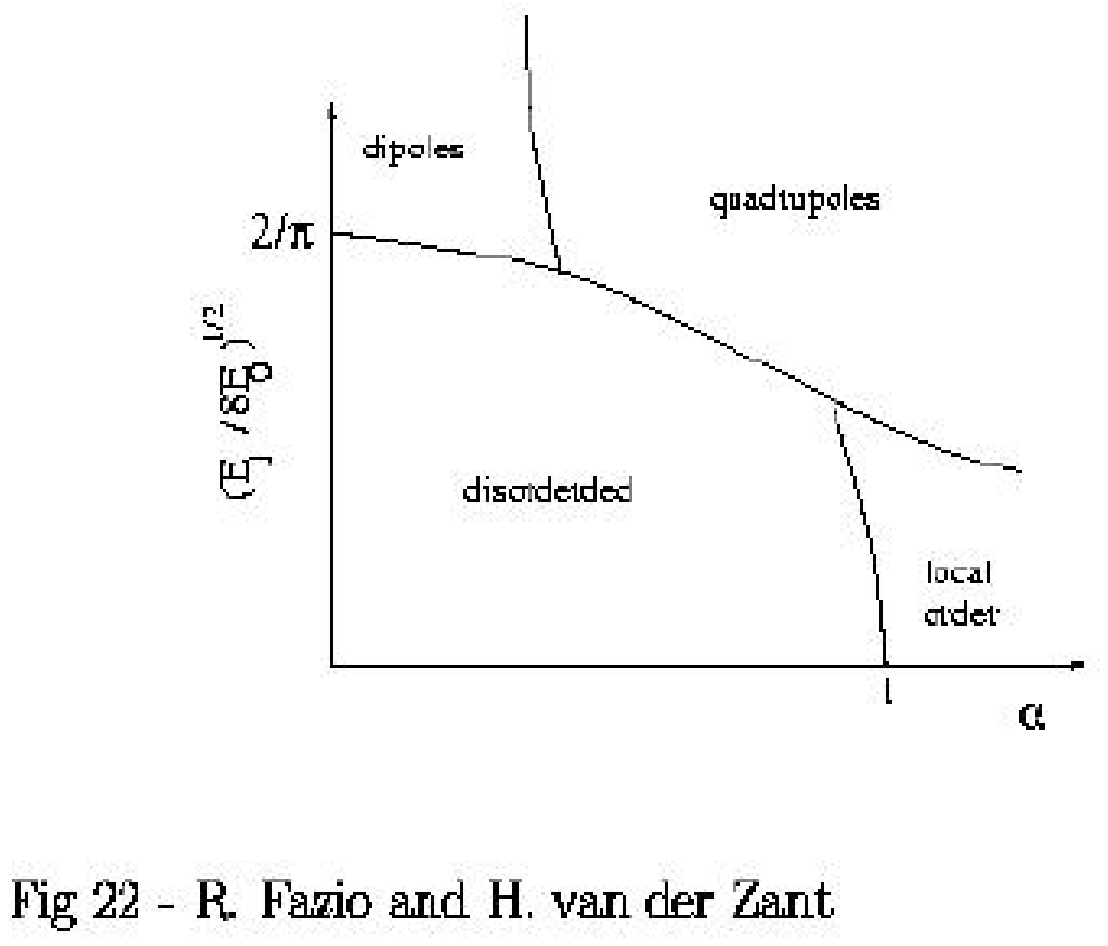}}}
\caption{Calculated phase diagram of a Josephson chain shown as a function of 
	the Josephson coupling and the dissipation strength. 
	The dissipative part of the action leads to two new phases. 
	For $\alpha > 1/2 $ the dipoles are bound in a gas of quadrupoles and moreover  
	for strong dissipation there is an additional phase transition which separate 
	the quadrupole phase from a phase in which the system shows local order.
	(From Ref.~\protect\cite{Bobbert90}).}
\label{1dpd}
\end{figure}

The presence of dissipation (see also Section \ref{dissip}) modifies the critical
behavior of the chain. The case of
a Josephson chain with Ohmic dissipation has been considered by several
authors~\cite{Zwerger89,Korshunov89,Bobbert90,Bobbert92} by means of dual
transformations and Monte Carlo simulations.
The main conclusions of this
series of works is the zero-temperature-phase diagram as a
function of dissipation strength and Josephson coupling as
shown in Fig.\ref{1dpd}. In addition to the S-I
phase boundary there are two new phases induced entirely
by dissipation:
\begin{itemize}
\item  for small Josephson coupling and large dissipation
	the chain is in a phase with local order. The phase difference at each junction
	is locked in time but the chain has no quasi long-range order 
\item for large
	dissipation and large Josephson coupling there is a new type of superconducting
	phase characterized by the phase slips bound in quadrupoles. 
\end{itemize}
The four different
phases can be measured by considering different setups 
as discussed in Ref.~\cite{Bobbert92}.

\begin{figure}
\centerline{{\epsfxsize=14cm\epsfysize=16cm\epsfbox{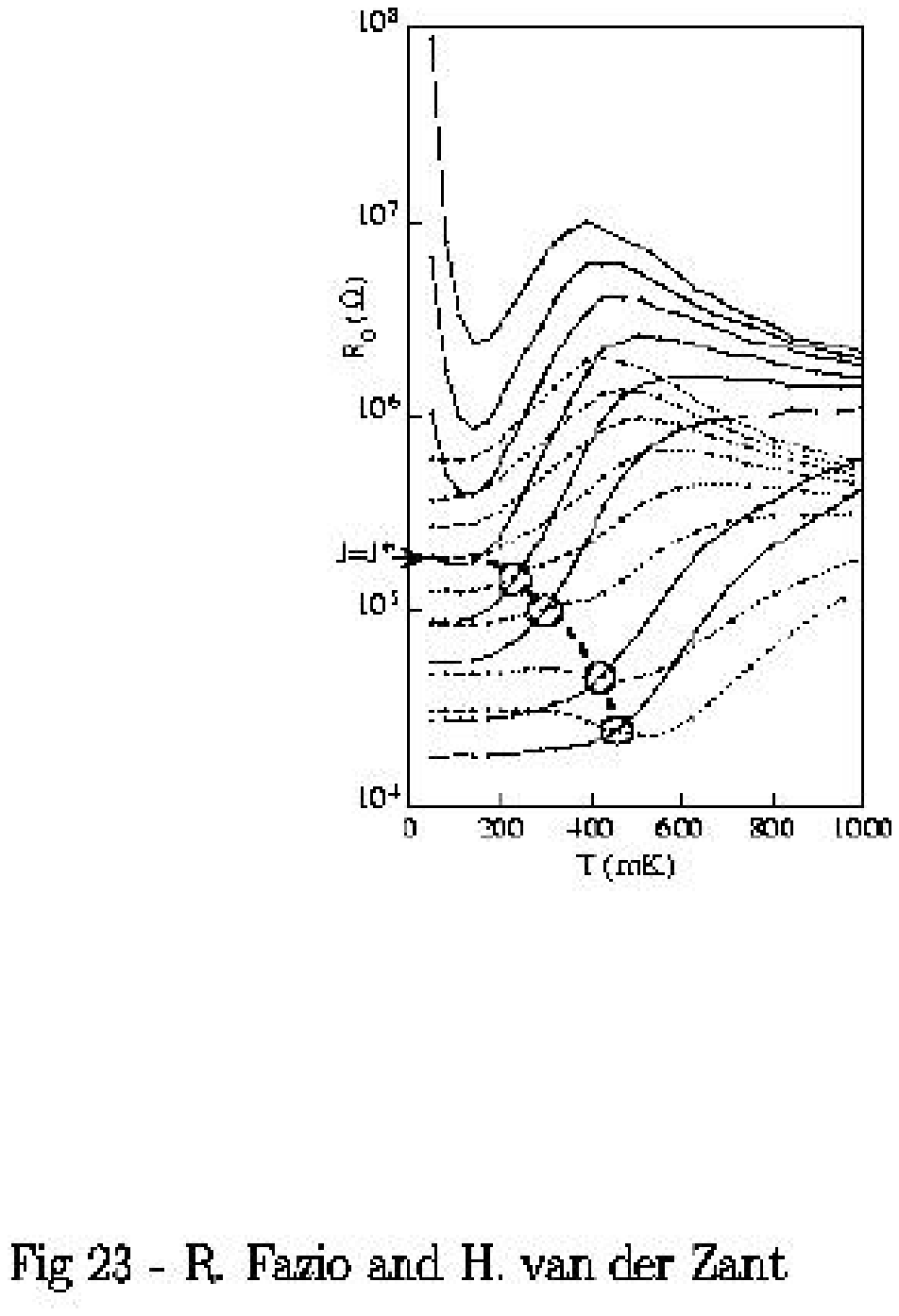}}}
\caption{Resistance vs. temperature for eight values of the magnetic field in the 
	range $0 - 64$~G. The two sets of curves correspond to two different 
	one-dimensional arrays with $N=255$ (solid lines) and $N=63$ (dashed lines) junctions.
	The longer array shows a sharper S-I transition. 
	At the point $J^{\star}$ ($J=\sqrt{E_J/E_C})$ the resistance is length independent. 
	(From Ref.~\protect\cite{Chow98}.)}
\label{figChow}
\end{figure}

We conclude this section by reviewing the experiments of Chow 
{\em et al.}~\cite{Chow98}. The dependence of the resistance on the temperature,
shown in Fig.\ref{figChow}, shows  a  non trivial scaling behavior.
The two set of curves (solid and dashed) refer to two chains of different
length. While in the insulating phase the resistance increases with the
number of junctions, in the superconducting phase the opposite trend is visible.
By identifying the scale-independent value of $R_0(T)$, Chow {\em et  al.}
were able to trace out the zero-temperature critical point (indicated
with $J^*$ in the figure). 

\begin{figure}
\centerline{{\epsfxsize=14cm\epsfysize=16cm\epsfbox{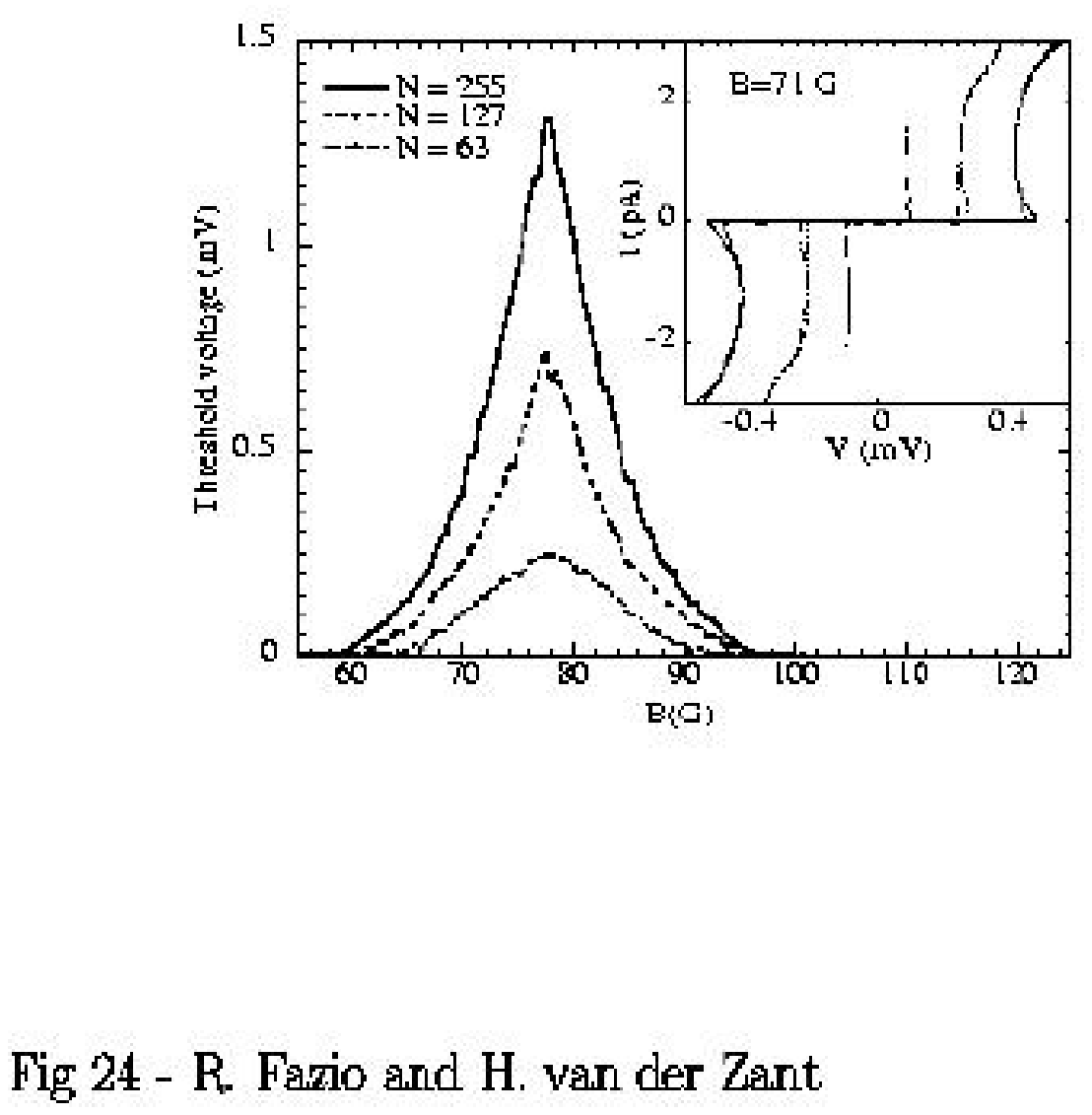}}}
\caption{The magnetic field dependence of the threshold voltage for one-dimensional 
	arrays of different length. The inset shows the corresponding $I-V$ curves. 
	(From Ref.~\protect\cite{Chow98}.)}
\label{Bgap}
\end{figure}

The resistance in the
superconducting chains can be explained in terms of formation of phase slips.
The flat tails in the curves are due to a finite-size effect 
and occur for temperatures of the order of the effective Coulomb gap.
In this region the probability of a phase slip event obtained
in the Coulomb gas picture presented above scales with the number of junctions 
as $N_x^{2-\pi \sqrt{E_J/8E_0}}$. Quantum
phase fluctuations are suppressed by increasing the system size.
In the insulating regime the $I-V$ curves show Coulomb blockade with a
threshold voltage which depends on the magnetic flux piercing the
SQUID~\cite{Haviland96} as shown in Fig.\ref{Bgap}. 
Thus, there is reasonable agreement with the theory, but finite-size effects make a 
quantitative analysis difficult because of the rapidly diverging correlation lengths.

\subsubsection{1D arrays as Luttinger Liquids}

The interest in one-dimensional arrays goes further as they can 
be described in terms of the
Luttinger liquid (LL) model~\cite{LLreviews}.
The low-energy excitations of the interacting electron gas in
one dimension are long-wavelength spin and charge oscillations,
rather than fermionic quasi-particle excitations.
Accordingly, the transport properties cannot be described in terms of the
conventional Fermi-liquid approach.
The density of states shows asymptotic power-law behavior at low energies.
Depending on the sign of the interaction an arbitrarily weak barrier 
in a quantum wire leads to perfectly reflecting (for
repulsive interactions) or transmitting behavior at low voltages~\cite{Kane92}.
It is customary to characterize this interaction by a parameter
$g$ such that $g=1$ in the noninteracting situation while $g > (<) 1$ in the
attractive (repulsive) case.

A Josephson chain seems an ideal system to explore LL correlations~\cite{Fazio96}.
In the limit of large Josephson coupling $g = \sqrt{E_J/8E_0} \gg 1$; i.e., the
chain behaves as an attractive LL. Glazman and Larkin~\cite{Glazman97} showed that
in a certain region of parameters (close to $q_x=1/2$) between the Mott lobe
and the superconducting region, there is a new intermediate phase which is
equivalent to the chain behaving as a {\em repulsive} LL. In order to characterize
this repulsive behavior one should consider 
a Josephson chain with a defect. One of the junctions could for example be made with 
a Josephson
coupling much smaller than the charging energy~\cite{Glazman97,Fazio96}.
The different phases in the phase diagram can be characterized by the dependence
of the Josephson current on the chain length. In the
superconducting phase, the defect has no effect as the number of
junctions increases. On the contrary, in the repulsive LL phase there
is a strong dependence on the number of junctions. 
The LL phase can be studied also by means 
of Andreev tunneling spectroscopy along the lines discussed in Ref.~\cite{Falci95}.

Repulsive LL behavior is also present in a Josephson ladder as discussed in 
Ref.~\cite{Choi98}.
The possibility  of repulsive LL behavior is related to a normal
phase of interacting bosons at zero temperature. In one-dimensional systems
such a possibility cannot be excluded and Monte Carlo simulations on a Josephson
chain~\cite{Baltin97} show a phase in which there is neither crystalline nor
superfluid order. The existence of a normal phase has been questioned
in Ref.~\cite{kuehner98} through Density Matrix Renormalization Group of 
the BH model. One should
note however that phase boundaries are non-universal and therefore the QPM and
BH system can lead to different results. 

\subsection{Field-tuned transitions}
\label{fieldtsection}
In arrays which are in the superconducting state at $f=0$ but
have an $E_J/E_C$ ratio close to the critical value,
a magnetic field can be used to drive the array into the insulating
state. This field-tuned transition has been
considered theoretically by M.P.A. Fisher \cite{fisher90a} in disordered 
systems and has first been observed by Hebard and Palaanen~\cite{hebard90,paalanen92} 
in thin InO$_x$ films. The interplay between disorder and vortex-vortex interactions 
plays an essential role. At low magnetic fields vortices at $T=0$ are pinned (by 
disorder) but for higher fields, the vortex density increases and at some
critical density, vortices Bose-condense (a vortex superfluid
leads to an infinite resistance). By employing duality arguments 
this transition can also be thought as a Bose condensation of vortices 
that occurs by changing the  applied magnetic field. 

The general characteristic of the field-tuned S-I transition is that when $f$ is 
increased from zero, the temperature derivative of the resistance changes sign at 
critical values $\pm f_c$. Fisher's analysis~\cite{fisher90a} leads to the 
following scaling  for the resistivity tensor close to the field tuned transition 
\begin{equation}
\rho_{\alpha,\beta} = \frac{h}{4e^2}\tilde{\rho}_{\alpha,\beta}(\frac{f-f_c}{T^{1/z\nu}})
\label{fieldtscaling}
\end{equation}
where $\nu$ is the exponent which controls the divergence of the correlation 
length at the transition and $z$ is the dynamical critical exponent (with $z\nu \ge 1$).
The resistivities are predicted  to be universal at the transition and should satisfy 
the relation 
\begin{equation}
	\sqrt{ \rho_{x,x}^2 + \rho_{x,y}^2 } = \frac{h}{4e^2}
\label{universal_resistance}
\end{equation}
These predictions were tested in Josephson arrays by the Delft~\cite{zant96,zant92a} 
and the Chalmers groups~\cite{chen95}.

\begin{figure}
\centerline{{\epsfxsize=14cm\epsfysize=16cm\epsfbox{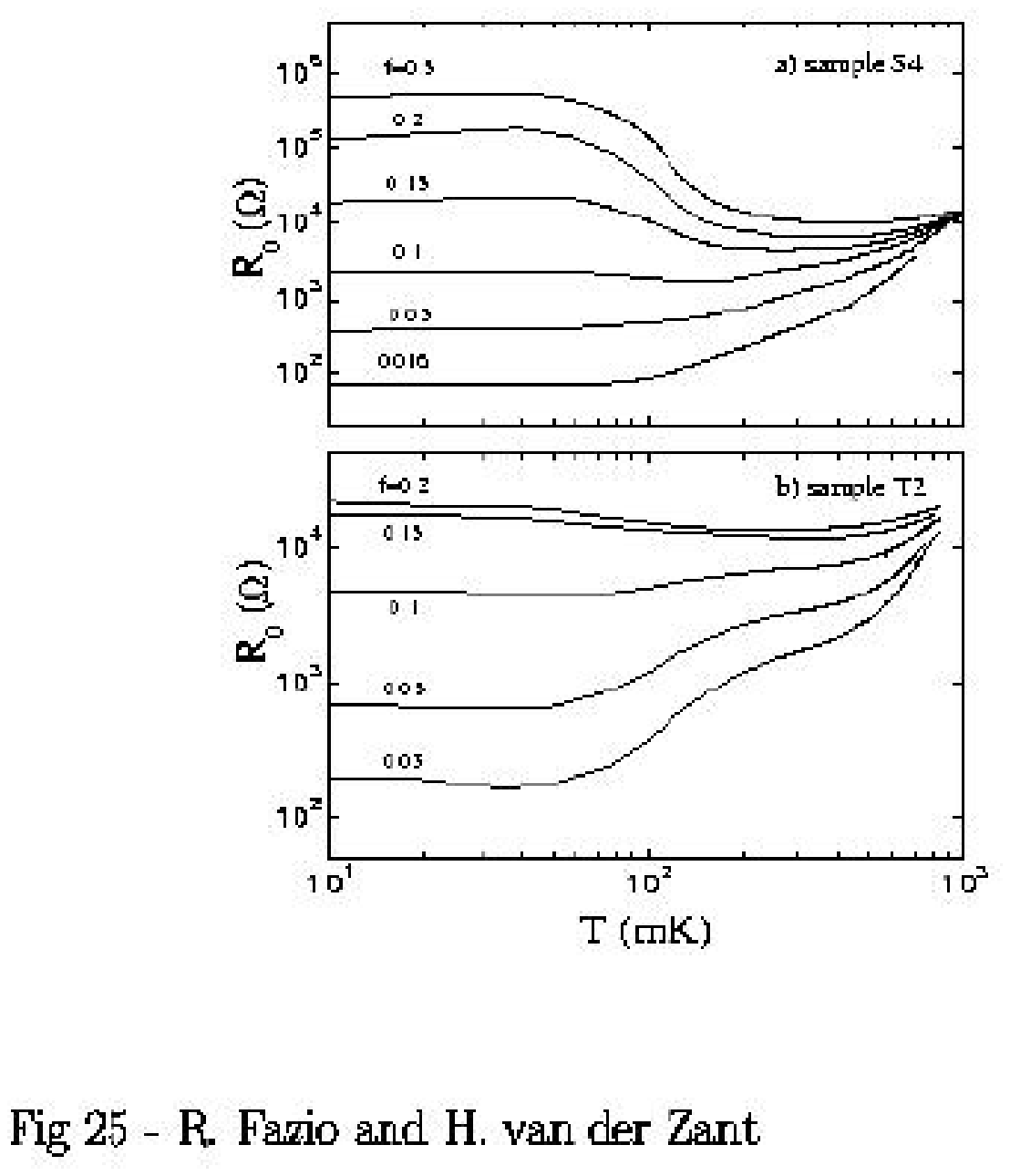}}}
\caption{Temperature dependence of the resistance for a square (a) and triangular (b) 
	2D Josephson array measured at different applied magnetic fields. The field 
	tuned S-I transition occurs at that frustration where the temperature 
	dependence of the resistance changes sign. 
	In both cases this change occurs at $f= 0.10-0.15$.
	(From Ref.\protect\cite{zant96}.)}
\label{rf_temp}
\end{figure}

\begin{figure}
\centerline{{\epsfxsize=14cm\epsfysize=16cm\epsfbox{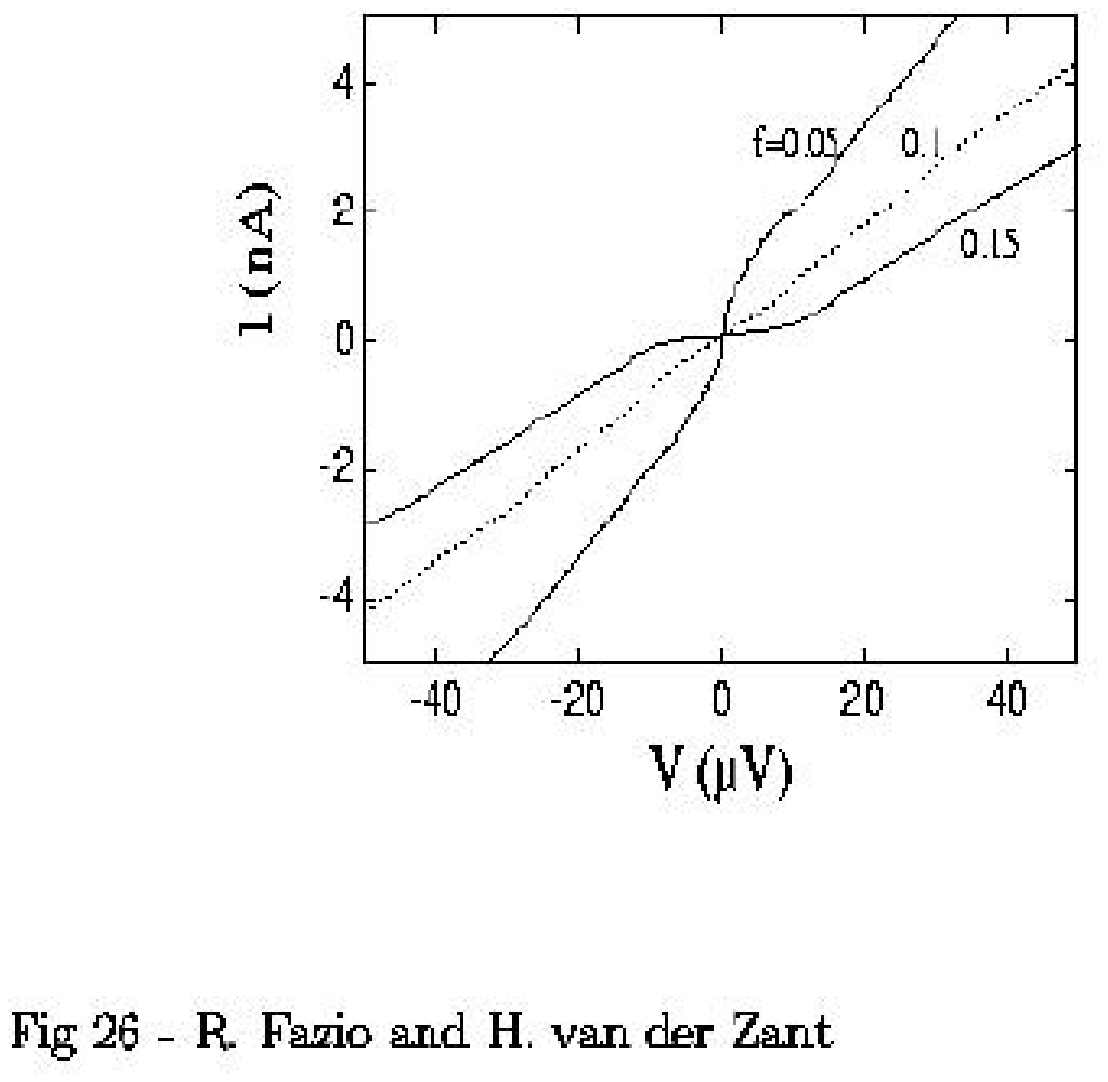}}}
\caption{$I-V$ characteristics measured at low temperature ($10$~mK) for three values 
	of the applied field. The square 2D array has an ($E_C/E_J$ ratio of 1.25. 
	The crossover from the superconducting to the insulating behavior is related 
	to the field tuned S-I transition. (From Ref.\protect\cite{zant96}.)}
\label{IV_fieldt}
\end{figure}

For several values of the frustration, the resistance as a function of temperature is 
shown in Fig.\ref{rf_temp}. 
Below a critical value $f_c$, the resistance decreases upon
cooling down ($dR_0/dT>0$). Above $f_c$ the resistance increases ($dR_0/dT<0$)
and for low temperatures reaches a value that might be orders of magnitudes higher 
than the normal-state resistance. This sign change in the temperature dependence 
corresponds to a change in the $I$-$V$ characteristics shown in Fig.\ref{IV_fieldt}. 
For $f<f_c$, a critical current is observed in the $I$-$V$ characteristics, 
whereas above $f_c$ a charging gap develops. 
Note that at low temperatures, the resistance flattens off. This is most likely
a finite size effect involving quantum tunneling of vortices. Finite-size effects are 
expected to play a more prominent role in JJAs as compared to films because arrays 
are typically 100 cells wide. In units of the coherence length,
films are much larger.

\begin{figure}
\centerline{{\epsfxsize=14cm\epsfysize=16cm\epsfbox{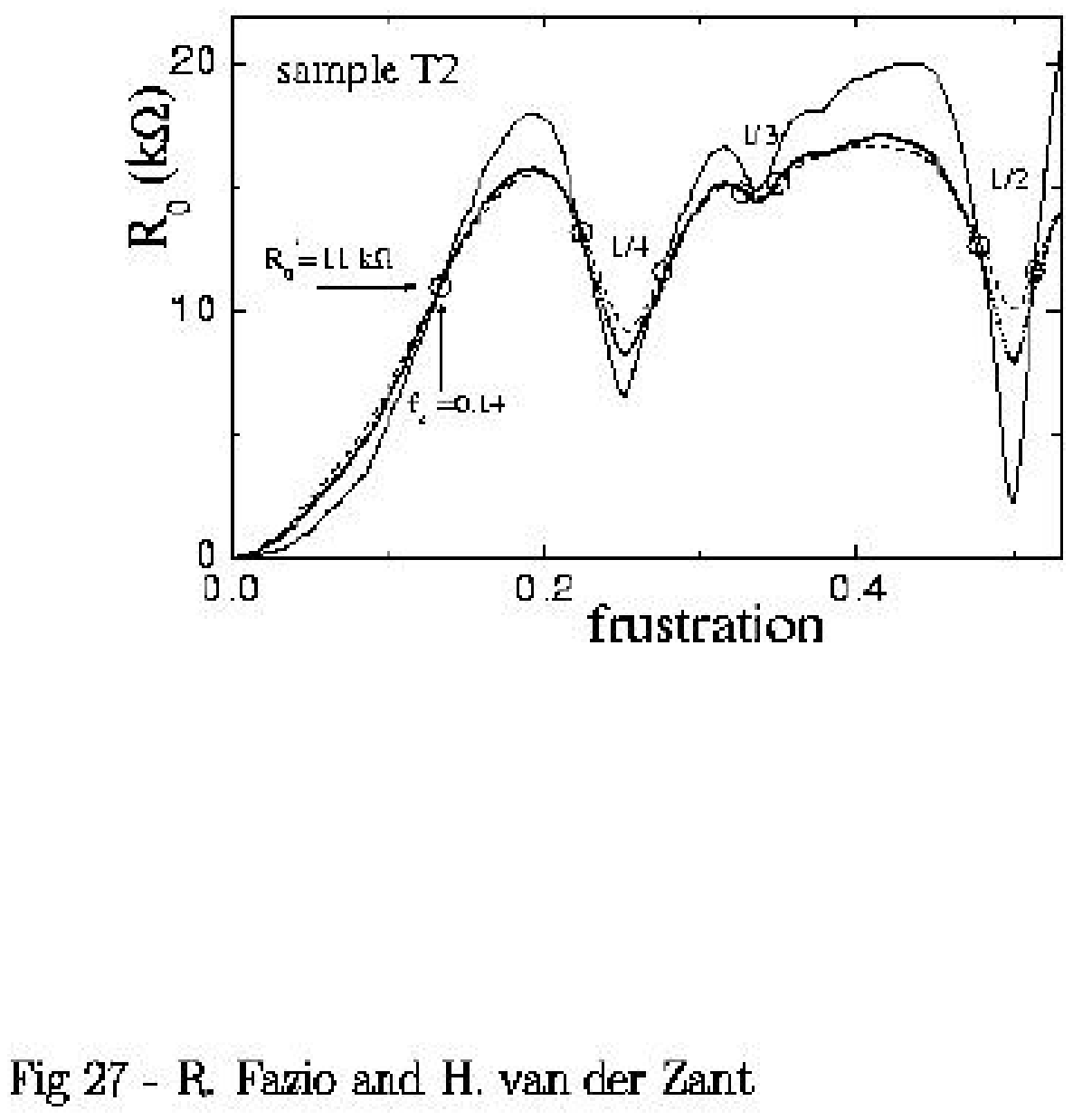}}}
\caption{Field dependence of the resistance of a triangular 2D array measured 
	at different temperatures, $T=50$ mK (solid line), $T=120$ mK (dotted line), 
	and $T=160$ mK (dashed line). The field-tuned transition is observed around 
	different fractional values of the frustration indicated by the open circles. 
	(From Ref.\protect\cite{zant96}.)}
\label{rt_temp}
\end{figure}

A more detailed way of observing the field-tuned S-I transition is obtained by
measuring the resistance versus magnetic field for different temperatures (see
Fig.\ref{rt_temp}).
In the range $0<f<1/3$, the $R(f)$ curves are very similar
to the ones measured in thin films.
Below the critical field $f_c$=0.14 the resistance becomes
smaller when the temperature is lowered and above
$f_c$ the resistance increases. 

\begin{figure}
\centerline{{\epsfxsize=14cm\epsfysize=16cm\epsfbox{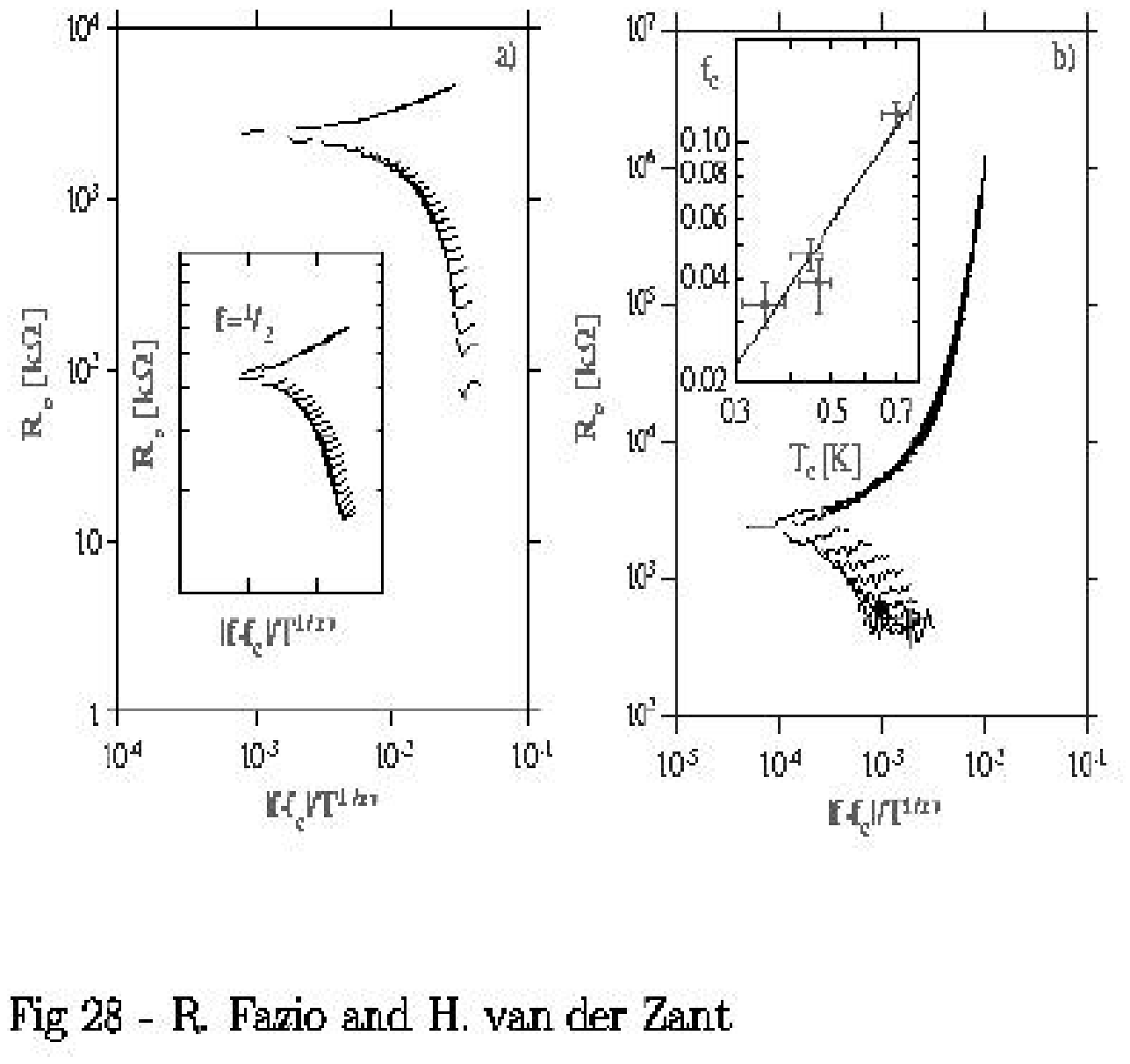}}}
\caption{Resistance as a function of the scaling parameter $|f-f_{c}| T^{-1/z_{B}\nu _{B}}$ 
	for two different arrays (data taken in the range $0<f<0.2$). 
	The data collapse onto a single curve: the upper part for the insulating 
	transition and the lower part for the superconducting transition.
	The inset in the figure on the left shows the scaling for the array close 
	to full frustration (data taken in the range $0.5<f<0.6$).  
	The inset in the figure on the right shows a log-log plot of the critical 
	frustrations $f_{c}$ as a function of the BKT transition temperature for the four 	
	measured samples. The line through the data yields a critical exponent $z=1.05$. 
	(From Ref.\protect\cite{chen95}.)}
\label{fieldtscal}
\end{figure}

From the scaling analysis~\cite{fisher90a}, it follows that the slopes of the 
$R(f)$ curves at $f_c$ should follow a power-law dependence on $T$ with power 
$-1/(z\nu)$. When on a double logarithmic plot the slopes of the $R(f)$
curves at $f_c$ are plotted versus $1/T$, one finds straight lines
in the temperature range $50<T<500$ mK.
>From the reciprocal of the slope, the product $z\nu$ can be determined. 
Values in JJAs range from 1.2 to 2 for the Delft data and from 1.5 to 8.2
for the Chalmers data, in agreement with
the theoretical expectations $z=1$ and $\nu \geq 1$.
The scaling resulting from Eq.(\ref{fieldtscaling}) is best seen by 
plotting the resistance as a function $(f-f_c)/T^{1/z\nu}$ as illustrated 
in Fig.\ref{fieldtscal}: A universal function is obtained 
by plotting the resistance as a function of $E_J(f-f_c)/(E_C T^{1/z\nu})$.
The tails on the superconducting side (bottom curve) correspond to
the finite-size effect mention above.

The exponent $z$ can also be obtained form the measurements by plotting
$f_c$ as a function of the zero-field BKT
transition temperature:
$$
	f_c \propto T_{J}^{2/z} \;\;\; . 
$$
The Delft-data points on the triangular arrays yield a rough estimate
of $z \approx 0.34$ and their two data points
on the square arrays of $z \approx 1.4$. The Chalmers data provide a more
accurate fit yielding $z =1.05$, in good agreement with the
theoretical expectation.  

Measurements on different thin films
show that the resistance right at the transition is of the order
of $R_Q$ but measurements are not conclusive regarding the universality. 
In arrays, this resistance is again of order $R_Q$,
yet in different arrays it varies between 1.6 and 12.5~k$\Omega$.
The Chalmers group has also measured the Hall resistance in order to 
check the validity of Eq.\ref{universal_resistance}. 
For two arrays, the Hall resistance at the critical point is of the order of
30~$\Omega$, but again $\sqrt{ \rho_{x,x}^2 + \rho_{x,y}^2 }$ 
is not a universal quantity.

A new feature introduced by JJAs is the existence of
field-tuned transitions near commensurate values of the
applied field, i.e., at $f_{comm} \pm f_c$~\cite{chen96,zant92a}. 
Studying the $R(f)$ curves of JJAs in more detail, 
critical behavior is not only seen around $f=0$, but also around
$f=\pm 1/4, \; \pm 1/3, \; \pm 1/2$, $\pm 2/3$, and $\pm 3/4$.
For each commensurate $f$-value $z\nu$
can be determined as described above and the
values of $z\nu$ are close to one. The sample-dependent 
critical resistances are of the order of a few k$\Omega$. 
Calculations on the
Bose-Hubbard model in a magnetic field~\cite{cha93}
show that the product $z\nu$ at $f=1/2$ is close to 1 in 
agreement with the measurement.

\section{Quantum Vortex Dynamics}
\label{qvd}

A vortex (antivortex) is a topological excitation. When going around its 
center in a closed loop, the phases of the  islands wind up to $2\pi$ (-$2\pi$).
Vortices have extensively been studied in classical arrays where they both 
determine both the phase diagram and the dynamical properties.
To a large extent, this still holds for quantum arrays where the interplay 
between vortex and charge dynamics plays the central role. 
The field tuned transition for instance can be understood as a Bose 
condensation of vortices or charges and the S-I transition in zero field 
can be analyzed using the duality between charge and
vortex excitations. In this section we show that vortices characterize 
the quantum dynamics of arrays as well. We concentrate on the 
superconducting side of the S-I transition where vortices are well defined 
excitations. 

As discussed before, quantum arrays have flux penetration depths that are 
larger than the array sizes. 
The magnetic field is almost uniform over the whole array area indicating 
that there is not one flux quantum in particular cell. 
\footnote
{Self-field effects may play a role in classical arrays since critical 
currents are substantially larger. Generally, speaking self-field effects 
manifest themselves in two ways~\cite{phillips93}. First, there are 
self-inductance effects which are short-ranged and caused by the self-inductance 
of the cell loop. It turns out that the cell-to-cell energy barrier is 
dominated by these short-range interactions. Second, there are 
mutual-inductive effects which have a longer range. For example, the current
distribution around a vortex changes from an exponential fall-off for 
self-inductances to an algebraic fall-off when the mutual inductances 
between all cell pairs are included.}

\begin{figure}
\centerline{{\epsfxsize=14cm\epsfysize=16cm\epsfbox{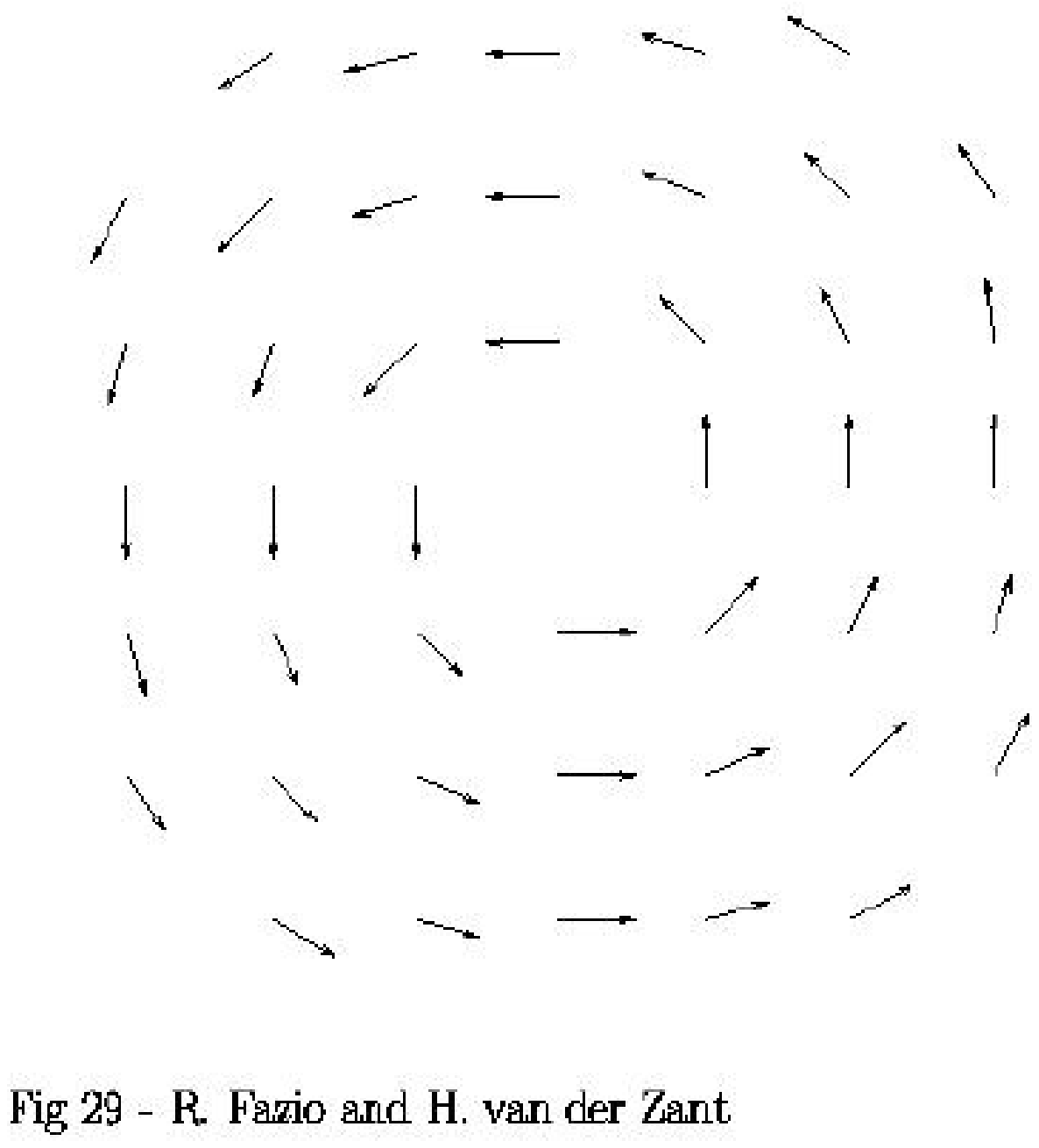}}}
\caption{Phase configuration of a vortex configuration. The arrows indicate the 
	phase of each island with respect to a given reference direction.}
\label{vortex}
\end{figure}

The essential aspect of vortices in junction arrays is therefore not the flux, 
but the distribution of phases. 
The phase configuration of a vortex (shown in Fig.\ref{vortex}) in a large 2D 
array can be approximated by the following analytical expression
\begin{equation}
\phi_{i}= \pm \arctan \left(\frac{y_{i}-y}{x_{i}-x}\right) 
\label{eq:arctan}
\end{equation}
where the site $i$ has coordinates $ {\bf r}_{i} = [x_{i},y_{i}]$ and the vortex 
center is placed at (${\bf r} =[x,y]$). 
The $\pm$ sign refers to the vortex (antivortex) configuration.
For most purposes Eq.(\ref{eq:arctan}) is accurate even very close to the 
vortex center. As we will see, the approximate arc-tan solution is very useful 
since it allows to express the action in terms of the coordinates of the vortex 
center ${\bf r}$ only (instead of in term of all the phases). 
This appears to be a reliable description as long as  the vortex can be 
considered as a rigid body. In most of the cases we discuss in this review, 
this turns out to be  a good approximation.

An important property of vortices is that supercurrents around them fall 
off inversely proportional to the distance $r$ from their core. Vortex-vortex 
interactions therefore have a long-distance character as they are proportional 
to $\ln r$. Eq.(\ref{eq:arctan}) is the solution for a single vortex in 
an infinite system. In finite systems, vortices interact with boundaries. 
The interaction of vortices with the open edges can be viewed as the attraction 
of a vortex with an image antivortex outside the array. Superconducting 
banks repel vortices; the interaction with these edges can be viewed as the 
one with an image vortex (of the same sign)  outside the array. The interaction 
between boundaries indicates that especially in small arrays the approximation  
given in Eq.(\ref{eq:arctan}) is no longer valid. Numerical calculations are 
then used to extract the quasi-static  phase configuration around a vortex. 

Experimentally, single-vortex dynamics is studied by applying a small magnetic 
field (low vortex densities) and performing transport measurements.
On the theoretical side, both numerical simulations and phenomenological models 
which lump the collective dynamics of the phases into the description of the 
motion of the vortex center, have been investigated. In this chapter, we first 
derive the classical equation of motion and show that vortices in underdamped 
arrays can be viewed as massive, point-like particles. We then continue with 
the quantum corrections to the equation of motion and discuss their consequences.
First, there is a renormalization of the vortex parameters (like the mass, 
damping, ...) due to quantum fluctuations. In addition there is a class of 
new phenomena which arise from the quantum dynamics of vortices which are 
treated in Sec.~\ref{quantumvortex} and \ref{localize}. They include macroscopic 
quantum tunneling of vortices in 2D arrays, quantum interference in a 
hexagon-shaped array,
Bloch oscillations of vortices in the periodic lattice potential 
and vortex localization in quasi one-dimensional arrays. In some experiments, 
evidence has been found that the (independent) single-vortex picture breaks 
down. In these cases, we briefly comment on the influence of vortex-vortex 
interactions.

\subsection{Classical equation of motion}
\label{eqmotion}
In this section, we review the steps to derive the equation of motion for a 
vortex. We analyze all contributions, i.e. its inertia, its dissipation, 
the external potential and the applied forces. To keep the notation simple 
we suppose, for the moment that the vortex moves along a given (say $\hat{x}$) 
direction with an average vortex velocity $v_x$. 

\underline{Vortex mass} - 
Moving vortices lead to phase changes across junctions and they therefore 
contribute to an electric energy. In principle all capacitances contribute, 
but as discussed before the main effect comes from the junction 
capacitance so that $E_{ch}=1/2\sum_{<ij>}^{}C V_{ij})$. (The contribution 
due to $C_0$ will be discussed in Section \ref{vaction}.) 
In a quasi-static approach this 
sum is calculated by comparing the phase differences across each junction 
at times $t$ and $t+1/v_x$ (in units of the lattice constant):  
$\Delta \phi_{ij} =  \phi_{ij}(t+1/v_x)-  \phi_{ij}(t)$.
The electric energy then acts like a kinetic energy term and the proportionality 
factor defines the vortex mass~\cite{simanek83,eckern89,larkin88,eckern90}:
\begin{equation} 
       E_{ch}= \frac{1}{2} M_v v_x^2 \; \; \; 
\end{equation}
with
\begin{equation}
M_v=  
               \frac{1}{8E_C} \;
               \sum_{<ij>}^{} (\Delta \phi_{ij})^2 \;.
\label{E_el}
\end{equation}
The problem of calculating the vortex mass is now reduced to finding
the phase differences across junctions at times
$t$ and $t+1/v_x$. Note, that Eq.(\ref{E_el}) 
can be applied to various array geometries if the phase
configuration around a vortex is known. 

In large 2D Josephson arrays the arctan form given in  
Eq.(\ref{eq:arctan})~\cite{lobb83} is used to evaluate the sum in 
Eq.~(\ref{E_el}). With the assumption that this arctan-phase
configuration remains the same when the vortex moves through the array, 
numerical evaluation of the phase differences 
in a large 2D square array yields the Eckern-Schmid value of the vortex mass
\begin{equation}
M_v = \frac{\pi^2}{4}  \;  E_C^{-1} \;\;\; . 
\label{mass_2D}
\end{equation}
In this calculation roughly half of the vortex mass is due to the 
junction the vortex crosses; the other half comes from all the 
other junctions in the array. For a triangular 2D array, a similar 
calculation can be done and the vortex mass is twice the mass of 
a square array. For typical arrays with $C=1$~fF and $a^2=10$~$\mu$m$^2$,
the vortex mass is 500 times smaller than the electron mass. This small value 
already indicates that quantum effects are likely to occur. 

It has been shown~\cite{orlando91} that near array edges the vortex mass 
vanishes when it approaches a free boundary of the array. 
These boundary effects are, however, negligible if the
vortex is a few lattice spacings away from the edge.
One can also include self-field effects. 
Currents now extend over a distance of the penetration depth $\lambda_{\bot}$ 
from the vortex center so that the arctan approximation can no longer be used.  
As the vortex is effectively reduced in size,
the sum of the $V_i$'s can be restricted to those junctions which are 
$\lambda_{\bot}$ from the vortex center. The result is a smaller vortex mass 
and its decrease with decreasing $\lambda_{\bot}$ is given in Ref.~\cite{trias96}.
The vortex mass is dependent on the $E_J/E_C$ ratio as well. Quantum 
corrections to the mass, on approaching the S-I transition~\cite{fazio94}
are discussed in Section \ref{vaction}.
To a good approximation  it has the value given in Eq.~(\ref{mass_2D}) 
for arrays that are not in the critical region $E_J \sim E_C$.

The vortex mass can also be calculated in geometries other than two dimensional 
arrays. In a purely 1D array ($N$ junctions in parallel connected by
two superconducting leads), 
the vortex phase configuration is given by 
$$
\phi_i=4\arctan[\exp((x_i-x)/\Lambda_J)], 
$$
where $x$ denotes the position of the vortex center. 
For $\Lambda_J<N$, the vortex has a kink-like shape, which extends over
a distance of the order of $\Lambda_J$; for $\Lambda_J>N$ 
the vortex is spread out equally over the whole system with 
$\phi_{x+1}-\phi_x \approx 2 \pi /N$. In this latter regime, 
$\Delta \phi_{t+1/v_x}(x+1,x)-\Delta \phi_{x+1,x}(t) = 2 \pi /N$ 
and $M_v=h^2/(8E_C N^2)$.
For $1<\Lambda_J<N$ the sum can be computed numerically or in
a continuum approximation. The sum over 
the phase differences squared is equal to $8/\Lambda_J$\cite{zant94}
and hence
\begin{equation}
M_v^{1D} = \frac{1}{ \Lambda_J} \; E_C^{-1} \; \; 
{\rm for} \; \; 1<\Lambda_J<N \; .
\end{equation}

In other geometries, the phase configuration of a vortex is not
exactly known. It can in principle be calculated. 
However, since a substantial contribution comes from the junction that 
the vortex crosses, Eq.(\ref{mass_2D}) can be used
as an estimate for the quasi-static vortex mass in these cases. 

The notion that the concept of the vortex mass is not new in the theory of 
superfluids. It has been discussed extensively for type-II superconductors and 
superfluid He~\cite{suhl65,simanek85,duan92}.

\underline{Dissipation} - As the vortex is a macroscopic object, it couples to 
the environment and experiences dissipation.
Quantum arrays generally have junctions that are underdamped, i.e., have a 
McCumber parameter $\beta_c>1$. 
One might therefore expect that vortex motion is underdamped as well.

In the simplest approximation one can assume that a moving vortex experience a 
viscous drag force characterized by a viscous coefficient $\eta$. In a 
Bardeen-Stephen like model $\eta$ is calculated using the following argument. 
The total power loss is the sum of all the resistive losses in the junctions. 
Assuming $R_e$ to be identical for all junctions,
the sum in the total power is the same as in the calculation of the vortex mass. 
For example, 
$$
	\eta=\frac{\Phi_0^2}{2R_e}
$$ 
for a large square 2D array. Here, $R_e$ is the 
effective voltage-bias resistance, i.e., the effective 
shunt resistance of each junction. At low temperatures $R_e$ is the subgap 
resistance which is many orders of magnitude larger than the normal state 
resistance, indicating the vortices can move through the medium with negligible 
damping. However, the simple model presented here does not take into account 
other sources of dissipation like the coupling to the low lying modes of the 
array (spin-waves) or quasi-particle tunneling (see next sections).

\underline{Lattice potential } -
The total Josephson energy associated to a vortex configuration (calculated 
for example by means of Eq.(\ref{eq:arctan})) depends on the vortex position.
The energy has a minimum value when the vortex is in the middle of a cell; the maximum value is reached when the vortex right on top of a junction. Neglecting vortex-vortex interaction and the influence of the array  
edges, vortices are only subject to a periodic lattice potential:  
$$
U_v(x) = \frac{1}{2} \gamma E_J \sin(2 \pi x) \;\; .
$$
Here, $\gamma$ is the energy barrier in units of $E_J$ a vortex has to overcome 
when moving from one cell to the next. 
In large 2D arrays with no self fields, 
$\gamma =0.2$ in a square geometry.  In a triangular arrays
the barrier is about a factor five lower, 
$\gamma =0.043$~\cite{lobb83}. 
Inclusion of self-field effects can be done and  $\gamma$
increases dramatically for $\lambda_{\bot}<1$~\cite{phillips93}.
In contrast, there is no energy barrier in 1D arrays if
$\Lambda_J>1$~\cite{bock94}.

\underline{Equation of motion for a single vortex} - 
A vortex in a Josephson array moves under influence of
a Lorentz force $\Phi_0 I $ in a direction perpendicular to
the current flow. The phenomenological damping term and the 
periodic lattice potential $U(x)$ provide additional forces.

Gathering all the ingredients discussed so the equation of motion 
can be written as:
\begin{equation}
   M_v \ddot{{\bf r}} \; + \; \eta \dot{{\bf r}} \; = 
   \; -{\bf \nabla }_{\bf r}U_v \; 
   - \; \Phi_0 \hat{{\bf z}} \times {\bf I}
\label{eq_mot}
\end{equation}
where ${\bf r} $ is the vortex position, ${\bf I}$ is the applied 
current per junction and $\hat{{\bf z}}$ is the unit vector perpendicular 
to the array~\cite{phunits}. 

The dynamics can be visualized as that of a massive particle moving
in a washboard potential analogous to the dynamics of a single junction in the  
RCSJ model. For the junction problem, the motion
is in artificial $\phi$-space; for the vortex problem the motion
is in real space. The mapping is exact for $2 \pi x \; \rightarrow \;\phi$ 
indicating that vortices in arrays produce the same dynamics as a 
single junction with a critical current of $\gamma I_c/2$ per junction, a 
McCumber parameter $\beta_{c,v}=\gamma \beta_{c}$ and a
plasma frequency $\omega_{p,v}=\sqrt{\gamma} \omega_{p}$.

Numerical evidence for massive vortices has been found by Hagenaars 
{\em et al.}~\cite{hagenaars96}. Their data show that a vortex may be 
reflected at an array edge, thereby changing its sign (i.e., it becomes an 
antivortex).
This behavior can be qualitatively understood within the model of a  
massive vortex interacting logarithmically with the image vortices
outside the array. In the same paper the authors also note that 
the way in which vortex inertia manifests itself depends on the dynamical 
situation considered. 

In the next three subsections, we first summarize the experimental details 
on classical arrays with $\beta_{c,v}>1$ and then discuss two phenomena that 
are not included in Eq.~(\ref{eq_mot}). First, experiments show that vortices 
in highly underdamped arrays experience more damping than can be expected from 
the simple approach we followed above. As it turns out coupling to spin waves 
becomes dominant in these arrays. Second, we review the theoretical and 
experimental results on the Hall effect. 

\subsubsection{Experiments on classical, underdamped arrays}

\begin{figure}
\centerline{{\epsfxsize=14cm\epsfysize=16cm\epsfbox{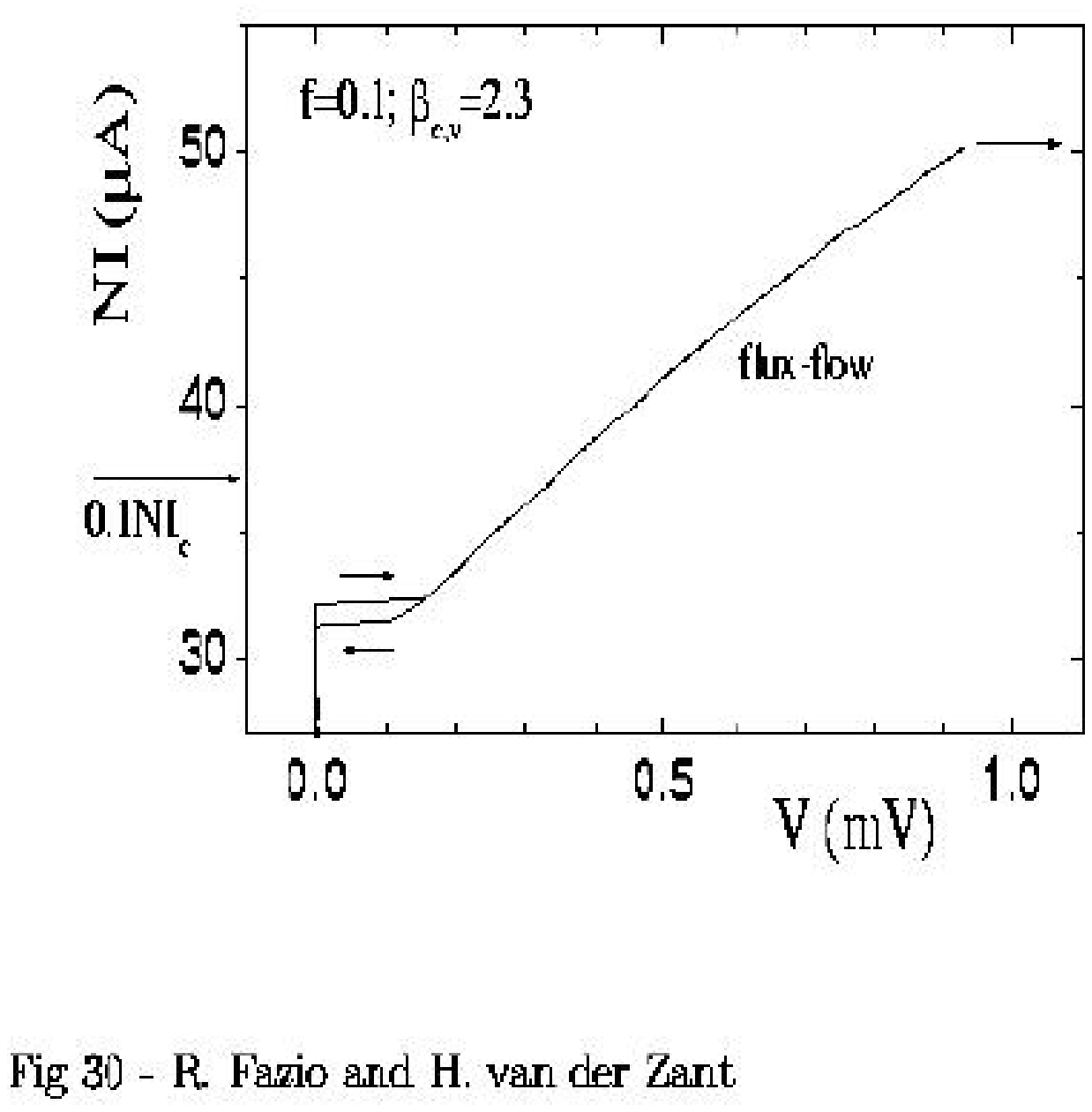}}}
\caption{A current-voltage characteristic of an underdamped 2D square, 
	aluminum array measured at low temperature ($10$~mK) in a magnetic 
	field of 0.1~$\Phi_0$  applied per cell ($f=0.1$).
	The arrow at the left indicates the expected depinning current of $0.1NI_c$ 
	with $N$ the number of junctions perpendicular to the direction of the current 
	flow. For small voltages hysteresis is seen. The flux-flow region is found above 
	the depinning current but below the current at which row switching sets in 
	(arrow at the left). (From Ref.\protect\cite{zant97}.)}
\label{underdamped}
\end{figure}

In Fig.\ref{underdamped} a typical example~\cite{zant91,zant97} of a current-voltage
$I$-$V$ characteristic of a classical 2D array is shown. The applied magnetic 
field corresponds to $f=0.1$. At low temperatures, hysteresis near the
depinning current indicates that $\beta_{c,v}>1$, consistent with the existence ]
of a mass term in the equation of motion. The depinning current itself is close 
to the expected value of $(\gamma/2)I_c=0.1I_c$ per junction.
From the analogy with the single junction problem, a RCSJ-like
$I$-$V$ characteristic is expected. 
This is generally not observed. The $I$-$V$ curves, instead, show 
a slight bending in the direction of the voltage axis opposite
to what is expected from the single-junction analogy.
This is also seen in simulations on
the properties of a single vortex in a 2D array with periodic 
boundaries~\cite{hagenaars94}.
Their numerical data points at a nonlinear viscous 
damping of the form:
$$
	\eta = \frac{A}{1+B v_{x}}
$$ 
where the $A$ and $B$ depend only on the McCumber parameter of the 
junctions. 

For currents well above depinning (above 50 $\mu$A and not visible in
Fig.\ref{underdamped}), the flux-flow state becomes unstable.
The $I$-$V$ enters a row-switched state~\cite{zant88} where
rows of junctions across the whole array width start to oscillate
coherently~\cite{phillips94}: all phases rotate 
continuously in time with a phase shift between them. 
In this regime, a description of the array in terms of 
vortex motion is no longer appropriate.

One should realize that the $I$-$V$ curve  was recorded at an applied
magnetic flux of 0.1~$\Phi_0$ per cell ($f=0.1$). Thus,
on average there is approximately one vortex per $1/f$ cells so that
the distance between vortices is only 
three cells. At such a short distance, vortices will interact with 
each other. The influence of these vortex-vortex interactions on the 
measured $I$-$V$ characteristics is not known in detail.

\subsubsection{Spin-wave damping}

As shown in Fig.\ref{underdamped} experimental $I$-$V$ characteristics 
in the flux-flow regime are generally more or less straight lines. Neglecting 
the influence of the
pinning potential $U_v$, Eq.(\ref{eq_mot}) indicates that for high bias the 
slope of this line should approach a conductance value corresponding to 
$1/(2fR_e)$ per junction. The $I$-$V$ curves therefore provide a way to 
estimate $R_e$ in 
the regime where vortices are driven with relatively large currents. 

A systematic study on highly underdamped arrays has been performed
by the Delft group~\cite{zant93}. The surprising result is that
for the most underdamped arrays $R_e$ is much lower
than the normal-state resistance. Such a low resistance cannot be explained
by the Bardeen-Stephen model. Apparently, vortices, 
when driven with a {\em large current}, experience
more damping than can be explained by ohmic dissipation alone.
A similar conclusion was drawn by Tighe {\em et al.}~\cite{tighe91}, who
concluded that in their underdamped arrays vortices moved in an overdamped 
manner. 

Several authors have suggested the possibility that energy can be lost
in the wake  of the moving vortex~\cite{zant91,zant93,tighe91}. 
The effective viscosity due to coupling to spin-waves can be calculated
in a semi-quantitative model~\cite{zant93} using the following argument: 
The oscillating part of the junction is modeled by an $L_J-C$ circuit
and the voltage drop across it is $V = \Phi_0 v \Delta \phi/(2\pi)$ which 
occurs for a time interval of the order of $v^{-1}$. As a response to this 
voltage step the phase difference starts to oscillate (note that at the same 
time the average phase advances when the vortex moves across the junction).
Equating the total power dissipated in the $L_J-C$ circuit to $\eta_{sw} v^2$, 
the result for the effective viscosity due to the plasma
oscillations (in square two-dimensional arrays) is
$$
\eta_{sw}=\frac{1}{\pi} \frac{\Phi_o^2}{2N} \frac{1}{\sqrt{L_J/C}} \;\;\; .
$$
When comparing this viscosity coefficient to the Bardeen-Stephen
viscosity coefficient, 
$$
\frac{\eta_{sw}}{\eta_{v}} \sim \frac{R_e}{\sqrt{L_J/C}}
$$
one sees that the more underdamped the arrays are, the more dominant 
the damping due to energy lost in the wake of the vortex becomes.

These observations have been confirmed by more systematic
calculations discussed in Section \ref{vaction}. In particular 
it is possible to obtain a self-consistent picture of vortex dynamics which 
includes the interaction with its 
environment. There are two main advantages for discussing the dynamics 
from this perspective. Firstly it is possible to evaluate quantum corrections
to the classical equation of motion and secondly it is possible to analyze the 
quantum dynamics in detail.

\subsubsection{The Hall Effect}
In addition to the Lorentz force which is due to the external current, a vortex 
is subject to a Magnus force which is transverse to vortex velocity. The study of 
the Magnus force in superfluids has a very long history. A detailed discussion
is outside the scope of this review (see Ref.~\cite{sonin97} and references 
therein). In Josephson arrays, in presence of a gate to the ground plane, 
particle-hole symmetry is broken.
A vortex feels a Magnus force~\cite{fazio92,fisher91} given by
\begin{equation}
        {\bf F} =  Q_x \Phi_0 \hat{{\bf z}} \times \dot{{\bf r}} \; .
\label{Magnus}
\end{equation}
Here we assumed for simplicity a homogeneous gate charge.
As a result of the combined effect of the Magnus force and
the Lorentz force, the vortices move at a certain angle, the Hall angle, 
with respect to the current. Its measurement yields informations on
the different dissipation sources in the system. Combining all the terms the 
equation of motion in the stationary limit the vortex moves at a constant 
velocity ${\bf v} = [v_x,v_y]= [v\cos \theta_H, v \sin \theta_H]$ obeying the 
following 
equations
\begin{eqnarray} 
	\eta v \cos \theta_H  & = & I_y \Phi_0 - Q_x v \Phi_0 \sin \theta_H
	\nonumber \\
	\eta v \sin \theta_H & = & Q_x v \Phi_0 \cos \theta_H
\end{eqnarray}
lead to the resistance tensor 
\begin{eqnarray}
	R_{xx} & = & R_{xx} = \frac{\Phi^2_0/ \eta}{1+ (Q_x  \Phi_0/ \eta )^2}
 	\nonumber \\
	R_{xy} & = & - R_{yx} = 
		\frac{Q_x\Phi^3_0/ \eta^2}{1+ (Q_x  \Phi_0/ \eta )^2}
\end{eqnarray}
and to the Hall angle $\theta_H$ 
\begin{equation}
	\tan \theta_H   =  \frac{Q_x  \Phi_0}{\eta }
\label{hall}
\end{equation}
The main consequence of the previous results is that the Hall effect should 
be larger in low resistance samples. 

In experiments on classical Josephson arrays the Hall angle is usually 
found to be very small (see e.g. Ref.~\cite{wees87}). Samples are usually 
characterized by random offset charges and as a
result the Magnus force averages to approximately zero. However, up to now 
there is no general agreement on this explanation. From a 
theoretical point of view there are questions related to the derivation 
of Eq.(\ref{Magnus}) from first principles. 

In Ref.\cite{fazio92} the Magnus force was obtained from the QPM, 
implying that only the  external charge enters in determining the Hall angle. 
A reexamination of the problem by Makhlin and Volovik~\cite{makhlin95} related 
the (apparent) absence of the Hall angle to the near exact cancellation of the 
Magnus force with the spectral-flow force. On deriving the effective action from 
the BCS Hamiltonian Volovik~\cite{volovik97} shows that the offset charges, 
contributing to the Hall angle, have two different physical origins. 
In addition to the one stemming from the 
coupling to the ground plane, there is an additional contribution which depends
on the particle-hole asymmetry of the spectrum. This latter term is of the order 
of the small factor $(\Delta/E_F)^2$, with $E_F$ being the Fermi energy. This 
confirms the expectation that the Hall angle should be small in Josephson 
arrays~\cite{halldis}. 

\begin{figure}
\centerline{{\epsfxsize=14cm\epsfysize=16cm\epsfbox{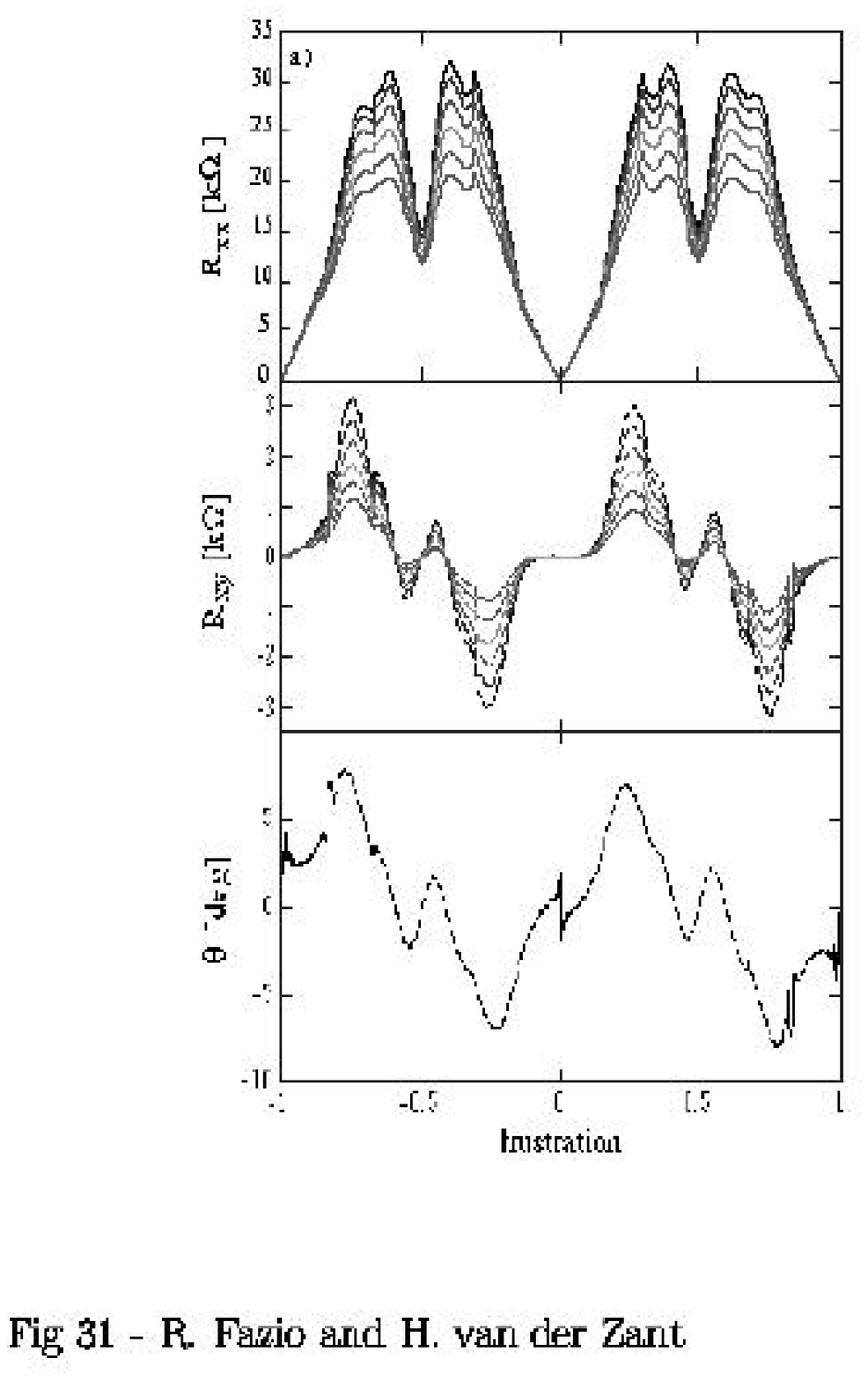}}}
\caption{Longitudinal resistance $R_{0xx}$ (a) and the Hall resistance $R_{0xy}$ (b), 
	and the Hall angle $\Theta$ (c) as a function of frustration. 
	$R_{0xx}$, and $R_{0xy}$ are shown 	for various temperatures ranging from  
	$T=20$ (top), 75, 100, 125, 150, 175~mK. 
	$R_{0xx}$ is symmetric around $f=0$ and $f=\pm 1/2$ whereas $R_{0xy}$ changes 
	sign upon passing through these frustrations. 
	(From Ref.~\protect\cite{delsing97}.)}	
\label{hall-exp}
\end{figure}

In {\em quantum} Josephson arrays Hall measurements have been performed by the 
Chalmers 
group~\cite{delsing97,chen95a}. Their results are shown in Fig.\ref{hall-exp}.
The transverse resistance is odd in the magnetic 
field. Combining the results for the longitudinal and transverse part the 
Hall angle can be extracted. Comparing the 
results with the classical results~\cite{wees87}, 
the Chalmers experiments indicate a larger Hall angle. The only apparent 
difference is the smaller ratio $E_J/E_C$ and this is consistent with the 
theoretical expectation that the offset charges are responsible for the 
Hall effect. It is reasonable to expect that these have a negligible effect 
on approaching the classical limit.

Finally, it is worth mentioning that the field-tuned transition discussed 
previously 
also manifests itself in $R_{xy}$. An interesting feature which still remains 
unexplained is that $R_{xx}$ and$R_{xy}$ are related by the following empirical 
relation
\begin{equation} 
	R_{xy} \sim \frac{\partial R_{xx}}{\partial f } 
\end{equation}
similarly to what happens in the Quantum Hall effect.

\subsection{Ballistic vortex motion}
\label{ball}
Besides the experimental verification of the mass term in the equation of 
motion~\cite{zant91,tighe91} a considerable interest was focused on the 
direct observation of the ballistic motion.
Ballistic vortex motion has not only been observed in long continuous
junctions~\cite{matsuda83}, where energy barriers for cell-to-cell motion and 
spin-wave coupling are absent, but also in discrete 1D 
arrays~\cite{fujimaki87,zant95,watanabe95} 
and in 2D aluminum arrays~\cite{zant92b}. 
The idea goes as follows: If vortices are massive particles, they should keep on 
moving if the current is turned off. In an experiment, this concept can be 
realized by accelerating vortices up to a high velocity $v_0$ so that their 
kinetic energy is much larger than the lattice potential.
With Eq.(\ref{eq_mot}) one finds that $v_0 \approx \Phi_0 I/ \eta$
if one neglects the lattice potential. 
Then, these fast-moving vortices can be launched into a force-free environment 
where voltages probes can be used to detect their path through this region.

The observation of ballistic motion is only possible in a velocity 
window ($ v_{min} < v < v_{max}$) bound from below by the presence of the pinning 
potential and from above by the various damping mechanisms which set in at high 
velocities.
The criterion $E_{kin} \geq E_{pot}$ translates into a lower bound for the vortex 
velocity required to observe ballistic motion
\begin{equation}
       v_{min} = \frac{\sqrt{\gamma} \omega_{p}}
                                {\pi} \; .
\label{u_min}
\end{equation}
Note, that for a 1D system with $\Lambda_J>1$, $\gamma \approx 0 $ 
so that the minimum vortex velocity is small.

The vortex velocity cannot be chosen arbitrarily
large. Fast moving vortices can trigger row
switching in the array~\cite{zant88,nakajima81}. 
Simulations~\cite{bobbert92} indicate that in 2D arrays the vortex
velocity must be limited to $v <\omega_p$. 
Another limitation comes from coupling to spin-waves.  
In 2D arrays, there is a threshold vortex velocity 
below which this coupling is weak. It has been 
shown~\cite{geigenmuller93,eckern93} that a moving vortex
only couples to spin-waves above $v_{max} \approx 0.1 \omega_p$. 
The requirement that 
$$
	\frac{\sqrt{\gamma} \omega_{p}}{\pi} < v  < 0.1 \omega_p
$$
indicates that ballistic motion is possible in triangular 
arrays just above depinning. (Quantum fluctuations make the window larger; 
details are discussed in the next section.)
 
Let us analyze the ballistic motion using Eq.~(\ref{eq_mot}).
With no current applied and for vortices launched at high enough velocity 
(to ignore 
pinning effects), the equation of motion reduces to 
$$
M_v {\dot v_x} \; + \; \eta v_x \; = \; 0 \;\;\; ,
$$
indicating that the vortex velocity decreases exponentially in time as
$v_x = v(0) \exp[-M_v t /\eta]$. A mean free vortex path can then be 
defined as
\begin{equation}
    \lambda_{free} \; = \; \frac{v(0)M_v}{ \eta} \; 
    = \; \pi^{-1} \frac{I} {I_c} \; \beta_{c} \; \; ,
\end{equation}
for a square 2D array.
The factor $I /\pi I_c$ is typically of order 0.1 so that at
high temperatures with $R_e=R_N$, $\lambda_{free} \approx 1$. 
At low temperatures with  
$R_e \gg R_N$ (the corresponding $\beta_c$ can be high as $10^7$), 
$\lambda_{free} \gg 1$. 

\begin{figure}
\centerline{{\epsfxsize=14cm\epsfysize=16cm\epsfbox{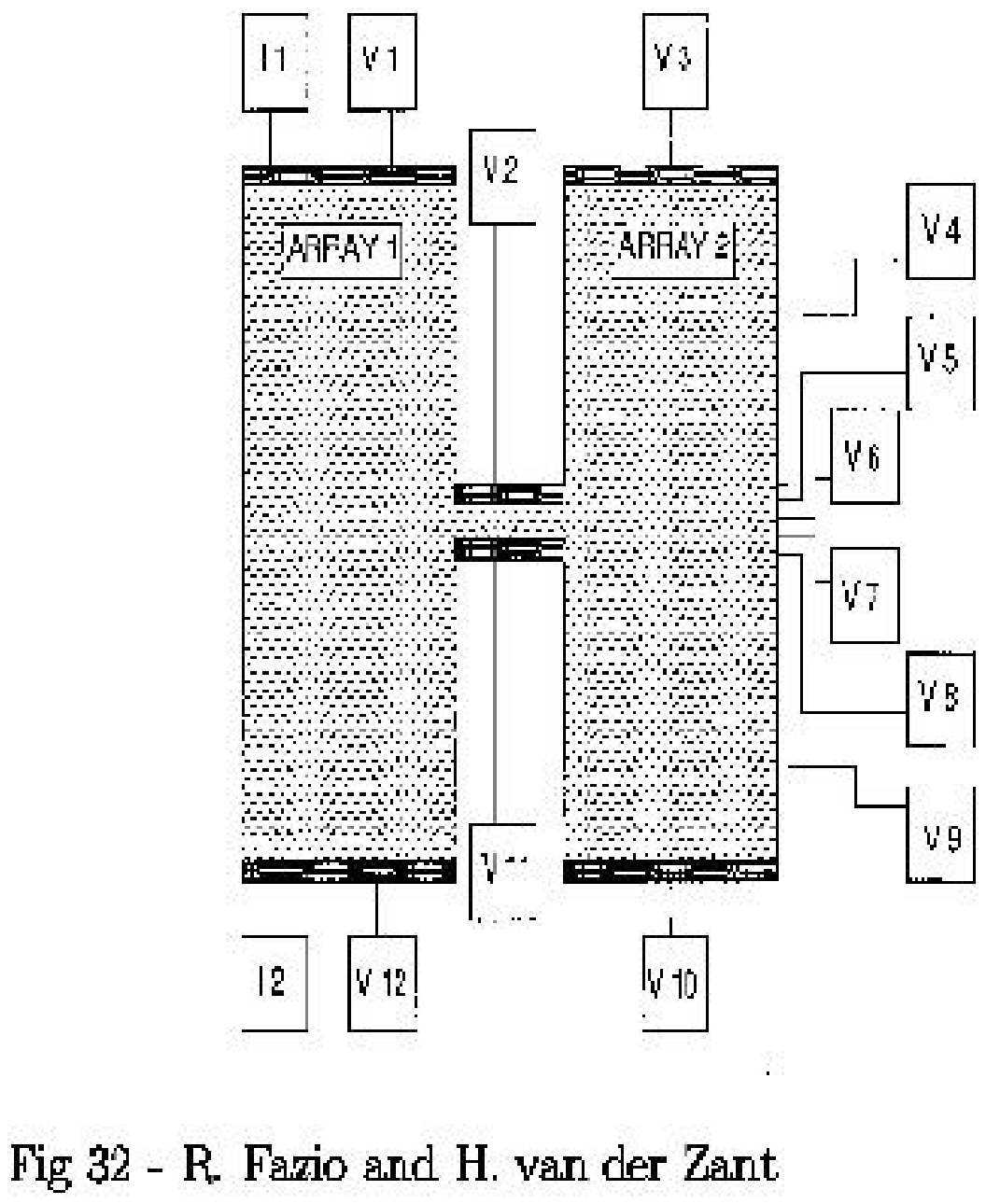}}}
\caption{ Sample lay-out used to measure ballistic vortices in 2D Josephson arrays. 
	(From Ref.~\protect\cite{zant92}.)}
\label{H-sample}
\end{figure}

The experiment with 2D arrays was performed by van der Zant 
{\em et al.}~\cite{zant92} using the sample configuration shown in 
Fig.\ref{H-sample}. 
It consists of two 2D arrays which are connected
by a narrow channel of 20 cells long and 7 cells wide.
Superconducting banks on both side of the channel confine the
vortices in the channel and, in order to reduce the influence
of the lattice potential, the arrays and the channel
were made in a triangular geometry.

\begin{figure}
\centerline{{\epsfxsize=14cm\epsfysize=16cm\epsfbox{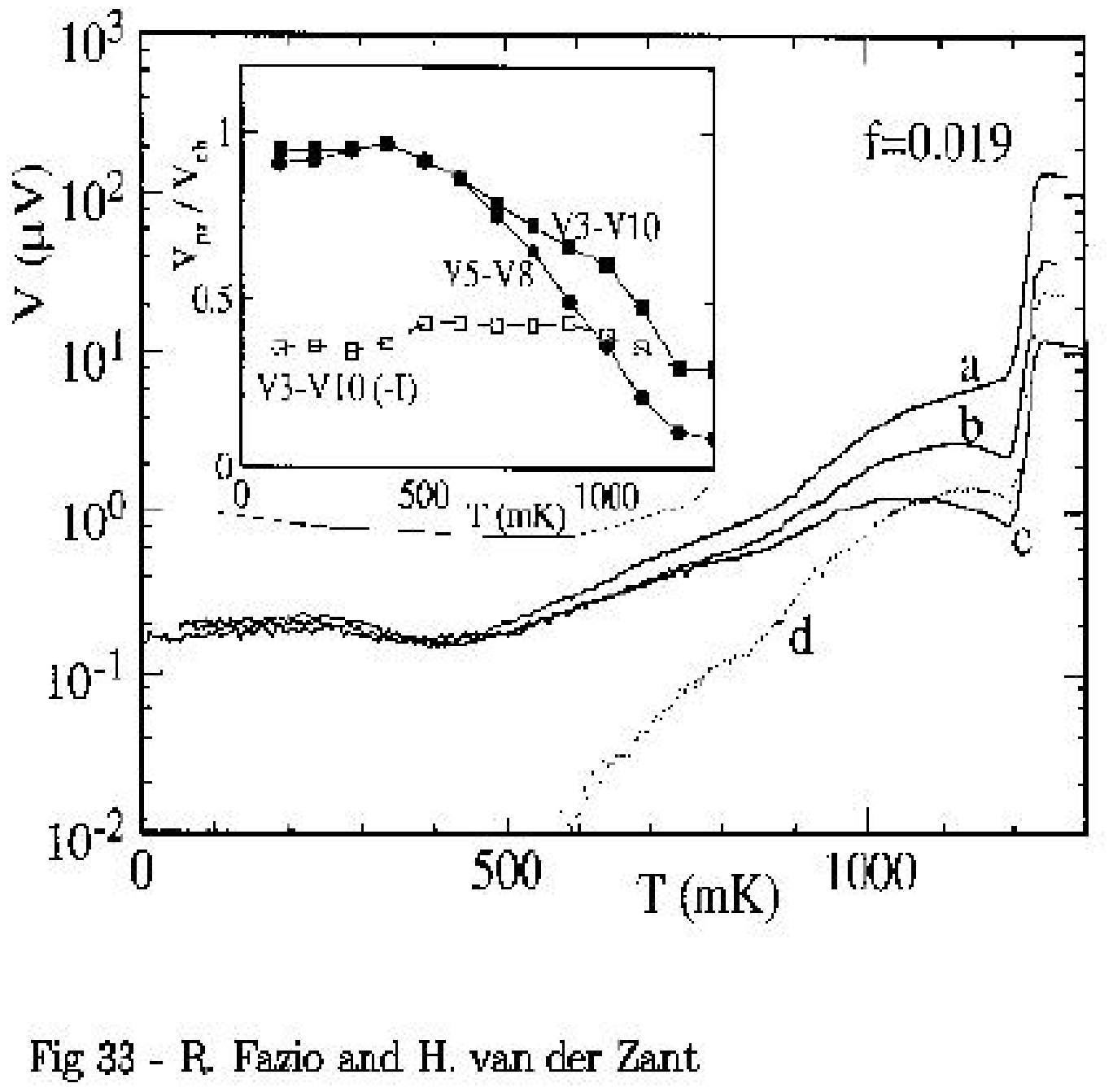}}}
\caption{Voltage vs. Temperature measured across the channel (a), between probes V3 and V10 (b), 
	between probes V5 and V8 (c), and between probes V3 and V4 (d). In the inset voltages 
	are plotted in units of the voltage across the channel. The dashed line corresponds to 
	the voltage across V3-V10 when the current direction is reversed.
	At low temperatures, the voltage across the two probes opposite from the channel 
	is almost equal to the voltage across the channel: all vortices that go through 
	the channel leave the array between V7 and V8. With reversed current direction 
	vortices are accelerated in the opposite direction and no ballistic motion is observed.
	(From Ref.~\protect\cite{zant92}.)}
\label{Vdrops}
\end{figure}

In the array on the left hand side, vortices generated by a small magnetic field,
are accelerated up to a high velocity. Some of these high-energetic
vortices will enter the channel and will then be launched
into the detector array (array on the right hand side). 
There is no driving current applied
to this array. A set of voltage probes around the detector array 
is used to detect the places where vortices leave the
force-free environment. 
(The voltage measured across two probes is proportional to the
number of vortices passing the probes per unit time.)
The results of the local voltage drops as a function of temperature it is shown 
in Fig.\ref{Vdrops}.
At high temperatures, vortices move diffusively and voltages are observed
between all voltage probes. As a consequence, the voltage across the probes V3 
and V10 is much smaller than the channel voltage. 

At low temperatures, the voltage measured between the two probes situated
just opposite to the channel is almost equal to the channel
voltage. Vortices  cross the second array in a narrow beam 
(see Fig.\ref{H-sample}).
This ballistic vortex motion is observed for $T<500$~mK, for small applied 
magnetic fields ($0.01<f<0.025$) and for currents just above depinning. For high
magnetic fields, vortex-vortex interactions start to play a role
when more than one vortex is in the channel at the same time and
for too high currents coupling to spin-waves probably starts to play
a role.

\subsection{Effective single vortex action}
\label{vaction}
In this Section we derive a more general approach to describe 
the single vortex dynamics  which incorporates the coupling to
spin-waves and which is also valid in the quantum regime where $E_J
\approx E_C$~\cite{otterlo94a,fazio94,otterlo94b}. After introducing the 
effective vortex action, we first consider the classical 
limit and show that all known results can be recovered. We then proceed to 
the quantum regime. The vortex mass is calculated as well as the velocity 
above which spin-wave damping starts to become effective.

As discussed in Appendix~\ref{single-vortex}, the effective action for a 
single vortex is given by
$$
	S_{eff}=\frac{1}{2} \sum_{a,b=x,y}
	\int d\tau d\tau' \dot{r}^{a}(\tau) M_{ab}[{\bf r}(\tau)-
	{\bf r}(\tau'),\tau-\tau']\dot{r}^{b}(\tau') ,
$$
\begin{equation}
	M_{ab}=\sum_{jk} \nabla_{a}\Theta({\bf r}(\tau)-{\bf r}_{j}) 
	\langle q_{j}(\tau)
	q_{k}(\tau')\rangle \nabla_{b}\Theta({\bf r}_{k}-{\bf r}(\tau'))
\label{eq:effect}
\end{equation}
($\Theta({\bf r}) =arctan(y/x)$).
Thus, {\it vortex dynamics is governed by the charge-charge 
correlation}, which depends on the full
coupled charge-vortex gas. The effective action
Eq.(\ref{eq:effect}) describes dynamical vortex properties in the whole 
superconducting region  and is therefore a good starting point for the 
investigation of vortex properties down to the S-I transition. 
The cumulant expansion that leads to Eq.(\ref{eq:effect}) is correct in 
the $E_{J}\gg E_{C}$ limit where the charges can be 
considered as continuous variables and where the vortex fluctuations can be 
disregarded.
In general the average defined in Eq.(\ref{eq:1ststep}) is far from
being gaussian so that one may argue that higher order cumulants should
be considered. Nevertheless nothing prevents us to analyze Eq.(\ref{eq:effect}) 
keeping in mind that a full description of vortex motion may require the 
analysis of a dynamical equation that
contains also terms proportional to higher powers of the vortex velocity.

The expression given in Eq.(\ref{eq:effect}) reproduce the known results
in the classical limit where $E_{J}\gg
E_{C}$. In this region of the phase diagram the charges may be considered
to be continuous variables and vortex fluctuations may be neglected so that
the charge-charge correlation reads 
$$
	\langle q q \rangle_{k,\omega_{\mu}}=
	E_{J}k^{2}\frac{1}{(\omega^{2}_{\mu}+\omega^{2}_{k})} 
$$
and 
$$	\omega^{2}_{k}= \frac{4e^2E_J}{C_0}\frac{k^2}{1+\lambda^2k^2} \;\;\; .
$$ 
The spin-wave dispersion is described by $\omega_{k}$. It is optical,
i.e. $\omega_{k}=\omega_{p}$, for long range Coulomb interactions,
whereas
for on-site interactions we have $\omega_{k}=\bar{\omega}_{p} k$.
Here $\bar{\omega}_{p}=\sqrt{4e^2E_{J}/C_{0}}$ is the plasma frequency
for the case of on-site Coulomb interactions.

The action (\ref{eq:effect}) reduces to that of a free particle in the
limit of small velocities $\dot{r}(\tau)$.
The corresponding adiabatic vortex mass $M_{v}$ is
$$
	M_{v}=\int_0^{\beta} d\tau M_{xx}(0,\tau) \; ,
$$
which reduces in the classical limit to 
$$
M_{v}= \frac{\pi^{2}}{4E_{C}}+\frac{\pi C_0}{4e^2} \ln(L)
$$
with $L$ the array dimension. Thus both $C_0$ and $C$
yield a contribution to the mass. The self-capacitance contribution
depends on the system size $L$. For generic sample
sizes and capacitance ratio's the size-dependent contribution is 
smaller than the Eckern-Schmid mass (the value in Eq.~(\ref{mass_2D})). 
The effect of a uniform background charge on the 
vortex mass was considered by Luciano {\em et al.}\cite{luciano95,luciano96}. 
The frustration  of charging leads to a renormalization of the mass towards the 
classical value.

The spin-wave damping that a moving vortex experiences may also be
calculated from Eq.(\ref{eq:effect}). Varying the vortex coordinate 
$r^{a}(\tau)$ in
Eq.(\ref{eq:effect}) yields the equation of motion 
\begin{equation}
	2\pi \epsilon_{ab} I_{b}/I_{c}= \partial_{\tau}
	\int d \tau' M_{ab}(r(\tau)-r(\tau'),\tau-\tau')\dot{r}^{b}(\tau')
\label{eq:eom}
\end{equation}
($\epsilon_{xx}=\epsilon_{yy}=0,\epsilon_{xy}=-\epsilon_{yx}=1 $).
Its constant velocity solutions in the presence of an external current determine
the non=linear relation between driving current and vortex velocity (i.e. the 
current-voltage characteristics), once
the charge-charge correlation is analytically continued to real frequencies 
(i.e., if $i\omega_{\nu}\rightarrow \omega +i \delta$)~\cite{geigenmuller93}.
The relevant information is contained in the real part of Eq.(\ref{eq:eom}), which
reads in Fourier components and for a constant vortex velocity
\begin{equation}
	I^{y}/I_{cr}= \frac{v}{4} \int d\omega\int d^{2}k
	\frac{k^{2}_{y}}{k^{2}}
\delta(\omega-vk_{x})[\delta(\omega-\omega_{k})+\delta(\omega-\omega_{k})]\; .
\label{eq:reom}
\end{equation}
The delta functions express the spin-wave dispersion (from the analytic 
continuation of the charge-charge correlation) and
the vortex dispersion respectively. The overlap integral determines
the amount of dissipation a moving vortex suffers from coupling to
spin-waves. Adopting the smooth momentum integration cut-off, 
introduced in Ref.\cite{eckern89}, one recovers in the classical
limit the results of Refs.\cite{geigenmuller93,eckern93}. 

While the static damping is zero for vortex velocities below threshold 
(which implies the possibility of ballistic motion), a dynamical friction due 
to the 
coupling to the plasma oscillations is always present for frequencies higher 
than a 
given frequency threshold~\cite{luciano97}. The latter contribution approaches 
to zero when the velocity increases to the threshold velocity. 
However, radiative dissipation of the vortex affects the
threshold for ballistic motion.  What is important in this analysis is that by 
changing the frequency of the applied current, one is able to extract the 
domain of validity where a vortex can be defined as a macroscopic object.

Inclusion of quantum effects contributes to the opening
of a more robust velocity interval where ballistic motion can be
observed. When the ratio $E_{J}$/$E_{C}$ decreases the charge-charge correlation
must be  calculated beyond the classical approximation: The discreteness of 
charge transfer has to be taken into account, which is in particular important 
at short distances. 
For long range Coulomb interactions and in the absence of vortex 
fluctuations the charge-charge correlation function may be rewritten as
\begin{equation}
	\langle q q \rangle_{k,\omega_{\mu}}=
	\frac{k^{2}E_{J}}{\omega^{2}_{\mu}+\bar{\omega}^{2}_{k}} ,
	\;\;\bar{\omega}^{2}_{k}= \omega^{2}_{k} + 4\pi^{2} E_{J}\xi^{2}
	k^{2}.
\label{eq:aha}
\end{equation}
where  the correlation length is
$$
	\xi^{2} \sim \sqrt{\frac{E_{C}}{E_J}}\left(\pi
	e^{-\sqrt{E_{C}/E_J} c/\pi} \right)^{\frac{1}{1-\delta}} \!\!,
	\;\; \delta=\sqrt{\frac{E_{C}}{E_J\pi^2}}
$$
and the constant $c$ is of order one. The S-I phase transition 
takes place at  $E_{J}/E_{C}=1/\pi^{2}$. Thus, without
vortex fluctuations the phase transition is at a smaller $E_{J}/E_{C}$
value than the $2/\pi^{2}$ that follows from a duality argument \cite{fazio91a}.
The spin-wave dispersion is affected at small distances (large $k$) and the 
mass is now given by:
$$
	M= \frac{\pi }{16E_C\xi^{2}}\ln\left[
	1+4\pi\xi^{2}\right]
$$
In the limit of small $\xi$
the Eckern-Schmid mass is recovered. An extrapolation to the S-I
transition where $\xi \rightarrow \infty$ yields a mass that vanishes at the
transition.

With the charge charge correlation given in Eq.(\ref{eq:aha}) we may
calculate the spin-wave damping of vortex motion due to the coupling to spin-waves
beyond the classical limit. Replacing $\omega_{k}$ by $\bar{\omega}_{k}$ in
Eq.(\ref{eq:reom}),
the overlap integral over the delta functions only contributes for
vortex velocities
that are higher than a threshold velocity
$$
	v_{max} \sim \xi\sqrt{8 E_{J}E_C}
$$ 
By taking into account quantum effects the spin-wave spectrum enlarges
the velocity window in which vortices move over the lattice
potential without emitting spin-wave.
An extrapolation to the S-I transition yields a diverging threshold
Velocity so that vortices and spin-waves decouple.
Note, that the outcome of this calculation also has 
consequences for the classical equation of motion. For instance, 
it shows that coupling to spin-wave dissipation is reduced for velocities
$< 0.5 \omega_p$. This value is a factor five higher than
expected from classical considerations~\cite{geigenmuller93,eckern93}. 

Spin-wave damping is not the only source of dissipation. Vortices, in their 
motion, can excite quasi-particles as well if the local voltage drop (due to the 
finite velocity) exceeds the quasi-particle gap.
The effect of quasi-particles damping on vortex motion was considered in 
Ref.~\cite{choi98b}. In this case the equation of motion takes the form 
\begin{equation}
\ddot{{\bf r}}
+ \eta \dot{{\bf r}}
- \eta \Delta
  \int^tdt'\;{\rm{sinc}}[2\Delta(t-t')]
  \frac{\dot{{\bf r}}(t')}{\pi|{{\bf r}}(t)-{{\bf r}}(t')|^2}
= - \Phi_0 \hat{{\bf z}} \times {\bf I}
\end{equation}
where ${\rm{}sinc}x\equiv(2/\pi{}x)\sin{x}$. Despite the nonlinear form of the 
damping kernel, Choi {\em et al.}~\cite{choi98b} showed that the frictional 
force on a vortex is linear in the vortex velocity for any practical 
purpose ( in particular in the long-time and long-wavelength limit, 
where the semiclassical equation of motion is mostly concerned).

\subsection{Quantum vortices}
\label{quantumvortex}
If vortices are massive particles that move ballistically,
one can think of them as quantum mechanical objects.
Like an electron, a vortex in a periodic potential will have a
Bloch wave function with momentum $p = \hbar k$ and thus a wavelength of $h/vM_v$.
At present, many experiments have verified the concept of a quantum 
vortex~\cite{zant96,chen96,zant91,tighe91,oudenaarden96a,oudenaarden96b,oudenaarden98}.
In this chapter, we discuss three examples: macroscopic quantum tunneling of 
vortices~\cite{zant96,chen96,zant91,tighe91}, the observation of vortex 
interference in a hexagon-shaped array~\cite{elion93}
and Bloch oscillations of vortices in the periodic lattice 
potential~\cite{oudenaarden96}.

\subsubsection{Macroscopic quantum tunneling of vortices}

In a classical description vortices oscillate in the 
minima of the washboard potential with frequency $\omega_{p,v}$.
In quantum arrays, these oscillations are
quantized. To estimate when quantum fluctuations in the vortex
position become important, we compare the zero-point energy
$\frac{1}{2} \hbar \omega_{p,v}=
\frac{1}{2}\sqrt{8\gamma E_JE_C}$ to the 
energy barrier $U_{\rm bar}=\gamma E_J$.
The two energies are equal if $E_J/E_C = \sqrt{2 / \gamma}$.
In this quantum vortex regime, the zero-point fluctuations are large enough 
to allow for quantum tunneling of vortices. 

\begin{figure}
\centerline{{\epsfxsize=14cm\epsfysize=16cm\epsfbox{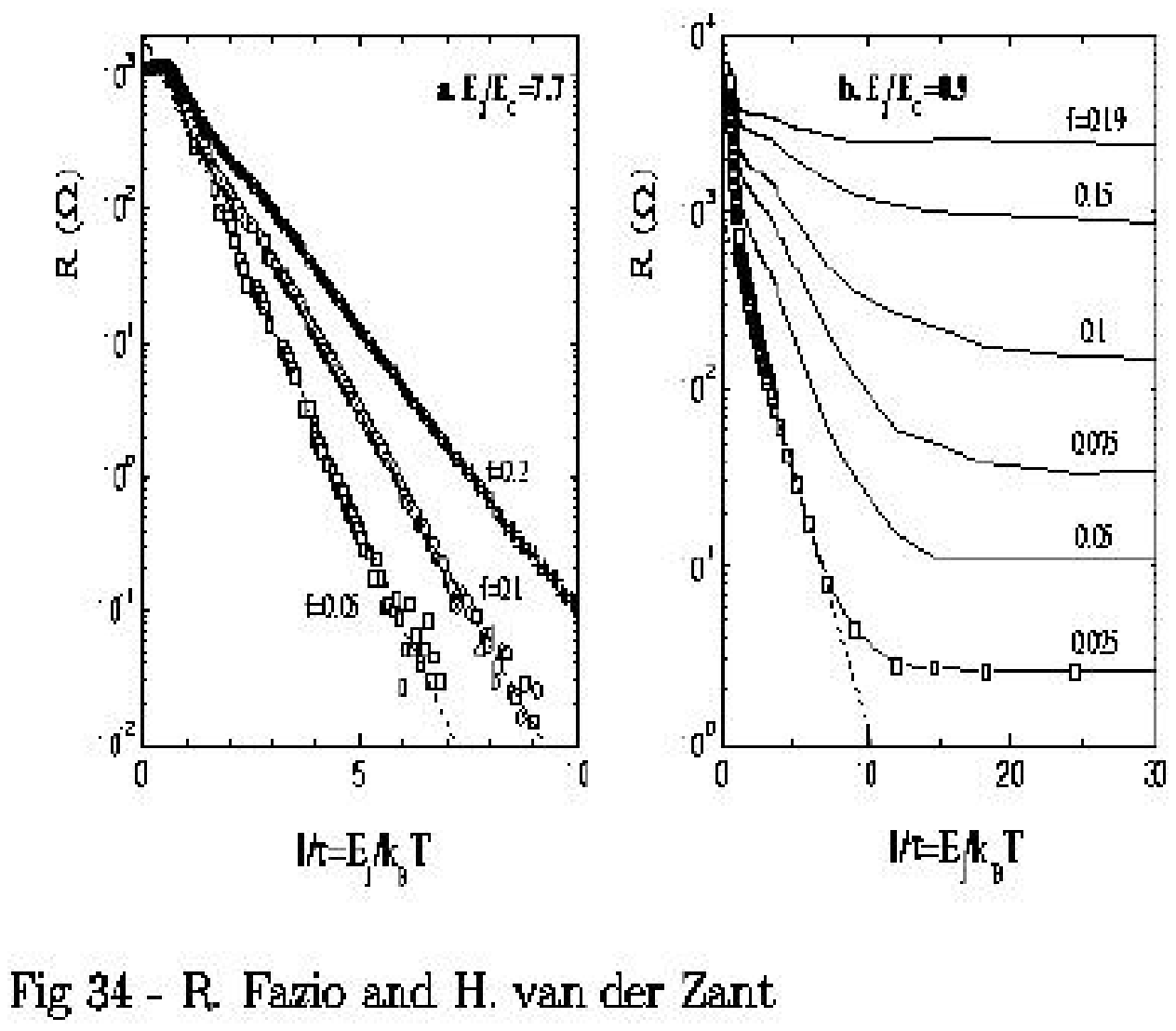}}}
\caption{Measured zero-bias resistance per junction versus the inverse normalized temperature 
	measured for two different square arrays.
	At low temperatures the resistance of the sample with the smaller $E_J/E_C$ ratio (b) 
	is temperature independent indicative for quantum tunneling of vortices. 
	(From Ref.\protect\cite{zant97}.)}
\label{mqt1fig}
\end{figure}

In Fig.\ref{mqt1fig}, the resistance per junction (linear response)
as a function of temperature is given for two square arrays~\cite{zant96}; 
in the classical (a) and in the quantum regime (b). The resistance  of the 
classical array decreases exponentially all the
way down to the lowest temperatures. The slopes define
the barrier for this thermally activated process:
$R \propto \exp{[-U_{\rm bar}/k_BT]} = \exp{[-\beta \gamma E_J]}$. 
In contrast, the resistance of the quantum array levels
off at a temperature $T_{cr}$ below which it
remains constant. We denote this constant value with $R_c$. 
Above $T_{cr}$, again thermally activated
behavior is observed. Similar resistance curves have been reported by the
Chalmers group~\cite{chen96}.

A first estimate of the tunnel rates and of  $R_c$ can obtained from the analogy 
with the single-junction problem.
In the moderate damping regime\cite{martinis87}:
\begin{equation}
       R_c \approx 140 R_Q f \sqrt{S} {\rm e}^{-S}
\label{eq:rc}
\end{equation}
where the action $S$ is given by
\begin{equation}
       S = \sqrt{0.95 \frac{ \gamma E_J}{E_C}} \;
           \left( 1 + \frac{0.87}{\sqrt{\beta_{c,v}}} \right) \; .
\label{eq:s}
\end{equation}
The $\sqrt{E_J/E_C}$ dependence of the critical resistance, implied by the 
previous equation, has indeed been reported by the Chalmers group~\cite{chen96}. 

An estimate for $T_{cr}$ can be obtained by equating $S$ 
to $\gamma E_J/T_{cr}$.
Neglecting the term with $\beta_{c,v}$, the result is:
\begin{equation}
     E_J/T_{cr} = \frac{2.5}{\sqrt{\gamma}} \; 
                    \sqrt{ \frac{E_J}{E_C}} \; .
\label{rate}
\end{equation}
With $E_J$ of the order of
$E_C$, the inverse normalized critical temperature ($E_J/T_{cr}$) is typically 
somewhat larger than 2.5 in agreement with the data.

Comparing the measured values of $R_c$ with the 
estimates given in Eq.(\ref{eq:rc}) and Eq.(\ref{eq:s}), the tunnel rates in the
measurements are lower than
expected even when taking $R_N$ as the resistance determining
$\beta_{c,v}$.
A smaller $R_c$ 
is consistent with a single vortex model in which the vortex mass is an
order of magnitude larger than the one calculated in the
quasi-static approximation. It is likely that vortices do not
move as rigid objects and calculations have shown
that the dynamic band mass of a vortex can be an order of
magnitude larger \cite{geigenmuller91}.
However, considering this uncertainty, no definite conclusion about the 
validity of the single-junction model can be drawn for the
observed flattening of the resistance. Other models like 
collective tunneling cannot be excluded.

A surprising result is that the array (a) in Fig.\ref{mqt1fig} does not 
show any signature of quantum tunneling. Our simple argument
given above indicates that for this array the zero-point
energy is of order $U_{\rm bar}$. The absence of
quantum tunneling is explained by the fact that
in Fig.\ref{mqt1fig}  the measured energy barriers are of order $E_J$, 
instead of $0.2E_J$. The Delft group has reported~\cite{zant96} a systematic 
increase of the measured energy barrier in the range $2<E_J/E_C<20$.
This increase is not yet understood.  

The theoretical analysis of inter-site vortex tunneling in Josephson arrays 
was first formulated by Korshunov~\cite{korshunov87,korshunov88} who evaluated 
the instanton action $S_{inst}$ associated to vortex tunneling between adjacent 
plaquettes. $S_{inst}$, related to a hop from one plaquette to a neighboring one, 
determines tunnel  rates and the depinning current. It can be obtained in the 
language of the Coupled Coulomb gas approach~\cite{fazio91b} by evaluating 
the action associated with the trajectory 
$$
\dot{v}_{i,t}=
v_{i,\tau+\tau_{\epsilon}}-v_{i,\tau}=\delta_{\tau,t}
[\delta_{i,x+1}-\delta_{i,x}]
$$ for a hop from
$x,\tau \rightarrow x+1,\tau+\tau_{\epsilon}$, with the result 
(see Eq.(\ref{eq:effect}))
$$
	S_{inst}=\frac{1}{2} M_{xx}(0,0) .
$$
In the limit of large Josephson coupling one recovers all known results, i.e.
for general capacitance matrix
\begin{equation}
	S_{inst}=\frac{\pi E_{J}}{4\omega_{p}}
\left[\sqrt{\pi}\sqrt{\lambda^{2}+4\pi}+
	\frac{\lambda^{2}}{2} \ln\left(\frac{2\sqrt{\pi}}{\lambda}+
	\sqrt{1+\frac{4\pi}{\lambda^{2}}}  \right) \right] \;\;\ .
\label{eq:act}
\end{equation}
It reduces to $S_{inst}=\pi^{3/2}E_{J}/4\bar{\omega}_{p}$ and
$S_{inst}=\pi^{2}E_{J}/2\omega_{p}$ for $C=0$ and $C_{0}=0$
respectively.

Korshunov pointed out that instanton-instanton  interaction cannot be neglected. 
Vortex tunneling is incoherent in the temperature range 
$E_J \gg T \gg \sqrt{E_J E_0}$. 
In this case the tunneling probability $W$ is given by
$$
	W \sim \frac{E_J^{1/2}e^{-S_{inst}}}{T^{3/2}} e^{2\pi \ln2 T/E_J}
$$
up to the crossover temperature $T \sim\sqrt{E_J E_0}$ where the activated 
behavior takes place. 

Dissipation associated to vortex tunneling was discussed by Ioffe and 
Narozhny~\cite{ioffe98}. Since the  the time associated to vortex tunneling
is slow compared to $\Delta ^{-1}$, the dissipation which accompanies 
this process arises from rare processes when a vortex excites a quasi-particle 
above the gap. These authors  find that this source of dissipation can be 
significant even in the adiabatic limit.

\subsubsection{Vortex interference: the Aharonov-Casher effect}

In 1984 Aharonov and Casher~\cite{aharonov84} studied
the interference of particles with a magnetic moment moving 
around a line charge. This AC effect is the dual of the 
Aharonov-Bohm effect~\cite{aharonov59} that describes quantum interference
of charged particles moving around magnetic flux.
The AC effect has first been observed using neutron 
beams in Ref.~\cite{cimmino89}. 
The concept of vortex-interference in superconductors  
has been introduced by Reznik and Aharonov~\cite{reznik89} and
their ideas have been adapted to ring-shaped Josephson arrays by
van Wees~\cite{wees91} and by Orlando and Delin~\cite{orlando91a}. 
Although there are similarities with the
conventional AC effect, there are important differences. In JJAs vortices
do not carry a flux unlike the Abrikosov vortices in a bulk superconductor.
Moreover, as already stressed, JJAs form an artificial 2D space.
The observation of the AC effect for vortices 
has been reported by Elion {\em et al.}~\cite{elion93}, only two years after the 
first theory papers~\cite{wees91,orlando91a} appeared.

\begin{figure}
\centerline{{\epsfxsize=14cm\epsfysize=16cm\epsfbox{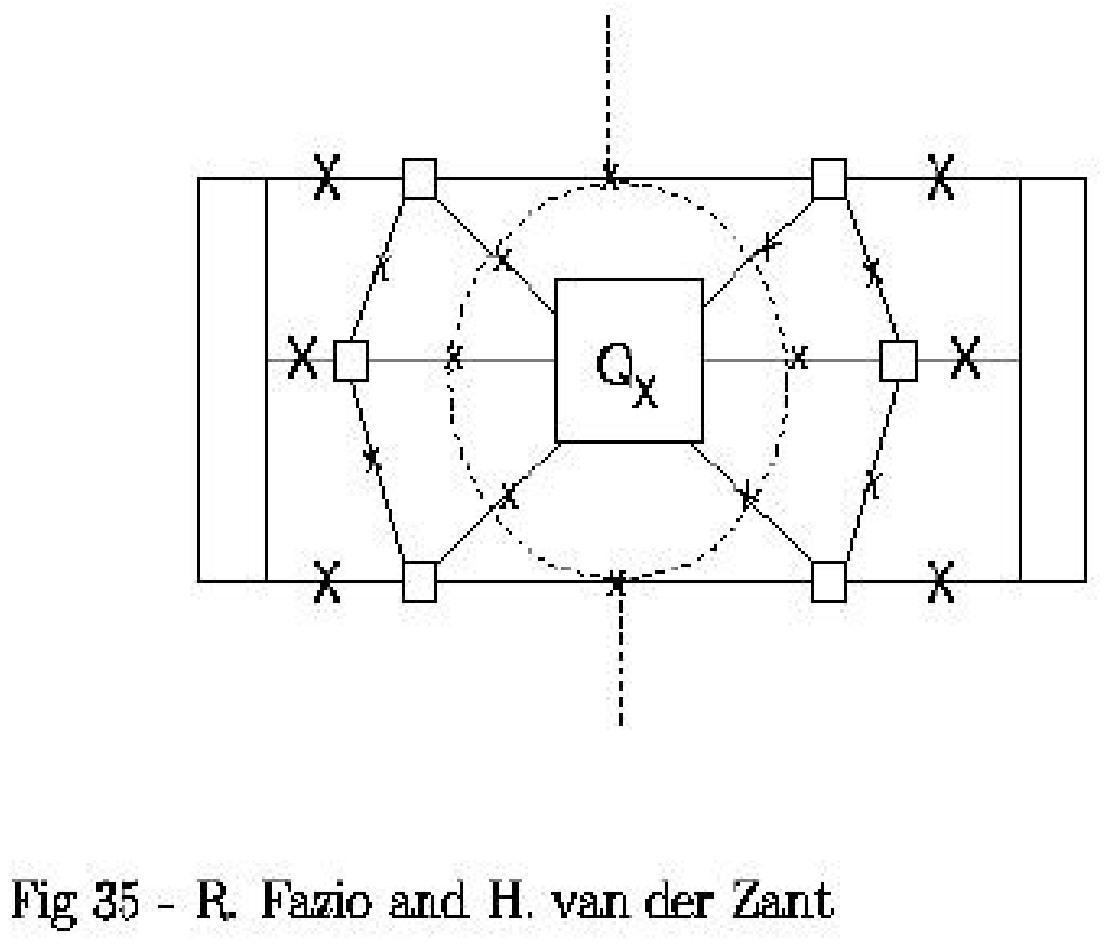}}}
\caption{Schematic drawing of the hexagon-shaped Josephson array to measure vortex 
	interference. (From Ref.~\protect\cite{elion93}.)}
\label{fig_ac}
\end{figure}

In the array of Elion {\em et al.}, vortices follow trajectories indicated as 
dotted lines in the Fig.\ref{fig_ac}.
The sample consists of a hexagon-shaped array with six triangular cells. 
Large-area junctions couple the hexagon 
to superconducting banks so that only two paths for vortex 
motion are possible. The large-area junctions confine vortices
to the hexagon, but the coupling to the superconducting
banks is not so strong that phases of the islands are set
by the banks. A gate controls the charge on the superconducting
island in the middle of the hexagon. 

In the experiment the differential resistance in the
flux-flow regime has been measured as a function of the gate voltage. 
When fixing the current through the array, clear oscillations
in the differential resistance are observed. The measured period, however,
corresponds to $e$, half the value expected from theory.
The factor of two arises from tunneling of quasi-particles which
effectively limit the quantum phase difference to
values in the range $-\pi/2$ to $\pi/2$. 
(Quasi-particle tunneling changes the vortex phase difference by
$\pi$ and becomes favorable as soon as the induced charge
equals $e/2$, i.e. when the phase difference equals $\pi/2$.)

It is possible to derive the AC effect from the Quantum Phase Model.
In a more transparent way, although less rigorous, one starts from the
representation of the partition function as a path integral over the phases
and charges (see Eq.(\ref{Sqphi5})) with the inclusion of a uniform background 
charge, i.e. $q \longrightarrow q - q_x$.
The term in the action of the QPM 
which is relevant for the AC effect is the one which is linear in
$\dot{\phi }_i$ in Eq.~(\ref{Sqphi5}).  
If a vortex is present, the configuration of the phases 
will be related to its position $\:\:{\bf r}(\tau )\:\:$
and, by going over the same steps that lead to the single vortex action in 
Eq.~\ref{eq:effect}, there is an additional term to the effective action equal to
\begin{equation}
S_{AC} = - i \sum_{i}\frac{q_{x,i}}{2e}
 \int _{0}^{\beta} d \tau
{\bf \dot{r}} \cdot {\bf \nabla} \Theta \left [
 {\bf r}_i -{\bf r}(\tau ) \right ]
\label{sac}
\end{equation}
Eq.(\ref{sac}) defines a pseudo-charge gauge field for the vortex 
$$
 2e {\bf A}_Q({\bf r}) = q_{x} {\hat{\bf z}} \times {\bf r} / r^2
$$
which is singular at the origin of the vortex.
Thus the  external charge act like a vector potential for the vortex. 
The phase factor $\chi$ implied by Eq.(\ref{sac}) is
\begin{equation}
\chi = 2 \pi \sum _{i  
\subset \Gamma } \frac{q_{x,i}}{2e}  \nonumber
\end{equation}
where the sum extends to all islands enclosed by  the trajectory
$\;\: \Gamma \:\:$.

\subsubsection{Bloch oscillations}

Electrons in metals move in the periodic potential created by the 
positively charged ions. The electron wave functions overlap
and energy bands are formed. A constant electric field accelerates
electrons, 
but in the absence of scattering, electrons would be Bragg reflected 
at the Brillouin zone edges.
Electrons then undergo an oscillatory motion in space (Bloch
oscillations). No charge would be transported. 
In metals scattering takes place before
the electrons can reach the zone edge so that 
Bloch oscillations do not appear.
In semiconductor superlattices~\cite{bloch_semi} Bloch oscillations
have been observed because of the larger superlattice period and because of 
less scattering in the controlled fabricated structures. Coherent Cooper pair 
tunneling in current bias Josephson junctions leads to a 
phenomenon analogous to  Bloch oscillations~\cite{kuzmin91}.

\begin{figure}
\centerline{{\epsfxsize=14cm\epsfysize=16cm\epsfbox{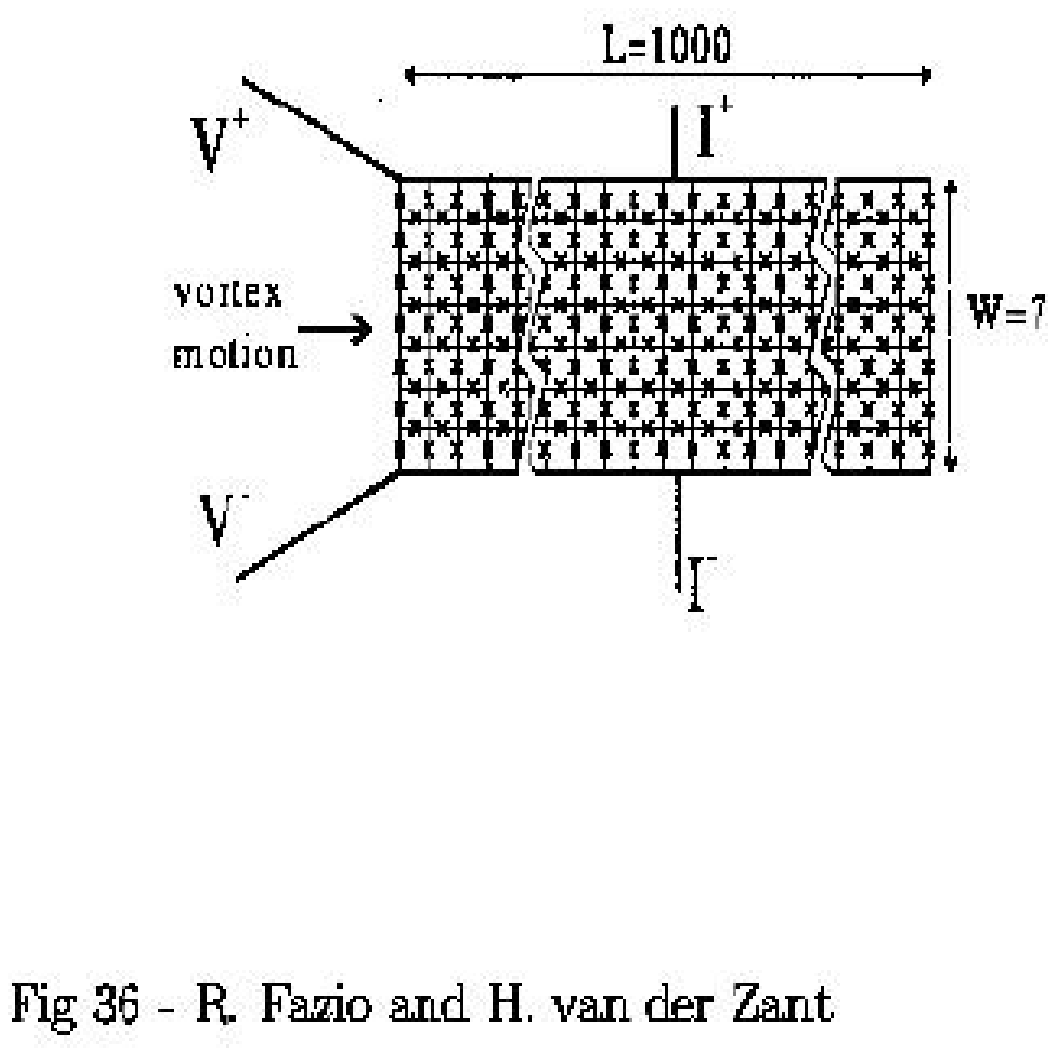}}}
\caption{A sketch of the sample layout of a quasi one-dimensional Josephson array. 
	The current is injected in the middle while the voltage probes are situated at 
	the end of the busbars. 
	(From Ref.~\protect\cite{oudenaarden96a}.)}
\label{alexlayout}
\end{figure}

\begin{figure}
\centerline{{\epsfxsize=14cm\epsfysize=16cm\epsfbox{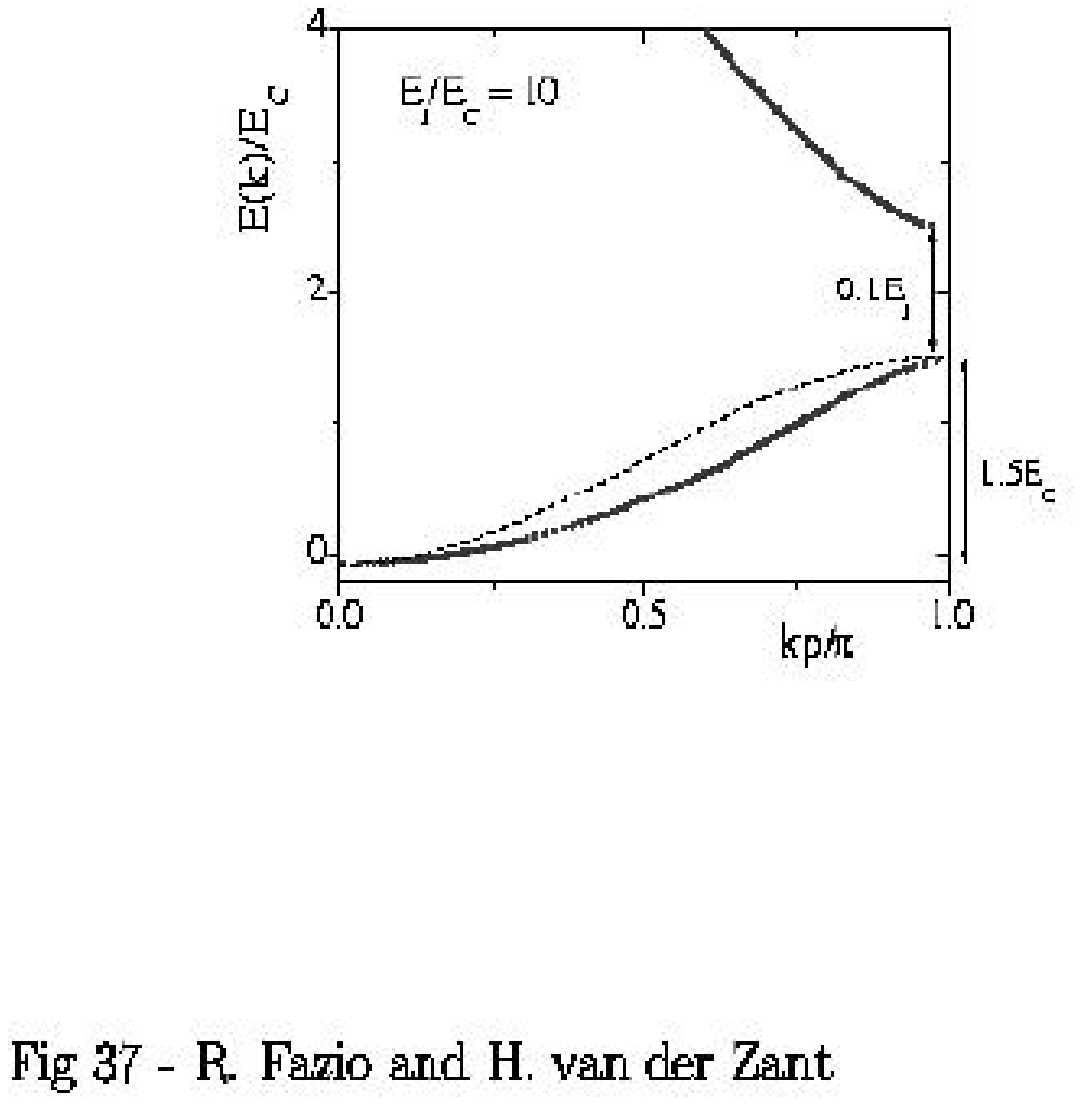}}}
\caption{Schematic drawing of the energy bands for a vortex moving in a quasi-1D Josephson array. 
	Dots: numerical calculated energy bands starting from Schr\"odingers equation with a 
	cosine potential. The dashed line shows the first band of a cosinusoidal dispersion 
	relation with the same band width.
	(From Ref.~\protect\cite{zant97}.)}
\label{vorbands}
\end{figure}

Vortices in a periodic potential should also from energy bands. It is possible  
to study Bloch oscillation for vortices~\cite{oudenaarden96} in  
a quasi-1D Josephson array that is a few cells wide and 1000
cells long. A sketch of the sample layout is shown in Fig.\ref{alexlayout}. 
For low densities, the bus-bars force the vortices to move in the middle
row so that they experience a purely 1D sine potential. 
For a free vortex the energy depends 
quadratic on the wave vector $k$: $E(k)=k^2/2M_v$, which equals
$E(k)=2E_C$ at the Brillouin zone edge. In a periodic potential, energy gaps
open up at the zone edges. The gap is equal to the Fourier coefficient
of the lattice potential~\cite{kittel}. 
For a sine potential $\frac{1}{2} \gamma E_J
\sin(2 \pi x)$,
the gap is then $\gamma E_J$. Thus, vortices in arrays form energy
bands with a bandwidth of the order $E_C$ and an energy gap 
of $ \frac{1}{2} \gamma E_J$ as illustrated in Fig.\ref{vorbands}.
Assuming the lowest band to be cosine-shaped ($E(k)
=\frac{1}{2}E_C(1-\cos(k))$),
the equation of motion $F=  \hbar {\rm d}k/{\rm d} t$ is:
\begin{equation}
   \hbar \frac{{\rm d}k}{{\rm d} t} = \Phi_0 I - \eta v(k) \;
\;
   \; \; {\rm where} \; \; \; \;  \eta u(k) = \frac{1}{\hbar} \frac{{\rm
d}E(k)}{{\rm d}k} =
   \frac{E_C}{2\hbar} \sin (k).
\label{blocheq}
\end{equation}
As defined before, $I$ denotes the applied current per
junction, $\eta$ the phenomenological viscosity and $v(k)$ the average vortex 
velocity.       

In the absence of damping, with a small current applied, the wave vector
changes linearly in time: the vortex thus reaches the Brillouin zone
edge where it will be Bragg reflected. This Bragg reflection results in 
an oscillatory motion in $k$-space. 
On average the vortex velocity  is zero and the time it takes
the vortex to complete one oscillation follows from 
$\Delta t = \Delta k /<{\rm d}k/{\rm d}t>$ with $\Delta k = 2 \pi $. 
The corresponding Bloch oscillation frequency ($\nu_B$) is:
\begin{equation}
\nu_B \; = \; \frac{I}{2e} \; ,
\label{blochfreq}
\end{equation}
and the amplitude of the oscillation is
\begin{equation}
	x \; = \; \int \frac{1}{\phi_0 I} \; =
  \; \frac{E_C}{E_J} \; \frac{I_c}{ \pi I}. 
\end{equation}
When biasing an array with 1000 junctions with currents of the order of
$\mu$A, Bloch frequencies are in the range 1-10~GHz.
Since $E_C \approx E_J$ and $I_c/I$ is typically 100, the 
Bloch oscillations extend over 10 cells.

\begin{figure}
\centerline{{\epsfxsize=14cm\epsfysize=16cm\epsfbox{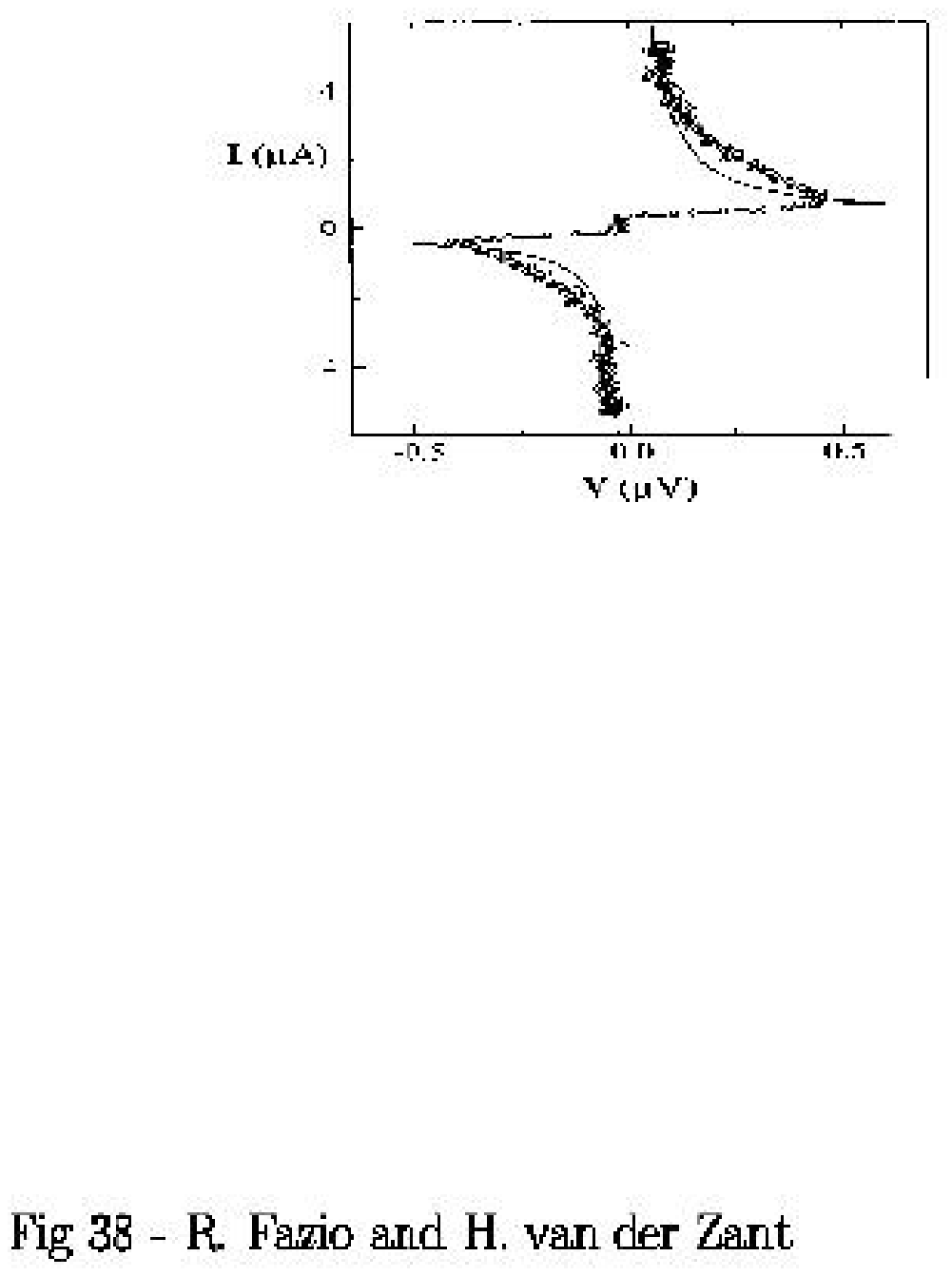}}}
\caption{A nose-shaped $I$-$V$ characteristic indicative for Bloch
	oscillating vortices. Data (circles) has been obtained in a quasi-1D array with size 7 by 
	1000 and has been measured at 10~mK for a 1D vortex density ($n$) of 0.04. 
	The solid line is the analytical result discussed in the text. 
	(From the Ph.D. thesis of A. van Oudenaarden, Delft, 1998, unpublished.)}
\label{fig_bloch}
\end{figure}

A characteristic feature of Bloch oscillating vortices is a nose-shaped
form of the dc current-voltage characteristic (see Fig.\ref{fig_bloch}).
For very small bias, there is a small supercurrent because 
vortices need to overcome 
the energy barriers near the array edges (finite-size effect). 
Just above the depinning
current, any amount of dissipation prevents vortices from reaching the
zone edges. Bloch oscillations do not exist in this regime and 
an increase of the current yields an increase of the measured voltage
across the array. 

When increasing the current beyond some point,
dissipation is not strong enough to prevent the vortices from 
reaching the zone edges. Bloch oscillations are now possible. In the
$I$-$V$
characteristic a sudden decrease of the voltage is then expected with
a negative differential resistance: the
oscillating vortices do not contribute to the net transport of
vortices through the array. 
Eqs.(\ref{blocheq}) can be solved with the result
\begin{equation}
   V(I) = n \pi \frac{E_C}{e} \; \frac{I}{I_0} \; \; \; \;  \; \; \; \; 
   \; \; \; \;  \; \; \; \; {\rm if } \; \; \; \; I<I_0
\end{equation}
\begin{equation}
   V(I) = n \pi \frac{E_C}{e} \; 
\frac{I}{I_0}\sqrt{(1-\sqrt{1-\left(\frac{I_o}{I}\right)^2})} 
   \; \; \; \; \; \; \; \; {\rm if } \; \; \; \; I>I_0,
\end{equation}
where $I_0=\eta E_CW/(2\Phi_0)$ and $n$ is the one-dimensional 
vortex density (=number of vortices divided by the number of cells 
in the direction of motion). 
The solid line in Fig.\ref{fig_bloch} is a fit to these equations.
(The line is offset by a small positive current to correct for
the depinning current).  Although the overall shape 
of the experimental curve resembles that of the theoretical prediction,
the experimental value of the band width ($E_C$ is the previous discussion) 
is one order of magnitude smaller than expected.
This discrepancy is not understood, but is consistent with a vortex mass
that is larger than the calculated, quasi-static vortex mass. Note that
the  data extracted from the study of the macroscopic quantum tunneling 
also indicated the same trend for the experimental vortex mass. 

Additional information on Bloch vortices can be obtained by irradiating 
the sample with a microwave signal. Steps occur in the $I$-$V$
characteristics
when the external frequency locks to the Bloch frequency. Surprisingly, 
the experiments show that the Bloch frequency depends on the vortex
density.
This dependence is not accounted for by the independent-vortex model
presented 
above (see Eq.~(\ref{blochfreq})) and suggests a collective oscillation of
the 1D vortex chain (see next section). We will come back to this issue when 
discussing the formation of a Mott insulator in these quasi-1D arrays.

\subsection{One-dimensional vortex localization}
\label{localize}
In Section~\ref{onedi} , we already introduced quasi-one dimensional
arrays as model systems for the study of interacting bosons in one dimensions.
For an ideal periodic potential and in the absence of interactions between 
the particles, the solution of the Schr\"odinger equation consists of Bloch 
waves that extend throughout the whole chain. 

In quasi-1D Josephson arrays, however, vortex localization can occur in
several ways. We already discussed the
existence of Bloch oscillations when an external force is exerted on a
vortex. One can view this oscillatory motion in space as
localization of individual bosons analogous to Wannier-Stark
localization of electrons~\cite{wannier-stark}:
the extend of the wave function is decreased when the external
force on the quantum particle is increased.

In the next two subsections, we treat two other cases where  boson localization 
occurs.
Commensurability effects~\cite{kardar86} may lead to localization of vortices 
in quasi-1D arrays. The repelling interaction between vortices plays a
crucial role in this so-called Mott-localization\cite{fisher89b}.
Another mechanism to localize Bloch waves is disorder and this
phenomenon is known as Anderson localization\cite{anderson58}.
Two important remarks should be made at this ponit. First, vortex localization,
as discussed  in the next subsection, has been studied by measuring the 
zero-bias resistance. 
In this case only a very small current is applied in contrast to the
experiment showing the Bloch oscillations.
Second, one should realize that in contrast to localization in
electronic systems,
vortex localization leads to a zero resistive state. In superconductors
motion of vortices is the cause of dissipation. If they are localized, the 
sample is superconducting.

\subsubsection{Mott insulator of vortices}

More than ten years ago, localization of bosonic particles with a
short-range repulsive interaction has been
studied theoretically by Fisher {\it et al.}\cite{fisher89b}.
They found a Mott insulating phase
for commensurate filling and a superfluid phase for incommensurate filling.
Strong disorder destroys the Mott phase and there is the possibility to have
a Bose glass (for theoretical studies of the phase-diagram
of bosons on a chain see Refs.~\cite{freericks94,batrouni90,niyaz94} 
and references therein).

Experimentally, Mott localization of vortices has been studied by van 
Oudenaarden {\it et al.}\cite{oudenaarden96a,oudenaarden98}. They
explored the influence of the interaction strength, the bandwidth, the sample
geometry and temperature on the stability of the Mott states. First, we
will summarize the experimental results. Then, we
discuss their experiment in the context of a recent theory
of Bruder {\it et al.}\cite{bruder99}.

\begin{figure}
\centerline{{\epsfxsize=14cm\epsfysize=16cm\epsfbox{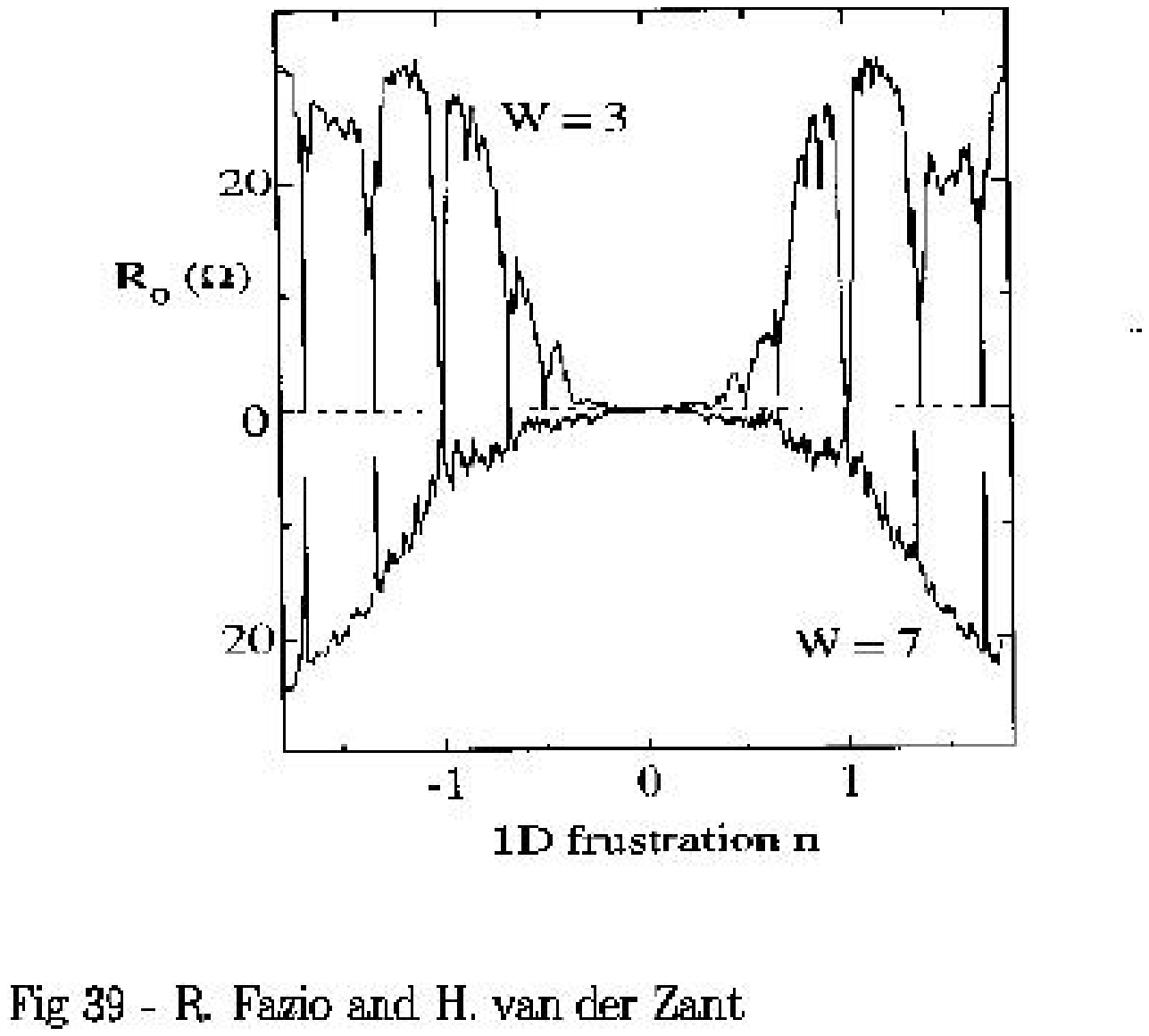}}}
\caption{Zero-bias resistance $R_0$ as a function of the one-dimensional 
	vortex density ($n=Wf$) for two samples with different widths $W$. 
	The sample length is 1000 cells and data has been obtained at 30~mK.
	The bottom curve ($W=7$) is mirrored with respect to the $x$-axis for clarity.
	(From Ref.~\protect\cite{oudenaarden98}.)}
\label{alexmott1}
\end{figure}

The experiment consists in measuring the zero-bias resistance vs.
magnetic field for quasi-one dimensional arrays of different lengths and 
$E_J/E_C$ ratios. It is convenient to define a 1D frustration $n$, 
which is associated with magnetic field piercing through a cell
of area $W$ (in units of the lattice constant): $n=WB/\Phi_0 = Wf$. 
In Fig.\ref{alexmott1} the results are shown for
arrays of three and seven cells wide, i.e., $W=3$ and 7 respectively.
The plot for the $W=7$-sample is mirrored with respect to the $x$-axis
for clarity. For both samples, clear dips occur for certain values of the 2D
frustration index $f$, i.e., for certain values of the vortex density.
When plotted vs. this 1D frustration index, the 1D nature of the sharp dips 
becomes visible. Dips are found at the same rational values of $n=1/3, 1/2, 
1, 2$. 

More detailed measurements of the resistance dips show that they are not
infinitely sharp. There is a certain window of $n$ in which the resistance 
vanishes and the vortex chain is pinned. 
The interaction energy proportional to $E_J$
(see below) dominates the bandwidth proportional to $E_C$ in this
regime. Beyond this window the resistance increases sharply, indicating
that the vortex chain is depinned and that the bandwidth dominates the interaction 
energy. From this consideration, one expects the window to be larger for
samples with a larger $E_J/E_C$ ratios. In the experiment, this dependence
has indeed been observed.

Around commensurate filling, the system is incompressible: small
changes of the magnetic field (i.e. $n$) do not lead to a change in the 
number of vortices in the chain. This process costs a certain energy, 
called the Mott gap. By analyzing thermally activated transport 
in the Mott states, the value of this gap can be deduced from the experiments 
and values in the order of Kelvins are reported.
The influence of the array length was also studied. No significant
differences were observed in arrays with lengths larger than 200 cells. This 
observation demonstrates that edge effects do not play an important role
and that the long arrays are indeed one-dimensional systems.

A quantitative theory of the commensurate-incommensurate transition of 
a vortex chain in  quasi-one-dimensional Josephson arrays has been formulated by 
Bruder {\em at al.}~\cite{bruder99}. They showed that the transition to the 
incommensurate state is due to the proliferation of soliton excitations
of the vortex chain. Since the range of the interaction 
between the vortices is much longer than the inter-vortex distance, 
solitons consist of many vortices, and possess a large effective mass. 
The transfer of one flux quantum between the array edges is then due to soliton 
propagation through the sample. The number of solitons necessary to
transfer one vortex is equal to the ratio of the periods of the vortex
lattice and the junction array.

The analysis of Bruder  {\em at al.}~\cite{bruder99}  focuses on determining the 
energy barrier $E_R$ of the observed thermal activation in the resistance vs. 
temperature curves. Although this approach is of semi-classical nature, quantum 
effects are crucial in relating the parameters of the effective theory 
(expressed in terms of soliton excitations) to the microscopic couplings 
($E_C$ and $E_J$) of the Josephson array. 

\begin{figure}
\centerline{{\epsfxsize=14cm\epsfysize=16cm\epsfbox{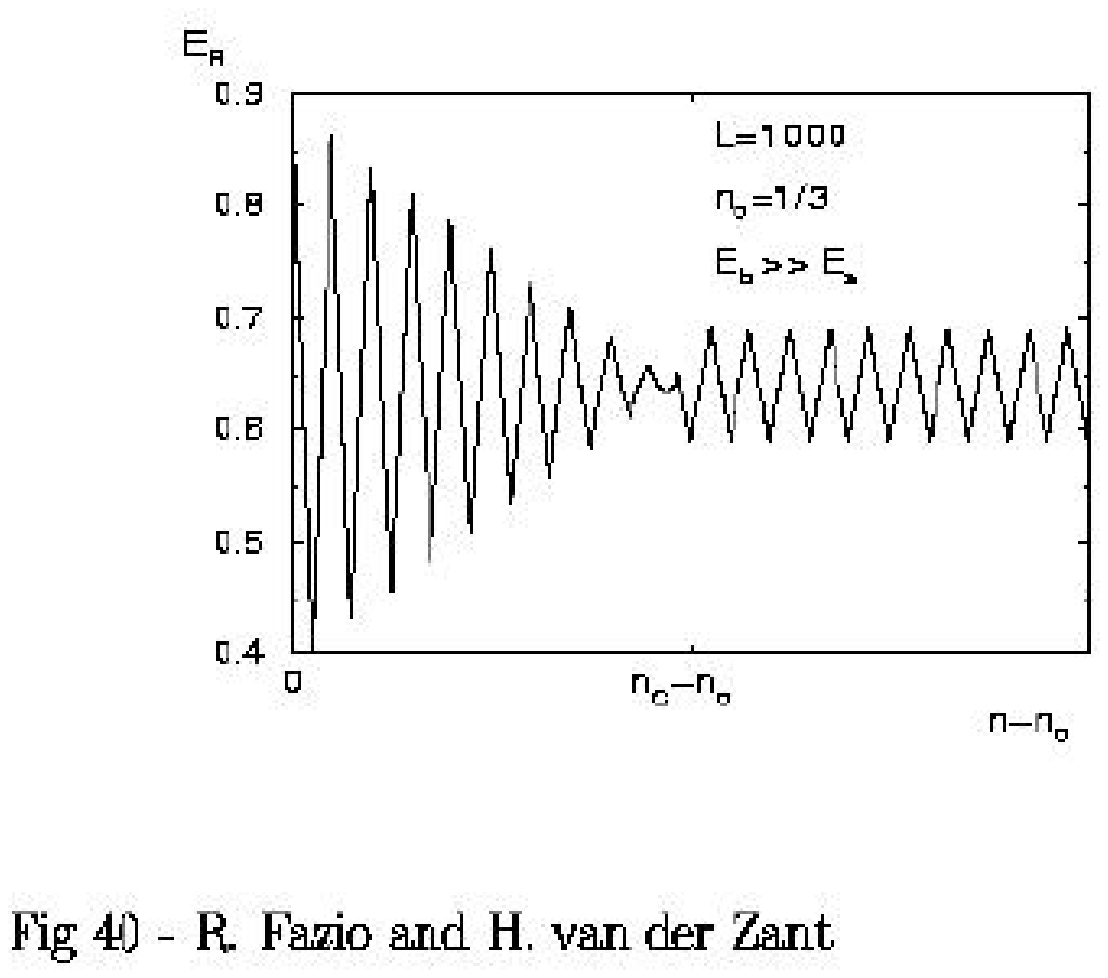}}}
\caption{Activation energy as a function of $n-n_0$ ($n_0$ is the commensurate density)
	in the case that the  
	boundary pinning $E_b$ dominates over the soliton formation energy $E_s$. 
	On the incommensurate side of the transition, $n \: > \: n_C $, 
	solitons form spontaneously and the physics is determined by boundary 
	pinning and the elastic energy. 
	(From Ref.~\protect\cite{bruder99}.)}
\label{brudermott1}
\end{figure}

\begin{figure}
\centerline{{\epsfxsize=14cm\epsfysize=16cm\epsfbox{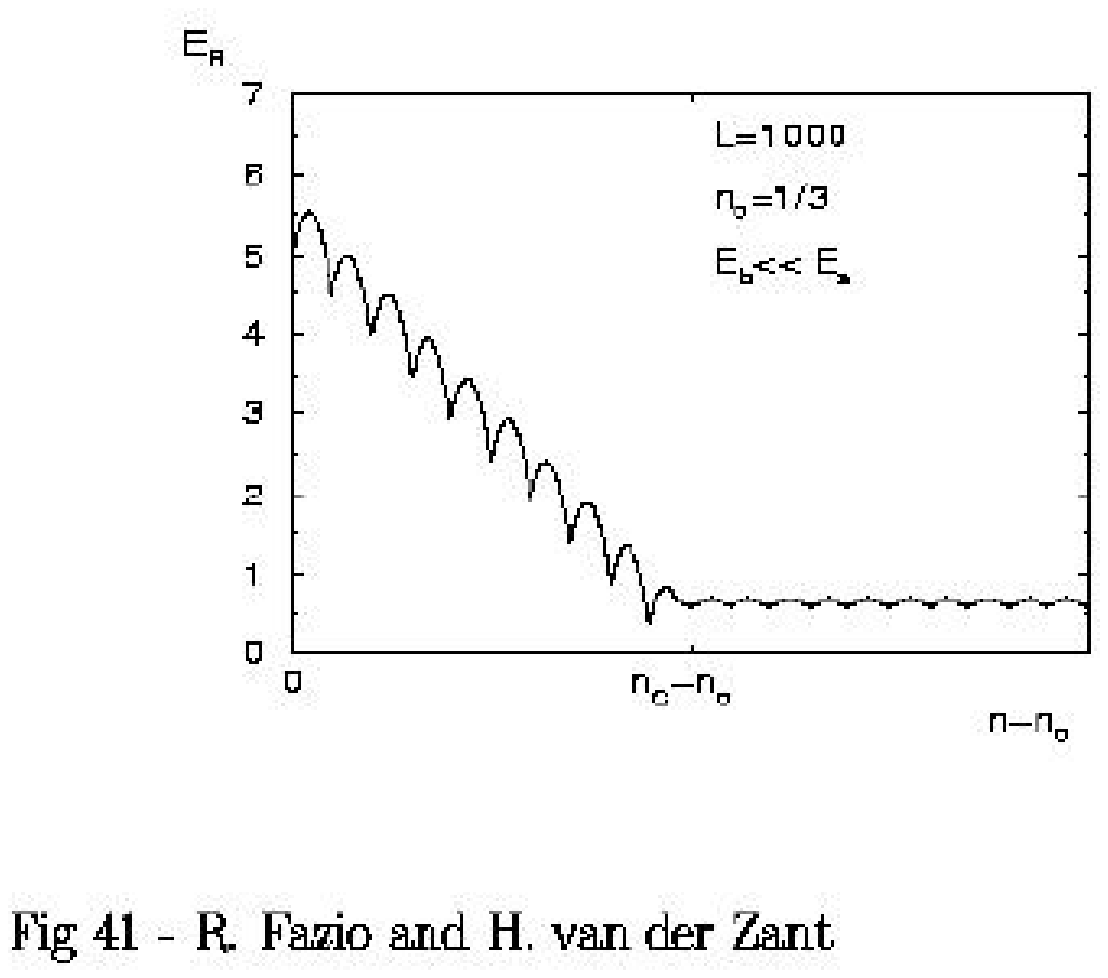}}}
\caption{Activation energy as a function of $n-n_0$ when the soliton 
	formation energy $E_s$dominates over the  boundary pinning $E_b$ (opposite 
	limit as considered in Fig.~\ref{brudermott1}).
	(From Ref.~\protect\cite{bruder99}.)}
\label{brudermott1b}
\end{figure}

In the commensurate phase, there are two contributions to $E_R$. 
The first contribution comes from the activation energy of a soliton 
and the second term summarizes the boundary pinning energies. This 
boundary effect can be understood as follows: Because of commensurability, 
the process of vortex flow through the array can be viewed as the motion
of a rigid vortex chain. Therefore, the vortex chain
cannot adjust itself to the boundary pinning potential. The potentials
produced by the two array ends both contribute to $E_R$ and
the relative phase of these two contributions depends on whether the total flux 
piercing the junction array equals an integer number of flux quanta or not. 
Consequently, the second -boundary pinning- term to $E_R$ oscillates with the 
magnetic flux piercing the array. In Fig.\ref{brudermott1} $E_R$ is plotted 
for the case of small boundary pinning while the opposite situation
is shown in Fig.\ref{brudermott1b}. The short period oscillations 
are determined by boundary-pinning term while the vanishing of $E_R$ at the 
edge of Mott lobe is driven by the soliton energy.

In the incommensurate state, the vortex chain is compressible, and can adjust 
itself to the boundaries of the array. As
a result, the main contribution to the activation energy is due to the
boundary pinning potential and the elastic energy.  

\begin{figure}
\centerline{{\epsfxsize=14cm\epsfysize=16cm\epsfbox{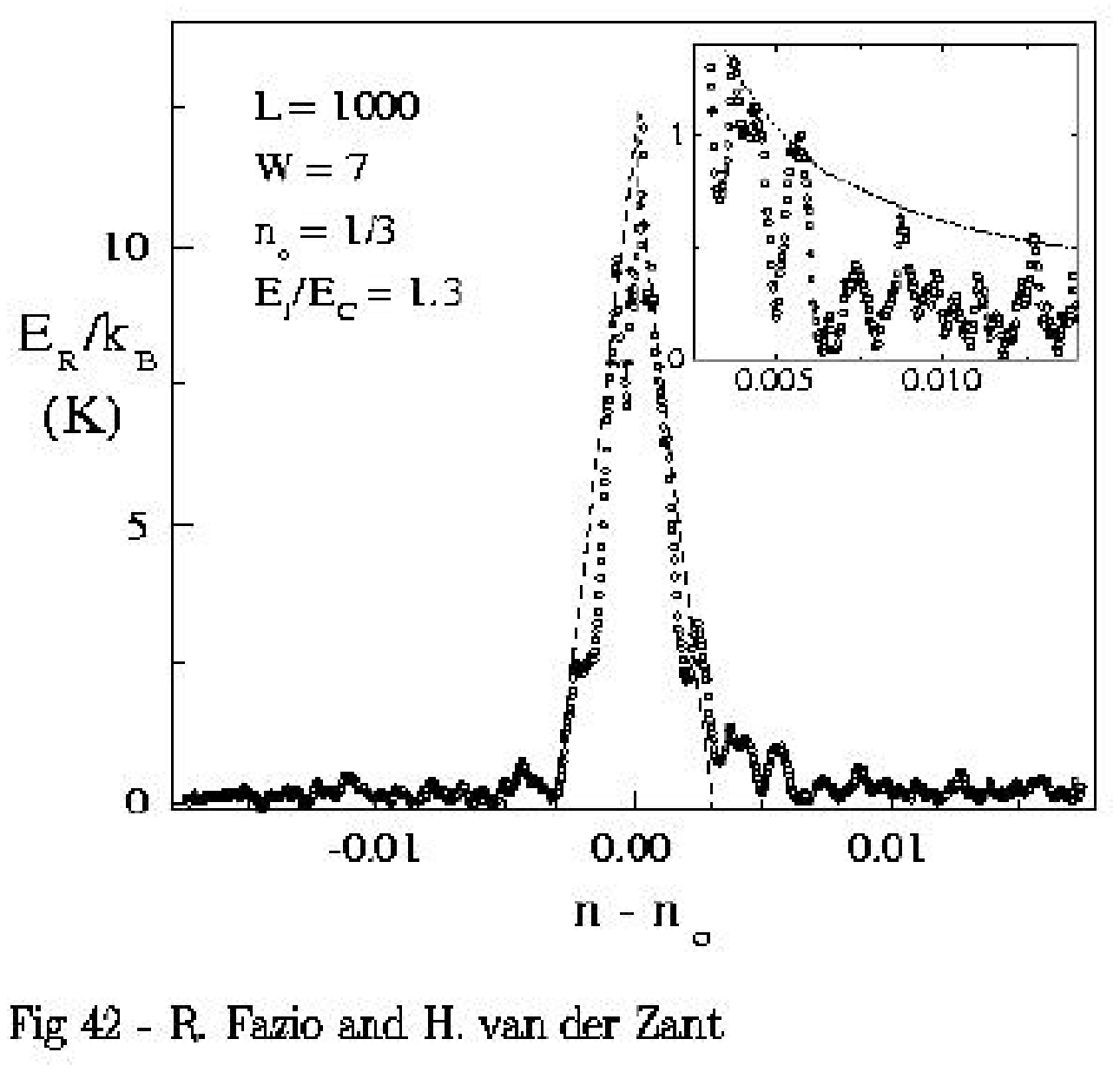}}}
\caption{Measured activation energy of an array of $1000 \times 7$ cells
with $E_J=0.9$ K and $E_C=0.7$ K. The dashed line is a fit to
the data yielding the width of the Mott region. The inset
shows $E_R$ inside the Mott phase.
(From Ref.~\protect\cite{bruder99}.)}
\label{brudermott2}
\end{figure}

The comparison between theory and experiment is shown in 
Fig.\ref{brudermott2}. 
The theoretical results lead to an estimate for the soliton 
energy much larger than the boundary pinning which is of the order of 
$\sim 0.5K$. The theoretical value of the activation energy
$$
	E_R \sim 8 K
$$
is in very good agreement with the experiments.

Two observations provide additional support for the theory of Bruder  
{\it et al.}. First, the regions of $n$ corresponding to the Mott 
phase are extremely narrow suggesting, at least in the conventional 
non-interacting picture, a weak interaction between the particles.
Consequently, within the Mott phase the activation energies for particle 
transport are expected to be small as well. 
However, the observed value of $E_R$ is one order of magnitude larger
than the energies $E_C$ and $E_J$, which determine the single-vortex
band spectrum.
A second feature in favor is the strong oscillating behavior of the 
resistance (with a period proportional to the inverse length of the 
chain) outside the Mott region. These 
oscillations would not be expected in a model of almost-free quasi-particles 
within the delocalized phase. 


\subsubsection{Anderson localization of vortices}

In 1958, Anderson\cite{anderson58} showed that disorder has a
dramatic influence
on transport properties. Disorder reduces the spatial extent of wave
functions
to such an extent that transport can completely be blocked.
Three years later
Mott and Twose\cite{mott61} showed that 1D systems
are particularly susceptible
for disorder: even weak disorder leads to strong localization.
Many studies on Anderson localization have been performed on 3D and 2D
samples. One-dimensional model systems are harder to find.
Josephson-junction
arrays have the great advantage that disorder can be introduced in a
controlled
way. Experimentally, Anderson localization of vortices has been studied
by the Delft group \cite{oudenaarden96b}. Their results will be outlined
in this subsection.

In the experiment by the Delft group disorder has been introduced
by constructing superlattice structures. The superlattice is formed by
replacing
all the junctions of a column by junctions that are twice as large.
Consequently,
these barrier junctions have a Josephson energy that is two times larger
than that
of the adjacent junctions.
The barrier junctions yield a peak in the potential landscape for
vortices
traveling through the array. Numerical calculations
show that this barrier is $1.7E_J$, which
is about one order of magnitude larger than the energy barrier for
cell-to-cell
motion.

A perfect superlattice structure is made by introducing columns of
barrier junctions on a distance of exactly $10$ lattice cells. 
For an array of length 1000, this
means that there are 100 columns that have been changed. Disorder is now
introduced by changing the distance between two barriers Samples
with different
amounts of disorder have been fabricated. In the least disordered 
samples (labeled with $\delta=1$) the barriers were separated by 
$9$, $10$ or $11$ lattice cells with equal probabilities.
In other disordered samples (labeled with $\delta=2$) barriers placed at distances
$8$, $9$, $10$, $11$, or $12$ lattice constants again with equal probability.
All samples contained 100 barriers.

\begin{figure}
\centerline{{\epsfxsize=14cm\epsfysize=16cm\epsfbox{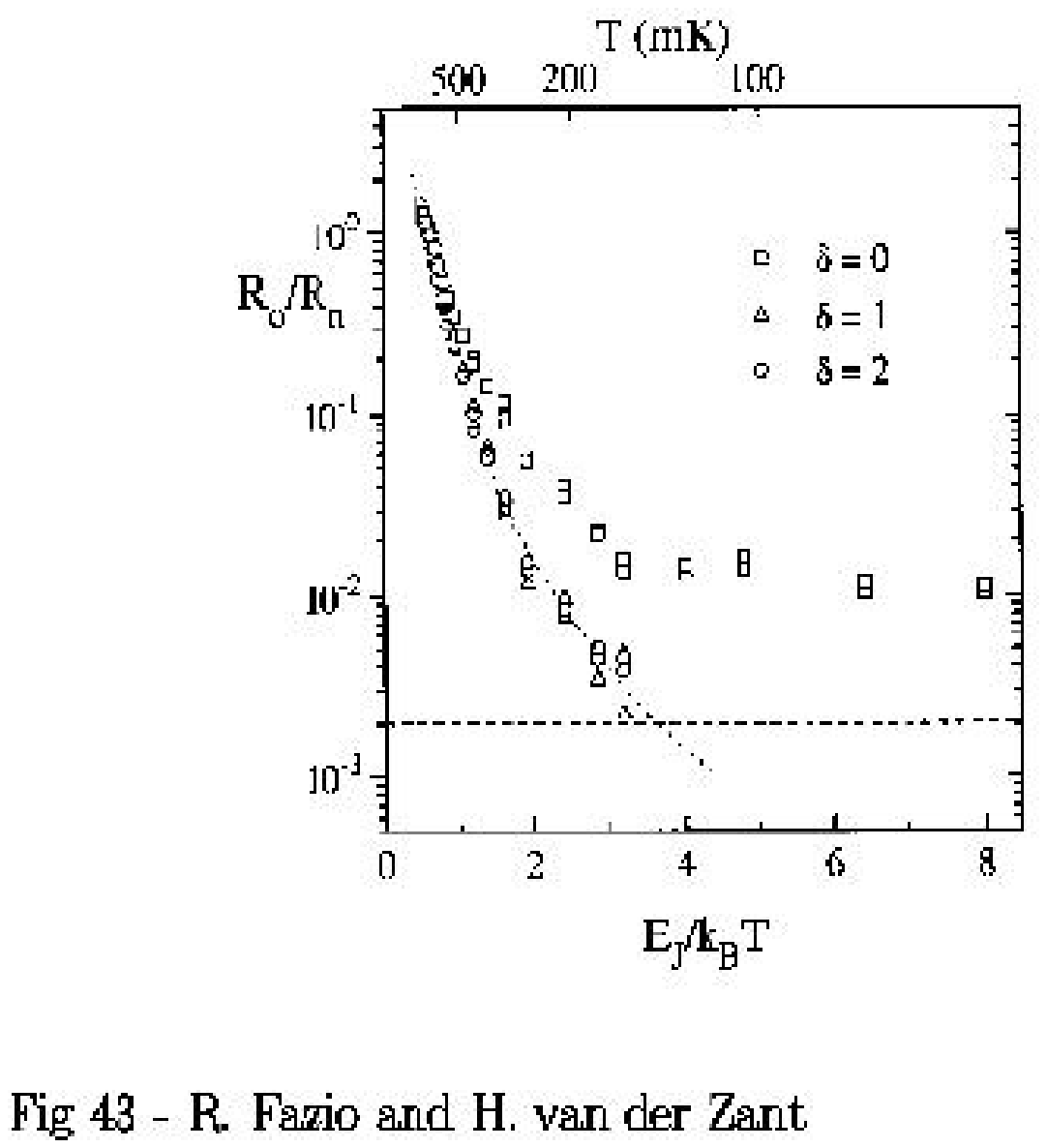}}}
\caption{Arrenius plot of the linear resistance for the ordered (squares)
	and for the disordered (triangles, circles) arrays. 
	At low temperatures (right hand side of the figure), the resistance of the disordered 
	arrays (triangles and circles) has dropped below measuring accuracy (dashed line) 
	indicating vortex localization. (From Ref.~\protect\cite{oudenaarden96b}.)}
\label{alexloc1}
\end{figure}

The vortex quantum properties are probed by measuring the zero-bias
resistance as a function of temperature for the perfect periodic array  as
well as for the
disordered arrays. Since the topic of interest is the study of quantum
transport,
the vortex density needs to have a non-commensurate value to avoid the
Mott
state as discussed in the previous subsection.
The result for $n=0.44$ is shown in Fig.\ref{alexloc1}. At high 
temperatures all three
arrays show the same behavior: transport is thermally activated with an
energy
barrier of $3E_J$, a factor of two larger than the expected value.
When the temperature is lowered, however, a significant difference is
observed
between the periodic sample and the two disordered samples. For the
perfect periodic
sample a finite resistance is measured at the lowest temperatures. In
this regime
the resistance is independent on temperature indicating vortex transport
by
quantum tunneling. In contrast, the resistances of the disordered
samples have dropped below
the measuring accuracy of the set-up. Thus, in the perfect array
vortices
are mobile whereas they are localized in the disordered arrays.

The zero-bias resistance has been studied for several values of the
vortex density.
For $n<0.3$, the zero-bias resistance is too small to be resolved for
the periodic array. In the range $0.3<n<0.8$, the resistance of the
periodic array
is significantly larger than that of the disordered arrays and this is
the
region where vortex localization occurs.
But for even larger vortex densities the behavior of all three arrays is
almost the
same showing a flattening of the resistance at the lowest temperatures.
In all
three arrays vortices are now mobile. At these high vortex densities
the distance between them is small and their repulsive interaction can
no longer be neglected. The experiment
shows that in this case delocalization occurs.

Above we have discussed the experiment in terms of
Anderson localization which strictly speaking occurs when the
interaction between
the bosons is very weak; i.e., when the bosons act as independent
particles that
are localized for arbitrarily weak disorder. In the experiment, vortex
densities are
large. Disorder now competes with the interaction strength. A
sufficiently
strong interaction can delocalize the particles, whereas strong disorder
will localize
them again in a Bose-glass phase. To distinguish between Anderson
localization and
localization in a Bose glass, more measurements are needed.
Experiments should be performed on arrays with fixed disorder but
different $E_J$ to clarify the role of the interaction strength.

\section{Future directions}
\label{fd}

In this last chapter of the review, we discuss some future directions for 
research on Josephson networks. Of course, additional information on the 
quantum nature of Josephson networks can be obtained by using new measuring 
techniques (e.g. the ac-measuring method that has successfully been applied 
to classical arrays) and better samples (e.g. arrays with a well-controlled 
environment). Here, we outline the concepts behind two experiments -persistent 
vortex currents and vortex quantum Hall effect- of which some theoretical 
calculations are available. We summarize the main ideas as well as the 
experimental requirements/improvements for observation of these effects. 
Future lines of research may also involve the use of Josephson networks 
as model systems in unexplored areas of physics. Biophysics may be such 
a field and an interesting example in this respect is vortex transport 
in ratchet arrays. Undoubtedly, the most exciting new line of future 
research is quantum computation. 
In this paper, we do not have space to treat quantum computation with 
Josephson circuits in great depth. We therefore only present the concept 
and summarize the current status of art.

\subsection{Persistent vortex currents}

\begin{figure}
\centerline{{\epsfxsize=14cm\epsfysize=18cm\epsfbox{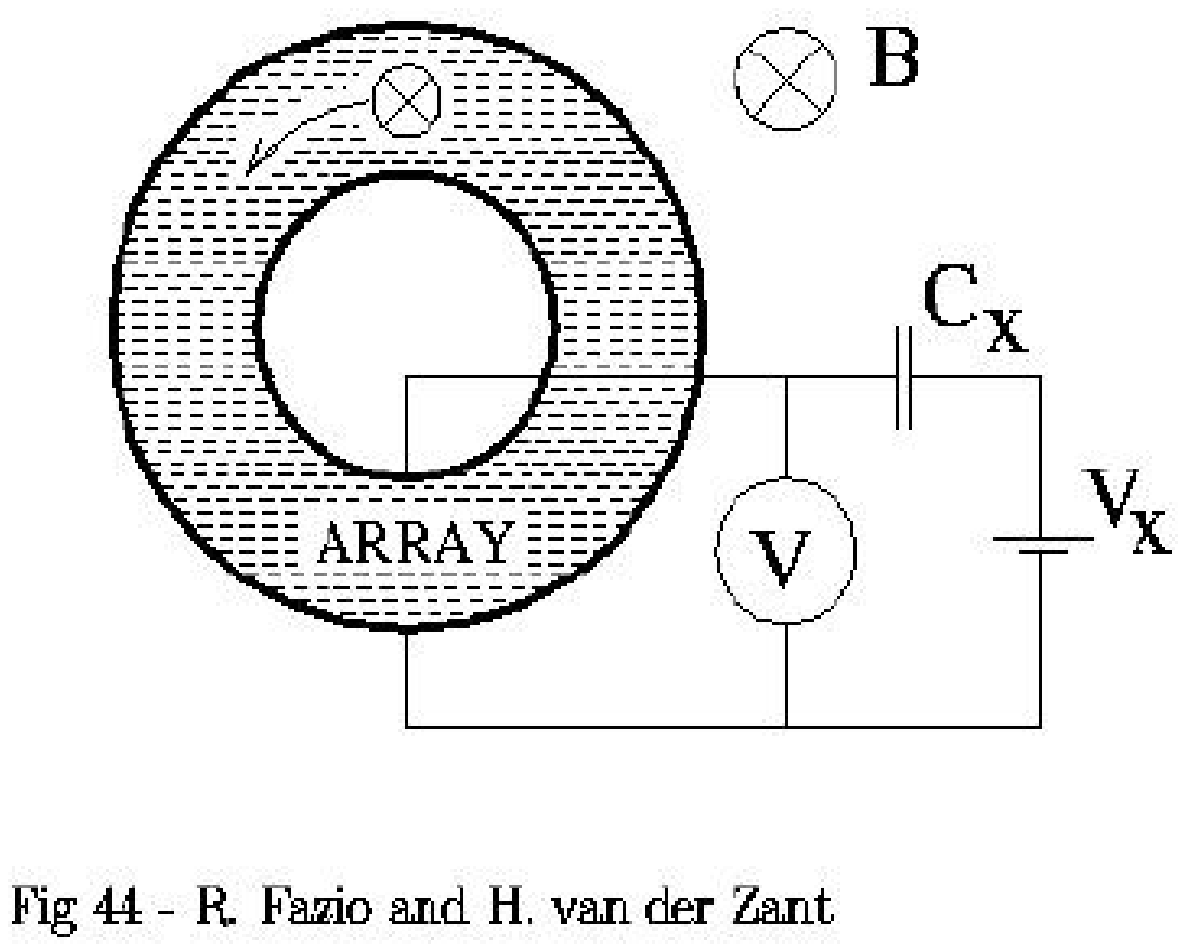}}}
\caption{The experimental setup to detect interference effects of vortices: The Corbino 
	disk (From Ref.~\protect\cite{wees91}.)}
\label{Corb}
\end{figure}

Persistent currents in rings made of normal metals are a manifestation 
of the Aharonov-Bohm effect for quantum coherent electrons in a multiply 
connected geometry. In the absence of any driving current, the flux 
through the ring induces the motion of charge carriers that can be 
detected at low temperatures~\cite{buettiker}.
Charge-flux duality indicates that in Josephson rings a persistent current 
of vortices may be expected generated by the charge induced on the 
inner island.
A persistent vortex current leads to a persistent voltage across the ring 
and as such would be manifestation of the Aharonov-Casher effect discussed 
in Sec.~\ref{quantumvortex}. Although vortex interference in an 
open Josephson circuit has been reported by Elion {\em et al.}~\cite{elion93}, 
there are no experiments reporting the existence of persistent vortex 
currents in Josephson Corbino circuits. The proposed setup to measure 
persistent vortex current is shown in Fig.\ref{Corb}.

When the mean free vortex path is long enough, we can neglect the
dissipation in the dynamics and the Hamiltonian of a vortex in 
a discrete Josephson ring (with $N$ junctions) can be written as
\begin{equation}
\label{eq:hamil}
       H=\frac{1}{2M_v} \left( P-\frac{\Phi_0}{N} Q_x \right)^2
\end{equation}
where $P$ is the vortex momentum and $Q_x$ the charge on the inner 
island of the ring.
The vortex dynamics is quantized and the set of discrete energy
values is given by 
\begin{equation}
\label{eq:cap}
       E_\nu =\frac{(2e\nu-Q_x)^2}{2C_{eff}} \;  \; \; \; {\rm with } \; \;  
           \; \; C_{eff}=\left( \frac{N}{\Phi_0} \right)^2 \; M_v
\end{equation}
where $\nu$ is an integer and where $C_{eff}$ agrees
with the continuum result of Ref.~\cite{hermon95} and with
$C_{kin}$ in Ref.~\cite{orlando91a}. 
Thus, a Josephson ring with a vortex trapped inside acts like a perfect capacitor.
A similar situation arises in a
single classical junction in the absence of screening. With exactly one 
vortex trapped, 
its critical current is zero. The difference lies in the value of the effective 
capacitance, which in quantum rings differs from the geometrical one. For example, 
for a ring consisting of a square
2D array  $C_{eff}= (N^2/2) C$ and for a 1D ring with $\Lambda_J < N$,
$C_{eff}= (2N/\pi^2 \Lambda_J) NC$.

Although this change in capacitance can in principle be measured, quantum
rings should exhibit a more interesting phenomenon: persistent vortex currents.
This can easily been seen from the Hamiltonian of Eq.(\ref{eq:hamil}) since
it has the same from as the 
Hamiltonian for electrons in a metal ring. Duality arguments then indicate the 
existence of a persistent 
voltage due to the persistent motion~\cite{orlando91,wees91,choi93} of a 
vortex induced by $Q_x$.
The basic reasoning is as follows: With some disorder, small gaps open up in 
the energy spectrum  of Eq.(\ref{eq:cap}) and energy bands form. 
With no current applied, the charge $Q_x$ dictates the quantum dynamics of 
the ring. It determines the vortex velocity and therefore the voltage across 
the ring because the vortex speed is 
proportional to $\partial{E_n} / \partial{Q_x}$. As a result, a 
persistent voltage appears across the ring which  
is periodic in $Q_x$ with period $2e$. 

Within the free-particle model, the maximum voltage can be estimated.
The voltage $V$ due to a circling vortex with velocity $v$ is equal to 
$\Phi_0 v/N$. The velocity is equal to $(N/\Phi_0) 
\partial {E_\nu} / \partial{Q_x}$. Then for $\nu=0$ and $Q_x=e$, 
the maximum voltage equals 
$e/C_{eff} = e/M_v \;(\Phi_0/N)^2$. For a ring consisting of a square
2D array  $V_{max}= (2e/ CN^2)$.  
With $N=10$, $C=1$~fF, $V_{max} \sim 3$~$\mu$V.

In the picture presented above, the vortex is treated as a free, non-interacting 
particle.
A detailed description of vortex dynamics, however, is only possible if 
one considers its dissipative environment. 
As discussed before, an important dephasing mechanism is the existence of linear 
spin-waves. 
This coupling leads to damping and hence to a finite phase-coherence length for 
vortices.
This problem was studied in Ref.~\cite{fazio96b}.
For the persistent voltage one obtains
\begin{equation}
	\langle 2eV\rangle=\frac{4\pi}{\beta}\frac{
	\sum^{\infty}_{\nu=1}n\sin(2\pi \nu Q_{x}/2e)\exp(-S_{\nu})}
	{1+2\sum^{\infty}_{\nu =1}\cos(2\pi \nu Q_{x}/2e)\exp(-S_{\nu})} 
\label{pvolt}
\end{equation}
where $S_{\nu}$ corresponds to the action in the sector of $\nu$ winding 
numbers.
In the adiabatic mass approximation in the limits $E_{J}\gg E_{C}$ and
$R_{N}\stackrel{<}{\sim} R_{K}$, the action  
reduces to that of a free particle with mass $ M_{v}=\pi^{2}/4E_{C}$. 
The saddle point action is
\begin{equation}
 	S^{0}_{\nu}=\frac{M_v}{2\beta}(\nu N)^{2}
\label{adiab}
\end{equation}
and the persistent voltage $V$ is a
sawtooth function of $Q_{x}$ at zero temperature and a sine-like function at 
higher temperatures. It depends on the system size
through the ratio of the radius $N/\pi$ to the thermal wavelength of the vortex
$\lambda_{T}= \sqrt{2\pi/M_{v}T}= \sqrt{8\beta E_{C}/\pi}$.

To go beyond the adiabatic mass approximation, the
actions $S_{\nu}$ can be calculated with the full non-local kernel
(see Eq.~\ref{eq:effect}) 
that incorporates the effect of inelastic processes due to the interaction with 
spin-waves. An important consequence of the quantum corrections is that the 
vortex dephasing length
\begin{equation}
	L_{\varphi}=\omega_{p}\xi/T
\end{equation}
increasing with the ratio $E_{C}\sim E_{J}$ ($\xi$ gets larger on approaching the 
S-I transition). For $E_{C}\sim E_{J}\sim 1K$,
$L_{\varphi}$ may be as large as 100 lattice constants at $T$=
10~mK, which is much larger than the thermal wavelength $\lambda_{T}$ for the 
vortex in the adiabatic mass approximation. 
In the limit $\xi\ll N$ and $\omega_{\nu }\ll\omega_{p}$ the adiabatic result
is recovered.
If $\xi\sim 1$  a larger persistent voltage is found. If the ratio
$E_{J}/E_{C}$ is reduced, the coherence length $\xi$ grows beyond the radius
of the system, and the persistent voltage grows even more. 
As compared to the adiabatic mass limit, the persistent
voltage including the vortex-spin-wave coupling is always larger. For $E_{J}$=
$E_{C}$=1~K, $T$=10~mK and  $N=10$, a persistent voltage in
the microvolt range is expected, which is observable.

Other interesting issues involve the effect of the underlying lattice and of
disorder on the persistent voltage. 
They both induce backscattering and the opening
of gaps in the band structure $E(Q_{x})$. One expects Zener tunneling across
the band gaps to occur if the voltage $V_{x}$ is switched on fast, yielding
a higher transient current that relaxes to the persistent current after some
relaxation time.

Experimentally, fabrication of quantum rings is difficult, but does not seem 
to be impossible. 
First of all vortices should have long mean free paths. 
As we have seen, 2D arrays exhibit only a small window
for which free propagation of vortices is possible. Purely 1D discrete 
Josephson rings with one vortex trapped~\cite{zant94} seem to be more promising 
in this respect since for $\Lambda_J>1$ there is no energy barrier for vortex 
motion. 
Coupling to spin-waves can be small and long mean free vortex paths are 
expected~\cite{zant95a}. 
Secondly, Josephson rings should be decoupled from their environment. This can be 
achieved by placing high-Ohmic resistors or alternatively arrays of junctions 
in the leads close by. 

A third restriction comes from the charging energy.
Calculations on the continuous Josephson system~\cite{hermon94}  show that 
the temperature has to be smaller than $e^2/(2 \pi^2 C_{eff})$ in
order to observe quantum vortex dynamics.
Temperatures of the order of 100~mK therefore require $C_{eff}$ to be
1~fF. This requirement indicates that aluminum junctions have to be 
smaller than 0.01~$\mu$m$^2$ and that the whole ring structure can not be 
made too large. The capacitances to ground would otherwise be too dominant. 
The fourth restriction comes from the shadow-evaporation 
method itself. Both the wire connecting the central island in the middle of 
the ring and the gate capacitor -preferable situated underneath or on top of 
the central island- have to be made in separate fabrication steps. Alignment 
and good electrical isolation between the different layers have to be established.

\subsection{The Quantum Hall effect}
Quantum electron transport in two-dimensional systems in the presence of 
an applied magnetic field is one of the most intensively investigated 
areas in condensed matter. When the filling factor (ratio between the 
electron density and the density of flux quanta ) is of the order of one, the 
Quantum Hall effect (integer and fractional) occurs~\cite{prange}.
In the previous sections we showed that charges and vortices behave 
as quantum particles hopping coherently in the artificial two-dimensional space 
created by the Josephson array. The duality between charges and vortices 
can be shown also in  the presence of external frustration; magnetic field 
for charges and offset 
charges for vortices. In a series of papers~\cite{nazarov94,choi94,stern94}
it was suggested that a Quantum Hall effect could be observed in Josephson 
arrays both for charges and vortices. The three proposals address different 
regimes: Nazarov and Odintsov~\cite{nazarov94} 
describe the possibility of Hall states for Cooper pairs while the authors in 
Refs.\cite{choi94,stern94} consider the Hall 
fluid of vortices.

In the limit $E_C>>E_J$ and in the case of very low density, 
charges behaves as a dilute Bose gas with 
strong repulsion. Under these conditions, analytic and numerical 
calculations~\cite{nazarov94} support the idea that in a magnetic field 
Cooper pairs form Laughlin-type incompressible states (a Cooper pair fluid):
the charge density 
changes in a stepwise function by changing the external parameters. The 
incompressible states give rise to the quantization of the Hall 
conductance 
$$
	\sigma_{xy} = \frac{4e^2 \nu}{h}
$$
where the filling factor $\nu=q/f$ is given by the ratio between the charge 
density $q$ and the magnetic frustration $f$. According to Odintsov and Nazarov 
two sets of Hall plateaus exist. One corresponds to the fractional quantum Hall 
effect with $\nu= 2m$ ($m$ integer). 
The other corresponds to the integer quantum  Hall effect with $\nu= l/2$ 
($l$ integer).

The opposite limit in which the Josephson energy dominates, has been considered 
in Refs.~\cite{choi94,stern94}. 
In this case vortices condense to form a 
quantum Hall fluid and the transverse conductivity is now given by:
$$
	\sigma_{xy} = 2m \frac{4e^2 }{h} \;\;\; .
$$

Despite the similarities highlighted here, there are important differences 
as compared to the ``electronic'' case. 
Both charges and vortices are bosons, moreover they are interacting particles.
Interactions modify the results. For instance, in the 
vortex case the logarithmic interaction changes the 
longitudinal response~\cite{stern94} as well. 

Up to now there is no experimental evidence for the quantum Hall effect in 
Josephson arrays. 
For the Cooper-pair fluid, the random offset charges form a serious obstacle
for observation of the Quantum Hall states.
For the vortex fluid the situation is less clear.
Additional theoretical work is required to 
locate the region in parameter space where the Hall fluid is the 
ground state (as opposed to the Abrikosov lattice for example). 
Effects related to disorder, quantum correction of the mass, and dissipation 
should all be taken into account.

\subsection{Quantum Computation with Josephson junctions}

Quantum Computation (QC) has recently excited many scientists from 
various different areas of physics, mathematics and computer science. 
In contrast to its classical counterpart, quantum information processing
is based on the controlled unitary evolution of quantum mechanical
systems. The great interest in this field is certainly related
to the fact that some problems which are intractable with classical
algorithms can be solved much faster with QC. Factorization of large numbers
as proposed by Shor is probably the best known example in this respect. 
This section briefly reviews the  recent work in this field
using Josephson nano-circuits.  For excellent reviews devoted to QC
we refer to Refs.\cite{ekert96,steane98}. Furthermore, many elementary
books on quantum mechanics treat the physics of two-level
systems. A review devoted to the implementation of quantum computation 
by means of Josephson nano-circuits just appeared~\cite{makhlinrev}.

The elementary unit of any quantum information process is the
{\it qubit}. The two values of the classical bit are replaced by
the ground state ($|0>$) and excited ($|1>$) state of a two-level
system. (Note that it common to adopt the spin-1/2 language as we 
will do here.) Already at this stage a fundamental difference between 
classical and quantum bits emerges. While information is stored either 
in $0$ or in $1$ in a classical bit, any state $|\psi(t)> = a(t)|0> + b(t)|1>$
can be used as a qubit. 

Manipulations of spin systems have been widely  studied and nowadays NMR 
physicists can prepare the spin system in any state and let it evolve 
to any other state.
Controlled evolution between the two degenerate states $|0>$ and $|1>$
is obtained by applying resonant microwaves to the system but 
state control can also be achieved with a fast DC pulse of high amplitude. 
By choosing the appropriate pulse widths, the NOT operation can be established 
\begin{eqnarray}
        |0>  & \longrightarrow & |1> \\
                \nonumber
        |1>  & \longrightarrow & |0> \\
                \nonumber
\end{eqnarray}
or the Hadamard transformation
\begin{eqnarray}
        |0>  & \longrightarrow & \frac{1}{\sqrt{2}}(|0> + |1>) \\
                \nonumber
        |1>  & \longrightarrow & \frac{1}{\sqrt{2}}(|0> - |1>) \; .
\end{eqnarray}

These unitary operations alone do not make a quantum computer yet.
Together with one-bit operations it is of fundamental importance to 
perform two-bit quantum operations; i.e., to control the unitary evolution
of entangled states. Thus, a universal quantum computer needs both one and 
two-qubit gates (it has been shown that most of two-qubit gates
are universal\cite{barenco95}). 
One example of a two-qubit gate is the Control-NOT operation:
\begin{eqnarray}
        |00>  & \longrightarrow & |00> \\
                \nonumber
        |01>  & \longrightarrow & |01> \\
                \nonumber
        |10>  & \longrightarrow & |11> \\
                \nonumber
        |11>  & \longrightarrow & |10>
\end{eqnarray}
The unitary single-bit operations and this 
Control-NOT operation are sufficient for performing all tasks of a 
quantum computer. Therefore, quantum computers can be viewed as programmable 
quantum  interferometers. Initially prepared in a 
superposition of all the possible input states, the computation evolves in 
parallel along all its possible paths, which interfere constructively 
towards the desired output state. 
It is this intrinsic parallelism in the evolution of quantum systems that 
allows for exponentially more efficient ways of performing computation.

It is of crucial importance that qubits are protected from the environment, 
i.e., from any source that could cause decoherence~\cite{decoherence}.
This is a very difficult task because at the same time one also has to 
control the evolution of the qubits, which inevitably means that the qubit 
is coupled to the environment. In quantum optics experiments, single atoms 
are manipulated which are almost decoupled from the outside world. 
Large-scale integration (needed to make a quantum computer useful) seems to be 
on the other hand impossible. Qubits made out of solid-state devices 
(spins in quantum 
dots or superconducting nano-devices), may offer a great advantage in 
this respect because fabrication techniques allow for scalability to 
a large number of coupled qubits. 

At present different proposals have been put forward to use superconducting 
nano-circuits~\cite{makhlin97-99,averin98,mooij99,ioffe99,fazio99} for the 
implementation of quantum algorithms. Depending on the operating regime, they 
are commonly referred to as charge~\cite{makhlin97-99,averin98,fazio99,falci00}
and flux~\cite{mooij99,ioffe99} qubits. 
We briefly discuss both approaches and summarize the experimental 
advances made so far.

\begin{figure}
\centerline{{\epsfxsize=14cm\epsfysize=18cm\epsfbox{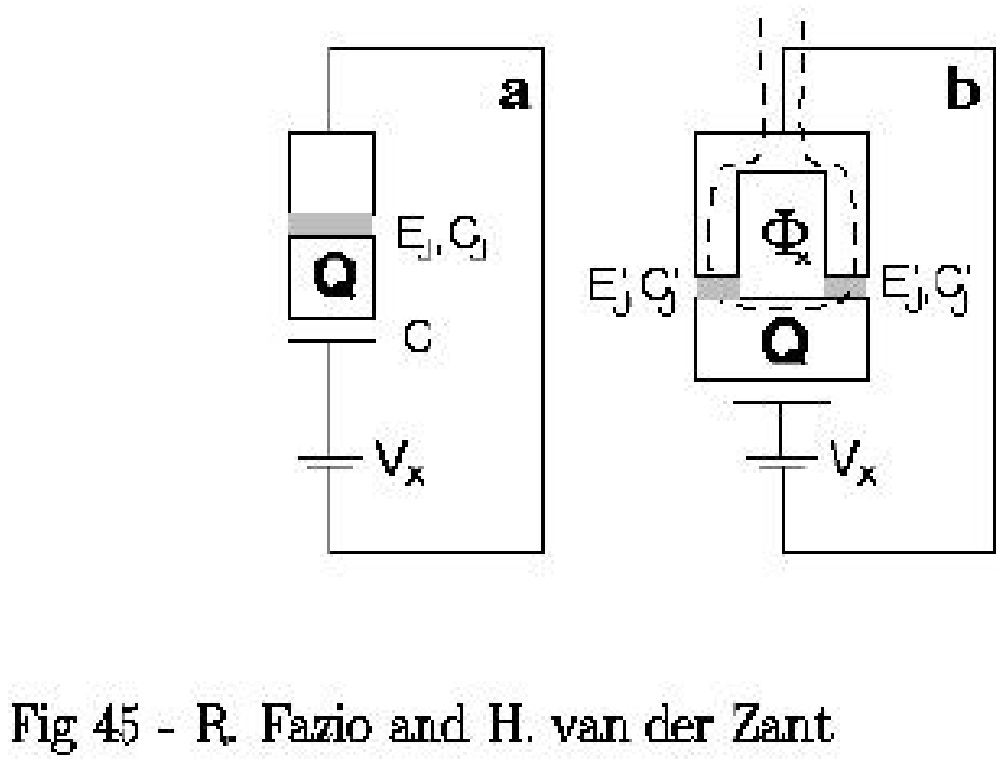}}}
\caption{a) A charge qubit.
	b) Improved qubit design as proposed by the Karlsruhe group. 
	The island is coupled to the
	circuit via two Josephson junctions with parameters $C'_{\rm J}$
	and $E_{\rm J}'$. This dc-SQUID is tuned by the external flux 
	which is controlled by the current through the inductor loop 
	(dashed line). The setup allows switching the effective Josephson \
	coupling to zero. 	
	(Reprinted by permission from  Nature {\bf 398}, 305 (1999 copyright 1999
	Macmillan Magazines Ltd.)}
\label{onebit}
\end{figure}

\underline{Charge qubits\cite{makhlin97-99,averin98}} -
In this case the qubit is realized by the two nearly degenerate charge
states of a single electron box as shown in Fig.\ref{onebit}.
They represent the states $|0 \rangle $, $|1 \rangle $ 
of the qubit. In the computational Hilbert space 
the ideal evolution of the system is governed by the Hamiltonian
\begin{equation}
	H  = 
		-\Delta E_{{\rm ch},i} (
	        (|0><0| - |1><1|)		 
	- \frac{E_{J}}{2} (|0><1| + |1><0| ) 
\label{oneb}
\end{equation}
where $\Delta E_{{\rm ch},i} = E_{\rm ch}(n_{x} - 1/2)$. Any one bit
operation can be realized by varying the external
charge $n_x$ and, in the proposal of Ref.~\cite{makhlin97-99} 
by varying the Josephson coupling as well. Modulation of $E_J$ is achieved 
by placing the Cooper-pair box in SQUID geometry. 
The advantage of this choice is that
during idle times the Hamiltonian can be "switched off" completely
eliminating any trivial phase accumulation which should be subtracted
for computational purposes.

As discussed before, a quantum computer can be realized
once two bit gates are implemented. The Karlsruhe group has proposed an
inductive coupling between qubits which lead to a coupling of the type
\begin{equation}
	H_C  = 
		-E_{L} \sigma_y^{(1)}\sigma_y^{(2)} \;\;\; .
\label{twob1}
\end{equation}
This type of coupling is very close in spirit to the coupling
used in the ion-trap implementation of QC. The main advantage of this
choice is that qubits are coupled via an infinite range coupling and
that the two bits can easily be isolated.  
A different scheme has been proposed in Ref.\cite{averin98}. 
They emphasize the adiabatic aspect of conditional dynamics and suggest 
to use capacitive coupling between gates as to reduce unwanted transitions 
to higher charge states.  The coupling reads $H_C  = -E_{C}
\sigma_z^{(1)}\sigma_z^{(2)}$ and the qubit is now defined as 
a finite one-dimensional array of junctions. By means of gate voltages
applied at different places in the array the bit-bit coupling can
be modulated in time and a control-NOT can be realized. 

The experiments on the superposition of charge states in Josephson 
junctions~\cite{Matters95,bouchiat98} and the recent achievements in 
controlling the  coherent evolution of quantum states in
a Cooper pair box~\cite{nakamura99} render superconducting nanocircuits 
interesting candidates to implement solid state quantum computers.
The experiment by Nakamura {\em et al.}~\cite{nakamura99} goes as follows.
Initially, the system is prepared in the ground state.
Appropriate voltage pulses bring the system in resonance so that the two 
charge states are in a coherent superposition $a(t)|0> + b(t)|1>$.
The final state is measured by detecting a tunneling current through 
an additional probe-junction. 
For example, zero tunneling current implies that the system ended up in the 
$|0>$ state, whereas a maximum current indicates that the final state 
corresponds to the excited one. 
In the experiment the tunneling current shows an oscillating behavior as 
a function of pulse length, thereby demonstrating the evolution of a 
coherent quantum state in the time domain.

Nakamura {\em et al.} also estimate the dephasing time and report it to be 
of the order of few nanoseconds.
The probe junction and 1/f noise presumably due the motion to trapped 
charges are the main source of decoherence. In their absence, the main 
dephasing mechanism is thought to be spontaneous photon emission to 
the electromagnetic environment. 
Decoherence times of the order of 1~$\mu$s should then be possible. 

\underline{Phase qubits\cite{mooij99,ioffe99}} - 
A qubit can also be realized with superconducting nano-circuits in 
the opposite limit $E_J \gg E_C$. An rf-SQUID (a superconducting loop 
interrupted by a Josephson junction) provides the prototype of such a device.  
The Hamiltonian of this system reads
\begin{equation}
{\cal H}=-E_J\cos\left(2\pi\frac{\Phi}{\Phi_0}\right)
+\frac{(\Phi-\Phi_x)^2}{2L} +\frac{Q^2}{2C} \;.
\end{equation}
Here, $L$ is the self-inductance of the loop and the phase difference 
across the junction ($2\pi\Phi/\Phi_0$) is related the flux $\Phi$ in 
the loop. 
The externally applied flux is denoted by $\Phi_x$.
The charge $Q$ is canonically conjugated to the flux $\Phi$.  
In the limit in which the self-inductance is large, the two first terms 
in the Hamiltonian form a double-well potential near $\Phi=\Phi_0/2$.  
Also in this case the Hamiltonian can be reduced to that of a two-state system. 
The term proportional to $\sigma_z$ measure  the asymmetry of 
the double well potential and the off-diagonal matrix elements depend on 
the tunneling amplitude 
between the wells. By controlling the applied magnetic field, all 
elementary unitary operations can be performed.

\begin{figure}
\centerline{{\epsfxsize=14cm\epsfysize=16cm\epsfbox{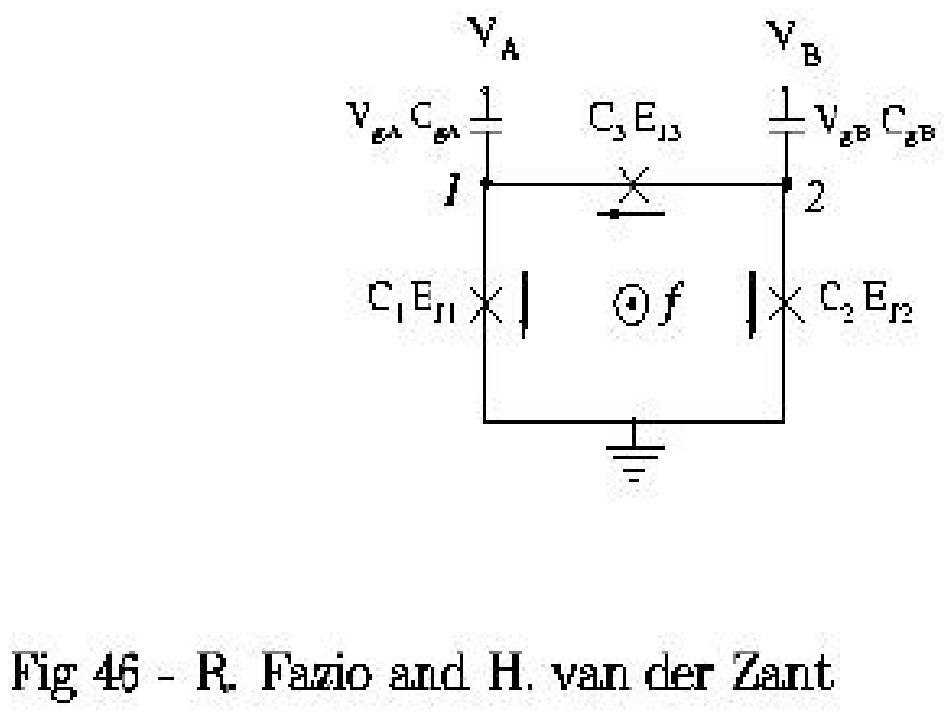}}}
\caption{The three-junction flux qubit.   
Josephson junctions~1 and 2 have the same Josephson energies $E_J$ and capacitance $C$; 
Josephson junction~3 has a Josephson energy and capacitance that are $\alpha$ times larger. 
The islands are coupled by gate capacitors $C_g= \gamma C$ to gate voltages $V_A$ and $V_B$. 
The arrows define the direction of the currents.  
(From Ref.~\protect\cite{mooij99}.)}
\label{fqubit}
\end{figure}

In order to fulfill various operational requirements more refined 
designs should be used.
In the proposal of Mooij {\em et al.}~\cite{mooij99}, qubits are formed 
by three junctions (see Fig.~\ref{fqubit}). Flux qubits are coupled 
by means of flux transformers which provide inductive coupling between them. 
Any loop of one qubit can be coupled to any loop of the other, but 
to turn off this coupling, 
one would need to have an ideal switch in the flux transformer.
This switch is to be controlled by high-frequency pulses and the related
external circuit can lead to decoherence effects.  

At present, both the Stony Brook
group (Friedman {\em et al.}~\cite{friedman00}) and the Delft group 
(Van der Wal {\em et al.}~\cite{wal00}) have demonstrated superposition 
of two magnetic flux states in superconducting loops. One state 
corresponds to the magnetic moment of $\mu$A-currents flowing clockwise 
whereas the other corresponds the same moment but of opposite sign due 
to the current flowing anti-clockwise. Coherent quantum oscillations have 
not yet been detected. 
To probe the time evolution, pulsed microwaves instead of continuous 
ones have to be applied. Observation of such oscillations would imply 
the demonstration of macroscopic quantum coherence (MQC). 
It is called macroscopic because the 
currents are built of billions of electrons coherently circulating 
within the superconducting ring. 

There are two main differences between the approaches of the two groups.
The Stony Brook group uses the excited states of an RF SQUID. 
The Delft circuit consists of a three-junction system and the continuous 
microwaves induce transitions between the ground state and the first 
excited state only. The three-junction geometry has the advantage that 
it can be made much smaller so that it is less sensitive to noise 
introduced by inductive coupling to the environment. Nevertheless, 
recent insights indicate that in all designs put forward so far the measuring 
equipment destroys quantum coherence. The meter is 
believed to be main obstacle to study the 'intrinsic' decoherence 
times.
Future work must evaluate the role of the measuring equipment and 
new measuring schemes should be developed in order to study MQC in 
Josephson loops.

\acknowledgments
Special thanks go to J.E. Mooij and G. Sch\"on for many years of 
fruitful collaboration, guidance and for sharing their insight and 
knowledge with us. 
\\
With C. Bruder, G. Falci, B. Geerligs, G. Giaquinta, T.P. Orlando, 
A. van Otterlo we had a 
long-standing fruitful and pleasant collaboration, we would like 
to thank them for this. We also thank C. Bruder for a careful 
reading of the manuscript.
\\
We acknowledge L. Amico, R. Baltin, P.A. Bobbert, Ya. Blanter, A. Kampf,
W. Elion, A. Tagliacozzo, K.-H. Wagenblast, G. T. Zaikin, 
D. Zappal\`a,  and G.T. Zimanyi for valuable collaboration on these topics.
\\
Finally we thank O. Buisson, J. Clarke, P. Delsing, A. Fubini, L. Glazman,
D.B. Haviland, P. Martinoli, Yu. Nazarov, A. Odintsov, A. van
Oudenaarden, B. Pannetier, 
M. Rasetti, P. Sodano, L. Sohn, M. Tinkham, V. Tognetti for many 
useful conversations.
\\
R.F is supported by the European Community under TMR and IST 
programmes and by INFM-PRA-SSQI, H.v.d.Z. is supported by the Dutch
Royal Academy of Arts and Sciences (KNAW).

\begin{appendix}
	\section{Array fabrication and experimental details}

\label{fabrication}
There are two types of junctions arrays: {\it proximity}
coupled arrays and arrays made of Josephson {\it tunnel}
junctions. The proximity coupled arrays consist of
superconducting islands (Nb or Pb) on top of a normal metal film (Cu)
and are solely used for the study of classical phenomena.
The main reason for this is the low (less than 1 Ohm)
normal-state resistance of the junctions.

At present there are two technologies to fabricate arrays with Josephson
{\it tunnel} junctions. Commercial niobium junctions are
fabricated using a trilayer with aluminum oxide as insulating barrier.
Reliable niobium junctions, however, are too large to observe the
quantum effects discussed in this review.
Quantum arrays are built up of all aluminum, tunnel junctions.
These Josephson junctions are fabricated with a shadow-evaporation
technique\cite{fulton87}. We will now outline the most recent
fabrication technique as used in the Delft group.

Samples are fabricated on silicon substrates with an insulating
SiO$_2$ top layer. In the first step two resist layers are spun
on the substrate. The lower layer is a solution of PMMA/MAA copolymer
in acetic acid and the upper layer is a solution of PMMA in
chlorobenzene. The resist sandwich is baked at 180~$^{\rm o}$C
for one hour. Then the sample mask is written by high-resolution
electron-beam lithography at 100~kV. After writing, the exposed resist
is developed in a 1:3 mixture of MIBK and 2-propanol for one minute.
The solubility of the lower resist is larger than the upper layer,
which leads to an undercut in the
lower resist layer. This undercut is necessary for the formation
of free-hanging bridges below which the junctions are formed
in the shadow evaporation technique.

After mask definition, a 24 nm thick aluminum layer is evaporated
under a given angle. Then the aluminum layer is exposed
to pure oxygen at a controlled pressure. By changing the pressure
the thickness of the aluminum oxide barrier is varied. In the second
evaporation step (the sample is not taken out of the vacuum)
a 40 nm thick aluminum layer is evaporated under the opposite angle.
After this step, the tunnel junction is formed. The remaining
resist layers with the unwanted aluminum on top are
removed by rinsing the sample in acetone.

We end this appendix with some remarks on the measuring set-up.
Arrays are measured in a dilution refrigerator inside $\mu$-metal and
lead magnetic shields at temperatures
down to 10$\thinspace$mK.
To protect the arrays from high energy photons generated by
room-temperature noise and radiation,
extensive filtering and placing the arrays in a closed copper box
are minimum requirements. Therefore, a typical set-up for the
measurements of quantum arrays has the following characteristics.
At the entrance of the cryostat, electrical leads
are filtered with radio-frequency interference (RFI)
feedthrough filters.
Arrays are placed inside a closed, grounded
copper box (microwave-tight).
All leads leaving this box are filtered with RC
filters for low-frequency filtering
($R=1$~k$\Omega$ and $C=470$~pF)
and with microwave filters.
A microwave filter consists of a coiled manganin wire
(length $\sim 5$~m), put inside a grounded copper tube that is
filled with copper powder (grains $< 30$~$\mu$m). The resistance of
the wire in combination with the capacitance to ground via the
copper grains provide an attenuation over 150~dB at frequencies
higher than 1~MHz.
The copper box with the RC and microwave filters is
situated in the inner vacuum
chamber and is mounted on the mixing chamber in good thermal
contact.

	\section{Triangular arrays and geometrical factors}

 \label{geometrical-factors}

In comparing properties of 
square and triangular arrays some care is necessary.
The energy required to store an additional electron on an island
is ${e^2 / 2C_{\Sigma}}$, where $C_{\Sigma}$ is the sum of the capacitances
to other islands and to ground. As in triangular arrays
all islands are coupled with $z=6$ instead of $z=4$ junctions, 
the required energy is 2/3 times smaller than that 
of an island in a square lattice.
Similarly, the freedom of the phase on a particular island is determined
by the Josephson coupling energy of all junctions connected to the island
and therefore it seems reasonable to assume that
in a triangular array the effective Josephson coupling energy is
3/2 times that of a square array.
Summarizing, we come to the conclusion that the effective $E_C/E_J$-ratio for 
a triangular array is a factor 4/9 lower than that of a square array.

	\section{Phase correlator}

\label{phase-correlator}
In this Appendix we show how to evaluate the phase-phase 
correlator $g_{ij}$ introduced in Section~\ref{phase0} (see also 
Ref.~\cite{bruder92}). The starting point is:
\begin{eqnarray}
g_{i0}(\tau)
	&=&\langle \exp[ \phi_{i}(\tau ) -\phi_{0}(0) \rangle_{ch}
	\nonumber \\
	&=&\frac{1}{Z_{ch}}\sum_{\{n_j\}}
\prod_j\int d\phi_{j0}\int_{\phi_{j0}}^{\phi_{j0}+2\pi n_j}
{\cal D} \phi_j e^{-S_{ch}} e^{ 2\pi i \sum_j q_{xj}n_j +
i[\phi_i(\tau)-\phi_0(0)]}
\label{eq:A1}
\end{eqnarray}
By making use of the parametrization 
$$
\phi_i(\tau) = \phi_{i0} + 2\pi i n_i \frac{\tau}{\beta} +\theta_i(\tau)
$$ 
it is possible to verify  that all the off-diagonal elements of the
correlation function, viz. $g_{i0}(\tau)$ for $i\ne 0$ vanish because of the
integrations over $\phi_{j0}$. The reason is that $S_{ch}$ does not depend on
the phase $\phi_0(\tau)$ itself but only on its time derivative.
It is therefore sufficient to calculate the on-site correlation function at
site
\begin{equation}
g(\tau) \equiv g_{00}(\tau)=\frac{1}{Z_{ch}}\sum_{\{n_j\}} \exp\Big(2\pi i
\sum_j q_{xj}n_j - T\sum_{ij}\frac{4\pi^2}{8e^2}n_i C_{ij} n_j\Big)
\exp(-2\pi i T n_0 \tau ) \, g_c(\tau)
\end{equation}
where $g_c(\tau)$, the correlation function for the case of continuous
charges, results from the remaining integral over $\theta(\tau)$. 
It is given by
\begin{equation}
g_c(\tau) = \exp\big[-2e^2 C_{00}^{-1} \tau(1-\tau T)\big] \; .
\end{equation}
By using the Poisson resummation formula
\begin{equation}
\sum_{\{n_i\}}\exp(-\sum_{ij}n_i A_{ij} n_j +2 \sum_i z_i n_i)=
\sqrt{\frac{\pi^N}{det A}}\sum_{\{q_i\}}\exp[-\sum_{ij}(\pi q_i+z_i)A_{ij}^{-1}
(\pi q_j+z_j)]
\end{equation}
and performing the Fourier transform 
$$
	g(\omega_\nu)=\int_0^\beta d \tau \exp(i\omega_\nu\tau) g(\tau)
$$
one obtains
\begin{equation}
g(\omega_\nu)=\frac{1}{Z_{ch}}\sum_{\{q_i\}}
e^{-\frac{2e^2}{T}\sum_{ij}(q_i-q_x)C_{ij}^{-1}(q_j-q_x)}
\frac{4e^2C_{00}^{-1}}{[2e^2C_{00}^{-1}]^2-
[4e^2\sum_jC_{0j}^{-1}(q_j-q_x)-i\omega_\nu]^2} \;\; .
\label{appcorr}
\end{equation}
The expressions for the coefficients $\epsilon$, $\gamma$, $\zeta$ and 
$\lambda$ in the 
coarse-graining approach follow from an expansion at small frequencies of 
Eq.(\ref{appcorr}).

	\section{Derivation of the coupled Coulomb gas action}
\label{coupled-coulomb-gas}
In this Appendix we briefly discuss the steps leading  to Eq.(\ref{Sqv11}). 
following the calculation in Ref.~\cite{fazio91a,fazio91b}. First a path integral
representation is introduced for the island charges. In terms of phase 
trajectories $\phi_i(\tau)$ and
charges $q_i(\tau) = Q_i(\tau)/2e$ the partition function takes the form
\begin{equation} \label{part11}
	Z  =  \prod_j \int dq_{j0} \int Dq_j \prod_{i} 
	  \sum_{\{m_i\}} 
	  \int D\phi_i (\tau) \exp (-S\{\phi,q\}),
\end{equation}
where the phases obey the boundary conditions 
$\phi_{j}(0) = \phi_{j}(\beta)+2\pi n_j$,
while the charge paths are periodic, 
$q_j(0) = q_j(\beta) = q_{j0}$. 
The effective action in the mixed representation is
\begin{equation}
	S\{\phi,q\} = \int_{0}^{\beta} d\tau 
	\left\{  \vphantom{\frac{}{{}_{}}}
	2e^2 \sum_{i,j} q_i(\tau) C^{-1}_{ij} q_j(\tau)
	+ i \sum_i q_i(\tau) \dot\phi_i(\tau) 
	- E_{\rm J}\sum_{\langle ij \rangle} \cos (\phi_{i} - \phi_{j})
					 \vphantom{\frac{}{{}_{}}} \right\}. 
\label{Sqphi5}
\end{equation}
Summation over winding numbers $\{n_i\}$ fixes the
charges $q_i$ to be integer-valued.
Starting from the partition function (\ref{part11}), we first introduce
the vortex degrees of freedom. This can be done by means of the Villain
transformation~\cite{villain75} (see also \cite{jose77}); the
time-dependent quantum problem requires some additional steps
\cite{fazio91a,fazio91b}. We introduce the lattice with spacing $\epsilon$ in
time direction; this spacing is of order of inverse Josephson
frequency: $\epsilon \sim (8E_{\rm J}E_{\rm C})^{1/2}$. In the Villain
approximation one replaces 
\begin{equation}
	\exp \Big\{ -\epsilon E_{J} 
	\sum_{\langle i j \rangle, \tau} 
	[1 - \cos(\phi_{i,\tau} - \phi_{j,\tau})] \Big\} 
	\rightarrow \sum_{\{
	{\bf m}_{i\tau}\}} \exp \Big\{ -\frac{\epsilon E_{\rm J} F(\epsilon
	E_{ J})}{2} \sum_{i,\tau} \vert \nabla \phi_{i\tau} - 2\pi
	{\bf m}_{i\tau} \vert^2 \Big\}. 
\label{coup1}
\end{equation}
Here, we have introduced a two-dimensional vector field
${\bf m_{i\tau}}$, defined on the dual lattice (alternatively, it can be
considered as a scalar field defined on bonds). The function
$$
	F(x) = \frac{1}{2x \ln({J_0(x)/J_1(x)})} \to \frac{1}{2x \ln(4/x)}, 
\;\;\;\; x \ll 1,
$$ 
has been introduced to ``correct'' the Villain transformation
for small $E_J$. 
As we see, its entire effect is to
renormalize (increase) the Josephson coupling $E_{J} \to E_{J}
F(\epsilon E_{J})$, but it does not affect the physics. In
the following we will use only the renormalized constant. 

The rhs.\ of Eq.(\ref{coup1}) can be rewritten as
$$
	\sum_{\{ {\bf J}_{i\tau}\}} \exp 
	\Big\{ -\frac{1}{2\epsilon E_{J}}
	\sum_{i,\tau} \vert {\bf J}_{i\tau} \vert^2 - i {\bf J_{i\tau}}
	\nabla \phi_{i\tau} \Big\} 
$$
Now the Gaussian integration over the phases can be easily performed,
yielding 
\begin{equation}
	Z = \sum_{q_{i\tau}} \sum_{{\bf J}_{i\tau}} \exp \Big\{ -2e^2
	\epsilon \sum_{i,j,\tau} q_{i\tau} C^{-1}_{ij}
	q_{j\tau}  - \frac{1}{2\epsilon E_{J}} \sum_{i,\tau} \vert
	{\bf J}_{i\tau} \vert^2 \Big\}, 
\end{equation}
and the summation is constrained by the continuity equation, 
$$
	\nabla \cdot {\bf J}_{i\tau} - \dot{q}_{i\tau} = 0 \;.
$$
The time derivative stands for a discrete derivative
$\dot{f} (\tau) = \epsilon_{\mu}^{-1} [f(\tau + \epsilon_{\mu}) -
f(\tau)]$.
The constraint is satisfied by the parameterization~\cite{elitzur79}
$$
	J^{(\mu)}_{i\tau} = 
	n^{(\mu)} ({\bf n}\nabla)^{-1} \dot{q}_{i\tau} +
	\epsilon^{(\mu\nu)} \nabla_{\nu} A_{i\tau}.
$$ 
Here the operator $({\bf n}\nabla)^{-1}$ is the line integral on the
lattice (in Fourier space it has the form $i(k_x + k_y)^{-1}$),
$\epsilon^{(\mu \nu)}$ is the antisymmetric tensor, while
$A_{i\tau}$ is an unconstrained integer-valued scalar field. 
It is important to mention that the continuity equation can be solved 
in different ways which can be mapped onto each other by gauge transformations
(see Ref.\cite{fisher89a}).  

With the use of the Poisson resummation (which requires introducing  a new
integer scalar field $v_{i\tau}$) the partition function can be
rewritten as
$$
	Z = \sum_{[q_{i\tau},v_{i\tau}]} \exp - S \{q,v\} \;.
$$
The effective action for the integer charges $q_{i}$ and vorticities
$v_{i}$ is
\begin{eqnarray}
	S[q,v] &=&  2e^2 \epsilon \sum_{ij\tau} q_{i\tau}
	C^{-1}_{ij} q_{j\tau} 
 	- \frac{1}{2\epsilon E_{\rm J}}
	\sum_{i\tau} \Big[ n^{(\mu)}
	({\bf n} \nabla)^{-1} \dot{q}_{i\tau} \Big]^2   
        \nonumber  \\
	&- & \frac{\epsilon
	E_{\rm J}}{4\pi} \sum_{ij\tau} \Big[ 2\pi v_{i\tau} 
	-
	\frac{i}{\epsilon E_{\rm J}} 
	\epsilon^{(\mu \nu)} \nabla_{\nu} n^{(\mu)}
	({\bf n} \nabla)^{-1} \dot{q}_{i\tau} \Big] G_{ij}
	\Big[ 2\pi
	v_{i\tau} - \frac{i}{\epsilon E_{\rm J}} 
	\epsilon^{(\mu \nu)} \nabla_{\nu}
	n^{(\mu)} ({\bf n} \nabla)^{-1} \dot{q}_{j\tau} \Big]	\nonumber 
\end{eqnarray}
The kernel $G_{ij}$ is the lattice Green's function,
i.e., the Fourier transform of $k^{-2}$. Finally, after some algebra
\cite{fazio91b} one arrives to the effective action of Eq.(\ref{Sqv11}), 
which we rewrote, for simplicity, in the continuous notations.

	\section{Effective single vortex action}

\label{single-vortex}
It is possible to derive  from 
the coupled Coulomb gas action an effective action for a single vortex of 
vorticity $v=\pm 1$. The single vortex effective action  includes the effect 
of the interaction with fluctuating charges and the other vortices which 
are present in the system due to quantum fluctuations. 
The desidered effective action is formally obtained by performing 
the sum in the partition function
over all charge and vortex configurations excluding the vortex whose dynamics
is to be studied. We introduce the vortex trajectory 
$$
v_{i,\tau}=v\delta({\bf r}_{i}-{\bf r}(\tau)) \;\;\;\; .
$$
The single vortex effective action can then be written as
\begin{eqnarray}
\nonumber
	S_{eff}= -\ln{\Big \langle} \epsilon 2\pi E_{J} v\sum_{ij,\tau}
	v_{i,\tau}G_{ij}\delta({\bf r}_{j}-{\bf r}(\tau))+\\
	+iv\sum_{ij,\tau} \dot{q}_{i,\tau}\Theta_{ij}
	\delta({\bf r}_{j}-{\bf r}(\tau))
	+iv\epsilon\sum_{ij,\tau} {\bf I}_{i,\tau}\cdot {\bf \nabla}\Theta_{ij}
	\delta({\bf r}_{j}-{\bf r}(\tau)){\Big \rangle} \;\; ,
\label{eq:1ststep}
\end{eqnarray}
where the average is to be taken with the full coupled Coulomb gas action. 
The first term describes the static interaction with other vortices, 
whereas the second describes the dynamical interaction with charges.
This expression, although exact, cannot be evaluated explicitly because of
the nonlinearity of the action. To proceed we expand the dynamical part of 
the average in Eq.(\ref{eq:1ststep}) in cumulants up to second order in the 
vortex velocity $\dot{r}(\tau)$. 
For a uniform external current distribution the result is 
$$
	S_{eff}=\frac{1}{2}\sum_{\tau\tau'} \dot{{\bf r}}^{a}(\tau) 
	{\cal M}_{ab}({\bf r}(\tau)-
	{\bf r}(\tau'),\tau-\tau')\dot{{\bf r}}^{b}(\tau') +
	2\pi i v\epsilon\sum_{\tau}\epsilon_{ab}I^{a}_{\tau}{\bf r}^{b}(\tau)\; ,
$$
\begin{equation}
	{\cal M}_{ab}=\sum_{jk} \nabla_{a}\Theta({\bf r}(\tau)-{\bf r}_{j}) 
	\langle q_{j\tau}
	q_{k\tau'}\rangle \nabla_{b}\Theta({\bf r}_{k}-{\bf r}(\tau')) \; ,
\end{equation}
where $a,b=x,y$ and $\epsilon_{ab}$ is the anti-symmetric 
tensor~\cite{chargemass}. 

\newpage
	\section{List of symbols}
\begin{tabular*}{12cm}[t]{ll}
 
Josephson Energy \hspace{3cm} & $E_J$ \\

Junction capacitance \hspace{3cm} & $C$ \\

Ground capacitance \hspace{3cm} & $C_0$ \\

Capacitance matrix \hspace{3cm} & $C_{ij}$ \\

Charging energy (junction) \hspace{3cm} & $E_C = e^2/(2C)$ \\

Charging energy (ground) \hspace{3cm} & $E_0 = e^2C_{00}^{-1}/2$ \\

Index labels for the islands in the array \hspace{3cm} & $i,j$ \\

Superconducting order parameter \hspace{3cm} & $ \Delta  e^{i\phi}$\\

Charge on the island \hspace{3cm} &  $Q$\\

External charge \hspace{3cm} &  $Q_x$\\

Vector potential \hspace{3cm} &  $A_{ij}= \int_i^j {\bf A}\cdot d{\bf l}$\\

Flux quantum  \hspace{3cm} &  $\Phi_0 = h/2e$\\

Magnetic frustration per plaquette \hspace{3cm} &  $f$\\

External current \hspace{3cm} &  $I$\\

Junction critical current \hspace{3cm} &  $I_c$\\

Quantum of resistance \hspace{3cm} &  $R_Q = h/(4e^2)$\\

Dissipation strength  \hspace{3cm} & $\alpha$ \\

Josephson plasma frequency \hspace{3cm} & $\omega_p = \sqrt{8E_JE_C}$ \\ 

BCS transition temperature  \hspace{3cm} &  $T_c$\\

Vortex unbinding transition \hspace{3cm} &  $T_J$\\

Charge unbinding transition \hspace{3cm} &  $T_{ch}$\\

Universal conductance at the S-I transition \hspace{3cm} &  $\sigma^*$\\

Vortex mass \hspace{3cm} &  $M_v$\\

McCumber parameter \hspace{3cm} &  $\beta _c$\\

Josephson Junctionn Array \hspace{3cm} &  JJA\\

Superconductor - Insulator \hspace{3cm} & S-I \\

Berezinskii - Kosterlitz - Thouless   \hspace{3cm} & BKT \\

Aharonov - Bohm \hspace{3cm} & AB \\

Aharonov - Casher \hspace{3cm} & AC \\

Bose - Hubbard  \hspace{3cm} & BH \\

Quantum Phase Model \hspace{3cm} & QPM \\ 
 \end{tabular*}

\end{appendix}

\newpage
\newpage

\newpage

\newpage

\newpage

\newpage

\newpage

\newpage
\newpage

\newpage

\newpage

\newpage

\newpage

\newpage

\newpage

\newpage

\newpage

\newpage

\newpage

\newpage

\newpage

\newpage

\newpage

\newpage

\newpage

\newpage

\newpage

\newpage

\newpage

\newpage

\newpage

\newpage

\newpage

\newpage

\newpage

\newpage

\newpage

\newpage

\begin{references}
\bibitem{IBM}
	R.F. Voss and R.A. Webb, Phys. Rev. B {\bf 25}, 3446 (1982);
	R.A. Webb, R.F. Voss, G. Grinstein and P.M. Horn,
	Phys. Rev. Lett. {\bf 51}, 690 (1983).
\bibitem{nato86}
	{\it Percolation, Localization, and Superconductivity}, 
    	M. Goldman and S.A. Wolf Eds., NATO ASI {\bf 108} (1986). 
\bibitem{delft88}
	{\it Coherence in Superconducting Networks},
    	J.E Mooij and G. Sch\"{o}n Eds., Physica {\bf B 152} pp 1-302 (1988).
\bibitem{trieste95}
	{\it Josephson Junction Arrays}, H.A. Cerdeira and
    	S.R. Shenoy Eds., Physica {\bf B 222}, pp 253-406 (1996).
\bibitem{frascati95}
	{\it Macroscopic Quantum Phenomena and Coherence in
    	Superconducting Networks}, C. Giovanella and M. Tinkham Eds., (World
    	Scientific, Singapore, 1995).
\bibitem{simanek94}
	E. Simanek, {\it Inhomogeneous Superconductors,
    	Granular and Quantum Effects}, (Oxford University Press, Oxford, 1994).
\bibitem{Berezinskii70}
	V.L. Berezinskii, Zh. Eksp. Teor. Fiz. 
    	{\bf 59}, 907 (1970) [Sov. Phys. JETP {\bf 32}, 493 (1971)].
\bibitem{Kosterlitz73}
	J.M. Kosterlitz and D.J. Thouless,
    	J. Phys. C {\bf 6}, 1181 (1973).
\bibitem{btkexp}
	J. Resnick, J. Garland, J. Boyd, S. Shoemaker, and
	R. Newrock, Phys. Rev. Lett. {\bf 47}, 1542 (1981);
	P. Martinoli, P. Lerch, C. Leemann, and H. Beck,
	J. Appl. Phys. {\bf 26}, 1999 (1987).
\bibitem{schoen90}
	G. Sch\"{o}n and A.D. Zaikin,
    	Phys. Rep. {\bf 198}, 237 (1990).
\bibitem{averin91}
	D.V. Averin and K.K. Likharev, {\it Mesoscopic
    	Phenomena in Solids},   B.L. Altshuler, P.A. Lee, and R.A. Webb 
	Eds., (North-Holland, Amsterdam, 1991).
\bibitem{sondhi97}
	S.L. Sondhi, S.M. Girvin, J.P. Carini, D. Shahar,
    	Rev. Mod. Phys. {\bf 69}, 315 (1997).
\bibitem{sachdev99}
	S. Sachdev, {\it Quantum Phase Transition}, (Cambridge
    	University Press, Cambridge, 1999).
\bibitem{goldman00}
	A. Goldman and N. Markovic, Physics Today November issue p. 39
	(1998).
\bibitem{weiss99}
	U. Weiss {\it Quantum Dissipative Systems}, (World Scientific,
	Singapore, 1999).
\bibitem{anderson64}
	P.W. Anderson, {\it Lectures on the Many Body
	Problem}, Caianiello Ed., (Academic Press, New York, 1964), p.113.
\bibitem{elion94}
	W.J. Elion, M. Matters, U. Geigenm\"uller, and J.E.
	Mooij, Nature {\bf 371}, 594 (1994).
\bibitem{matters95}
	M. Matters, W.J. Elion, and J.E. Mooij,
	Phys. Rev. Lett. {\bf 75}, 721 (1995).
\bibitem{tinkham96}
	M. Tinkham, {\it Introduction to Superconductivity},
    	(McGraw-Hill, New York, 1996).
\bibitem{barone82}
	A. Barone and G. Paterno, {\it Physics and Applications
    	of the Josephson Effect}, (J. Wiley, New York, 1982).
\bibitem{beenakker91}
	C.W.J. Beenakker and H. van Houten, Solid
    	State Physics {\bf 44}, 1 (1991).
\bibitem{imry97}
	Y. Imry, {\it Introduction to Mesoscopic Physics},
    	(Oxford University Press, Oxford, 1997).
\bibitem{karlsruhe94}
	{\it Mesoscopic Superconductivity}, F.W.J. Hekking,
    	G. Sch\"{o}n, and D.V. Averin Eds., Physica {\bf B 203} pp 
    	201-537 (1994).
\bibitem{curacao97}
	{\em Mesoscopic Electron Transport}, L.P. Kouwenhoven, L.L.
	Sohn and G. Sch\"on Eds., NATO ASI series E, Vol. 345, 
	Kluver (1997) 
\bibitem{doniach83}
	S. Doniach in {\em Percolation, Localization and 
	Superconductivity}, A.M. Goldman and S.A. Wolf Eds. 
	(Plenum Press, New York, 1983). 
\bibitem{mooij92}
	J.E. Mooij and G. Sch\"{o}n in {\it Single Charge
        Tunneling} H. Grabert and M.H. Devoret Eds., NATO ASI series 
        Vol.294 (Plenum, NY 1992), p. 275.
\bibitem{katsumoto95}
	S. Katsumoto, J. Low. Temp. Phys. {\bf 98}, 287 (1995).
\bibitem{ambegaokar63} 
	V. Ambegaokar and A. Baratoff,
	Phys. Rev. Lett. {\bf 10}, 486 (1963); {\bf 11}, 104(E) (1963).	
\bibitem{juncres}
	For small junctions with low $R_N$, the oxide barrier is of 
	the order of one atomic layer. Such thin layers
	may produce leaky junctions. We find that for our 1~fF junctions, 
	this lower limit is about 1~k$\Omega$. Thus, our aluminum tunnel 
	junctions may become leaky when $R_NC<10^{-12}$~s. 
	For very small junctions with $C<0.1$~fF, this
	criterion indicates that $R_N>R_Q \; (=h/4e^2=6.45$~k$\Omega$).
\bibitem{local}	
	The local rule instead of the global 
	rule\protect{\cite{schoen90}}
	to describe the coupling of tunneling processes to the 
	environment seems to be more appropriate because  the voltage 
	offset is measured at high-bias currents. 
\bibitem{lu98}
	J.G. Lu, J. M. Hergenrother, and M. Tinkham,
	Phys. Rev. B {\bf 57}, 4591 (1998). 
\bibitem{conduct}
	$C_0$ can be increased considerably when placing the array 
	on a conducting ground plane as done 
	by A.J. Rimberg et {\em al}, Phys. Rev. Lett. {\bf 78}, 2632 (1997).
\bibitem{jackson75}
	J. D. Jackson, {\it Classical Electrodynamics}, 
	(John Wiley $\&$  Sons, New York 1962).	
\bibitem{env}Quasi-particles may be generated by the 
	environment (e.g. by photons). 
	In a single small junction, the high-frequency coupling to the 
	environment determines the effective damping, yielding an effective 
	impedance of the order of $100~\Omega$ (G. Ingold and Yu. V. 
	Nazarov in {\it Single Charge tunneling} H. Grabert and M.H. 
	Devoret Eds., NATO ASI series vol.294 (Plenum,NY 1992)). 
	This impedance can be increased, i.e., a single junction can be 
	decoupled from its  environment by placing high-ohmic resistors 
	or arrays of small junctions in the leads  close to the junction. 
	From the latter we expect that junctions inside a 2D array 
	are decoupled from the leads.	
\bibitem{mccumber}
	In classical arrays the scale of damping in junctions 
	is commonly defined through the McCumber parameter 
	$\beta_c(T)=2\pi I_c(T) CR_e^2/\Phi_0$, 
	where $R_e$ is the effective 
	damping resistance for each junction.
\bibitem{caldeira81} 
	A.O. Caldeira and A.J. Leggett,
    	Phys. Rev. Lett. {\bf 46}, 211 (1981).
\bibitem{caldeira83} 
	A.O. Caldeira and A.J. Leggett, 
    	Ann. Phys. (N.Y.) {\bf 149}, 347 (1983).
\bibitem{ambegaokar82} 
	V. Ambegaokar, U. Eckern, and G. Sch\"{o}n,
    	Phys. Rev. Lett. {\bf 48}, 1745 (1982).
\bibitem{beloborodov00} 
	Recently Beloborodov {\em et al.} considered 
	normal arrays and found a new effective action that in the low 
	temperature limit yields the dynamically screened 
	Coulomb interaction of a normal metal, whereas at 
	high temperatures recovers the standard quantum dissipative 
	action; I.S. Beloborodov, K.B. Efetov, 
	A. Altland, F.W.J. Hekking, cond-mat/0006337.
\bibitem{fisher89b} M.P.A. Fisher, B.P. Weichman, G. Grinstein, and 
    	D.S. Fisher, Phys. Rev. B {\bf 40}, 546 (1989).
\bibitem{liu73}	
	K. Liu and M. Fisher, J. Low. Temp. Phys. 
    	{\bf 10}, 655 (1973).
\bibitem{bruder93} 
	C. Bruder, R. Fazio, G. Sch\"{o}n,
    	Phys. Rev. B {\bf 47}, 342 (1993).
\bibitem{amico00}
	L. Amico and V. Penna, Phys. Rev. B {\bf 62}, 1224 (2000).
\bibitem{das99}
	D. Das, S. Doniach, Phys. Rev. B {\bf 60}, 1261 (1999).
\bibitem{herbut99}
	I.F. Herbut, Phys. Rev. B {\bf 60}, 14503 (1999).
\bibitem{kopec00}
	T.K. Kopec and J.V. Jos\`e, 
	Phys. Rev. Lett. {\bf 84}, 749 (2000).
\bibitem{amico00a}
	L. Amico, Mod. Phys. Lett. B, {\bf 14}, 759 (2000).
\bibitem{fazio88}
	R. Fazio, G. Falci, and G. Giaquinta,
	Physica B {\bf 152}, 257 (1988).
\bibitem{lerner00}
	I. V. Yurkevich, I. V. Lerner, cond-mat/0007317.
\bibitem{geerligs89} 
	L.J. Geerligs, M. Peters, L.E.M. de Groot, 
    	A. Verbruggen, and J.E. Mooij, 
	Phys. Rev. Lett. {\bf 63}, 326 (1989).
\bibitem{zant96} 
	H.S.J. van der Zant, W.J. Elion, L.J. Geerligs, and 
    	J.E. Mooij, Phys. Rev. B {\bf 54}, 10081 (1996).
\bibitem{chen95} 
	C.D. Chen, P. Delsing, D.B. Haviland, Y. Harada, 
    	and T. Claeson, Phys. Rev. B {\bf 51}, 15654 (1995).
\bibitem{chen96} 
	C.D. Chen, P. Delsing, D.B. Haviland, Y. Harada,
    	and T. Claeson, Phys. Rev. B {\bf 54}, 9449 (1996).
\bibitem{strongin70} 
	M. Strongin, R.S. Thompson, O.F. Kammemer, and 
	J.E. Crow, Phys. Rev. B {\bf 1}, 2078 (1970).
\bibitem{deutscher80}
	G. Deutscher, B. Bandyopadhay, T. Chui, P. Lindenfeld, W.L. 
	McLean, and T. Worthington, Phys. Rev. Lett. {\bf 44}, 1150 (1980).
\bibitem{kobayashi80} 
	S. Kobayashi, Y. Tada, and W. Sasaki,
	J. Phys. Soc. Jpn. {\bf 49}, 2075 (1980).
\bibitem{orr86} 
	B.G. Orr, H.M. Jaeger, A.M. Goldman, and C.G. Kuper,
    	Phys. Rev. Lett. {\bf 56}, 378 (1986).
\bibitem{jaeger89} 
	H.M. Jaeger, D.B. Haviland, B.G. Orr, and 
\bibitem{kobayashi92} 
	S. Kobayashi, Surf. Sci. Reports {\bf 16}, 1 (1992).
    	A.M. Goldman, Phys. Rev. B {\bf 40}, 182 (1989).
\bibitem{yagi96} 
	R. Yagi, T. Yamaguchi, H. Kazawa, and S. Kobayashi,
	J. Phys. Soc. Jpn. {\bf 65}, 36 (1996).
\bibitem{haviland89} 
	D.B. Haviland, Y. Liu, and A.M. Goldman, 
    	Phys. Rev. Lett. {\bf 62}, 2180 (1989).
\bibitem{liu91} 
	Y. Liu, K.A. Greer, B. Nease, D.B. Haviland, G. Martinez,
    	J.W. Haley, and A.M. Goldman, Phys. Rev. Lett. {\bf 67}, 
	2068 (1991).
\bibitem{yazdani95} 
	A. Yazdani and A. Kapitulnik,
    	Phys. Rev. Lett. {\bf 74}, 3037 (1995).
\bibitem{zant90} 
	H.S.J. van der Zant, H.A. Rijken and J.E. Mooij,
	J. Low Temp. Phys. {\bf 79}, 289 (1990).
\bibitem{efetov80} 
	K.B. Efetov, Sov. Phys. JETP {\bf 51}, 1015 (1980).
\bibitem{fazekas84} 
	P. Fazekas, B. Muelshlegel, M. Schroter, 
    	Z. Phys. B {\bf 57}, 193 (1984). 
\bibitem{fazio86} 
	R. Fazio, and G. Giaquinta, 
    	Phys. Rev. B {\bf 34}, 4909 (1986).
\bibitem{fishman88a}
	R.S. Fishman, Phys. Rev. B {\bf 38}, 4437 (1988).
\bibitem{feigelman97} 
	M.V. Feigel'man, S.E. Korshunov, and A.B. Pugachev,
    	JETP Lett. {\bf 65}, 566 (1997). 
\bibitem{simanek81} 
	E. Simanek, Phys. Rev. B {\bf 23}, 5762 (1982).
\bibitem{lozovik81} 
	Yu.E. Lozovik and S.G. Akopov, J. Phys. C {\bf 14},
	L31 (1981).
\bibitem{wood82} 
	D.M. Wood and D. Stroud,
    	Phys. Rev. B {\bf 25}, 1600 (1982).
\bibitem{cuccoli99}
	A. Cuccoli, A. Fubini, V. Tognetti, R. Vaia
	Phys. Rev. B {\bf 61}, 11289 (2000).
\bibitem{fishman88b} 
	R.S. Fishman and D. Stroud,
    	Phys. Rev. B {\bf 37}, 1499 (1988);
    	Phys. Rev. B {\bf 38}, 280 (1988).
\bibitem{jacobs88} 
	L. Jacobs, J.V. Jos\`e, M.A. Novotny, and A.M. Goldman,   
    	Phys. Rev. B {\bf 38}, 4562 (1988).
\bibitem{jose94} 
	J.V. Jos\`e and C. Rojas, Physica B {\bf 203}, 481 (1994);
	Phys. Rev. B {\bf 54}, 12361 (1996).
\bibitem{fazekas80} 
	P. Fazekas, Z. Phys. B {\bf 45}, 215 (1980).
\bibitem{kim97}
	B.-J. Kim, J. Kim, S.-Y. Park, M.Y. Choi
	Phys. Rev. B {\bf 56}, 395 (1997).
\bibitem{jose84} 
	J.V. Jos\`e, Phys. Rev. B {\bf 29}, 2836 (1984).
\bibitem{doniach81} 
	S. Doniach, Phys. Rev. B {\bf 24}, 5063 (1981).
\bibitem{kissner93}
	J. Kissner and U. Eckern, Z. Phys. B 
	{\bf 91}, 155 (1993).
\bibitem{herz76}
	J. Herz, Phys. Rev. B {\bf 14}, 1175 (1976).
\bibitem{parisibook} 
	G. Parisi, {\em Statistical Field Theory},
	Addison Wesley (1988).
\bibitem{fisher88} 
	M.P.A. Fisher and G. Grinstein, Phys. Rev. Lett. 
    	{\bf 60}, 208 (1988).
\bibitem{savit80}
	R. Savit, Rev. Mod. Phys. {\bf 52}, 453 (1980)
\bibitem{villain75}  
	J. Villain, J. Physique {\bf 36}, 581 (1975).
\bibitem{jose77} 
	J.V. Jos\'{e}, L.P. Kadanoff, S. Kirkpatrick, 
    	and D.R. Nelson, Phys. Rev. B {\bf 16}, 1217 (1977).
\bibitem{elitzur79} 
	S. Elitzur, R. Pearson, and J. Shigemitzu,
	Phys. Rev. D {\bf 19}, 3638 (1979).  
\bibitem{nienhuis87} 
	B. Nienhuis, in: {\em Phase transitions and critical
	phenomena}, Vol.~11, ed. by  C.~Domb and J.~L.~Lebowitz (Academic
	Press, London, 1987), p.1.
\bibitem{fisher89a} 
	M.~P.~A. Fisher and D-H. Lee, Phys. Rev. B 
    	{\bf 39}, 2756 (1989).
\bibitem{fazio91a} 
	R. Fazio and G. Sch\"{o}n, Phys. Rev. B {\bf 43}, 5307  (1991).
\bibitem{fazio91b} 
	R. Fazio, U. Geigenm\"uller, and G. Sch\"on, in {\em
	Quantum Fluctuations in Mesoscopic and Macroscopic Systems},
	H.A. Cerdeira, {\it et al.} eds. (World Scientific, 1991), p.\ 214.  
\bibitem{mooij90} 
	J.E. Mooij, B.J. van Wees, L.J. Geerligs, M. Peters, 
    	R. Fazio, and G. Sch\"{o}n, Phys. Rev. Lett. {\bf 65}, 645 (1990).
\bibitem{blanter96} 
	Ya.M. Blanter, and G. Sch\"{o}n, Phys. Rev. B
    	{\bf 53}, 14534 (1996).
\bibitem{rojas95}
	C. Rojas, J.V. Jos\'e, and A.M. Tikofsky,
	Bull. Am. Phys. Soc. {\bf 40}, p.68, B11-7 (1995).
\bibitem{blanter97} 
	Ya.M. Blanter, R. Fazio, and G. Sch\"{o}n, Nucl. Phys. B
    	{\bf S58}, 79 (1997).
\bibitem{zant92} 
	H.S.J. van der Zant, L.J. Geerligs, and J.E. Mooij,
    	Europhys. Lett. {\bf 19}, 541 (1992).
\bibitem{yagi97}
	R. Yagi, T. Tamaguchi, H. Kazawa and S. Kobayashi, 
	J. Phys. Soc. Jpn. {\bf 66}, 2429 (1997)
\bibitem{v-BKT-self}
	In the opposite limit with $C \ll C_0$ the vortex BKT transition 
	temperature is	
	$T_{J} =\frac{\pi E_{J}}{2}\left(1 - \frac{1}{3\pi}\frac{E_{C}}{E_J}\right)$
	\protect{\cite{jose94}}.
\bibitem{normal-BKT1}
	In the case of normal arrays, the transition temperature is 
	$E_C/4\pi$. This implies that, in the charge regime, an array 
	has a lower resistance in the normal case compared to the 
	superconducting one. (The factor 4 reduction of $T_{ch}$ is 
	related to the charge $e$ of the quasi-particles 
	as compared to $2e$ of the Cooper pairs). 
	A higher resistance in 'superconducting' charging arrays has indeed 
	been observed in Ref.\protect{\cite{mooij90}}.
\bibitem{tighe93} 
	T.S. Tighe, M.T. Tuominen, J. M. Hergenrother, and M. Tinkham,
	Phys. Rev. B {\bf 47}, 1145 (1993).
\bibitem{delsing94} 
	P. Delsing, C.D. Chen, and D.B. Haviland,
	Phys. Rev. B {\bf 50}, 3059 (1994).
\bibitem{normal-BKT2}
	The case of normal arrays was investigated, for instance, by
	R. Yamada, S. Katsumoto, and S. Kobayashi, 
	J. Phys. Soc. Jpn. {\bf 62}, 2229 (1993).
\bibitem{tuominen92}M.T. Tuominen, J. M. Hergenrother, T.S. Tighe,and 
	M. Tinkham, Phys. Rev. B {\bf 69}, 1997 (1992).
\bibitem{zant88} 
	H.S.J. van der Zant, C.J. Muller, L.J. Geerligs, C.J.P.M.
	Harmans, and J.E. Mooij, Phys. Rev. B {\bf 38}, 5154 (1988). 
\bibitem{frustration}  
	W.Y. Shih and D. Stroud, Phys. Rev. B {\bf 30}, 6774
	(1984); M.Y. Choi and S. Doniach, Phys.Rev. B {\bf 31}, 4516 (1985);
	T.C. Halsey, Phys. Rev. B {\bf 31}, 5728 (1985).
\bibitem{martinoli00} 
	P. Martinoli and C. Leemann,
	J. Low Temp. Phys. {\bf 118}, 699 (2000).
\bibitem{hofstadter76} 
	D.R. Hofstadter, Phys. Rev. B {\bf 14}, 2239 (1976).
\bibitem{fishman87}
	R.S. Fishman,  and D. Stroud, Phys. Rev. B {\bf 37} 
	1499, (1987).  
\bibitem{kim98} 
	B.-J. Kim, G.-S. Jeon, M.-S. Choi, M. Y. Choi
	Phys. Rev. B {\bf 58}, 14524 (1998). 
\bibitem{niemeyer99}
	M. Niemeyer, J. K. Freericks and H. Monien,
	Phys. Rev. B {\bf 60}, 2357 (1999). 
\bibitem{jose96}
	J.V. Jos\'e, T.K. Kope\'c, and C. Rojas,
	Physica B {\bf 222}, 353 (1996).
\bibitem{choi85}
	M. Y. Choi and S. Doniach, Phys. Rev. B {\bf 31}, 4516 (1985).
\bibitem{cha93}	
	M.-C. Cha and S. M. Girvin, Phys. Rev. B {\bf 49}, 9794 (1994).
\bibitem{granato93-94}
	E. Granato, Phys. Rev. B {\bf 48}, 7727 (1993);
	J. Appl. Phys. {\bf 75} 6690 (1994).
\bibitem{kruperin00}
	V.A. Krupenin, D.E. Presnov, A.B. Zorin, J. Niemeyer
	J. Low Temp. Phys. {\bf 118}, 287 (2000).
\bibitem{lafarge95} 
	P. Lafarge, J.J. Meindersma, J.E. Mooij, in
    	{\it Macroscopic Quantum Phenomena and Coherence in
    	Superconducting Networks}, C. Giovanella and M. Tinkham Eds., (World
    	Scientific, Singapore, 1995), pag. 94. 
\bibitem{freericks94} 
	J.K. Freericks and H. Monien,
    	Europhys. Lett. {\bf 26}, 545 (1994); Phys. Rev. B {\bf 53}, 
	2691 (1996).
\bibitem{batrouni90} 
	G.G. Batrouni, R.T. Scalettar, and G.T. Zimanyi,
    	Phys. Rev. Lett. {\bf 65}, 1765 (1990).
\bibitem{scalettar91}  
	R.T. Scalettar, G.G. Batrouni, and G.T. Zimanyi,
    	Phys. Rev. Lett. {\bf 66}, 3144 (1991).
\bibitem{bruder92} 
	C. Bruder, R. Fazio, A.P. Kampf, A. van Otterlo, and
    	G. Sch\"{o}n, Physica Scripta T {\bf 42}, 159 (1992).
\bibitem{amico98} 
	L. Amico and V. Penna, Phys. Rev. Lett. {\bf 80}, 2189 (1998).
\bibitem{grignani00} 
	G. Grignani, A. Mattoni, P. Sodano, and A. Trombettoni, 
	Phys. Rev. B {\bf 61}, 1676 (2000).
\bibitem{andreev69}
	A.F. Andreev and I.M. Lifshitz, 
	Sov. Phys. JETP {\bf 29}, 1107 (1969). 
\bibitem{leggett70} 
	A.J. Leggett, Phys. Rev. Lett. {\bf 25}, 1543 (1970). 
\bibitem{matsuda70} H. Matsuda and T. Tsuneto,
    	Suppl. Prog. Theor. Phys {\bf 46}, 411 (1970).
\bibitem{lengua90} 
	G.A. Lengua and J.M. Goodkind,
	J. Low. Temp. Phys. {\bf 79}, 251 (1990).
\bibitem{meisel92}  
	M.W. Meisel, Physica {\bf 178}, 121 (1992);
	and references therein.
\bibitem{roddick93}  
	E. Roddick and D.H. Stroud, Phys. Rev. B 
    	{\bf 48}, 16600 (1993).
\bibitem{otterlo94a} 
	A. van Otterlo and K-H. Wagenblast,
    	Phys. Rev. Lett. {\bf 72}, 3598 (1994).
\bibitem{otterlo95} 
	A. van Otterlo and K-H. Wagenblast, R. Baltin, C. Bruder,
    	R. Fazio, and G. Sch\"on, Phys. Rev. B {\bf 52}, 16176 (1995).
\bibitem{batrouni95} 
	G.G. Batrouni, R.T. Scalettar, G.T. Zimanyi, and A.P. Kampf,
    	Phys. Rev. Lett. {\bf 74}, 2527 (1995);
	R.T. Scalettar, G.G. Batrouni, A.P. Kampf, and G.T. Zimanyi,
	Phys. Rev. B {\bf 51}, 8467 (1995).
\bibitem{frey97} 
	E. Frey and L. Balents, Phys. Rev. 
	B {\bf 55}, 1050 (1997).
\bibitem{pich98}
	C. Pich and E. Frey,  Phys. Rev. B {\bf 57}, 13 712 (1998).
\bibitem{amico97} 
	L. Amico, G. Falci, R. Fazio, and G. Giaquinta, 
    	Phys. Rev B {\bf 55}, 1100 (1997).
\bibitem{murthy97} 
	G. Murthy, D.P. Arovas, and A. Auerbach,
	Phys. Rev B {\bf 55}, 3104. (1997).
\bibitem{gabay93}
	M. Gabay and A. Kapitulnik,
	Phys. Rev. Lett. {\bf 71}, 2138 (1993).
\bibitem{zhang93}
	S.C. Zhang, Phys. Rev. Lett. {\bf 71}, 2142 (1993).
\bibitem{balents95a} 
	L. Balents, D. R. Nelson, 
	Phys. Rev. B {\bf 52}, 12951 (1995).
\bibitem{mullen94}
	K. Mullen, H.T.C. Stoof, M. Wallin, and S.M. Girvin,
	Phys. Rev. Lett. {\bf 72}, 4013 (1994).
\bibitem{balents95} 
	L. Balents, Europhys. Lett. {\bf 33}, 291, (1996).
\bibitem{roddick95}E. Roddick and D.H. Stroud,
	Phys. Rev. B {\bf 51}, 8672 (1995).
\bibitem{nelson88}
	D.R. Nelson, Phys. Rev. Lett. {\bf 60}, 1973 (1988).
\bibitem{feigelman93}
	M.V. Feigel'man, V.B. Geshkenbein, L.B. Ioffe,
	and A.I. Larkin, Phys. Rev. B {\bf 48}, 16641 (1993).
\bibitem{blatter94} 
	G. Blatter, M.V. Feigel'man, V.B. Geshkenbein,
	A.I. Larkin, and V.M. Vinokur,
	Rev. Mod. Phys. {\bf 66}, 1125 (1994).
\bibitem{frey94}  
	E. Frey, D.R. Nelson, and D.S. Fisher,
	Phys. Rev. B {\bf 49}, 9723 (1994).
\bibitem{schmid83} 
	A. Schmid, Phys. Rev. Lett. {\bf 51}, 1506 (1983).
\bibitem{penttila99} 
	J. S. Penttila, U. Parts, P. J. Hakonen, M. A. Paalanen, 
	E. B. Sonin, Phys. Rev. Lett. {\bf 82}, 1004 (1999).
\bibitem{impedence}
	see M. Devoret and H. Grabert in {\it Single Charge
        Tunneling} H. Grabert and M.H. Devoret Eds., NATO ASI series 
        Vol.294 (Plenum, NY 1992).
\bibitem{Fisher90b} 
	M.P.A. Fisher, G. Grinstein, and S.M. Girvin,
    	Phys. Rev. Lett. {\bf 64}, 587 (1990).
\bibitem{Wen90} 
	X.G. Wen and A. Zee, Int. J. Mod. Phys. (1990). 
\bibitem{takahide00} 
	Y. Takahide, R. Yagi, A. Kanda, Y. Ootuka, and 
	S. Kobayashi, Phys. Rev. Lett. {\bf 85}, 1974 (2000).
\bibitem{rimberg97} 
	A.J. Rimberg, T.R. Ho, C. Kurdak, J. Clarke, K.L. 
    	Campman, and A.C. Gossard, Phys. Rev. Lett. {\bf 78}, 2632 (1997).
\bibitem{chakravarty86} 
	S. Chakravarty, S.A. Kivelson, G.T. Zimanyi, 
	and B.I. Halperin, Phys. Rev. B {\bf 35}, 7256 (1986).
\bibitem{ferrell88}
	R.A. Ferrell and B. Mirhashem,
	Phys. Rev. B {\bf 37}, 648 (1988).
\bibitem{kampf88a} 
	A. Kampf and G. Sch\"{o}n, Physica {\bf 152}, 239 (1988).
\bibitem{kampf87} A. Kampf and G. Sch\"{o}n, Phys. Rev. B {\bf 36}, 
    	3651 (1987).
\bibitem{simanek88}
	E. Simanek and R. Brown, Phys. Rev. B {\bf 35}, 7256 (1988).
\bibitem{falci91} 
	The effect of the phase-dependent renormalization of the
	capacitance was discussed in G. Falci, R. Fazio, V. Scalia and  G. 
	Giaquinta, Phys. Rev. B  {\bf 43}, 13053 (1991).
\bibitem{choi89}
	J. Choi and J.V. Jos\`e, Phys. Rev. Lett. {\bf 62}, 1989 (1989).
\bibitem{delsing97} 
	P. Delsing, C.D. Chen, D.B. Haviland, T. Bergsten, 
	and T. Claeson, in {\em Superconductivity in Networks and Mesoscopic 
	Structures}, C. Giovannella and C.J. Lambert Eds., 
	American Institute, of Physics (1997).
\bibitem{zwerger88}
	W. Zwerger, J. Low Temp. Phys. {\bf 72}, 291 (1988).
\bibitem{panyukov89}
	S.V. Panyukov and A.D. Zaikin,
	J. Low Temp. Phys. {\bf 75}, 365 (1989); 
	{\em ibid} {\bf 75}, 389 (1989).
\bibitem{kampf88b} A. Kampf and G. Sch\"{o}n, Phys. Rev. B {\bf 37}, 
    	5954 (1988).
\bibitem{chakravarty86b} 
	S. Chakravarty, G.L. Ingold, S.A. Kivelson, 
	and A. Luther, Phys. Rev. Lett. {\bf 56}, 2303 (1986).
\bibitem{chakravarty88} 
	S. Chakravarty, S.A. Kivelson, and G.T. Zimanyi, 
	Phys. Rev. B {\bf 37}, 3283 (1988).
\bibitem{falci91b} 
	G. Falci, R. Fazio, and G. Giaquinta, 
    	Europhys. Lett. {\bf 14}, 145 (1991).
\bibitem{cuccoli00}
	A. Cuccoli, A. Fubini, V. Tognetti, and R. Vaia,
	cond-mat/0002072.
\bibitem{fisher87} 
	M.P.A. Fisher, Phys. Rev. B{\bf 36}, 1917 (1987).
\bibitem{zimanyi88}
	G.T. Zimanyi, Physica B {\bf 152}, 233 (1988).
\bibitem{Wagenblast97} 
	K-H., Wagenblast, A. van Otterlo, G. Sch\"{o}n, 
    	and G.T. Zimanyi, Phys. Rev. Lett. {\bf 78}, 1779(1997).
\bibitem{beck94}
	H. Beck, Phys. Rev. B {\bf 49}, 6153 (1994).
\bibitem{korshunov94}
	S.E.Korshunov, Phys. Rev. B {\bf 50}, 13616 (1994).
\bibitem{wagenblast98} 
	K-H. Wagenblast, A. van Otterlo, G. Sch\"{o}n, 
    	and G.T. Zimanyi, Phys. Rev. Lett. {\bf 79}, 2730 (1998).
\bibitem{kravchenko94}
	S.V. Kravchenko, G.V. Kravchenk, J.E. Furneaux, V.M. Pudalov,
	and M. D'Iorio, Phys. Rev. B {\bf 50}, 8039 (1994).
\bibitem{castellani95} 
	C. Castellani, C. Di Castro and M. Grilli,
	Phys. Rev. Lett. {\bf 75}, 4650 (1995).
\bibitem{footnotetr}
	In two-dimensional systems the resistance per
	square is scale invariant.  
\bibitem{Wen92} 
	X.G. Wen,  Phys. Rev. B  {\bf 46}, 2655 (1992).
\bibitem{Girvin92}
	S.M. Girvin, M. Wallin, M.-C. Cha, M.P.A. Fisher,
    	and A.P. Young, Prog. Teor. Phys. Supp. {\bf 107}, 135 (1992).
\bibitem{Cha91} 
	M.-C. Cha, M.P.A. Fisher, S.M. Girvin, M. Wallin,
    	and A.P. Young, Phys. Rev. B {\bf 44}, 6883 (1991).
\bibitem{Fazio96a} 
	R. Fazio and D. Zappal\`a, Phys. Rev. B {\bf 53}, R8883 (1996).
\bibitem{Sorensen92} 
	E.S. S\o rensen, M. Wallin, S.M. Girvin and 
    	A.P. Young, Phys. Rev. Lett {\bf 69}, 828 (1992).
\bibitem{Batrouni93}
	G.G. Batrouni, B. Larson, R.T. Scalettar,
	J. Tobochnik, and J. Wang, Phys. Rev. B {\bf 48}, 9628 (1993).
\bibitem{Makivic93}
	M. Makivic, N. Trivedi, and S. Ullah,
	Phys. Rev. Lett {\bf 71}, 2307 (1993).
\bibitem{Wallin94} 
	M. Wallin, E.S. S\o rensen, S.M. Girvin and 
    	A.P. Young, Phys. Rev. B {\bf 49}, 12115 (1994).
\bibitem{Runge92} 
	K. Runge, Phys. Rev. B {\bf 45}, 13136 (1992)
\bibitem{Herbut97} 
	I. Herbut, Phys. Rev. Lett. {\bf 79}, 3502 (1997).
\bibitem{Otterlo93} 
	A. van Otterlo, K-H. Wagenblast, R. Fazio and 
    	G. Sch\"{o}n, Phys. Rev. B {\bf 48}, 3316 (1993).
\bibitem{Kampf93} 
	A.P. Kampf and G.T. Zimanyi, Phys. Rev. B {\bf 47}, 
    	279 (1993).
\bibitem{Herbut98} 
	I. Herbut, Phys. Rev. Lett. {\bf 81}, 3916 (1998).
\bibitem{Damle97} 
	K. Damle and S. Sachdev, Phys. Rev. B {\bf 56}, 8714 (1997).
\bibitem{Damle98} 
	K. Damle and S. Sachdev, Phys. Rev. B {\bf 57}, 8307 (1998).
\bibitem{Sachdev98} 
	S. Sachdev, Phys. Rev. B {\bf 57}, 7157 (1998).
\bibitem{Kim91} 
	K. Kim and P.B. Weichman, Phys. Rev. B {\bf 43}, 13583 (1991).
\bibitem{granato90}
	E. Granato and J.M. Kosterlitz, 
	Phys. Rev. Lett., {\bf 65}, 1267 (1990).
\bibitem{dalidovich00}
	D. Dalidovich, P. Phillips,
	Phys. Rev. Lett. {\bf 84}, 737 (2000).
\bibitem{Chow98} 
	E.D. Chow, P. Delsing, and D.B. Haviland,
    	Phys. Rev. Lett. {\bf 81}, 204 (1998);
	D. B. Haviland, K. Andersson, P. Agren,
	J. Low Temp. Phys., {\bf 118}, 733 (2000).
\bibitem{Minnhagen87} 
	P. Minnhagen, Rev. Mod. Phys. {\bf 59}, 1001 (1987).
\bibitem{Bradley84} 
	R.M. Bradley and S. Doniach,
	Phys. Rev. B {\bf 30}, 1138 (1984).
\bibitem{1dfootnote}
	This choice corresponds to  choose equal lattice
	constants in space and time directions and therefore should not
	modify the critical properties of the transition.
\bibitem{Odintsov95}
	A.A. Odintsov, Phys. Rev. B {\bf 54}, 1228 (1996).
\bibitem{Zwerger89}
	W. Zwerger, Europhys. Lett. {\bf 9}, 421 (1989).
\bibitem{Korshunov89}
	S.E. Korshunov, Europhys. Lett. {\bf 9}, 107 (1989).
\bibitem{Bobbert90}
	P.A. Bobbert, R. Fazio, G. Sch\"on, and
	G.T. Zimanyi, Phys. Rev. B {\bf 41}, 4009 (1990).
\bibitem{Bobbert92}
	P.A. Bobbert, R. Fazio, G. Sch\"on, and
	A.D. Zaikin, Phys. Rev. B {\bf 45}, 2294 (1992).
\bibitem{Haviland96}
	D.B. Haviland and P. Delsing,
	Phys. Rev. B {\bf 54}, R6857 (1996).
\bibitem{LLreviews}
	J. S\'olyom, Adv. Phys. {\bf 28}, 201 (1979).
\bibitem{Kane92}
	C.L. Kane and M.P.A. Fisher,
        Phys. Rev. Lett. {\bf 68}, 1220 (1992);
\bibitem{Fazio96}
	R. Fazio, K.-H. Wagenblast, C. Winkelholz,
	and G. Sch\"on  Physica B {\bf 222}, 364 (1996) 
\bibitem{Glazman97}
	L.G. Glazman and A.I. Larkin,
	Phys. Rev. Lett. {\bf 79}, 3736 (1997).
\bibitem{Falci95}
	G. Falci, R. Fazio, A. Tagliacozzo, and G. Giaquinta,
        Europhys. Lett. {\bf 30}, 169 (1995).
\bibitem{Choi98}
	M.-S. Choi, M.Y. Choi, T. Choi and S.-I. Lee,
 	Phys. Rev. Lett. {\bf 81}, 4240 (1998).
\bibitem{Baltin97}R. Baltin and K.-H. Wagenblast,
	Europhys. Lett. {\bf 39}, 7 (1997).
\bibitem{kuehner98}
	T. Kuehner and H. Monien, Phys. Rev. B {\bf 58}, 
	R14741, (1998). 
\bibitem{fisher90a} 
	M.P.A. Fisher, Phys. Rev. Lett. {\bf 65}, 
    	923 (1990).
\bibitem{hebard90}
	A.F. Hebard and M.A. Paalanen,
    	Phys. Rev. Lett. {\bf 65}, 927 (1990).
\bibitem{paalanen92} 
	M.A. Paalanen, A.F. Hebard, and R.R. Ruel,
    	Phys. Rev. Lett. {\bf 69}, 1604 (1992).
\bibitem{zant92a} 
	H.S.J. van der Zant, F.C. Fritschy, W.E. Elion, 
    	L.J. Geerligs, and J.E. Mooij, Phys. Rev. Lett. 
	{\bf 69}, 2971 (1992).
\bibitem{phillips93}
	J.R. Phillips, H.S.J. van der Zant, J. White, and 
	T.P. Orlando, Phys. Rev. B {\bf 47}, 5219 (1993).
\bibitem{simanek83} 
	E. Simanek, Solid State Comm. {\bf 48}, 1023 (1983).
\bibitem{eckern89} U. Eckern and A. Schmid, 
    	Phys. Rev. B {\bf 39}, 6441 (1989).
\bibitem{larkin88} 
	A.I. Larkin, Yu. Ovchinnikov and A. Schmid, Physica B
    	{\bf 152}, 266 (1988).
\bibitem{eckern90} 
	U. Eckern, in {\em Applications of Statistical and Field 
	Theory Methods to Condensed Matter}, 
	Edited by R. Bishop (Plenum, New York, 1990).
\bibitem{lobb83} 
	C.J. Lobb, D.W. Abraham and M. Tinkham,
    	Phys. Rev. B {\bf 27}, 150 (1983).
\bibitem{orlando91} 
	T.P. Orlando, J.E. Mooij, and H.S.J. van der Zant,
    	Phys. Rev. B {\bf 43}, 10218 (1991).
\bibitem{trias96} 
	E. Trias, T.P. Orlando and H.S.J. van der Zant
 	Phys. Rev. 54{\bf 54}, 6568 (1996).
\bibitem{fazio94} 
	R. Fazio, A. van Otterlo, and G. Sch\"{o}n,
    	Europhys. Lett. {\bf 25}, 453 (1994).
\bibitem{zant94} 
	H.S.J. van der Zant, T.P. Orlando, S. Watanabe and S.H. Strogatz
	in Proc. of the NATO ARW on  {\em Mesoscopic Superconductivity},
	F. Hekking, G. Sch\"on, and
	D.V. Averin Eds, Physica B {\bf 203}, 490 (1994).
\bibitem{suhl65} 
	H. Suhl, Phys. Rev. Lett. {\bf 14}, 226 (1965).
\bibitem{simanek85} 
	E. Simanek, Phys. Rev. B {\bf 32}, 500 (1985).
\bibitem{duan92} 
	J.-M. Duan and A.J. Leggett, Phys. Rev. Lett. 
    	{\bf 68}, 1216 (1992).
\bibitem{bock94} 
	R.D. Bock, J.R. Phillips, H.S.J. van der Zant, and 
	T.P. Orlando, Phys. Rev. B {\bf 49}, 10009 (1994).
\bibitem{phunits}
	It might be  useful to express all the quantities characterizing 
	the vortex motion including all the dimensional constants:
	Vortex mass   $M_v = \frac{h}{16 a^2}  \;  E_C^{-1}$, dissipation strength
	$\eta=\Phi_0^2/ 2R_e a^2$, lattice potential   
	$U_v(x) = \frac{1}{2} \gamma E_J \sin(2 \pi x/a)$, 
	Lorentz force $\Phi_0 I/a $. 
\bibitem{hagenaars96}
	T.J. Hagenaars, P.H.E. Tiesinga, J.E. van Himbergen, and 
	J.V. Jos\`e,	in {\it Quantum Dynamics of Submicron Structures}, 
	eds. H.A. Cerdeira {\em et al.}, (Kluwer, Dordrecht, 1995), p.617.
\bibitem{zant91} 
	H.S.J. van der Zant, F.C. Fritschy, T.P. Orlando,
    	and J.E. Mooij, Phys. Rev. Lett. {\bf 66}, 2531 (1991).
\bibitem{zant97}
	H.S.J. van der Zant, in {\it Superconductivity in Networks and 
	Mesoscopic Structures}, C. Giovannella and C.J. Lambert Eds., 
	American Institute of Physics (1997). 
\bibitem{hagenaars94} 
	T.J. Hagenaars, P.H.E. Tiesinga, J.E. van Himbergen, and 
	J.V. Jos\`e, Phys. Rev. B {\bf 50}, 1143 (1994).
\bibitem{phillips94}
	J.R. Phillips, H.S.J. van der Zant, and 
	T.P. Orlando, Phys. Rev. B {\bf 50}, 9380 (1994). 
\bibitem{zant93} 
	H.S.J. van der Zant, F.C. Fritschy, T.P. Orlando,
    	and J.E. Mooij, Phys. Rev. B {\bf 47}, 295 (1993).
\bibitem{tighe91} 
	T.S. Tighe, A.T. Jonson, and M. Tinkham,
	Phys. Rev. B {\bf 44}, 10286 (1991).
\bibitem{sonin97}
	E.B. Sonin, Phys. Rev. B {\bf 55}, 485 (1997).
\bibitem{fazio92} 
	R. Fazio, A. van Otterlo, G. Sch\"{o}n, H.S.J. 
    	van der Zant, and J.E. Mooij, Helv. Phys. Acta {\bf 65}, 228 (1992).
\bibitem{fisher91} 
	M.P.A. Fisher, Physica A {\bf 177}, 553 (1991)
\bibitem{wees87}
	B.J. van Wees, H.S.J. van der Zant, and J.E. Mooij,	
	Phys. Rev. B {\bf 35}, 7291 (1987).
\bibitem{makhlin95} 
	Yu.G. Makhlin and G.E. Volovik, Pis'ma
	Zh. Eksp. Teor. Fiz. {\bf 62}, 923 (1985) [JETP Lett. {\bf 62}, 941
	(1995)]. 
\bibitem{volovik97}
	G.~E.~Volovik, cond-mat 9707136.
\bibitem{halldis} 	
	There were some controversy in the literature on this point. 
	It has been suggested that the offset charge in the equation of motion 
	should be replaced  with the electron number (F.~Gaitan and S.~R.~Shenoy, 
	Phys. Rev. Lett. {\bf 76}, 4404 (1996)). We do not share this point of 
	view since only the offset charges, which are responsible for a local 
	deviations 
	from charge neutrality in the array, lead to the Magnus force.
\bibitem{chen95a} 
	C. D. Chen, P. Delsing, D. B. Haviland, and T. Claeson
	in {\em Macroscopic Quantum Phenomena and Coherence in Superconducting 
	Networks}, C. Giovanella, and M. Tinkham Eds., 
	(World Scientific, 1995), pag. 121.
\bibitem{matsuda83}
	A. Matsuda and T. Kawakami, Phys. Rev. Lett. {\bf 51}, 694 (1983).
\bibitem{fujimaki87}
	A. Fujimaki, K. Nakajima, and Y. Sawada,
	Phys. Rev. Lett. {\bf 59}, 2985 (1987).
\bibitem{zant95} 
	H.S.J. van der Zant, T.P. Orlando, S. Watanabe,
    	and S.H. Strogatz, Phys. Rev. Lett. {\bf 74}, 174 (1995).
\bibitem{watanabe95} 
	S. Watanabe, S.H. Strogatz, H.S.J. van der Zant,
    	and T.P. Orlando, Phys. Rev. Lett. {\bf 74}, 379 (1995).
\bibitem{zant92b} 
	H.S.J. van der Zant, F.C. Fritschy, T.P. Orlando,
    	and J.E. Mooij, Europhys. Lett. {\bf 18}, 343 (1992).
\bibitem{nakajima81} 
	K. Nakajima and Y. Sawada, J. App. 
	Phys.{\bf 52}, 5732 (1981).
\bibitem{bobbert92} 
	P. Bobbert, Phys. Rev. B {\bf 45}, 7540 (1992).
\bibitem{geigenmuller93} 
	U. Geigenm\"{u}ller, C.J. Lobb and C.B. Whan,
    	Phys. Rev. B {\bf 47}, 348 (1993).
\bibitem{eckern93} 
	U. Eckern and E.B. Sonin, 
    	Phys. Rev. B {\bf 47}, 505 (1993).
\bibitem{otterlo94b} 
	A. van Otterlo, R. Fazio, and G. Sch\"{o}n, Physica B
    	Physica B {\bf 203}, 504 (1994);
        A. van Otterlo, R. Fazio, and G. Sch\"{o}n, 
    	Physica B {\bf 194-196}, 1153 (1994).
\bibitem{luciano95} 
	G. Luciano, U. Eckern, and J.G. Kissner,
    	Europhys. Lett. {\bf 32}, 669 (1995).
\bibitem{luciano96} 
	G. Luciano, U. Eckern, J.G. Kissner, and A. Tagliacozzo,
    	J. Phys: Condens. Matter {\bf 8}, 1241 (1996).
\bibitem{luciano97} 
	G. Luciano, U. Eckern, and A. Tagliacozzo,
	Phys. Rev. {\bf B} 56, 14686 (1997)
\bibitem{choi98b} 
	M-S. Choi, S-I. Lee, and M.Y. Choi, 
    	Phys. Rev. B {\bf 57}, 2720 (1998).
\bibitem{oudenaarden96a} 
	A. van Oudenaarden and J.E. Mooij,
    	Phys. Rev. Lett. {\bf 76}, 4947 (1996).
\bibitem{oudenaarden96b}
	A. van Oudenaarden, S.J.K. V\'{a}rdy and J.E. Mooij,
	Phys. Rev. Lett. {\bf 77}, 4257 (1996).
\bibitem{oudenaarden98} 
	A. van Oudenaarden, B. van Leeuwen,
    	M.P.P. Robbens, and J.E. Mooij,
    	Phys. Rev. B {\bf 57}, 11684 (1998).
\bibitem{elion93} 
	W.J. Elion, I.I. Wachters, L.L. Sohn, and J.E. Mooij,
    	Phys. Rev. Lett. {\bf 71}, 2311 (1993).
\bibitem{oudenaarden96} 
	A. van Oudenaarden S.J.K. V\'ardy, and J.E. Mooij,
    	Czhec. J. Phys. {\bf 46}, 707 (1996). 
\bibitem{martinis87}
	J.M. Martinis, M.H. Devoret, and J. Clarke,
	Phys. Rev. B {\bf 35}, 4682 (1987).
\bibitem{geigenmuller91}
	U. Geigenm\"{u}ller in {\it Macroscopic 
    	Quantum Phenomena}, T.D. Clark et al. Eds, (World Scientific, 1991).  
\bibitem{korshunov87} 
	S.E. Korshunov, JETP Lett. {\bf 46}, 484 (1987).
\bibitem{korshunov88} 
	S.E. Korshunov, Physica B {\bf 152}, 261 (1988).
\bibitem{ioffe98} 
	L. B. Ioffe, B. N. Narozhny, Phys. Rev. B {\bf 58}, 11449 (1998).
\bibitem{aharonov84} 
	Y. Aharonov and A. Casher,
    	Phys. Rev. Lett. {\bf 53}, 319 (1984).
\bibitem{aharonov59} 
	Y. Aharonov and D. Bohm, 
    	Phys. Rev. {\bf 115}, 485 (1959).
\bibitem{cimmino89}
	A. Cimmino, G.I. Opat, A.G. Klein, H. Keiser, S.A. Werner,
	M. Arif, and R. Clothier, Phys. Rev. Lett. {\bf 63}, 380 (1989).
\bibitem{reznik89} 
	B. Reznik and Y. Aharonov,  Phys. Rev. D {\bf 40}, 4178 (1989).
\bibitem{wees91} 
	B.J. van Wees, Phys. Rev. Lett. {\bf 65}, 255 (1990).
\bibitem{orlando91a} 
	T.P. Orlando and K.A. Delin, Phys. Rev. B {\bf 43}, 8717 (1991).
\bibitem{bloch_semi}
	V.G. Lyssenko, G. Valusis, F. L\"oser, T. Hasche, 
	K. Leo,M.M. Dignam, and K. K\"ohler, Phys. Rev. Lett. 
	{\bf 79}, 301 (1997), 
	and references therein.
\bibitem{kuzmin91} 
	L.S. Kuzmin and D.B. Haviland,
    	Phys. Rev. Lett. {\bf 67}, 2890 (1991).
\bibitem{kittel} 
	C.L. Kittel, {\em Introduction to Solid State Physics},
	(John Wiley \& Sons, New York, 1986), Chapter 7.
\bibitem{wannier-stark}
	for a review see:
	E. Mendex and G. Bastard, Physics Today, June, page 34 (1993).
\bibitem{kardar86} 
	For other effect of commensurability in Josephson arrays see
	M. Kardar, Phys. Rev. B {\bf 33}, 3125 (1986).
\bibitem{anderson58} 
	P.W. Anderson, Phys. Rev. {\bf 109}, 1492 (1958).
\bibitem{niyaz94} 
	P. Niyaz, R.T. Scalettar, C.Y. Fong, and
	G.G. Batrouni, Phys. Rev. B {\bf 50}, 362 (1994).
\bibitem{bruder99} 
	C. Bruder, L.I. Glazman, A.I. Larkin,
    	J.E. Mooij, and A. van Oudenaarden,
    	Phys. Rev. B {\bf 59}, 1383 (1999).
\bibitem{mott61}
	N.F. Mott and W.D. Twose, Adv. Phys. {\bf 10}, 107 (1961).
\bibitem{buettiker}  
	M. B\"{u}ttiker, Y. Imry, and R. Landauer,
	Phys. Lett. A {\bf 96}, 365 (1983).
\bibitem{hermon95} 
	Z. Hermon, A. Shnirman, and  E. Ben Jacob, 
    	Phys. Rev. Lett. {\bf 74}, 4915 (1995).
\bibitem{choi93}  
	M.Y. Choi, Phys. Rev. Lett. {\bf 71}, 2987 (1993).
\bibitem{fazio96b} 
	R. Fazio, A. van Otterlo, and A. Tagliacozzo,
    	Europhys. Lett. {\bf 36}, 135 (1996).
\bibitem{zant95a}
	H.S.J. van der Zant, T.P. Orlando, S. Watanabe and S.H. Strogatz 
	in {\it Quantum Dynamics of Submicron Structures}, 
	eds. H.A. Cerdeira {\em et al.}, (Kluwer, Dordrecht, 1995), p. 587.
\bibitem{hermon94}	
	Z. Hermon, A. Stern, E. Ben-Jacob,  Phys. Rev. B {\bf 49},
	9757 (1994).
\bibitem{prange} R. Prange and S. Girvin, {\em The Quantum Hall effect},
	Springer Verlag,Berlin (1987).	
\bibitem{nazarov94}
	Yu.V. Nazarov and A.A. Odintsov, Physica B {\bf 194-196}, 
	1737 (1994); A.A. Odintsov and Yu.V. Nazarov, 
	Phys. Rev. B {\bf 51}, 113 (1995).
\bibitem{choi94} 
	M.Y. Choi, Phys. Rev. B {\bf 50}, 10088 (1994).
\bibitem{stern94} 
	A. Stern, Phys. Rev. B {\bf 50}, 10092 (1994).
\bibitem{ekert96}
	A. Ekert and R. Jozsa, Rev. Mod. Phys.\
	{\bf 68}, 733 (1996).
\bibitem{steane98}
	A. Steane, Rep. Prog. Phys. {\bf 61},
        117 (1998).
\bibitem{makhlinrev}
	Y. Makhlin, G. Sch\"on and A. Shnirman, to be published in
	Rev. Mod. Phys. .
\bibitem{barenco95}
	A. Barenco, Proc. R. Soc. London A {\bf 449}, 679 (1995).
\bibitem{decoherence}
	G.M. Palma, K.-A. Suominen and A.K. Ekert,
        Proc. R. Soc. London A {\bf 452}, 567 (1996);
        W. Zurek, Physics Today {\bf 44}, 36 (1991).
\bibitem{makhlin97-99}
	A. Shnirman, G. Sch\"on and Z. Hermon,
	Phys. Rev. Lett. {\bf 79}, 2371 (1997); Y. Makhlin, 
	G. Sch\"on and A. Shnirman, Nature {\bf 398}, 305 (1999).
\bibitem{averin98} 
	D.A. Averin, Sol. State Comm. {\bf 105} 659 (1998).   
\bibitem{mooij99}
	J.E. Mooij, T.P. Orlando, L. Tian, C. van der Wal, 
        L. Levitov, S. Lloyd, and J.J. Mazo, Science {\bf 285}, 1036 (1999)
\bibitem{ioffe99} 
	L.B. Ioffe, V.B. Geshkenbein, M.V. Feigelman, 
        A.L. Faucher, and G. Blatter, Nature {\bf 398}, 679 (1999).
\bibitem{fazio99}
	R. Fazio, G.M. Palma and J. Siewert,
	Phys. Rev. Lett. {\bf 81}, 5385 (1999).
\bibitem{falci00}
	G. Falci, R. Fazio, G.M. Palma, J. Siewert, and
	V. Vedral, Nature {\bf 403}, 869 (2000).
\bibitem{Matters95}M.\ Matters, W.\ Elion, and J.E.\ Mooij,
	 Phys.\ Rev.\ Lett.\ {\bf 75}, 721 (1995).
\bibitem{bouchiat98} 
	V. Bouchiat, D. Vion, P. Joyez, D. Esteve, and 
        M. Devoret, Physica Scripta {\bf T76}, 165 (1998).
\bibitem{nakamura99} 
	Y. Nakamura, Yu.A. Pashkin, J.S. Tsai, 
        Nature {\bf 398}, 786 (1999).
\bibitem{friedman00} 
	J.R. Friedman, V. Patel, W. Chen, S.K. 
	Tolpygo and J.E. Lukens, Nature {\bf 406}, 43 (2000).
\bibitem{wal00} 
	C.H. van der Wal, A.C.J. ter Haar, F.K. Wilhelm, R.N. Schouten,
	C.J.P.M. Harmans, T.P. Orlando, S. Lloyd, J.E. Mooij,
	to be published in Science.
\bibitem{fulton87}
	T.A. Fulton and G.J. Dolan,
	Phys. Rev. Lett. {\bf 59}, 109 (1987).
\bibitem{chargemass} 
	Similar considerations for $E_{\rm C} \gg E_{\rm J}$ 
	leads to the effective action of charges in JJA. In particular 
	in the adiabatic limit it is possible to obtain the charge mass 
	$M_{\rm q} = (a^2 E_{\rm J})^{-1}$ 
	Thus, in two dimensional arrays of Josephson junctions a charge-vortex 
	duality exists. In the limiting case $E_{J} \gg E_{C}$ the 
	vortices are well-defined. They form Coulomb gas, and can be 
	considered as
	particles with masses. In the opposite limit, 
	$E_{C} \ll E_{J}$, the charges
	are the relevant excitations. The properties of charges are the same as
	the properties of vortices in the corresponding limiting case.
	An analogous situation occurs in 1D Josephson chains, in the 
	presence of extra 
	inductive terms charge solitons with a larger mass may appear as
	topological excitations (see Z.Hermon, E.Ben-Jacob, and G.Sch\"on,
     	Phys. Rev. B {\bf 54}, 1234 (1996)).  
\end{references}
\end{document}